\documentclass[eqsecnum,11pt]{article}

\usepackage{color}
\usepackage{cite}
\usepackage{amsmath}
\usepackage{amsfonts}
\usepackage{bm}
\usepackage{bbm}
\usepackage{graphicx}

\newcommand{\Ent}[1]{[\mkern - 2.5 mu [#1] \mkern - 2.5 mu ]}
\def\buildrel#1\under#2{\mathrel{\mathop{\kern0pt #2}\limits_{#1}}}

\def\corresponds{{\lower.2ex\hbox{=}}{\rm\kern-.75em^\triangle}}
\def\succsim{\succ\kern-.9em_\sim\kern.3em}
\def\precsim{\prec\kern-1em_\sim\kern.3em}
\def\slantfrac#1#2{\kern1em^{#1}\kern-.3em/\kern-.1em_{#2}}
\def\lfrac#1#2{{}^{#1\!}\kern-.0em/_{#2}}

\def\R{{\mathbbm{R}}}

\def\N{{\mathbbm{N}}}
\def\C{{\mathbbm{C}}}
\def\t{\theta}
\def\H{{\cal H}}
\def\G{{\cal G}}
\def\Rl{R(\lambda,H)}
\def\Re{{\rm Re}\,}
\def\Im{{\rm Im}\,}
\def\Sc{Schr\"o\-dinger o\-pe\-ra\-tor}
\def\RSPT{Rayl\-eigh-Schr\"o\-dinger per\-turbation theory}
\def\Sc{Schr\"o\-dinger}

\newtheorem{theorem}{Theorem}
\newtheorem{definition}[theorem]{Definition}

\newcommand{\nn}{\nonumber\\}

\newcommand{\gammat}{\Gamma^{\left(2,0\right)}}
\newcommand{\gammaf}{\Gamma^{\left(4,0\right)}}
\newcommand{\dgammat}{{\partial\Gamma^{\left(2,0\right)}/{\partial p^2}}}
\newcommand{\gammato}{\Gamma^{\left(2,1\right)}}

\begin{document}

\bibliographystyle{myprsty}

\vspace{-0.05cm}

\begin{center}
\begin{tabular}{c}
{\Large \bf From Useful Algorithms for Slowly Convergent Series}\\[1.2ex]
{\Large \bf to Physical Predictions Based on Divergent 
Perturbative Expansions}
\end{tabular}
\end{center}

\begin{center}
E. Caliceti,$^{a,*}$ 
M. Meyer--Hermann,$^{b}$ 
P. Ribeca,$^{c}$ 
A. Surzhykov,$^{d}$ 
and U. D. Jentschura$^{d,*}$ 
\end{center}

\vspace*{-1.0cm}

\scriptsize

\begin{center}
$^{a}${\it Dipartimento di Matematica,
Universit\`{a} di Bologna, 40127 Bologna, Italy}\\[1ex]
$^{b}${\it Frankfurt Institute for Advanced Study, Max von Laue-Stra\ss{}e 1,
60438 Frankfurt am Main, Germany}\\[1ex]
$^{c}${\it Gesellschaft f\"{u}r Schwerionenforschung
(GSI), Planckstra\ss{}e 1, 64291 Darmstadt}\\[1ex]
$^{d}${\it Max--Planck--Institut f\"{u}r Kernphysik,
Saupfercheckweg 1, 69117 Heidelberg, Germany}
\end{center}

\vspace*{1.0cm}

\normalsize

\begin{center}
\begin{minipage}{16.0cm}
{\bf This review is focused on 
the borderline region of theoretical physics and mathematics. 
First, we describe numerical methods for the acceleration of the 
convergence of series. These provide a 
useful toolbox for theoretical physics
which has hitherto not received the attention it actually deserves. 
The unifying concept for convergence acceleration methods
is that in many cases,
one can reach much faster convergence than by adding 
a particular series term by term. In some cases, it is even possible
to use a divergent input series, together with a 
suitable sequence transformation, for the construction of 
numerical methods that can be applied to the calculation of special
functions. This review both aims to 
provide some practical guidance as well as a
groundwork for the study of specialized literature.
As a second topic, we review some recent developments in the field
of Borel resummation, which is generally 
recognized as one of the most versatile methods for the 
summation of factorially divergent (perturbation) series.
Here, the focus is on algorithms which make optimal use of
all information contained in a finite set of perturbative coefficients.
The unifying concept for the various aspects of
the Borel method investigated here
is given by the singularities of the Borel transform,
which introduce ambiguities from a mathematical point of view
and lead to different possible physical interpretations.
The two most important cases are:
(i) the residues at the singularities correspond to the
decay width of a resonance, and (ii) the presence of the singularities
indicates the existence of nonperturbative contributions
which cannot be accounted for on the basis
of a Borel resummation and require generalizations toward
resurgent expansions. Both of these cases are illustrated by examples.}
\end{minipage}
\end{center}

\vspace{0.2cm}

quantum electrodynamics (specific calculations),\\
computational techniques,\\
numerical approximation and analysis.
\vfill

\hrule
$^*$ Corresponding authors. E-mail: caliceti@dm.unibo.it
and jentschura@mpi-hd.mpg.de.

\newpage

\tableofcontents

\newpage

%
%
\section{Background and Orientation}
\label{theory}

The purpose of this section is that of being a starting point and a 
reference for the reader. Its contents are organized as follows:
\begin{itemize}
\item In Sec.~\ref{Sec:IntroductionToAcceleration}, a general 
introduction to convergence acceleration and resummation is provided.

\item In Sec.~\ref{Sec:mathbasis}, some basic terminology is 
introduced. This terminology is a basis for the general 
familiarity with the notions that govern the construction of 
convergence acceleration methods.
\end{itemize}
The concepts contained in this section constitute to a large extent
a prerequisite for the material presented in the following sections.

%
%
\subsection{Introduction to Convergence Acceleration and Resummation}
\label{Sec:IntroductionToAcceleration}

The basic idea of convergence acceleration is to extract hidden 
information contained in partial sums of a specific slowly 
convergent series, and to use that information in order to make, 
loosely speaking, an educated guess about higher-order partial sums 
which eventually converge to some limit. In many cases, the 
``educated guesses'' can lead to spectacular numerical results which 
represent a drastic improvement over a term-by-term summation of the 
series, even if the series is formally convergent.

Convergence acceleration methods have a long history. E.g., as mentioned of
p.~{\em ix} of Ref.~\cite{LiLuSh1995}, such methods were
already used by the Chinese mathematicians Liu Hui (A.D. 263) and
Zhu Chongzhi (429---500) for obtaining better approximations to
$\pi$. It is thus clear that, at least in rudimentary form,
convergence acceleration methods are much older than calculus.
Another reference to early successful attempts can be found on
pp.~90---91 of a book by Brezinski~\cite{Br1991a} where it is
mentioned that convergence acceleration methods have already been
used in 1654 by Huygens and in 1674 by Seki Kowa, the probably most
famous Japanese mathematician of his time. Both Huygens and Seki
Kowa were trying to obtain better approximations to $\pi$. Huygens
used a linear extrapolation scheme which is a special case of what
we now call Richardson extrapolation~\cite{Ri1927}, and Seki Kowa
used the so-called $\Delta^2$ process, which is usually attributed
to Aitken \cite{Ai1926}.

Sequence transformations are principal numerical tools to overcome 
convergence problems. In this approach, 
a slowly convergent or divergent sequence $\{ s_n 
\}_{n=0}^{\infty}$, whose elements may for instance be the partial 
sums
$s_n \; = \; \sum_{k=0}^{n} \, a_k$
of an infinite series, is converted into a new sequence 
$\{ s^{\prime}_n \}_{n=0}^{\infty}$ with hopefully better numerical 
properties. Before the invention of electronic computers, mainly linear
sequence transformations were used. These matrix transformations compute
the elements of the transformed sequence $\{ s_n^{\prime}
\}_{n=0}^{\infty}$ as weighted averages of the elements of the input
sequence $\{ s_n \}_{n=0}^{\infty}$ according to the general scheme
$s_n^{\prime} \; = \; \sum_{k=0}^{n} \, \mu_{n k} \, s_k$.
The theoretical properties of such matrix transformations are very 
well understood \cite{Kn1964,Ha1949,Pe1966,%
Pe1969,ZeBe1970,Wi1981,Wi1984,PoSh1988,Bo2000}. Their main 
appeal lies in the fact that for the weights ${\mu}_{n k}$ 
 some necessary and sufficient conditions could be 
formulated which guarantee that the application of such a matrix 
transformation to a convergent sequence $\{ s_n \}_{n=0}^{\infty}$ 
yields a transformed sequence $\{ s_n^{\prime} \}_{n=0}^{\infty}$ 
converging to the same limit $s = s_{\infty}$. Theoretically, 
this regularity is extremely desirable, but from a practical 
point of view, it is a disadvantage. This may sound paradoxical. 
However, according to a remark on p.~X of 
Ref.~\cite{Wi1981}, the size of the domain of regularity of a 
transformation and its efficiency seem to be inversely related. 
Accordingly, regular linear transformations are in general at most 
moderately powerful, and the popularity of most linear 
transformations has declined considerably in recent years.

Modern nonlinear sequence transformations as for instance 
the $\epsilon$ algorithm~\cite{Wy1956eps} and 
the $\rho$ algorithm~\cite{Wy1956b} or
the $\vartheta$ algorithm~\cite{Br1971} have largely
complementary properties: They are nonregular, which means that that the
convergence of the transformed sequence is not guaranteed, let alone to
the correct limit. In addition, the theoretical properties of nonlinear
transformations are far from being completely understood. Nevertheless,
they often accomplish spectacular results when applied to appropriate
problems. Consequently, nonlinear transformations now dominate both
mathematical research as well as practical applications, as documented
in a large number of recent books \cite{Ba1975,Br1977,Br1978,%
BeOr1978,Br1980pade,Wi1981,MaSh1983,CuWu1987,De1988,%
BrRZ1991,LiLuSh1995,Wa1996,BaGr1996, 
Si2003,BoLaWaWa2004} 
and review articles~\cite{Gu1989,We1989,Ho2000,Br2000} on this topic. 

We here attempt to present these transformations in both a
special context (with a focus on applications) as well
as in the general context given by an underlying principle of 
construction exemplified by the $E$ algorithm.
One of the recurrent themes in the construction of sequence
transformations is the separation of the series into a 
partial sum $s_n = \sum_{k=0}^n a_n$ and a remainder
$r_n = - \sum_{k = n+1}^\infty a_n$ so that $s_n = s + r_n$.
The remainder term $r_n$ is to be approximated by remainder estimates and 
eliminated from the input data $s_n$ by suitable transformations.

We restrict the discussion to the acceleration of sequences.
{\color{black} Quite naturally, convergence acceleration methods
can be proposed and applied to the pointwise evaluation
of partial-wave expansions which occur 
ubiquitously in physics (see, e.g.,~Ref.~\cite{JeMoSoWe1999,JeMoSo1999}),
and to Fourier expansions and other orthogonal series~\cite{Br2004}.
Here, the sequences are typically composed of the partial sums
of the respective expansions.}
For applications to vector or matrix problems, the reader
is referred to the books~\cite{BrRZ1991,Br1997} and 
numerous articles, e.g.~\cite{SiFoSm1986,HoRaKr1995}, 
and for applications to numerical quadrature, in particular
with respect to oscillatory functions to be integrated over
semiinfinite intervals, we refer to the books~\cite{Ev1993,Si2003}
and several articles which describe corresponding 
applications, e.g.~\cite{SaHo2003}. 

Forming a
bridge between convergence acceleration and resummation, one may
observe that some acceleration algorithms may also be used for the
resummation of (mildly) divergent series, but are in general less
versatile than the Borel method. The resummed value of a divergent
series, which is sometimes called antilimit, can be attributed,
according to Euler, to the particular finite mathematical entity
whose expansion gives rise to the original series. L.~Euler wrote in
a letter to Goldbach (1745): {\it `Summa cuiusque seriei est valor
expressionis illius finitae, ex cuius evolutione illa series
oritur,''} which may be translated from the Latin language as:
``The sum of any given series [whether convergent or divergent] is
the value of the specific finite expression whose expansion gave
rise to that same series.''
One can mention a number of accepted methods for series
(re-)summation: e.g., there are averaging processes for oscillatory
partial sums like the
C\'{e}saro method~\cite{Ha1949} or sophisticated mathematical constructions
like zeta function regularization~\cite{El1995}, which
depend on the availability of closed-form analytic expressions
for all terms that need to be summed, and permit the variation
of a free parameter (the ``argument'' of the generalized zeta
function) from the physically relevant to a mathematically
favourable domain, where the defining series becomes convergent.
Later, in order to allow for the determination of the physical
property of interest, one performs an analytic continuation of the
parameter back to the physical domain. In the same context,
one can mention the Mellin-Barnes transform~\cite{PaKa2001}.

Divergent series result from any eigenvalue perturbation theory, unless
the perturbation is Kato-bounded with respect to the 
unperturbed Hamiltonian (see Ref.~\cite{ReSi1978}).
In particular, 
the application of simple, ``naive'' Rayleigh--Schr\"{o}dinger 
perturbation theory to a harmonic oscillator with an $x^4$ perturbation 
(where $x$ is the position operator) already leads to a divergent 
series in higher orders of perturbation theory. 
While the degree of acceptance of divergent series 
within the mathematical 
scientific community underwent some undeserved oscillations during the 
last two centuries, we are today in the fortunate 
situation that their properties are routinely investigated
and harvested in the context of physical problems that originate from 
perturbation theory in quantum mechanics and field theory.

One of the most versatile and most relevant resummation
algorithms, especially those originating 
from perturbation theory, is the Borel method~\cite{Bo1899,Bo1928,Bo1975}. 
The reason is that many physical divergences are factorial,
and for these, the Borel summation is the most powerful one.
Already in 1952 Dyson \cite{Dy1952} had argued that
perturbation expansions in quantum electrodynamics should diverge,
and it later turned out~\cite{Li1976Lett,Li1976,Li1977Lett,Li1977,%
BrKr2000,BrKr2001} that the 
divergence should be factorial. Also, 
Bender and Wu \cite{BeWu1969,BeWu1971,BeWu1973} demonstrated that
Dyson's speculation was not a mere mathematical sophistication: They
proved that the coefficients of the standard
Rayleigh--Schr\"odinger perturbation expansions
for the energy eigenvalues of the anharmonic oscillators with
anharmonicities $x^{2 m}$ with $m \ge 2$ grow at least factorially 
with the index $n$ as $(m \, n)!$.
A new subfield of theoretical physics emerged which is nowadays
called large-order perturbation theory
(see for example \cite{ArFeCa1990,LGZJ1990,Ad1994}
and references therein). Following these works, many other quantum
mechanical systems described by perturbation expansions were investigated,
and in the overwhelming majority of all cases very similar divergence
problems were observed. Moreover, nowadays the factorial divergence of
perturbation expansions in quantum field theory and in related
disciplines now almost has the status of a law of nature (see for
example the discussion in \cite{Fi1994}).

The Borel resummation process replaces the divergent 
series, after analytic continuation of the Borel transform,
by a Laplace-type integral, which is sometimes called 
the Laplace--Borel integral for obvious reasons.
A famous ambiguity results in the case of singularities of the 
transform along the integration contour. 
In principle, one may imagine three possible interpretations
for the ambiguity introduced by the singularities,
and it is interesting to remark that all of them might 
be realized in nature. They are marked here as
(i), (ii) and (iii) and will be briefly discussed now.
(i) In some cases (e.g., the cubic anharmonic
oscillator and the Stark effect), the resummation is 
achieved by using a complex integration contour, which goes 
around the singularity on a half circle and leads to an 
imaginary part for the (resonance) energy eigenvalues of the 
Hamiltonians under investigation. Of course, the sign of the 
imaginary part depends on the clockwise or anticlockwise
circulation around the pole and cannot be determined on the 
basis of the purely real perturbation series alone: additional
physics input is needed and provided by the very nature of a 
resonance. These cases find a natural mathematical
interpretation via the concept of distributional Borel summability.
(ii) There are situations
in which the singularities of the Borel transform indicate the 
presence of further contributions that cannot be obtained
on the basis of perturbation theory alone, even if perturbation 
theory is combined with the most sophisticated resummation methods.
Rather, one has to generalize perturbation theory to a 
resurgent expansion~\cite{DeDi1991,Ph1989,CaNoPh1993,Bo1994}. 
In some cases, there are subtle cancellations
of the imaginary parts created by the singularities 
of the Borel transform of the perturbation series against
explicit imaginary parts of higher-order instanton effects which 
are characterized by nonanalytic factors, leading to purely
real eigenenergies for, e.g., quantum mechanical 
potentials of the double-well type. The instanton
effects, in turn, are decorated again by divergent nonalternating 
perturbation series, which give to yet further imaginary parts,
but these are again compensated by instanton effects of 
even higher order. The generalized perturbative expansion which 
entails all of these effects is a so-called resurgent 
expansion. 
(iii) One may easily construct mathematical model problems
where the correct resummation is achieved by a principal-value
prescription for the evaluation of the Borel integral.
It has been conjectured that this prescription might be 
the correct one for a number of physical problems as well.
Summarizing, we may state that the 
ambiguity of the Borel resummation process definitely exists.
This ambiguity provides for an interpretation of the Borel transform 
even in cases where the complete physical answer cannot be 
obtained on the basis of perturbation theory alone.

The resummation methods allow us to determine eigenenergies and 
other physical properties on the basis of a finite number of 
perturbative coefficients alone. One can now even go one step further
and observe that, on the one hand, the large-order behaviour of perturbation 
theory is connected to the underlying perturbative potential or 
interaction Lagrangian (in the case of a field theory), and yet,
on the other hand, the potential also determines the 
large-coupling expansion for the potential or the field
theory under investigation. So, one can even try to 
guess the large-coupling expansion from a 
finite number of terms of the weak-coupling perturbative
expansion, because these already give an impression of the 
large-order behaviour. This paradoxical endeavour has
been pursued in recent years, in a series of papers, 
by Suslov~\cite{Su2001phi4,Su2001,Su2002, Su2005rev}.

Let us now attempt to form a bridge between convergence
acceleration and resummation via sequence transformations,
which can be used for the resummation of (mildly)
divergent series in addition to their use as
convergence accelerators. We recall
that the convergence acceleration of the sequence
$\{s_n \}_{n = 0}^\infty$ is achieved by eliminating the remainders
$r_n$ from a necessarily finite set of input data $\{s_n\}_{n=0}^m$, in
order to obtain a better estimate $s'_m$ to $s$ than the obvious
estimate which would be provided by $s_m$. Clearly, the ``elimination
of remainders'' is a concept which can be transported to the
case of a divergent series. The only difference is that for a convergent
series, $r_n \to 0$ as $n \to \infty$, whereas for a divergent series,
$\lim_{n\to\infty} r_n$ does not exist. In some cases,
the sequence transformations provide us with numerically favourable,
highly efficient algorithms for the computation of the antilimit
of the divergent input series, so that they constitute competitive
numerical algorithms for the actual computation of the antilimit,
which could otherwise only obtained by direct evaluation of the 
expression which gave rise to the divergent series in the first place,
or by the Borel method in connection with 
Pad\'{e} approximants for the analytic continuation
of the Borel transform. The latter method, although versatile, 
is numerically less efficient because it entails, at least, the numerical
evaluation of the Laplace--Borel integral.

%
%
\subsection{Mathematical Basis}
\label{Sec:mathbasis}

%
%
\subsubsection{Special Mathematical Symbols}
\label{Sec:SpecSymSpecFun}

In order to fix ideas, it is perhaps useful to recall a few
mathematical symbols. In particular, 
\begin{equation}
\mathbb{N} = \{ 1, 2, 3, \ldots \}
\end{equation}
denotes the set of positive integers, $\mathbb{N}_0 = \mathbb{N} \cup \{ 0 \}$ 
stands for the set $\{ 0, 1, 2, \ldots \}$ of nonnegative integers, and
\begin{equation}
\mathbb{Z} = \{ 0, \pm 1, \pm 2, \ldots \}
\end{equation}
is the set of
positive and negative integers. Moreover, $\mathbb{R}$ and $\mathbb{C}$
denote the sets of real and complex numbers, respectively. 

The symbol $\Ent{x}$
is the integral part of $x \in \mathbb{R}$, i.e., the largest integer $m \in 
\mathbb{Z}$ satisfying the inequality $m \le x$.

The nonstandard notation 
\begin{equation}
\{ s_n \}_{n=0}^{\infty}
\end{equation}
is used to 
denote a sequence of elements $s_n$ 
with $n \in \mathbb{N}_0$. In this work,
we usually assume that every element $s_n$ represents the partial 
sum of an infinite series: 
\begin{equation}
\label{basics}
s_n = \sum_{k=0}^n a_k \,.
\end{equation}
Indeed, such a definition implies that the 
sequence elements $s_{-1}, s_{-2}, s_{-3},
\ldots$ with negative indices are empty sums and thus 
equal to zero. 

As just indicated, empty sums are always interpreted as zero:
\begin{equation}
\sum_{k=m}^n \, a_k \; = \; 0 \, ,
\qquad \text{if} \quad m > n \, . 
\end{equation}
Similarly, an empty product is always interpreted as one:
\begin{equation}
\prod_{k=m}^n \, a_k \; = \; 1 \, ,
\qquad \text{if} \quad m > n \, . 
\end{equation}
 
Let $f(z)$ and $g(z)$ be two functions defined on some domain $D$ in
the complex plane and let $z_0$ be a limit point of $D$, possibly the
point at infinity. Then,
\begin{equation}
\label{defO}
f(z) \; = \; \mathrm{O} (g(z)) \, , \qquad z \to z_0 \, ,
\end{equation}
means that there is a positive constant $A$ and a neighborhood $U$ of
$z_0$ such that 
\begin{equation}
\vert f(z) \vert \; \le \; A \vert g(z) \vert  
\end{equation}
for all $z \in U \cap D$. If $g(z)$ does not vanish on $U \cap D$ this
simply means that $f(z)/g(z)$ is bounded on $U \cap D$. Similarly,
\begin{equation}
\label{defo}
f(z) \; = \; \mathrm{o} (g(z)) \, , \qquad z \to z_0 \, ,
\end{equation}
means that for any positive number $\epsilon \in \mathbb{R}$ there
exists a neighborhood $U$ of $z_0$ such that
\begin{equation}
\vert f(z) \vert \; \le \; \epsilon \vert g(z) \vert
\end{equation}
for all $z \in U \cap D$. If $g(z)$ does not vanish on $U \cap D$ this
simply means that $f(z)/g(z)$ approaches zero as $z \to z_0$.

%
%
\subsubsection{Finite Differences}
\label{Sec:FinDiff}

So--called finite difference operators~\cite{Jo1965,MT1981}
are essential in the construction of sequence transformations.
Let us consider a function $f(n)$ defined on the 
set $\mathbb{N}_0$ of nonnegative
integers. Then, the forward difference
operator $\Delta f(n)$ is defined
by
\begin{equation}
\label{difference_definition}
\Delta f(n) \; = \; f(n+1) - f(n), \qquad n \in \mathbb{N}_0 \, .   
\end{equation}
Higher powers of this operator can be defined recursively:
\begin{equation}
\Delta^k f(n) = \Delta [ \Delta^{k-1} f(n) ] \, ,
 \qquad k \in \mathbb{N} \, , 
\end{equation}
with $\Delta^0 f(n) = f(n)$.
Apart from the forward difference one may also introduce the shift
operator:
\begin{equation}
\label{shift_definition}
E f(n) \; = \; f(n+1) \, ,
\end{equation}
whose higher powers can again be defined recursively:
\begin{eqnarray}
E^k f(n) & = & E [ E^{k-1} f(n) ] \; = \; 
f(n+k), \qquad k \in \mathbb{N} \, ,
\end{eqnarray}
with $E^0 f(n) = f(n)$.
As seen from definitions (\ref{difference_definition}) and
(\ref{shift_definition}), the forward difference $\Delta$
and shift $E$ operators are connected by the relationship 
\begin{equation}
\Delta \; = \; E - 1 \, ,
\end{equation}
where $1$ is the identity operator. This relationship can be combined with 
the binomial theorem to give
\begin{equation}
\Delta^k f(n) \; = \; (-1)^k \, \sum_{j=0}^k \, (-1)^j \, 
{\binom {k} {j}} \, f(n+j) \, , \qquad k \in \mathbb{N}_0 \, .  
\end{equation}
In the following, we always tacitly assume that in the case
of quantities, which depend on several indices like
$\mathcal{A}_{k}^{(n)}$, the difference operator $\Delta$ and the shift
operator $E$ only act only upon $n$ and not on any
other indices: e.g., we have $\Delta \mathcal{A}_{k}^{(n)} =
\mathcal{A}_{k}^{(n+1)} - \mathcal{A}_{k}^{(n)}$.

Furthermore, we should clarify the notation
for sequences with $f(n) = s_n$: The quantity $\Delta s_n$ 
is to be understood as $\Delta s_n = s_{n+1} - s_n = a_{n+1}$, 
and we have $\Delta^2 s_n = s_{n+2} - 2 s_{n+1} + s_n = \Delta a_{n+1} = 
a_{n+2} - a_{n+1}$ and so on [we assume the structure 
given in Eq.~(\ref{basics}) for the $s_n$]. 

%
%
\subsubsection{Asymptotic Sequences and Asymptotic Expansions}
\label{Sec:AsySeqAsyExp}

A finite or infinite sequence of functions $\{ \Phi_{n} (z)
\}_{n=0}^{\infty}$, which are defined on some domain D of complex 
numbers on which all $\Phi_n (z)$ are nonzero except possibly at $z_0$,
is called an asymptotic sequence as $z \to z_0$ if for all 
$n \in \mathbb{N}_0$,
\begin{equation}
\Phi_{n+1} (z) \; = \; \mathrm{o} (\Phi_n (z)) \, , 
\qquad z \to z_0 \, . 
\end{equation}
The formal series  
\begin{equation}
f(z) \sim \sum_{n=0}^{\infty} a_n \, \Phi_n (z) \, , 
\label{f2Phi_asyExp}
\end{equation}
which need not be convergent, is called an asymptotic expansion
of $f(z)$ with respect to the asymptotic sequence $\{ \Phi_n (z)
\}_{n=0}^{\infty}$ in the sense of Poincar\'e
\cite{Po1886} if, for every $m \in \mathbb{N}_0$,
\begin{equation}
\label{asymptotic_sequence_Poincare}
f(z) - \sum_{n=0}^m a_n \, \Phi_n (z) \; = \; \mathrm{o} ( \Phi_m (z)) 
\, , \qquad z \to z_0 \, . 
\end{equation}
If such a Poincar\'e-type asymptotic expansion exists, it is unique, and
its coefficients $a_n$ can be computed as follows,  
\begin{equation}
a_m \; = \;
\lim_{z \to z_0} \; \Bigl\{ \bigl[ f(z) - \sum_{n=0}^{m-1} \; a_n
\Phi_n (z) \bigr] \, / \, \Phi_m (z) \Bigr\}
\, , \qquad m \in \mathbb{N}_0 \, .   
\end{equation}
The particularly important case of an asymptotic expansion
with respect to the asymptotic sequence 
$\{ z^{-n} \}_{n=0}^{\infty}$ about $z_0 = \infty$ has
the form of an asymptotic power series in $1/z$,
\begin{equation}
f(z) \; \sim \; \sum_{k=0}^{\infty} \; \alpha_k / z^k \, .
\qquad z \to \infty \,. 
\end{equation}
In the current article, we will often use the notation
\begin{equation} 
\label{DefP}
f (z) \; \sim \; \sum_{k=0}^{\infty} \alpha_k \, z^k\, ,
\qquad z \to 0 \,, 
\end{equation}
for asymptotic expansions or perturbative expansions
of a physical quantity $f(z)$ with 
respect to the asymptotic sequence $\{ z^n \}_{n=0}^{\infty}$ 
in the sense of Poincar\'e.

%
%
\subsubsection{Remainder Estimates}
\label{Sec:SpecSeq}

Let us assume that $\{ s_n \}_{n=0}^{\infty}$ either converges to some
limit $s$, or, if it diverges, can be summed by an appropriate summation
method to give $s$. Following \cite{Sh1955},
this generalized limit $s$ is frequently called
antilimit in the case of divergence. 
In either case, there is a partition of a sequence element
$s_n$ into the limit or antilimit $s$ and the remainder $r_n$ according to   
\begin{equation}
\label{definition_remainder}
s_n \; = \; s \, + \, r_n 
\end{equation}
for all $n \in \mathbb{N}_0$. If $s_n$ is the partial sum
\begin{equation}
s_n \; = \; \sum_{k=0}^n \, a_k
\end{equation}
of an infinite series, then the remainder $r_n$ obviously satisfies
\begin{equation}
   \label{remainder_estimate}
   r_n \; = \; - \sum_{k=n+1}^{\infty} \, a_k \, .  
\end{equation}
In particular, this implies that the remainders $r_n$ are
all negative if the series terms $a_n$ are all positive.

The dependence of the magnitude of the remainder $r_n$ on the index $n$
is a natural measure for the convergence of the sequences or series. Often,
it is of considerable interest to analyze the asymptotics of the
sequence of remainders $\{ r_n \}_{n=0}^{\infty}$ as $n \to \infty$. 
Let us assume that we have found an asymptotic
sequence $\{ \varphi_{k} (n) \}_{k=0}^{\infty}$ 
as $n \to \infty$ with $\varphi_0 (n) = 1$ for the function $r_n/\omega_n$, 
which is equivalent to identifying 
some dominant part (leading term) $\omega_n$ 
of $r_n$ with respect to the asymptotic sequence 
$\{ \varphi_{k} (n) \}_{k=0}^{\infty}$. We call $\omega_n$ the remainder 
estimate. In particular, this means that
the remainder term $r_n$ can be expanded asymptotically according to
\begin{equation}
\label{remainder_estimate_def}
r_n / \omega_n \; \sim \; \sum_{k=0}^{\infty} \, c_k \, \varphi_k (n)
\, , \qquad n \to \infty \, .
\end{equation}
Sequences of remainder estimates $\{ \omega_n \}_{n=0}^{\infty}$
will be of considerable importance in this report. The reason is that it
is often possible to obtain at least some structural information about
the behavior of the dominant term of a remainder $r_n$ as $n \to
\infty$. It will become clear later that convergence acceleration or
summation methods, which explicitly utilize the information contained in
the remainder estimates $\{ \omega_n \}_{n=0}^{\infty}$, frequently lead
to efficient algorithms.

%
%
\subsubsection{Types of Convergence}
\label{Sec:TypConv}

The ideas of ``linear'' and ``logarithmic'' convergence turn
out to be very useful concepts when analyzing the types
of convergence of series, and are used very often in the
literature on convergence acceleration. They are recalled in
the current section. 

We observe that 
the behaviour of the remainder (\ref{remainder_estimate}) as a function
of $n$ is a natural 
measure for the rate of convergence of the series. In this subsection, 
therefore, we discuss how different types of behaviour of the 
$r_n$ at $n \rightarrow \infty$ may be used to classify the types of 
convergence of a particular sequence. In order to perform such a
classification, we write
\begin{equation}
\label{convergence_classification}
\lim_{n \to \infty} \; \frac {s_{n+1} - s} {s_n - s} \; = \; \lim_{n
\to \infty} \; \frac {r_{n+1}}{r_n} \; =\; \rho \, ,
\end{equation}
where, of course, we assume that the sequence $\{ s_n \}_{n=0}^{\infty}$ 
converges to some limit $s$.
If, in this equation, $ 0 < \vert \rho \vert < 1$, we say that the 
sequence $\{ s_n
\}_{n=0}^{\infty}$ converges linearly, if $\rho = 1$ holds, we say
that this sequence converges logarithmically, and if $\rho = 0$
holds, we say that it converges hyperlinearly. Moreover, 
$\vert \rho \vert > 1$ implies that the sequence 
$\{ s_n \}_{n=0}^{\infty}$ diverges. 

In Ref.~\cite{ClGrAd1969}, it has been shown that if the terms $a_k$ of a
convergent series are all real and have the same sign, then
for $s_n = \sum_{k=0}^n a_k$, we can establish a connection
between the remainder $r_n$ and the series term $a_n$ as follows,
\begin{equation}
\label{conv_formula}
\lim_{n \to \infty} \; \frac{r_{n+1}}{r_n} \; = \;
\lim_{n \to \infty} \; \frac{a_{n+1}}{a_n} \; = \; 
\rho \, .  
\end{equation}
Effectively, this means that we can replace in the fraction
$r_{n+1}/r_n$ each of the $r_n$ by $\Delta s_n$,
\begin{equation}
\label{convergence_formula}
\lim_{n \to \infty} \; \frac{r_{n+1}}{r_n} \; = \;
\lim_{n \to \infty} \; \frac{\Delta s_{n+1} }{ \Delta s_{n} } \; = \; 
\rho \, .  
\end{equation}
Apart from the particular case of the real and nonalternating
$a_n$, Eq.~(\ref{convergence_formula}) can be also applied to 
arbitrary hyperlinearly and linearly convergent real sequences \cite{Wi1981}. 

We are ready now to illustrate the application of Eq.~(\ref{convergence_classification}) 
for the classification of the (type of) convergence of some particular
series. To this end, let us start from the Riemann zeta function,
defined by a Dirichlet series as given in Ref.~\cite{Ed1974},
\begin{equation}
   \label{zeta_general}
   \zeta(z) \, = \, \sum\limits_{n=0}^{\infty} \frac{1}{(n+1)^z} \, .
\end{equation}
We would like to determine the type of convergence of this series 
for a particular case of $z$ = 2. For this case, the reminder 
(\ref{remainder_estimate}) reads:
\begin{equation}
   \label{zeta_reminder}
   r_n = - \sum\limits_{k = n+1}^{\infty} \frac{1}{(k+1)^2} \, .
\end{equation}
By applying the Euler--Maclaurin formula [see, e.g., Eqs.~(2.01)---(2.02) on 
p. 285 of Ref.~\cite{Ol1974}] to the right side of this 
expression, we find the leading and the first sub--leading asymptotics
of the reminder $r_n$,
\begin{equation}
\label{zeta_reminder_asymptotics}
r_n \sim - \frac{1}{n + 2} - \frac{1}{2 (n + 2)^2} + 
\sum_{j=1}^\infty (-1)^j \, \frac{B_{2 j}}{(n+2)^{2j+1}}  \,, \qquad
n \to \infty \,. 
\end{equation}
Here, $B_{2j}$ is a Bernoulli number [p. 281 of Ref.~\cite{Ol1974}].
Together with Eq.~(\ref{convergence_classification}), this immediately
implies that the series (\ref{zeta_general}) at $z$ = 2 converges 
logarithmically, i.e. $\rho = 1$ for $n \to \infty$.

A more elaborate example concerns the convergence
properties of the alternating series with partial sums:
\begin{equation}
\label{alternating_zeta_general}
s_n = \sum\limits_{k=0}^{n} \frac{(-1)^k}{(k+1)^2} \,, \qquad
\lim_{n\to\infty} s_n = \frac{\pi^2}{12}\, .
\end{equation}
This series is similar to the Dirichlet series 
for the zeta function $\zeta(2)$,
but is strictly alternating in sign. The fact that it
converges to $\pi^2 / 12$ is shown,
for example, in Sec.~2.2 of Ref.~\cite{Ti1986}.
Via a somewhat rather lengthy calculation,
we find for $n \to \infty$:
\begin{equation}
\label{alternating_zeta_reminder_asymptotics}
r_n = - \sum\limits_{k = n+1}^{\infty} \frac{(-1)^k}{(k+1)^2}
\sim \frac{(-1)^n}{2} \, \left( \frac{1}{n^2} -
\frac{3}{n^3} + \frac{6}{n^4} - \frac{9}{n^5} +
{\rm O}(n^{-6}) \right)\,.
\end{equation}
By inserting this reminder into Eq.~(\ref{convergence_classification}),
we obtain that $\rho = -1$. This means that
while $|\rho| = 1$, we have $\rho \neq 1$ and so the series
is not logarithmically convergent. 
In this contribution, we will refer
to convergent series with remainder terms $r_{n+1}/r_n \to -1$
for $n \to \infty$ (i.e., series with $\rho = -1$) as 
alternating marginally convergent series. 

So far, we have discussed the application of the remainder estimates 
for the classification of the convergence of a particular single 
sequence. The estimates $r_n$, however, can be also be applied to compare the 
rate of convergence of two different series. To this end, let us assume 
that two sequences $\{ s_n \}_{n=0}^{\infty}$ and $\{
s^{\prime}_n \}_{n=0}^{\infty}$ both converge to the same limit $s$. We
shall say that the sequence $\{ s^{\prime}_n \}_{n=0}^{\infty}$
converges more rapidly than $\{ s_n \}_{n=0}^{\infty}$ if
\begin{equation}
\lim_{n \to \infty} \; \frac { s^{\prime}_n - s }{ s_n - s}
\; = \; \lim_{n \to \infty} \; \frac { r^{\prime}_n }{ r_n } \; = \;
0 \, .
\label{DefAccConv}
\end{equation}

A further natural question 
is whether one can replace in Eq.~(\ref{DefAccConv}) $r_n$ by $\Delta s_n$
and $r'_n$ by $\Delta s'_n$, in
analogy to Eq.~(\ref{convergence_formula}):
\begin{equation}
\lim_{n \to \infty} \; \frac { s^{\prime}_{n+1} - s^{\prime}_n }
{ s_{n+1} - s_n} \; = \;
\lim_{n \to \infty} \; \frac { \Delta s^{\prime}_n }
{ \Delta s_n} \; = \; 0 \, .    
\label{AltConAccConv}
\end{equation}
Note that in typical cases, the quantity 
$s^{\prime}_{n+1} - s^{\prime}_n$ which 
may result from the application of a sequence transformation to
$\{ s_n \}_{n=0}^\infty$ cannot simply be replaced by some known 
term $a_n$.

It seems~(see Ref.~\cite{We1989}) that the equivalence of
(\ref{DefAccConv}) and (\ref{AltConAccConv}) cannot be 
established universally without making some
explicit assumptions about how fast the sequences $\{ s_n
\}_{n=0}^{\infty}$ and $\{ s^{\prime}_n \}_{n=0}^{\infty}$ approach
their common limit $s$. 
In particular, if $\{ s_n \}_{n=0}^{\infty}$ 
converges linearly, then the transformed series 
$\{ s^{\prime}_n \}_{n=0}^{\infty}$ can 
converge more rapidly only if it converges at
least linearly or even faster. We can always write
\begin{equation}
\frac{ \Delta s^{\prime}_n }{ \Delta s_n } \; = \;
\frac{s^{\prime}_{n+1} - s^{\prime}_n} {s_{n+1} - s_n} \; = \;
\frac{ s^{\prime}_n - s } { s_n - s} \;
\frac{[(s^{\prime}_{n+1} - s) \, / \, (s^{\prime}_n - s)] \, - \, 1}
{[(s_{n+1} - s) \, / \, (s_n - s)] \, - \, 1}  \; = \;
\frac{ r^{\prime}_n } { r_n } \;
\frac{ r^{\prime}_{n+1} / r^{\prime}_n \, - \, 1}
{r_{n+1} / r_n - \, 1} \, . 
\end{equation}
Let us assume now that both $\{ s^{\prime}_n \}_{n=0}^{\infty}$ 
as well as $\{ s_n \}_{n=0}^{\infty}$ fulfill 
Eq.~(\ref{convergence_formula}) with limits $\rho'$ and $\rho$. 
We can then write
\begin{equation}
\label{reformulate_condition}
\lim_{n \to \infty} \frac{ \Delta s^{\prime}_n }{ \Delta s_n } \; = \;
\lim_{n \to \infty} \frac { r^{\prime}_n }{ r_n }  \, 
\frac{\rho' - 1}{\rho - 1}\,.
\end{equation}
The equivalence of
(\ref{DefAccConv}) and (\ref{AltConAccConv}) immediately follows
under the assumptions made, which include the 
linear convergence of the input series (this ensures $\rho \neq 1$).
If, however, $\{ s_n \}_{n=0}^{\infty}$ converges logarithmically, the
denominator of the second term on the right-hand side of
Eq.~(\ref{reformulate_condition}) approaches zero as
$n \to \infty$. In this case, some additional assumptions about the rate
of convergence of $\{ s^{\prime}_n \}_{n=0}^{\infty}$ to $s$ have to be
made.

Finally, we should perhaps remark that the analogous terminology of 
``exponential'' and ``algebraic'' convergence 
(instead of ``linear'' and ``logarithmic'' convergence) 
is often used in both the physical as well as the mathematical literature,
in particular in situations where the ``convergent entities'' are different
from partial sums of a series. An example would be
the rate of convergence of iterative solutions to dynamical problems.

%
%
\subsubsection{Stieltjes Functions and Stieltjes Series}
\label{FunStiSer}

Stieltjes series are, roughly speaking, paradigmatic examples
of asymptotic series for which it is possible to find remainder
estimates. These estimates characterize the remainder after the 
calculation of a partial sum and can be used for the construction
of sequence transformations which profit from the additional information
contained in the remainder estimates (see, for example,
Section II in \cite{BeWe2001} and references therein).  

Specifically,
a function $F(z)$ with $z \in \mathbb{C}$ is called a Stieltjes
function if it can be expressed as a Stieltjes integral according to
\begin{equation}
F (z) \; = \; \int_{0}^{\infty} \,
\frac{\mathrm{d} \Phi (t)} {1+zt} \, ,
\qquad \vert \arg (z) \vert < \pi \, .
\label{StieFun}
\end{equation}
Here, $\Phi (t)$ is a bounded, nondecreasing function taking infinitely
many different values on the interval $0 \le t < \infty$. Moreover, the
moment integrals
\begin{equation}
\mu_n \; = \; \int_{0}^{\infty} \, t^n \, \mathrm{d} \Phi (t) \, ,
\qquad n \in \mathbb{N}_0 \, ,
\label{StieMom}
\end{equation}                  
must be positive and finite for all finite values of $n$. 
If we insert the geometric series $1/(1+x) = \sum_{\nu=0}^{\infty}
(-x)^{\nu}$ into the integral representation (\ref{StieFun}) and
integrate term-wise using (\ref{StieMom}), we see that a Stieltjes
function $F (z)$ can be expressed by its corresponding
Stieltjes series:
\begin{equation}
F (z) \; = \;
\sum_{\nu=0}^{\infty} \, (-1)^{\nu} \, \mu_{\nu} \, z^{\nu} \, .
\label{StieSer}
\end{equation}
Whether this series converges or diverges depends on the behavior of the
Stieltjes moments $\mu_n$ as $n \to \infty$.       

If we insert the relationship
\begin{equation}
\sum_{\nu=0}^{n} \ (-x)^{\nu} \; = \; \frac{1}{1+x} \, - \, 
\frac{(-x)^{n+1}}{1+x} 
\end{equation}
for the partial sum of the geometric series into the integral
representation (\ref{StieFun}), we see that a Stieltjes function can be
expressed as a partial sum of the Stieltjes series (\ref{StieSer}) plus a 
truncation error term which is given as an integral (see for
example Theorem 13-1 of \cite{We1989}):
\begin{equation}
F (z) \; = \; \sum_{\nu=0}^{n} \, (-1)^{\nu} \, \mu_{\nu} \, z^{\nu}
\, + \, (-z)^{n+1} \, \int_{0}^{\infty} \,
\frac{t^{n+1} \mathrm{d} \Phi (t)} {1+zt} \, ,
\qquad \vert \arg (z) \vert < \pi \, .
\label{StieFunParSum}
\end{equation}
Moreover, the truncation error term in (\ref{StieFunParSum}) satisfies,
depending upon the value of $\theta = \arg (z)$, the following
inequalities (see for example Theorem 13-2 of \cite{We1989}):
\begin{equation}
\left\vert (-z)^{n+1} \, \int_{0}^{\infty} \, 
\frac{t^{n+1} \mathrm{d} \Phi (t)} {1+zt} \right\vert \; \le \;
\begin{cases}
\mu_{n+1} \, \vert z^{n+1} \vert \, ,
& \qquad \vert \theta \vert \le \pi /2 \, , \\[2.0ex]
\mu_{n+1} \, \left| \displaystyle \frac{z^{n+1}}{ \sin(\theta) } \right| \, ,
& \qquad \pi / 2 < \vert \theta \vert < \pi \, .
\end{cases}
\label{StieRem}
\end{equation}
Notice that the expression $\mu_{n+1} \, \vert z^{n+1} \vert$
emerges as a natural remainder estimate for the partial sum
$\sum_{\nu=0}^{n} \, (-1)^{\nu} \, \mu_{\nu} \, z^{\nu}$ of the 
input series. This observation will later be used in 
Sec.~\ref{NonlinearST}
for the construction of a special class of sequence transformations.
Detailed discussions of the properties of Stieltjes series and their
special role in the theory of summability can be found in 
Chap.~8.6 of~\cite{BeOr1978} or in Chap.~5 of~\cite{BaGr1996}.

%
%
\section{Sequence Transformations}
\label{seqtr}

In the current section, sequence transformations are introduced and discussed
in several steps. The goal is both to introduce basic
notions connected with the construction of convergence accelerators,
and to enable the reader to read and understand further, more
specialized literature sources. 
\begin{itemize}
\item In Sec.~\ref{Sec:QuickStart}, an overview of some of the 
most common convergence acceleration methods is given, and the
reader is presented with a crude quick-start guidance for the 
choice of suitable convergence acceleration methods.
\item In Sec.~\ref{Chap:ConvAccelMeth}, selected convergence
acceleration methods are discussed. Some of these profit 
from the availability of explicit remainder estimates and can 
sometimes lead to quite spectacular numerical results.
\item In Sec.~\ref{ConvAccelAppl}, the methods are applied 
to a number of concrete computational problems, for illustrative
purposes. The examples concern both straightforward
applications of some of the methods as well as more sophisticated 
cases, where methods have to combined in an adaptive way.
\end{itemize}
The Chapter thus provides both abstract as well as very concrete
information regarding convergence accelerators.

%
%
\subsection{Quick Start}
\label{Sec:QuickStart}

As a general preamble, it should be noticed that, as everything comes at a
price, there is unfortunately no magic method working in all situations: 
a famous proof in Ref.~\cite{DeGB1982} shows that a universal
transformation accelerating all numerical sequences cannot exist, even if we
restrict our attention to convergent sequences, neglecting the 
divergent case.

No matter which convergence acceleration method is finally 
selected for a particular problem at hand, 
some practical effort is always necessary to get the best results:
on the one hand, a straightforward usage of the
algorithms presented in this review, in the sense of computational recipes,
can often lead to misleading and nonsensical numerical results; 
on the other hand, quite spectacular results can often be achieved 
if careful use is made of all the available information.
A two-step procedure should be followed: before starting the actual numerical
investigations the known theoretical results should always be reviewed, to
understand if by chance the problem at hand satisfies the applicability
conditions of one or more of the existing methods. Only if the answer to this
question is negative the user should try out some algorithms blindly, and even
in this case a preliminary critical assessment of the nature of the problem
can often help in immediately ruling out a number of methods.
To illustrate these points, we lay out a possible 
rough guidance for choosing and
implementing the best sequence acceleration method among those which 
are presented in in the rest of the chapter.
A much more complete (and interactive) version
of this decision grid can be found in the form of computer program
named HELP as introduced and discussed in Ref.~\cite{BrRZ1991}.
Here, we attempt to present a rather compact discussion.

Before we turn our attention to specific cases, 
let us mention that it is possible to divide many of the algorithms
discussed in the literature into two classes.
In the case of ``more traditional methods'', an assumption
about the mathematical structure of the remainder term $r_n$ 
is implicitly coded into the algorithm. The algorithm is constructed
to eliminate remainder terms of a specific structure step by step.
The terminology is that the sequence transformation 
is exact for a specific subset of input series. 
In the case of ``modern nonlinear sequence transformations,''
an explicit remainder is coded directly and explicitly into the 
algorithm:  the calculation cannot proceed without the external
input of remainder estimates in addition to the terms of the 
series to be summed. In the case of the 
modern nonlinear sequence transformations, 
the algorithm only has to make an assumption 
regarding the correction term which relates the exact
remainder and the remainder estimate, but it is unnecessary to 
eliminate the entire remainder from the series (only the 
correction term needs to be dealt with). 

For both the ``modern'' as well as the ``traditional''
algorithms discussed here, one may indicate a certain 
unifying concept. Indeed, many of the convergence acceleration methods
can be cast into the general structure of the $E$ 
algorithm (see~\cite{Br1980E,Ha1979E} and Sec.~3.3
of Ref.~\cite{We1989}). By examination of the 
$E$ algorithm, the following fundamental paradigm,
on which all methods seem to be based,
can be exposed in a particularly clear fashion:
``Make an assumption about the remainder and eliminate 
it from the partial sums of the series as far as possible.'' 

In view of the ``non-universality theorem'' of Ref.~\cite{DeGB1982},
the task of finding a suitable acceleration transformation
acquires a certain heuristic component. Instead of giving 
an immediate answer, we should therefore attempt to 
formulate some useful questions which should be asked first:
\begin{itemize}
\item Is a remainder estimate $\omega_n$ for the 
remainder term $r_n$ known, or is no such information
available? 
In the first case, one has the option of using one of the 
nonlinear sequence transformations
discussed here in Secs.~\ref{NonlinearST} 
and~\ref{Relatedsequences} and in 
Secs.~7 and~8 of Ref.~\cite{We1989}. These transformations
profit from remainder estimates and make implicit assumptions about the 
correction term which corrects the remainder estimate in the 
direction of the explicit remainder.
If no remainder estimate is available, 
then one only has the option of one of the more traditional
methods covered in Secs.~\ref{PadeApproximation} 
and~\ref{OtherTraditional}, which do not need as input 
an explicit remainder estimate. Note, however, that the 
more traditional algorithms nevertheless make assumptions about the 
mathematical structure of the remainder term: the quality of the
results varies strongly depending on the validity of these
assumptions, and thus one can generally recommend to try more than 
one method on a particular given problem.
\item Is the convergence of the input series known to be 
linear or logarithmic, and is the 
sign pattern of the terms of the input series 
alternating or nonalternating? If this knowledge is available,
then the choice of applicable convergence accelerators
can be narrowed down further. Details can be found throughout
Chap.~\ref{seqtr}.
In general, the application of convergence accelerators 
to alternating series is numerically much more stable than 
to nonalternating series. Furthermore, the 
acceleration of logarithmic convergence is generally 
recognized as a more demanding task as compared to 
the acceleration of linear convergence. However, 
as discussed in this review, a few rather powerful algorithms 
are currently available.

\item Is it the aim to obtain a specific numerical 
result for a physical quantity under investigation,
e.g., an energy level or a cross section, to the highest accuracy 
based on a finite number of terms of an input series, or
is the goal to develop a high-performance algorithm, e.g.,
for a special function which is given by an asymptotic series expansion
whose terms are known? In the first case, there is no need 
for a computationally efficient implementation of the acceleration algorithm:
typically, a knowledge of its definition is sufficient, and in
order to avoid numerical loss in higher transformation
orders, one may use multi-precision
arithmetic, which is implemented in today's computer algebra
systems (e.g., Ref.~\cite{Wo1988}) in intermediate steps.
E.g., if there is no concern about computational
efficiency, then it is possible to directly key the definitions given in
Eqs.~(\ref{dLevTr}) and (\ref{dWenTr}) into a computer program,
without making use of the recursive schemes defined in
Eqs.~(\ref{LevRec}) and (\ref{WenRec}) below.
The same applies to prototyping applications for numerical
algorithms. In the second case, however, one typically has
highly powerful recursive schemes available for the calculation
of higher-order transformations which reduce the 
calculational demand by orders of magnitude. The latter
aspect is crucial to the applicability of convergence accelerators.
In order to appreciate this aspect, one should remember
that it was the discovery of the $\epsilon$ algorithm
in Ref.~\cite{Wy1956eps}
which sparked a huge interest in Pad\'{e} approximants
which in turn constitute a paradigmatic example of a
versatile convergence acceleration algorithm.
\end{itemize}
We can now turn our attention to the actual algorithms which
have a certain prominent status among the many methods available 
from the literature. Here, it is inevitable to name some algorithms 
which will be defined only in Chap.~\ref{seqtr} below. The authors
hope that this aspect will not lead to confusion.
In general, one may distinguish 
four cases, according to Eq.~(\ref{convergence_classification}): 
\begin{itemize}
\item Linear convergence, alternating series: 
In this case, a number of methods is 
available, e.g., the $\epsilon$ algorithm or the (iterated)
$\Delta^2$ process always lead to some acceleration.
These two algorithms are discussed in Secs.~\ref{PadeApproximation}
and~\ref{OtherTraditional}, respectively.
A further possibility is provided by polynomial 
extrapolation with geometrically convergent interpolation
points (see Sec.~\ref{OtherTraditional}, Sec.~6.1 of Ref.~\cite{We1989},
and p.~6 ff.~as well as pp.~36--42 of Ref.~\cite{Br1978},
and Sec.~3.1 of Ref.~\cite{PrFlTeVe1993}).
The mentioned algorithms fall into the ``traditional'' category 
discussed above, and are by construction exact for model 
sequences which approximate typical cases of linearly convergent
series. Moreover, the acceleration of alternating series typically
constitutes a numerically very stable process in higher transformation
orders. If a remainder estimate is available, then one 
may also study the behaviour of special sequence transformations,
typically the nonlinear $t$ and $d$, and the $\tau$
and $\delta$ transformations discussed here in 
Secs.~\ref{NonlinearST} and~\ref{Relatedsequences} 
(see also Table~\ref{master} in the current review) and in 
Secs.~7 and~8 of Ref.~\cite{We1989}. Observe, however, 
that the more traditional algorithms also make special 
assumptions about the mathematical structure of the remainder
terms, and that these may lead to numerically very favourable results
if the series matches the corresponding assumptions of the 
algorithm. 
\item Linear convergence, nonalternating series: 
From a theoretical point of view, there is no difference 
between the cases of linearly convergent, alternating series
and linearly convergent, nonalternating series.
However, from a practical point of view, the difference is 
huge: because any convergence accelerator automatically 
entails the calculation of higher weighted differences of 
elements of the input series, the calculation of higher-order
transformations of a nonalternating series can be a numerically
highly unstable process. The following question does not even 
have to be asked, as it usually penetrates practical considerations
immediately: Does the accelerated series converge fast enough to the 
required accuracy in order to be able to overcome the 
pileup of numerical loss of precision due to roundoff?
If the convergence is fast enough,
then the direct application of one of the 
above mentioned algorithms to the input series is possible. 
If the roundoff errors pile up too fast, then it becomes 
necessary to make additional efforts. One possibility is to take
the nonalternating series as input to a condensation transformation,
to generate an alternating series as a result of the 
condensation transformation, and then, to take the resulting 
series as input to an additional second acceleration
process: here, variants of the CNC transformation
(see Sec.~\ref{Sec:CNCT}) may be useful. 
\item Alternating marginally convergent series:
If remainder estimates are available, then
it is possible to use nonlinear sequence transformations,
e.g., the $u$, $d$ and $v$, and the $y$, $\delta$ and $\phi$
transformations (see Sec.~\ref{Relatedsequences} here and Secs.~7 and~8 
of Ref.~\cite{We1989}). Only for the alternating case,
the $\epsilon$ algorithm (see Sec.~\ref{PadeApproximation}) may still 
lead to some acceleration, but its numerical results in the case 
of logarithmically convergent series are in general very poor 
[see also Eq.~(\ref{oops}) below].
\item Logarithmic convergence, nonalternating series:
This is in general the most problematic case, because
numerical roundoff almost inevitably piles up in higher
transformation orders. 
If it is possible to clearly discern the mathematical 
structure of the input series, then it may still be 
possible to use a one-step convergence acceleration algorithm
with good effect. An example would be polynomial extrapolation
with interpolation points which are given as inverse powers
of the index, if the input series has a corresponding 
structure (see Sec.~\ref{OtherTraditional}, Sec.~6.1 of
Ref.~\cite{We1989} and Ref.~\cite{Br1978}). 
Another option is to transform first 
to an alternating series via a condensation transformation, 
and to use an efficient accelerator
for the resulting alternating series. The latter method leads to 
variants of the CNC transformation (see Sec.~\ref{Sec:CNCT}).
\end{itemize}
The above list of algorithms is incomplete.
E.g., the $\vartheta$ algorithm derived in Ref.~\cite{Br1971,Br1978}
is known to be quite an efficient accelerator for both 
linearly convergent as well as logarithmically convergent 
sequences. In some sense, the 
$\vartheta$ algorithm has a special status because it cannot be derived
as a variant of the $E$ algorithm. 
As remarked in Sec.~15.5 of Ref.~\cite{We1989}, 
the $\vartheta$ algorithm as well as its iterations 
${\cal J}^{(n)}_k$ are less robust than the $\epsilon$ and 
$\rho$ algorithms, and in addition, the $\vartheta$ algorithm
is outperformed by nonlinear sequence transformations 
(notably, the $u$ and $v$ transformations) in many cases
of logarithmically convergent sequences. Nevertheless,
because of its unique position as a rather robust,
and rather universal accelerator, the $\vartheta$ algorithm
definitely has to be mentioned in this review 
(see Sec.~\ref{OtherTraditional}). In that sense,
and at the risk of some over-simplification, we can say that
the $\vartheta$ algorithm interpolates between the rather universally
applicable, but moderately powerful $\epsilon$ and $\rho$ algorithms
on the one hand and the specialized, but potentially spectacularly
powerful nonlinear sequence transformation with remainder estimates
on the other hand. 

A rather heroic attempt at a universally applicable
accelerator is the COMPOS algorithm presented in
\cite{BrRZ1991}, which consists of many base sequence transformations
(up to $9$ in the original implementation) being performed in parallel.
The results of the parallel transformations 
are sorted according to their apparent convergence
via the SELECT algorithm
(here, the apparent convergence is given by the
relative change in the numerical values of the transforms as the
order of the transformation is increased).
The output sequences (or a subset of them) are COMPOSed 
by making use of the $E$-algorithm.
It has been proven that if one of the base transformation accelerates a
sequence, then also the composite transformation will accelerate the same
sequence.
In addition, the speed of convergence of the composite transformation
is usually not much worse than that of the best base transformation for the
sequence (and can even be better if many of the base transformations converge
quickly). A not so serious criticism of the COMPOS algorithm
is based on the observation that the algorithm fails
when one or more of the base transformations converge quickly to a wrong
answer. The interested reader can consult \cite{BrRZ1991} for further details,
but this drawback is of course shared by un-composed transformations as
well. A rather serious drawback of the COMPOS algorithm, however, lies
in the computational effort implied by the necessity to perform
the initial transformations in parallel: if the goal is to
develop an efficient numerical algorithm for a mathematical
entity defined by a slowly convergent series by implementing a
convergence accelerator, then the computation overhead implied by
the COMPOS algorithm can be quite severe.

Hints concerning the actual implementation of the methods discussed 
here can be found, e.g., in Refs.~\cite{We1989,FeFoSm1983} concerning
sequence transformations, including the $\epsilon$ and the $\rho$ 
algorithms, and in Ref.~\cite{BrRZ1991} concerning a number of 
rather sophisticated algorithms, including COMPOS.
Finally, let us remark that a complete implementation
of the CNC transform, adapted to the computation of the Lerch
transcendent, is available for download~\cite{JeHomeHD}.
With these preliminaries being discussed, let us know
go {\em in medias res}.
 
%
%
\subsection{Selected Methods}
\label{Chap:ConvAccelMeth}

%
%
\subsubsection{Pad\'{e} Approximants}
\label{PadeApproximation}

As already discussed in Section \ref{Sec:AsySeqAsyExp},
in many problems of theoretical physics, a physical quantity 
$f(z)$ can be presented in terms of its asymptotic or perturbation
expansion (\ref{DefP}) with respect to the asymptotic \textit{power}
series $\{ z^n \}_{n=0}^{\infty}$. Often, the convergence problems
with such power series can be overcome by making use of the 
Pad\'e approximants. The literature
on Pad\'e approximants is abundant. A classic reference 
is~\cite{Ba1975}, and further volumes 
(Refs.~\cite{Gi1973,Gi1978,BeOr1978,BaGr1996})
provide additional valuable information.
From among the many other articles that deal with the subject,
we only mention Refs.~\cite{Ba1965,BaGa1970,Ba1972,Br1976,BeOr1978,Ba1990,%
BaGr1996,Ma1997}.
A Pad\'e approximant produces a rational approximation to the
original input series which is given by the ratio of two
polynomials. For instance, the $[ l / m ]$ Pad\'e approximant to a 
power series (\ref{DefP}) 
\begin{equation}
f(z) = \sum_{k=0}^\infty \alpha_k \, z^k
\end{equation}
is given by the ratio of two polynomials $P_l(z)$ and 
$Q_m(z)$ of degree $l$ and $m$, respectively:
\begin{equation}
\label{pade_definition}
[ l / m ]_f(z) \;=\; \frac{P_l(z)}{Q_m(z)}\;=\;
\frac{p_0 + p_1 \, z + \ldots + p_{l} \,  z^{l}}
{q_0 + q_1 \, z + \ldots + q_{m} \, z^{m}} \;,
\end{equation}
where $q_0$ is canonically fixed to a unit value.
The polynomials $P_l(z)$ and $Q_m(z)$ are constructed so that the
Taylor expansion of the Pad\'e approximant (\ref{pade_definition}) agrees
with the original input series up to the
order of $l+m$
\begin{equation}
\label{padedef}
f(z) \;-\; [ l / m ]_f(z) \; = \; \sum_{k=0}^{\infty} \alpha_k \, z^k 
- [ l / m ]_f(z) \; = \;{\rm O}(z^{l+m+1})\,.
\end{equation}
This equation, therefore, can be used in order to find the coefficients
of the polynomials $P_l(z)$ and $Q_m(z)$. That is, by comparing coefficients
at the same orders of $z$ and assuming $q_0$ = 1, one gets a system
of $l+m+1$ coupled equations to determine $p_0, p_1, q_1, p_2, q_2,
\dots$.
In most cases it is quite inconvenient to solve this system of equations. A 
more attractive recursive prescription for the evaluation of Pad\'e
approximants is the $\epsilon$-algorithm given by the
formulas~\cite{Wy1956eps,We1989}
\begin{subequations}
\label{epsalg}
\begin{eqnarray}
\epsilon_{-1}^{(n)} & = & 0\,, \qquad \epsilon_0^{(n)} \; = \; s_n\,, \\[2ex]
\epsilon_{k+1}^{(n)} & = & \epsilon_{k-1}^{(n+1)} \;+\; 1/ [
\epsilon_k^{(n+1)} - \epsilon_k^{(n)} ]
\label{epsalgrec}
\end{eqnarray}
\end{subequations}
for $n, k \in \mathbb{N}_0$. For the case when the input data 
$\epsilon_0^{(n)} = s_n$ are the partial sums
\begin{equation}
\label{sntopower}
s_n \;=\; 
\sum_{k=0}^n \; a_k \; = \; 
\sum_{k=0}^n \; \alpha_k \; z^k 
\end{equation}
of the power series Eq.~(\ref{DefP}), the elements
$\epsilon_{2k}^{(n)}$ of the two-dimensional table of transformations
produce Pad\'e approximants according to
\begin{equation}
\label{EpstoPade}
\epsilon_{2k}^{(n)} \;=\; [ n+k / k ]_f (z) \,.
\end{equation}
The odd elements of the table of transformations 
$\epsilon_{2j+1}^{(n)}$ (with $j \in \mathbbm{N}_0$) are only auxiliary
quantities.

The above formulas would not be meaningful without some 
interpretation. First, we observe that the Pad\'e approximant 
$[ n / 0 ]_f (z)$ exactly reproduces the partial sum $s_n$
of the input series. For the computation of $[ n + 1 / 1 ]_f (z)$,
we need $s_n$, $s_{n+1}$ and $s_{n+2}$ (two elements
in addition to the start element $s_n$), and therefore
the approximant $[ n + 1 / 1 ]_f (z)$ constitutes a transformation
of second order. In order to compute
the Pad\'e approximant $[ n+k / k ]_f (z)$,  
we need the elements $s_{n+1}, \dots, s_{n+2 k}$ in addition
to $s_n$, and the transformation therefore is of order $2 k$.

In Eq.~(4.3-2) of Ref.~\cite{We1989}, it is remarked that
the elements of the $\epsilon$ table connected by the formula
(\ref{epsalgrec}) form a rhombus in the $(n,k)$ plane. The
general paradigm of the $\epsilon$ algorithm is to use the 
recursive scheme in order to increase the transformation order as
far as possible for a given number of elements of the input
sequence. If the elements $s_n, \dots, s_{n+m}$ are
known, then highest-order transform which can be calculated 
is $\epsilon_{2 \Ent{m/2}}^{(n)}$. Rather sophisticated
techniques are available for a computationally 
efficient implementation of recursive schemes like the 
one given in Eq.~(\ref{epsalg}). Typically, in most cases the surprising 
conclusion is that a single, one-dimensional array of elements
is sufficient in order to perform the transformation,
whereas the structure of the recursion relation would
otherwise suggest that at least a two-dimensional array is needed.
Details on the latter point can be found, e.g., in Secs.~4.3, 
7.5 and 10.4 of Ref.~\cite{We1989}.

An important point, which has certainly not escaped the
reader's attention, is that for $z=1$, the two forms of the
input data in Eq.~(\ref{sntopower}) become equivalent with
$a_k = \alpha_k$, and in this case, the Pad\'e approximant is nothing
but a (nonlinear) sequence transformation of the input series
$\{ s_n \}_{n=0}^\infty$. In fact, the $\epsilon$ algorithm 
makes no difference between input data which are partial sums
of a power series of term $\alpha_k z^k$ 
and partial sums of a simple input series of terms $a_k$. 
It might also be helpful to remark that the 
$\epsilon$ algorithm is exact for sequences of the form
(see e.g. Ref. \cite{Wy1966})
\begin{equation}
\label{padeexact}
s_n = \sum_{j=0}^{k-1} c_j \, \lambda^n_j \,, \qquad
|\lambda_0| > |\lambda_1| >  \ldots >  |\lambda_{k-1}|\,,
\end{equation}
which is a terminating series composed of, loosely speaking,
``linearly convergent terms'' (if $|\lambda_0| < 1$).

It is time now for a very brief historical detour.
The modern history of convergence 
started with the seminal paper~\cite{Sh1955},
in which the Shanks transformation $e_k(s_n)$ was introduced.
This transformation later turned out
to be equivalent to the calculation of Pad\'e approximants. 
In fact, the transformation $e_k(s_n)$ and the $\epsilon$
algorithm are related by [see, e.g., Eq.~(4.2-2) of Ref.~\cite{We1989}]:
\begin{equation}
\epsilon^{(n)}_{2 k} = e_k(s_n)\,.
\end{equation}
It was the article~\cite{Wy1956eps} which elucidated 
that a highly efficient recursive scheme for the calculation
of the approximants $e_k(s_n)$ can be devised. The Shanks transformation,
from our point of view, constitutes a way of expressing a
diagonal Pad\'e approximant $[ n+k/k]_f(z)$ to a function.

The entries of the $\epsilon$ table with odd lower index
satisfy
\begin{equation}
\epsilon_{2j+1}^{(n)} \;=\; \frac{1}{e_j(\Delta s_n)} \, = \;
\mathrm{O}(z^{-n-2j-1}) .
\end{equation}
Let us interpret this relation. We have assumed that 
the sequence composed of the $s_n$ can be written
as $\left\{ \sum_{k=0}^n \alpha_k \, z^k \right\}_{n=0}^\infty$.
The sequence composed of the $\Delta s_n$ is thus 
$\left\{ \alpha_{n+1} \, z^{n+1} \right\}_{n=0}^\infty$.
The expression for the $n$th sequence element
$\alpha_{n+1} \, z^{n+1}$ can be written as the partial sum 
of a power series,
\begin{equation}
\label{partsumdelta}
\Delta s_n = \alpha_0 + \sum_{k=0}^n \left( \alpha_{k+1} \, z^{k+1} - 
\alpha_k \, z^k \right) \,.
\end{equation}
As $n \to \infty$, all coefficients of this power series vanish,
and the function $\Delta s_\infty$ is zero.
The quantities $e_j(\Delta s_n)$ can thus be identified 
as $[n + j/j]$-Pad\'e approximants 
to a specific function defined by a power series whose partial
sums are given by Eq.~(\ref{partsumdelta}), and since all 
coefficients of that function vanish as $n \to \infty$, 
we have $e_j(\Delta s_n) = \mathrm{O}(z^{n+2j+1})$ by 
construction of the Pad\'e approximant.

In writing expressions like $\epsilon^{(n)}_{2 k}$ or $e_k(s_n)$,
we encounter a certain problem in the notation. Indeed,
$n$ is an index which is normally associated to 
a specific element $s_n$, and once $n$ is fixed, $s_n$ is
fixed also. Here, as in many other literature sources
in the field, $n$ is often used in a double meaning:
first, to indicate the starting element of a transformation
(fixed index), and second, to indicate all the sequence
elements $s_m$ with $m \in \{ n, \dots, n + j \}$ which
are needed in order to construct a transformation  of a specified
order $j$. Observe that in order to calculate $e_k(s_n)$,
the elements $s_n, \dots, s_{n+2k}$ are needed, as $e_k(s_n)$
is a transformation of order $2k$. The simple notation $e_k(s_n)$
might otherwise suggest that the transformation takes as 
input only the element $s_n$, and this is simply not the case.
However, a certain compromise between the exactness and 
completeness of a notation and its compactness and practical
usefulness must be found, and the notation $e_k(s_n, \dots, s_{n + 2 k})$
undeniably looks rather clumsy. The customary notation
$e_k(s_n)$ is thus being adopted here.

Let us briefly review the behaviour of the $\epsilon$ algorithm
for a number of model series. We follow Ref.~\cite{Wy1966}. 
For a linearly convergent series
with remainders
\begin{equation}
s_n \sim s + \sum_{j=0}^\infty c_j \, \lambda_j^n \,, \qquad
1 > \lambda_0 > \lambda_1 > \dots > 0 \,, \qquad n \to \infty\,,
\end{equation}
the $\epsilon$ algorithm has the following behaviour 
as $n \to \infty$ for fixed $k$,
\begin{equation}
\epsilon^{(n)}_{2k} \sim s + 
c_k \left( \frac{\Pi_{j=0}^{k-1} (\lambda_k - \lambda_j)}{
\Pi_{j=0}^{k-1} (1 - \lambda_j)} \right)^2 \, \lambda^n_k \,,
\qquad n \to \infty \,.
\end{equation}
The $\epsilon$ algorithm thus accelerates convergence 
according to the definition in Eq.~(\ref{DefAccConv})
because $(\lambda_k/\lambda_0)^n \to 0$ for $n \to \infty$
(we have $\lambda_k < \lambda_0$ by assumption).
For a marginally convergent sequence
with strictly alternating remainders, 
\begin{equation}
s_n \sim s + (-1)^n \sum_{j = 0}^\infty \frac{c_j}{(\beta + n)^{j+1}} \,,
\qquad \beta > 0 \, , \qquad n \to \infty \,,
\end{equation}
the following estimate has been found in Ref.~\cite{Wy1966} for fixed $k$,
assuming $c_0 \neq 0$,
\begin{equation}
\label{oops}
\epsilon^{(n)}_{2k} \sim s + 
\frac{(-1)^n (k!)^2}{2^{2 k} \, (\beta + n)^{2 k + 1}} c_0 \,.
\quad n \to \infty \,.
\end{equation}
Also here, the $\epsilon$ algorithm accelerates convergence
because $(\beta + n)/(\beta + n)^{2 k + 1} \to 0$ for $n \to \infty$
for $k > 0$. However, the expression $(k!)^2$ in the numerator
suggests that the $\epsilon$ algorithm encounters problems 
when one tries to increase the order of the transformation $k$
for given index $n$ of the start element of the transformation.
For monotone (nonalternating) series,
\begin{equation}
\label{failure1}
s_n \sim s + \sum_{j = 0}^\infty \frac{c_j}{(\beta + n)^{j+1}} \,,
\qquad \beta > 0 \, , \qquad n \to \infty \,,
\end{equation}
the $\epsilon$ algorithm is unable to accelerate convergence,
\begin{equation}
\label{failure2}
\epsilon^{(n)}_{2k} \sim s + 
\frac{c_0}{(k+1) \, (\beta + n)} \,,
\quad n \to \infty \,.
\end{equation}
Indeed, the convergence theory of Pad\'e approximants
is quite well developed~\cite{Ba1975,BaGr1996}, and the above
examples are provided only for illustrative purposes.

{\color{black}
Besides the model series discussed above, we would like 
to recall briefly the convergence and acceleration results obtained
with the $\epsilon$ algorithm when applied to two important classes
of sequences: totally monotonic and totally oscillating sequences.
Here, we start our discussion with totally monotonic sequences
$s_n$ which by definition have the following property:
\begin{equation}
\label{totally_monotonic_definition}
(-1)^k \, \Delta^k s_n \geq 0 \qquad \forall k, n 
\in \mathbb{N}_0 \, .
\end{equation}
This condition implies that since $s_n \geq 0$
(set $k=0$ in Eq.~(\ref{totally_monotonic_definition})), we have 
$s_n \geq s_{n+1} \geq 0$ 
(set $k=1$ in Eq.~(\ref{totally_monotonic_definition})).
A totally monotonic sequence converges
in such a way that $\forall n$, we have $s_n \geq s \geq 0$.
Another, equivalent definition
of a totally monotonic sequence is possible. Namely, it can 
be proven that a necessary and sufficient condition that $s_n$ is a totally
monotonic sequence is given by the 
existence of a bounded and non-decreasing measure 
${\rm d}\alpha(x)$ in $[0, 1]$ such that:
\begin{equation}
   \label{totally_monotonic_definition_another}
   s_n = \int\limits_{0}^{1} x^n {\rm d}\alpha(x) \, . 
\end{equation}
The sequences $s_n = \lambda^n$ where $0 < \lambda < 1$ and
$s_n = 1/(n+1)$ are examples of totally monotonic sequences.

The properties of totally monotonic sequences have been discussed
in detail elsewhere (see, for example, Refs.~\cite{BrRZ1991,Br1975}).
In the present manuscript, we just mention the convergence properties. 
Namely, it has been proven in~\cite{Br1973}
that if the $\epsilon$ algorithm is applied to a 
sequence $s_n$ which converges to $s$ and if there exist
parameters $a$ and $b$
such that $a \, s_n + b$ is a totally monotonic sequence, then 
it is possible to prove the following properties which ensure
the convergence of the Pad\'{e} approximants given by the 
$\epsilon$ algorithm to the limit $s$, as both $n$ and/or 
$k$ are increased. Namely, we have $\forall k$ and $\forall n$
(see p.~89 of Ref.~\cite{BrRZ1991}),
\begin{subequations}
\label{totally_monotonic_conv_epsilon}
\begin{align}
& 0 \leq a \epsilon_{2k+2}^{(n)} + b \leq a \epsilon_{2k}^{(n)} + b \,, \\[2ex]
& 0 \leq a \epsilon_{2k}^{(n+1)} + b \leq a \epsilon_{2k}^{(n)} + b \,, \\[2ex]
& 0 \leq a \epsilon_{2k+2}^{(n)} + b \leq 
a \epsilon_{2k}^{(n+1)} + b \,, \\[2ex]
& 0 \leq a \epsilon_{2k+2}^{(n)} + b \leq 
a \epsilon_{2k}^{(n+2)} + b \,, \\[2ex]
& \forall k \quad {\rm fixed}, 
\quad \lim_{n \to \infty} \epsilon_{2k}^{(n)} = s \,, \\[2ex]
& \forall n \quad {\rm fixed}, 
\quad \lim_{k \to \infty} \epsilon_{2k}^{(n)} = s \,. 
\end{align} 
\end{subequations}
As expressed by Theorem 2.23 of Ref.~\cite{BrRZ1991},
all the columns and all the descending diagonals 
of the $\epsilon$ algorithm can be used for the acceleration of convergence 
of a totally monotonic series, provided 
the input series is a non-logarithmic totally 
monotonic series. We recall that the $\epsilon$ algorithm is 
unable to accelerate logarithmic convergence.

In a similar way as for a totally monotonic series, the $\epsilon$ algorithm
can also be useful for the summation of totally oscillating series. 
Sequences of that kind
are defined in the following way: a $s_n$ is totally
oscillating if the sequence $(-1)^n s_n$ is totally monotonic. Again, 
it has been proven that if the $\epsilon$ algorithm is applied to a 
sequence $s_n$ which converges to $s$ and if there exists $a$ and $b$
such that $a s_n + b$ is totally oscillating sequence,
then we have $\forall k$ and $\forall n$ according 
to Ref.~\cite{BrRZ1991}:
\begin{subequations}
\label{totally_monotonic_conv_epsilon2}
\begin{align}
& 0 \leq a \, \epsilon_{2k+2}^{(2 n)} + 
b \leq a \, \epsilon_{2k}^{(2 n)} + b \,, \\[2ex]
& a \, \epsilon_{2k}^{(2n+1)} + b \leq 
a \, \epsilon_{2k+2}^{(2n+1)} + b \leq 0 \,, \\[2ex]
& a \, \left( \epsilon_{2k}^{(2n+1)} - \epsilon_{2k}^{(2n)} \right) \leq
a \, \left( \epsilon_{2k+2}^{(2n+1)} - \epsilon_{2k+2}^{(2n)} \right) \leq 
0 \,, \\[2ex]
& 0 \leq a \left( \epsilon_{2k+2}^{(2n+2)} - \epsilon_{2k+2}^{(2n+1)} \right)
\leq a \left( \epsilon_{2k}^{(2n+2)} - \epsilon_{2k}^{(2n+1)} \right) \,, 
\\[2ex]
& 0 \leq a \, \epsilon_{2k+2}^{(2n)} + b \leq 
a \, \epsilon_{2k}^{(2n+2)} + b \,,
\\[2ex]
& a \, \epsilon_{2k}^{(2n+3)} + b \leq 
a \, \epsilon_{2k+2}^{(2n+1)} + b \leq 0 \,,
\\[2ex]
& \forall k \quad {\rm fixed}, \quad 
\lim_{n \to \infty} \epsilon_{2k}^{(n)} = s \,, \\[2ex]
& \forall n \quad {\rm fixed}, \quad 
\lim_{k \to \infty} \epsilon_{2k}^{(n)} = s \,. 
\end{align} 
\end{subequations}
}

%
%
\subsubsection{Further Convergence Accelerators without Remainder Estimates}
\label{OtherTraditional}

The $\epsilon$ algorithm takes as input only the elements of the series
to be accelerated. No additional remainder estimates are needed.
This property is shared by the $\Delta^2$ process, 
the $\rho$ algorithm, and the $\vartheta$ algorithm,
which are discussed in the current subsection. 

The $\Delta^2$ process is widely applied to 
the summation of slowly convergent series. It is 
usually attributed to Aitken~\cite{Ai1926}, but there are
indications that it may already have been known to 
Kummer~\cite{Ku1837}.
Within this process, one makes use of the forward 
difference operator (\ref{difference_definition}) in order to transform 
the original sequence $\{ s_n \}_{n=0}^{\infty}$ into a
new sequence $\{ \mathcal{A}_n \}_{n=0}^{\infty}$, whose elements are defined
by:
\begin{equation}
   \label{AitkensA}
   \mathcal{A}_{n} = 
   s_n - \frac{\left[ \Delta s_n \right]^2}{\Delta^2s_n} \, .
\end{equation}
The structure of this equation clearly explains why Aitken's method
is called $\Delta^2$ method. Before discussing the convergence
properties of the transformation (\ref{AitkensA}), we 
would like to mention that
originally Aitken's process was designed to deal with 
the model sequence of the following form \cite{Ai1926},
which is a truncated form of Eq.~(\ref{padeexact}):
\begin{equation}
   \label{Aitkens_sequence}
   s_n = s + c \, \lambda^n \, .
\end{equation}
As seen from this equation, each sequence element $s_n$ contains
two coefficients $c \ne 0$ and $|\lambda| < 1$ as well as the limit $s$. 
The first and the second forward differences of such a sequence read:
\begin{equation}
\label{Aitkens_sequences}
\Delta s_n \; = \; c \, \lambda^n \, \left( \lambda - 1 \right) \, , 
\qquad
\Delta^2 s_n  \; = \; c \, \lambda^n \, \left( \lambda - 1 \right)^2 \, .
\end{equation}
By inserting now these relations into Eq.~(\ref{AitkensA}), we may easily find:
\begin{equation}
   \label{AitkensA_model}
   \mathcal{A}_{n} = 
   (s + c \, \lambda) - c \, \lambda = s \, ,
\end{equation}
which implies that the $\Delta^2$ process is exact for the 
model sequence (\ref{Aitkens_sequence}) at every stage. 
Like any other algorithm of numerical analysis, 
this method has its own domain of 
validity. Because the (convergence) properties of the transformation 
(\ref{AitkensA}) have been discussed 
in detail elsewhere \cite{Br1977,Br1978,Wi1981}, here we just mention that 
the $\Delta^2$ process (i) accelerates linear convergence and 
(ii) is regular but not accelerative for logarithmically convergent 
sequences. This indicates that $\Delta^2$ process has
similar properties as the $\epsilon$--algorithm but is usually not as
powerful as the latter one \cite{We1989}. One may increase the 
accelerative power of the Aitken's method, however, by applying this
method iteratively. That is, it is first applied to the original
sequence $\{ s_n \}_{n=0}^{\infty}$, then to the transformed 
sequence $\{ \mathcal{A}_n \}_{n=0}^{\infty}$ and so on. The
detailed discussion on the iterated $\Delta^2$ process 
is given in the texts by Brezinski and Redivo Zaglia (see pp. 128--131
of Ref.~\cite{BrRZ1991}) and Weniger (see pp. 223--228 of 
Ref.~\cite{We1989}) and, therefore, will be omitted in the present
report. Apart from the iterative procedure, moreover, a large 
number of modifications to the $\Delta^2$ algorithm have been presented 
in Refs.~\cite{Dr1976,JaOB1978,Sa1987} and many others.
At the end of a rather brief discussion on the Aitken's $\Delta^2$
method, we shall note that this method has very close connection 
to the $\epsilon$ algorithm as well as to the Pad\'{e}
approximants. For instance, it immediately follows from the derivation
of the sequence transformation $\mathcal{A}_{n}$ via the model sequence
(\ref{Aitkens_sequence}) that $\mathcal{A}_{n} = \epsilon_2^{(n)} =
[ n + 1/1 ]_f(z)$ is effectively a Pad\'{e} approximant to the 
input sequence given by $s_n = \sum_{k = 0}^{n} \alpha_k z^k$.
    
Beside the $\Delta^2$ process, another very well-known sequence transformation 
is the $\rho$ algorithm~\cite{Wy1956b}, which is exact for 
logarithmically convergent sequences of the following structure,
\begin{equation}
s_n = \frac{s x^k_n + a_1 x_n^{k - 1} + \ldots + a_k}
{x^k_n + b_1 x_n^{k - 1} + \ldots + b_k} \, , \, \qquad 
k, n \in \mathbb{N}_0 , 
\end{equation}   
where the $x_n$ are usually chosen to be $x_n = \beta + n$
with $\beta \geq 0$. The $\rho$ algorithm is defined by the scheme
\begin{subequations}
\label{rho_transformation}
\begin{eqnarray}
\rho_{-1}^{(n)} & = & 0 \,, \qquad \rho_0^{(n)} \; = \; s_n \,, \\
\rho_{k + 1}^{(n)} & = & \rho_{k - 1}^{(n + 1)} + 
\frac{x_{n + k + 1} - x_n}{\rho_k^{(n + 1)} - \rho_k^{(n)}} \,.
\end{eqnarray}
\end{subequations}
The recursive schemes for the 
$\epsilon$ and $\rho$ methods are almost identical,
as seen from Eqs.~(\ref{epsalg}) and (\ref{rho_transformation}). 
The only difference is that $\rho$ algorithm also
involves a sequence of interpolation points $\{ x_n \}_{n=0}^{\infty}$,
which are usually chosen to satisfy the 
condition $0 < x_0 < x_1 < x_2 ....$ and 
$x_n \to \infty$ if $n \to \infty$ (see p.~231 of Ref. \cite{We1989}). 
Moreover, similar to the case of the $\epsilon$ algorithm, only
the elements $\rho^{(n)}_{2k}$ with even orders serve as approximations
to the limit while the elements $\rho^{(n)}_{2k+1}$ with odd orders are just 
auxiliary quantities which diverge if the whole process converges.
Despite their formal similarity, the $\epsilon$ and $\rho$ methods 
differ significantly 
in their ability of convergence acceleration. For example, 
the $\epsilon$ algorithm is not able to accelerate nonalternating
logarithmically convergent sequences [see Eqs.~(\ref{failure1})
and~(\ref{failure2}) and the discussion around Eq.~(4.2-8) and
(4.2-9) of Ref.~\cite{We1989}, as well as Ref.~\cite{Wy1966}].
In contrast, such
a logarithmic convergence can be accelerated by the $\rho$ algorithm,
which fails in the case of linear convergence. Thus, the two algorithms can 
be considered to some extent as complementary. 

A scheme which interpolates between the $\epsilon$ and $\rho$
algorithms is the $\vartheta$ algorithm~\cite{Br1980E,Ha1979E}.
It assumes a special place in the hierarchy of the algorithms 
as it cannot be derived on the basis of the otherwise
very general $E$ algorithm discussed below in Sec.~\ref{ThetaAlgorithm}.
The $\vartheta$-algorithm is defined by the scheme
\begin{subequations}
\begin{align}
\vartheta_{- 1}^{\left( n \right)} = & \; 0\,, \qquad
\vartheta_0^{\left( n \right)} \; = \; s_n\,, \\[2ex]
\vartheta_{2 k + 1}^{\left( n \right)} = & \;
\vartheta_{2 k - 1}^{\left( n + 1 \right)} + 
\frac{\displaystyle 1}{\displaystyle \vartheta_{2 k}^{\left( n + 1 \right)} - 
\vartheta_{2 k}^{\left( n \right)}} \,, \\[4ex]
\vartheta_{2 k + 2}^{\left( n \right)} = & \;
\vartheta_{2 k}^{\left( n + 1 \right)} +
\frac{\displaystyle 
\left(\vartheta_{2 k}^{\left( n + 2 \right)}- 
\vartheta_{2 k}^{\left( n + 1 \right)} \right) \,
\left(\vartheta_{2 k + 1}^{\left( n + 2 \right)}- 
\vartheta_{2 k + 1}^{\left( n + 1 \right)} \right) }
{\displaystyle 
\left( \vartheta_{2 k + 1}^{\left( n + 2 \right)} - 
2\,\vartheta_{2 k + 1}^{\left( n + 1 \right)} +
\vartheta_{2 k + 1}^{\left( n \right)} \right)}   \,.
\end{align} 
\end{subequations}
with $k, n \in \mathbb{N}_0$.
Here, we have written out all expressions which might otherwise
be formulated a little more compactly by using the 
difference operator $\Delta$.
The $\vartheta$ algorithm can be considered as an intermediate form of the
$\varepsilon$ and the $\rho$ algorithm, as it is able to accelerate both 
some linearly and some logarithmically convergent sequences. 
Although the
$\varepsilon$-algorithm and the $\rho$-algorithm are particular cases of the
$E$-algorithm, the $\vartheta$-algorithm is not, and the same can be said about
various sequence transformations which were obtained by a manual modification
or iterations of some other previously known transformations, e.g.,
of the $\vartheta$ algorithm. For a complete
list of such sequences we refer the interested reader to 
\cite{BrRZ1991,We1989}.

%
%
\subsubsection{Sequence Transformations with Remainder Estimates}
\label{NonlinearST}

As seen from Eq.~(\ref{pade_definition}), the coefficients 
$\alpha_0, \alpha_1, \ldots , \alpha_{l+m}$ 
of the formal power series (\ref{DefP}) suffice to 
determine the Pad\'{e} approximant $[l / m]_f (z)$ completely without 
the need for 
any additional information. The same can be said about the 
methods discussed in Sec.~\ref{OtherTraditional}: all
the input required is in the partial sums of the input series.
However, the estimate of a truncation error (remainder term)
can often be given as the first term of asymptotic power series 
not included in the partial sum 
(see Sec.~\ref{FunStiSer} and Refs.~\cite{Ol1974,Ca1977}). 
In this Section, we 
discuss the sequence transformations which can benefit from such an additional 
information via remainder estimates. 
One of these transformations gave
very good results in the case of the sextic anharmonic oscillator, and
there is strong numerical evidence that it is able to sum the extremely
violently divergent perturbation series for the octic anharmonic
oscillator \cite{WeCiVi1993}.
Nonlinear sequence transformations have been 
the subject of a number of monographs
\cite{Br1980pade,Br1977,Br1978,Br1997,BrRZ1991,De1988,LiLuSh1995,%
MaSh1983,Wa1996,Wi1981} and review articles
\cite{We1989,Br1985anm,GaGu1974,Gu1989}.

In order to
start the discussion on the sequence transformations with 
explicit remainder estimates, let us again
consider the sequence $\{ s_n \}_{n=0}^{\infty}$ of partial sums of an 
infinite series either converges to some limit $s$, or, if it diverges, can 
be summed by an appropriate summation method to give the generalized limit $s$. 
As seen from 
Eq.~(\ref{definition_remainder}), the summation of slowly convergent or
divergent series to the (generalized) limit $s$ requires the estimation of the 
remainders $r_n$ from the sequence elements $s_n$. Unfortunately, a 
straightforward elimination of the
remainders is normally not possible, since they are either unknown or
not easily accessible. Thus, in realistic situations one can only hope
to eliminate approximations to the actual remainders. If this
can be accomplished, $\{ s_n \}_{n=0}^{\infty}$ is transformed into a
new sequence $\{ s^{\prime}_n \}_{n=0}^{\infty}$, whose elements can for
all $n \ge 0$ be partitioned into the generalized limit $s$ of the
original sequence and a transformed remainder $r^{\prime}_n$:
\begin{equation}
s^{\prime}_n \; = \; s \, + \, r^{\prime}_n \, .
\end{equation}
This approximative elimination process was successful if the transformed
remainders $r^{\prime}_n$ vanish as $n \to \infty$, which implies that
the transformed sequence $\{ s^{\prime}_n \}_{n=0}^{\infty}$ converges
to $s$. Thus, a sequence transformation is a rule which
transforms a slowly convergent or divergent sequence $\{ s_n
\}_{n=0}^{\infty}$ into a new sequence $\{ s^{\prime}_n
\}_{n=0}^{\infty}$ with hopefully better numerical properties.

The question arises
how assumptions about the structure of the remainder estimate
can be translated into an efficient computational scheme. 
The direct
elimination of approximations to $r_n$ from $s_n$ can be very
difficult. A considerable simplification can often be achieved by means
of a suitable reformulation. Let us consider the following model
sequence (see Section 3.2 of \cite{We1989}):
\begin{equation}
s_n \; = \;s \, + r_n \; = \; s \, + \, \omega_n \, z_n \, .
\label{Mod_Seq_Om}
\end{equation}
Here, $\omega_n$ is a remainder estimate
[cf. Eq.~(\ref{remainder_estimate_def})], which has to be chosen
according to some rule and which may depend on $n$ in a very complicated
way, and $z_n$ is a correction term, which should be chosen in
such a way that it depends on $n$ in a relatively smooth way. Moreover,
the products $\omega_n z_n$ should be capable of producing sufficiently
accurate approximations to the actual remainders $r_n$ of the sequence
to be transformed.

The ansatz (\ref{Mod_Seq_Om}) has another 
advantage: One may now easily devise a systematic way of 
constructing a sequence
transformation which is exact for this model sequence, 
if one can find a linear operator ${\hat T}$ which annihilates
the correction term $z_n$. 
Just apply ${\hat T}$ to $[s_n - s] /
\omega_n = z_n$. Since ${\hat T}$ annihilates $z_n$ according to ${\hat
T} (z_n) = 0$ and is by assumption linear, we find that the following
sequence transformation ${\cal T}$ is exact for the model
sequence (\ref{Mod_Seq_Om}), according to Eq.~(3.2-11) of \cite{We1989}:
\begin{equation}
{\cal T}(s_n, \omega_n) \; = \; \frac
{{\hat T} (s_n / \omega_n )} {{\hat T} (1 / \omega_n )}
\; = \; s \, .
\label{GenSeqTr}
\end{equation}
Simple and yet very powerful sequence transformations are obtained if
the annihilation operators are based upon the finite difference operator
$\Delta$ defined in 
Eq.~(\ref{difference_definition}). As is well known, 
a polynomial $P_{k-1} (n)$ of
degree $k - 1$ in $n$ is annihilated by the $k$-th power of
$\Delta$. Thus, we now assume that the correction terms $\{ z_n
\}_{n=0}^{\infty}$ can be chosen in such a way that multiplication of
$z_n$ by some suitable quantity $w_k (n)$ yields a polynomial $P_{k-1}
(n)$ of degree $k-1$ in $n$:
\begin{equation}
w_k (n) \, z_n \; = \; P_{k-1} (n) \, .
\end{equation}
Since $\Delta^k P_{k-1} (n) = 0$, the weighted difference operator
${\hat T} = \Delta^k w_k (n)$ annihilates $z_n$, and the corresponding
sequence transformation (\ref{GenSeqTr}) is given by the ratio
\begin{equation}
{\cal T}_k^{(n)} \bigl( w_k (n) \big\vert s_n, \omega_n \bigr)
\; = \; \frac
{\Delta^k \{ w_k (n) s_n / \omega_n \}}
{\Delta^k \{ w_k (n) / \omega_n \}} \, .
\label{Seq_w_k}
\end{equation}
Indeed, 
Eq.~(\ref{Seq_w_k}) represents the general form of the nonlinear sequence
transformation, based on annihilation operator ${\hat T}$. The exact
form of this transformation depends on the particular choice of the $w_k (n)$. 
For instance, $w_k (n) = (n + \beta)^{k-1}$,
which annihilates correction terms given by inverse power series,
\begin{equation}
\label{inversepower}
z_n = \sum_{j=0}^{k-1} \frac{c_j}{(\beta + n)^j} \,,
\end{equation}
with $\beta > 0$, yields the following 
sequence transformation \cite{We1989,Le1973}:
\begin{eqnarray}
\lefteqn{
{\cal L}_{k}^{(n)} (\beta, s_n, \omega_n) \; = \; \frac
{ \Delta^k \, \{ (n + \beta)^{k-1} \> s_n / \omega_n\} }
{ \Delta^k \, \{ (n + \beta)^{k-1}  / \omega_n \} }
} \nonumber \\
& \; = \; \frac
{\displaystyle
\sum_{j=0}^{k} \; ( - 1)^{j} \; 
\left( \begin{array}{c} k \\ j \end{array} \right) \;
\frac{(\beta + n + j)^{k-1}}{(\beta + n + k )^{k-1}} \;
\frac {s_{n+j}} {\omega_{n+j}} }
{\displaystyle
\sum_{j=0}^{k} \; ( - 1)^{j} \; 
\left( \begin{array}{c} k \\ j \end{array} \right) \;
\frac{(\beta + n +j )^{k-1}} {(\beta + n + k )^{k-1}} \;
\frac{1}{\omega_{n+j}} } \, ,
\label{LevTr}
\end{eqnarray}
where the subscript $k$
denotes the {\em order\/} of the transformation, and $n$ denotes the
starting index of the input data ($k+1$ sequence elements $s_{n}$, \dots
$s_{n+k}$ and remainder estimates $\omega_{n}$ \ldots $\omega_{n+k}$ are
needed for the computation of the transformation). The shift parameter
$\beta$ has to be positive in order to admit $n = 0$ in (\ref{LevTr}).
The most obvious choice would be $\beta = 1$.

Apart from the transformation (\ref{LevTr}), we may also
adopt the unspecified weights $w_k (n)$ in
(\ref{Seq_w_k}) to be the Pochhammer symbols according to $w_k (n) = (n +
\beta)_{k-1} = \Gamma (n+\beta+k-1)/\Gamma (n+\beta)$ with $\beta > 0$. 
This transformation is exact for corrections terms of the 
following structure,
\begin{equation}
\label{inversefactorial}
z_n = \sum_{j=0}^{k-1} \frac{c_j}{(\beta + n)_j} \,.
\end{equation}
This yields the transformation (see Eq.\ (8.2-7) of
\cite{We1989}):
\begin{eqnarray}
\lefteqn{
{\cal S}_{k}^{(n)} (\beta , s_n, \omega_n) \; = \; \frac
{ \Delta^k \, \{ (n + \beta)_{k-1} \> s_n / \omega_n\} }
{ \Delta^k \, \{ (n + \beta)_{k-1}  / \omega_n \} }
} \nonumber \\
& \; = \; \frac
{\displaystyle
\sum_{j=0}^{k} \; ( - 1)^{j} \; 
\left( \begin{array}{c} k \\ j \end{array} \right) \;
\frac {(\beta + n +j )_{k-1}} {(\beta + n + k )_{k-1}} \;
\frac {s_{n+j}} {\omega_{n+j}} }
{\displaystyle
\sum_{j=0}^{k} \; ( - 1)^{j} \; 
\left( \begin{array}{c} k \\ j \end{array} \right) \;
\frac {(\beta + n +j )_{k-1}} {(\beta + n + k )_{k-1}} \;
\frac {1} {\omega_{n+j}} } \, .
\label{WenTr}
\end{eqnarray}
Again, the most obvious choice for the shift parameter is $\beta = 1$.
The numerator and denominator sums in (\ref{LevTr}) and (\ref{WenTr})
can be computed more effectively with the help of the three-term
recursions, according to Eq.~(7.2-8) of \cite{We1989},
\begin{equation}
\label{LevRec}
L_{k+1}^{(n)} (\beta) \; = \; L_k^{(n+1)} (\beta)
- \, \frac
{ (\beta + n) (\beta + n + k)^{k-1} }
{ (\beta + n + k + 1)^k } \, L_k^{(n)} (\beta)\,,
\end{equation}
and according to Eq.\ (8.3-7) of \cite{We1989},
\begin{equation}
\label{WenRec}
S_{k+1}^{(n)} (\beta) \; = \; S_k^{(n+1)} (\beta)
 - \frac
{ (\beta + n + k ) (\beta + n + k - 1) }
{ (\beta + n + 2 k ) (\beta + n + 2 k - 1) } \, S_k^{(n)} (\beta) \, .
\end{equation}
The initial values $L_0^{(n)} (\beta) = S_0^{(n)} (\beta) = s_n /
\omega_n$ and $L_0^{(n)} (\beta) = S_0^{(n)} (\beta) = 1 / \omega_n$
produce the numerator and denominator sums, respectively, of ${\cal
L}_{k}^{(n)} (\beta , s_n, \omega_n)$ and ${\cal S}_{k}^{(n)} (\beta ,
s_n, \omega_n)$.

The remainder estimate 
\begin{equation}
\omega_n \; = \; a_{n+1} \; = \; \Delta s_n \, ,
\label{d_Est}
\end{equation}
was proposed by Smith and Ford in Ref.~\cite{SmFo1979} for both
convergent as well as divergent series with strictly alternating
terms. The use of this
remainder estimate in (\ref{LevTr}) and (\ref{WenTr}) yields the
following sequence transformations, defined according 
to Eqs.~(7.3-9) and (8.4-4) of \cite{We1989}:
\begin{eqnarray}
\lefteqn{
d_k^{(n)} (\beta, s_n) \; = \;
{\cal L}_{k}^{(n)} (\beta , s_n, \Delta s_n)}
\nonumber \\
& \; = \; \frac
{\displaystyle
\sum_{j=0}^{k} \; (-1)^{j} \; 
\left( \begin{array}{c} k \\ j \end{array} \right) \;
\frac
{(\beta + n +j )^{k-1}} {(\beta + n + k )^{k-1}} \;
\frac {s_{n+j}} {\Delta s_{n+j}} }
{\displaystyle
\sum_{j=0}^{k} \; ( - 1)^{j} \; 
\left( \begin{array}{c} k \\ j \end{array} \right) \;
\frac
{(\beta + n +j )^{k-1}} {(\beta + n + k )^{k-1}} \;
\frac {1} {\Delta s_{n+j}} } \, ,
\label{dLevTr} 
\end{eqnarray}
and
\begin{eqnarray}
\lefteqn{
{\delta}_k^{(n)} (\beta, s_n) \; =  \;
{\cal S}_{k}^{(n)} (\beta , s_n, \Delta s_n)}
\nonumber \\
& \; = \; \frac
{\displaystyle
\sum_{j=0}^{k} \; ( - 1)^{j} \; 
\left( \begin{array}{c} k \\ j \end{array} \right) \;
\frac {(\beta + n +j )_{k-1}} {(\beta + n + k )_{k-1}} \;
\frac {s_{n+j}} {\Delta s_{n+j}} }
{\displaystyle
\sum_{j=0}^{k} \; ( - 1)^{j} \; 
\left( \begin{array}{c} k \\ j \end{array} \right) \;
\frac {(\beta + n +j )_{k-1}} {(\beta + n + k )_{k-1}} \;
\frac {1} {\Delta s_{n+j}} } \, .
\label{dWenTr}
\end{eqnarray}
In Appendix A of \cite{JeMoSoWe1999}, 
it was explained why the
transformations $d_{k}^{(n)}$ and ${\delta}_{k}^{(n)}$ are more
effective in the case of alternating series than in the case of 
nonalternating series. The failure of the transformation 
for cases where the remainder estimates are invalid is also illustrated
in Table 8.7 of Ref.~\cite{Je2003habil}. 
Other sequence transformations, which are also special cases of the
general sequence transformation (\ref{Seq_w_k}), can be found in
Sections 7---9 of \cite{We1989} and in Section 2.7 of 
Ref.~\cite{BrRZ1991}. A sequence transformation,
which interpolates between the sequence transformations (\ref{LevTr})
and (\ref{WenTr}), has been derived in \cite{We1992}.

%
%
\subsubsection{Further Sequence Transformations}
\label{Relatedsequences}

According to Eq.~(7.3-13) of Ref.~\cite{We1989}
(see also p.~19 of Ref.~\cite{Wi1981} and the elucidating
discussion in Ref.~\cite{We2004}), one can 
heuristically motivate 
different remainder estimate for specific types of input series.
Let us begin with series $s_n = \sum_{k=0}^\infty a_n$ where the 
terms have the following structure,
\begin{equation}
a_n \sim \alpha_0 \, \lambda^n \, n^\Theta \, ,
\qquad n \to \infty\,.
\end{equation}
Here, $\alpha_0$ is a constant.
For $|\lambda| < 1$, which can be seen as a paradigmatic 
case of linear convergence, we have in leading order
\begin{equation}
r_n \sim \alpha_0 \, \frac{\lambda^{n+1}\, n^\Theta}{\lambda - 1} \,,
\qquad n \to \infty\,.
\end{equation}
Hence, we have in this case 
\begin{equation}
r_n \sim \frac{\lambda}{\lambda - 1} \, a_n \,, \qquad n \to \infty\,.
\end{equation}
The prefactor is irrelevant because it is $n$-independent, and therefore
the choice 
\begin{equation}
\omega_n = a_n
\end{equation}
for a simple remainder estimate
is heuristically motivated for linearly convergent series.
This choice leads to the $t$ and the $\tau$ transformations
[see Eqs.~(7.3-7) and (8.4-3) in \cite{We1989} and Table~\ref{master}
in this review].

We now turn to the case
\begin{equation}
a_n \sim \alpha_0 \, n^\Theta\,,
\end{equation}
with ${\rm Re} \, \Theta < -1$, which can be seen as a paradigmatic
case of logarithmic convergence. The remainders fulfill
\begin{equation}
\label{remTheta}
r_n \sim \alpha_0 \, \frac{n^{ \Theta + 1 }}{\Theta + 1}\,,
\qquad n \to \infty\,.
\end{equation}
This can be related to the $a_n$ as follows,
\begin{equation}
r_n \sim \frac{1}{\Theta + 1} \, n \, a_n \qquad n \to \infty\,.
\end{equation}
Both remainders have to be negative if the input series
consists only of positive terms, assuming the 
series terms to be real. This follows from the definition
$s_n = s + r_n$, which can be reformulated as
$r_n = - \sum_{k=n+1}^\infty a_k$ for $s_n = \sum_{k=0}^n a_k$.
For the case (\ref{remTheta}), the negativity in ensured because of the
condition ${\rm Re} \, \Theta = \Theta < -1$ which is naturally imposed if the
input series is assumed to be convergent.
The prefactor is irrelevant because it is $n$-independent, and therefore
the choice $\omega_n = n \, a_n$, or a slight generalization,
\begin{equation}
\label{remuy}
\omega_n = (\beta + n) \, a_n\,,
\end{equation}
is heuristically motivated for logarithmically convergent,
nonalternating series. By shifting $n \to n + \beta$, we have
avoided potential problems in the non-asymptotic region 
(notably, the case $n=0$).
The choice (\ref{remuy}) leads to the $u$ and the $y$ transformations
[see Table~\ref{master} and Eqs.~(7.3-5) and (8.4-2) of Ref.~\cite{We1989}].

\begin{table}[tb!]
\begin{center}
\begin{minipage}{15cm}
\begin{center}
\caption{\label{master} Variants of the transformations (\ref{LevTr}) and 
(\ref{WenTr}) with different remainder estimates $\omega_n$.
The assumed structure of the correction term is either 
an inverse power series, according to Eq.~(\ref{inversepower}),
or an inverse factorial series, according to Eq.~(\ref{inversefactorial}).
The main type of convergence corresponds to a typical model 
series by which the choice of the remainder estimate is motivated.
However, according to Ref.~\cite{We1989}, the $u$, $y$, $v$ and 
$\varphi$ transformations have also been applied to linearly 
convergent series with numerically favourable results.
The $d$ and $\delta$ transformations are constructed 
according to an alternating marginally convergent model series.
However, the remainder estimate employed is also applicable 
to alternating linearly convergent series, and even to 
factorially divergent, alternating Stieltjes series,
following the discussion in Sec.~\ref{FunStiSer} 
[see also Eq.~(\ref{StieRem})]. The $c$ and $\chi$ transformations
are proposed here as examples of alternative transformations
which can easily be obtained as generalizations of other,
existing methods.}
\begin{tabular}{cccc}
\hline
\hline
 & \multicolumn{2}{c}{Assumed correction term:} & \\
Remainder estimate & Inverse powers & 
Inverse factorials & Main type of convergence \\
\hline
\rule[-3mm]{0mm}{8mm}
$\omega_n = (\beta + n) \, a_n$ & $u$ & $y$       & logarithmic  \\[-0.5ex]
                                &     &           & (linear) \\[2ex]
$\omega_n = \displaystyle \frac{a_n \, a_{n+1}}{a_n - a_{n+1}}$
                       & $v$ & $\varphi$ & logarithmic \\[-0.5ex]
                       &     &           & (linear) \\[2ex]
$\omega_n = a_n$       & $t$ & $\tau$    & linear \\[2ex]
$\omega_n = a_{n+1}$   & $d$ & $\delta$  & alternating marginal \\
  & & & (linear, factorially divergent) \\[2ex]
$\omega_n = \displaystyle \frac{a_n \, a_{n+2}}{a_{n+1}}$
                       & $c$ & $\chi$  & alternating marginal \\[-1.0ex]
  & & & (linear, factorially divergent) \\[2ex]
\hline
\hline
\end{tabular}
\end{center}
\end{minipage}
\end{center}
\end{table}

One may refine this choice a little bit, by 
observing that for $a_n \sim \alpha_0 \, n^\Theta$, we have
\begin{equation}
\frac{a_n \, a_{n+1}} {a_n - a_{n+1}} \sim 
\frac{1}{\Theta} \, n \, a_n \,.
\end{equation}
The $n$-independent prefactor is again completely irrelevant,
and the choice
\begin{equation}
\label{remvp}
\omega_n = \frac{a_n \, a_{n+1}} {a_n - a_{n+1}} 
\end{equation}
leads to the $v$ and $\varphi$ transformations
[see Table~\ref{master} and 
Eqs.~(7.3-11) and (8.4-5) of Ref.~\cite{We1989}]. The averaging 
over $n$ implied by the more sophisticated choice (\ref{remvp}) 
as compared to (\ref{remuy}) can be advantageous.

We now turn to the case
\begin{equation}
a_n \sim \alpha_0 \, (-1)^n \, n^\Theta\,,
\end{equation}
with ${\rm Re} \, \Theta < -1$, which can be seen as a paradigmatic
case of alternating marginal convergence (see also
Sect.~\ref{Sec:TypConv}). The remainders fulfill
\begin{equation}
r_n \sim \frac12 \, \alpha_0 \, (-1)^n \, (1+n)^\Theta \,, 
\qquad n \to \infty\,.
\end{equation}
This can be related to the $a_n$ as follows,
\begin{equation}
r_n \sim - \frac12\, a_{n + 1} \,, \qquad n \to \infty\,.
\end{equation}
The prefactor is irrelevant because it is $n$-independent, and therefore
the choice 
\begin{equation}
\label{remdd}
\omega_n = a_{n + 1}\,,
\end{equation}
is heuristically motivated for logarithmically convergent,
nonalternating series. 
The choice (\ref{remdd}) leads to the $d$ and the $\delta$ transformations
[again, see Table~\ref{master} and
Eqs. (7.3-9) as well as (8.4-4) of Ref.~\cite{We1989}].

The following ``averaged'' choice which generalizes the estimate 
$\omega_n = a_{n+1}$,
\begin{equation}
\label{remcc}
\omega_n = \frac{a_n a_{n + 2}}{a_{n+1}}\,,
\end{equation}
leads to transformations which we would like to refer to as $c$
and $\chi$,
\begin{equation}
c_k^{(n)} (\beta, s_n) \; = \;
\frac
{\displaystyle
\sum_{j=0}^{k} \; (-1)^{j} \; 
\left( \begin{array}{c} k \\ j \end{array} \right) \;
\frac
{(\beta + n +j )^{k-1}} {(\beta + n + k )^{k-1}} \;
\frac {a_{n+j+1} \, s_{n+j}} {a_{n+j} \, a_{n+j+2}} }
{\displaystyle
\sum_{j=0}^{k} \; ( - 1)^{j} \; 
\left( \begin{array}{c} k \\ j \end{array} \right) \;
\frac
{(\beta + n +j )^{k-1}} {(\beta + n + k )^{k-1}} \;
\frac {a_{n+j+1}} {a_{n+j} \, a_{n+j+2}}} \, ,
\label{ctrafo} 
\end{equation}
and
\begin{equation}
{\chi}_k^{(n)} (\beta, s_n) \; =  \;
\frac
{\displaystyle
\sum_{j=0}^{k} \; ( - 1)^{j} \; 
\left( \begin{array}{c} k \\ j \end{array} \right) \;
\frac {(\beta + n +j )_{k-1}} {(\beta + n + k )_{k-1}} \;
\frac {a_{n+j+1} \, s_{n+j}} {a_{n+j} \, a_{n+j+2}} }
{\displaystyle
\sum_{j=0}^{k} \; ( - 1)^{j} \; 
\left( \begin{array}{c} k \\ j \end{array} \right) \;
\frac {(\beta + n +j )_{k-1}} {(\beta + n + k )_{k-1}} \;
\frac {a_{n+j+1}} {a_{n+j} \, a_{n+j+2}} } \, .
\label{chitrafo}
\end{equation}
Not much has to be said about the $c$ and $\chi$ 
transformations since in our numerical studies we found 
that they appear to have very similar properties as compared to 
the $d$ and $\delta$ 
transformations [see Eqs.~(\ref{dLevTr}) and~(\ref{dWenTr}),
respectively]. Other recent developments with regard
to the construction of novel sequence transformations are
described in Refs.~\cite{CiZaSk2003,We2004}.
As possible further modifications of the nonlinear
sequence transformations, we should also mention 
multi-point methods (see, e.g., Ref.~\cite{RoBhBh1998}).

%
%
\subsubsection{Accuracy--Through--Order Relations}
\label{Accuracyrelations}

We would now like to insert a small
detour into the analysis of the sequence transformation,
by discussing so-called accuracy-through-order relations
fulfilled by the rational approximants.
On the basis of these relations,
predictions for unknown series coefficients can be made.

According to (\ref{padedef}), the difference 
between $f (z)$, defined according to Eq.~(\ref{DefP}),
and the Pad\'e approximant $[ l / m ]_f (z)$ is of
order $O \bigl( z^{l+m+1} \bigr)$ as $z \to 0$. 
This means that the
Taylor expansion of $[ l / m ]_f (z)$ around $z = 0$ reproduces
the first $l+m+1$ coefficients $\alpha_0$, $\alpha_1$, \ldots,
$\alpha_{l+m}$ of the perturbation series (\ref{DefP}). Consequently, the
Pad\'e approximant can be expressed as follows:
\begin{eqnarray}
[ l / m ]_f (z) \, = \, \sum_{k=0}^{l+m} \, \alpha_k \, z^k \, + \,
{\cal R}_{l+m+1} (z) \, = \, f_{m+l} (z) \, + \,
{\cal R}_{l+m+1} (z) \, ,
\label{ErrorPA_P}
\end{eqnarray}
where the truncation error term ${\cal R}_{l+m+1} (z)$ is of order 
${\rm O} \bigl( z^{l+m+1} \bigr)$ as $z \to 0$. By performing now 
a formal power series expansion of this error term around $z = 0$: 
\begin{equation}
{\cal R}_{l+m+1} (z) \; = \; z^{l+m+1} \, 
\{ \alpha_{l+m+1} + \alpha_{l+m+2} \,\, z +
\dots \} \, .
\end{equation}
we may find those coefficients of the
perturbation series (\ref{DefP}) that were not used for the construction
of the Pad\'e approximant $[ l / m ]_f (z)$.

Moreover, as it follows from the definition of the asymptotic
sequence in sense of Poincar\'e, the $f(z)$ can be expressed as the partial sum $f_{n+l}(z)$ of the 
perturbation series (\ref{DefP}) plus an error term that is also of 
order ${\rm O} \bigl( z^{l+m+1} \bigr)$ as $z \to \infty$:
\begin{equation}
\label{ErrorSer_P}
f (z) \; = \; \sum_{n=0}^{l+m} \alpha_n \, z^n\, + R_{l+m+1} (z) \; = \; 
f_{m+l} (z) \, + \, R_{l+m+1} (z) \, .
\end{equation}
Similarly to the Pad\'e approximant, the formal Taylor expansion of this error
term gives:
\begin{equation}
R_{l+m+1} (z) \; = \; z^{l+m+1} \, \{ \tilde{\alpha}_{l+m+1} +
\tilde{\alpha}_{l+m+2} \, z + \dots \} \, .
\end{equation}

Let us now assume that $[ l / m ]_f (z)$ approximates $f (z)$ for
sufficiently large values of $l$ and $m$ with
satisfactory accuracy,
\begin{equation} f (z) \; \approx \; 
[ l / m ]_f (z) \, .  
\label{ErrorParSumP} 
\end{equation} 
By making use of Eqs.~(\ref{ErrorPA_P}) and (\ref{ErrorSer_P}), this
expression implies that
\begin{equation}
R_{l+m+1} (z) \; \approx \; {\cal R}_{l+m+1} (z) \, ,
\label{ApproxErr}
\end{equation}
and, therefore:
\begin{equation}
\label{OTA_Pade}
\tilde{\alpha}_{l+m+1} \; \approx \; \alpha_{l+m+1} \, , \quad
\tilde{\alpha}_{l+m+2} \; \approx \; \alpha_{l+m+2} \, , \quad \ldots \, .
\end{equation}
As seen from these relations, at least the leading coefficients of the
expansion for ${\cal R}_{l+m+1} (z)$ should provide reasonably accurate
approximations to the analogous coefficients of the expansion for
$R_{l+m+1} (z)$.

The set of equations 
(\ref{OTA_Pade}) represents the order-through-accuracy
relations for the case of Pad\'e approximant. Similar relations can be obtained
for all transformations listed in Table~\ref{master}.
For the $d$ and $\delta$ transformations,
the final results read~\cite{WeCiVi1991}:      
\begin{eqnarray}
&&f(z)\;-\;  d_k^{(n)}(\beta,s_n) \;=\; {\rm O}(z^{k+n+2})\,, \nonumber\\
&&f(z)\;-\;  \delta_k^{(n)}(\beta,s_n) \;=\; {\rm O}(z^{k+n+2}) \;.
\end{eqnarray}

The following example illustrates the order-through-accuracy
relations.  First, we consider an alternating asymptotic series
\begin{equation}
{\cal A}(z) \; \sim \; \sum_{n=0}^\infty (-1)^n \, n! \, z^n\,,
\end{equation}
whose first terms read
\begin{equation}
\label{exact24}
{\cal A}(z) \; = \;
1 - z + 2 \, z^2 - 6 \, z^3 + 
\underbrace{24 \, z^4}_{\mbox{exact term}} + {\rm O}(z^5)\,.
\end{equation}
The second-order $\delta$ transformation is given by
\begin{equation}
\delta^{(0)}_2\;=\; \frac{1 + 3 z}{1 + 4 z + 2 z^2}\,.
\end{equation}
Upon re-expansion about $z=0$ we obtain
\begin{equation}
\label{pred20}
\delta^{(0)}_2 \;=\; 1 - z + 2 \, z^2 - 6 \, z^3 +
\underbrace{ 20 \, z^4}_{\mbox{by reexpansion}} + {\rm O}(z^5)\,.
\end{equation}
The series expansion for ${\cal A}(z)$ up to the order of $z^3$ is
reproduced ($k+1=3$), and the term of order $z^4$ has to be
interpreted as a prediction for the term in this order of perturbation
theory.  In higher order of transformation, the predictions become more
and more accurate.

We now consider the nonalternating series 
\begin{equation}
{\cal N}(z) \; \sim \; \sum_{n=0}^\infty n! \, z^n\,,
\end{equation}
which is generated by 
expansion of a certain integral, see Eq.~(\ref{defN}) below.
We attempt to illustrate that as long as the coupling is held variable, the
prediction does not depend on the sign pattern of the coefficients. 
Namely, we have
\begin{equation}
\label{exact24N}
{\cal N}(z) \; = \;
1 + z + 2 \, z^2 + 6 \, z^3 + 24 \, z^4 + {\rm O}(z^5)\,.
\end{equation}
The second-order $\delta$ transformation is given by
\begin{equation}
\delta^{(0)}_2\;=\; \frac{1 - 3 z}{1 - 4 z + 2 z^2}\,.
\end{equation}
Upon re-expansion about $z=0$ we obtain
\begin{equation}
\label{pred20N}
\delta^{(0)}_2 \;=\; 1 + z + 2 \, z^2 + 6 \, z^3 +
 20 \, z^4 + {\rm O}(z^5)\,,
\end{equation}
where again the coefficients of $z^4$ are seen to differ by $20\,\%$.
The prediction of perturbative coefficients may be possible even in
cases where the resummation of the series fails due to its
nonalternating character (see also Table 8.7 of Ref.~\cite{Je2003habil}).  
For nonalternating series, the
prediction of higher-order coefficients may be possible, but the
terms associated with these predictions cannot be used as remainder
estimates. These observations have been used 
for the prediction of higher-order perturbative
coefficients~\cite{ElStChMiSp1998,StEl1998,
ChElMiSt1999,ChElSt1999a,ChElSt1999b,ChElSt2000,ElChSt2000,%
SaLi1991,SaLiSt1993,SaLi1994c,SaLiSt1994,
SaLi1994a,SaLi1994b,SaElKa1995,Sa1995a,Sa1995b,
SaLiSt1995,ElGaKaSa1996,ElGaKaSa1996a,ElGaKaSa1996b,
Ga1997,ElKaSa1997,SaAbYu1997,JaJoSa1997,
ElJaJoKaSa1998,BuAbSaStMa1996,%
CvKo1998,Cv1998a,CvYu2000,JeWeSo2000,JeBeWeSo2000}.

%
%
\subsubsection{Combined Nonlinear--Condensation Transformations}
\label{Sec:CNCT}

Quite recently, the combined 
nonlinear-condensation transformation~\cite{JeMoSoWe1999}
(CNCT) has been proposed as a computational tool for the 
numerical evaluation of slowly convergent, nonalternating
series. The idea is to divide the acceleration process in 
two steps. The first step, which is a reordering process,
consists in a rearrangement of the terms
of the input series into an alternating series via 
a condensation transformation~\cite{vW1965}
(the condensation transformation was apparently
first mentioned on p.\ 126 of
Ref.~\cite{NP1961} and published only later~\cite{vW1965}).
In typical cases, the output of the first step is an alternating series
whose convergence is marginal (i.e. $\rho$ = -1).
It could appear that nothing substantial has been achieved in the first
step of the CNCT. The first step of the CNCT merely represents
a ``computational investment'' with the intention
of transforming the nonalternating input series into 
a form which is more amenable to the acceleration of convergence.
The second step, which represents a convergence acceleration process,
consists in the application of a powerful nonlinear sequence
transformation for the acceleration of the convergence of the 
alternating series which results from the first 
step of the CNCT.

Following Ref.~\cite{vW1965}, we transform the 
nonalternating input series [we have $a(k) \equiv a_k$]
\begin{equation}
\label{inputcnct}
\sum_{k=0}^{\infty} a(k)\,,
\quad a(k) \geq 0\,,
\end{equation}
whose partial sums are given by (\ref{basics}),
into an alternating series 
$\sum_{j=0}^{\infty} (-1)^j {\bf A}_j$.
After the first step of the transformation,
the limit of the input series is recovered according to
\begin{equation}
\label{VWSerTran}
\sum_{k=0}^{\infty} \, a (k) =
\sum_{j=0}^{\infty} \, (-1)^j \, {\bf A}_j \,.
\end{equation}
The quantities ${\bf A}_j$ are defined according to 
\begin{equation}
\label{A2B}
{\bf A}_j = \sum_{k=0}^{\infty} \, {\bf b}_{k}^{(j)} \, ,
\end{equation}
where
\begin{equation}
\label{B2a}
{\bf b}_{k}^{(j)} = 2^k \, a(2^k\,(j+1)-1) \, .
\end{equation}
Obviously, the terms ${\bf A}_j$ defined in Eq.~(\ref{A2B}) are all
positive if the terms $a (k)$ of the original series are all
positive. The ${\bf A}_j$ are referred
to as the condensed series~\cite{JeMoSoWe1999}, and the series
$\sum_{j=0}^{\infty} (-1)^j {\bf A}_j$ is referred to as the 
transformed alternating series, or alternatively as the Van
Wijngaarden transformed series.

The summation over $k$ in Eq.~(\ref{A2B}) dose not pose
numerical problems. Specifically, it can be easily 
shown in many cases of practical importance that the 
convergence of $\sum_{k=0}^{\infty} \, {\bf b}_{k}^{(j)}$ (in $k$)
is linear even if the convergence of $\sum_{k=0}^{\infty} a(k)$ is
only logarithmic. We will illustrate this statement
by way of two examples. As a first example, we consider a logarithmically
convergent input series whose terms behave asymptotically for 
$k \to \infty$ as
$a (k) \sim k^{-1 - \epsilon}$ with $\epsilon > 0$.
In this case, the partial sums
\begin{equation}
   \label{cond_ser_par}
   {\bf A}^{(n)}_j = \sum_{k=0}^{n} \, {\bf b}_{k}^{(j)}
\end{equation}
converge linearly with
\begin{equation}
\lim_{n \to \infty} \frac{{\bf A}^{(n+1)}_j - {\bf A}_j} 
  {{\bf A}^{(n)}_j - {\bf A}_j} \; = \; 
\frac{1}{2^\epsilon \, (j+1)^{1+\epsilon}} < 1 \,, \qquad
a(k) \sim k^{- 1 - \epsilon}\,, \quad k \to \infty.
\end{equation}
As a second example, we can give a series with $a (k) \sim k^{\beta} r^k$ 
where $0 < r < 1$ and $\beta$ real. Here, we have
$\rho = r < 1$, and the series
is (formally) linearly convergent. However, slow convergence may result 
if $\rho$ is close to one. In this case, the condensed series are
very rapidly convergent,  
\begin{equation}
\lim_{n \to \infty} \frac {{\bf A}^{(n+1)}_j - {\bf A}_j} 
  {{\bf A}^{(n)}_j - {\bf A}_j} \; = \; 0 \,, \qquad
a(k) \sim k^{\beta} r^k \,, \quad k \to \infty.
\end{equation}
Therefore, when summing over $k$ in evaluating the condensed series
according to Eq.~(\ref{A2B}), it is in many cases 
sufficient to sum the condensed series term by term,
and no further acceleration of the convergence is required.

As shown in~\cite{Da1969,JeMoSoWe1999}, the condensation transformation
defined according to Eqs.~(\ref{VWSerTran})--(\ref{B2a}) 
is essentially a reordering
of the terms of the input series $\sum_{k=0}^{\infty} a(k)$.
Furthermore, according to the Corollary on p.~92 of 
Ref.~\cite{Da1969}, for 
nonalternating convergent series whose terms decrease in magnitude
$\left( \vert a(k) \vert > \vert a(k+1) \vert \right)$, the equality 
(\ref{VWSerTran}) holds. 

Note that the property
\begin{equation}
{\bf A}_{2\,j-1} = \frac{1}{2} \,
\left[ {\bf A}_{j-1} - a(j-1) \right]\,, \qquad
j \in \mathbbm{N} \,,
\end{equation}
derived in~\cite{Da1969}, facilitates the numerical evaluation of a 
set of condensed series, by reducing the evaluation of condensed series 
of odd index to a trivial computation.

In the second step of the CNCT, the convergence of the series
which results from the condensation transformation,
\begin{equation}
\sum_{j=0}^{\infty} \, (-1)^j \, {\bf A}_j\,,
\end{equation}
and which on the right-hand side of Eq.~(\ref{VWSerTran}),
is accelerated by a suitable nonlinear sequence transformation,
e.g., by the $\delta$ transformation (\ref{dWenTr}).
The remainder estimates are constructed according to
\begin{equation}
s_n \to {\bf S}_n \,, \qquad 
\omega_n = \Delta s_n \to \Delta {\bf S}_n = (-1)^{n+1} \, {\bf A}_{n+1}\,.
\end{equation}
By making---if necessary---an appropriate shift of the 
index of the partial sums, it is possible to always
assume that ${\bf S}_0$ is the initial element of the 
second transformation. This leads to the CNC transforms
\begin{equation}
   \label{defcnc}
   {\cal T}_{\rm CNC}(n) = \delta_{n}^{(0)} \left( 1, {\bf S}_0 \right)\,.
\end{equation}
These require as input the elements
$\{ {\bf S}_0, \dots, {\bf S}_{n}, {\bf S}_{n+1} \}$
of the condensation transformed series. 

We here also indicate that according to Table~\ref{master},
it is possible to immediately generalize the CNC-type transformations
to the the CNC-$u$, CNC-$y$ and the 
CNC-$v$, CNC-$\varphi$ transformations.
The CNC-$d$ and CNC-$\delta$ have been studied in 
Refs.~\cite{PeKi1998,JeMoSoWe1999,JeAkMoSaSo2003}.

{\color{black}
Apart from the CNCT, as discussed here, a number of other summation
techniques are known which also make use of the division of the acceleration
process into two steps. For instance, the summation of logarithmic 
sequences can sometimes be accelerated by applying an
``extraction procedure.''
The main idea of this procedure is to extract from a given logarithmic sequence
a subsequence converging linearly (on the level of 
the partial sums $s_n$, not on the level of the series terms
$a_k$), and then to accelerate the sequence $\{ s_n \}_{n=0}^\infty$
by accelerating the extracted, linearly convergent subsequence
$\{ s_{{\cal N}(n)} \}_{n=0}^\infty$, where the 
${\cal N}(n)$ are the indices of the elements of the subsequence.
A detailed discussion
of this method is given, for example in 
Refs.~\cite{BrRZ1991,BrDeGB1983}.
In particular, it has been
proven that the acceleration of the extracted subsequence implies the 
acceleration of the initial (monotone logarithmic) sequence. 
Based on this result,
a few algorithms have been proposed for the extraction of (linearly 
convergent) subsequences, i.e.~for the 
determination of the ${\cal N}(n)$. 
Here, we present one of those algorithms whose aim is to find 
a linearly convergent subsequence where the parameter $\rho$
is understood as defined in Eq.~(\ref{convergence_classification}):
\begin{enumerate}
\item Let us choose 0$< \rho <$ 1, and set ${\cal N}(0) = 0$, $n  = 0$.
\item Increase $n \to n + 1$, and set 
${\cal N}(n+1) = {\cal N}(n) + 1$, then go to step 4. 
\item Increase ${\cal N}(n) \to {\cal N}(n) + 1$, then go to step 4.
\item If $|s_{{\cal N}(n)+1}-s_{{\cal N}(n)}| \leq 
\lambda^n |s_1 - s_0|$,
go to step 2, otherwise go to step 3.     
\end{enumerate}
For this algorithm, if applied to a monotone logarithmic sequence
with $\lim_{n \to \infty} s_n = s$ and 
$\lim_{n \to \infty} \Delta s_{n+1}/\Delta s_n = 1$,
the resulting subsequence is found to satisfy \cite{BrRZ1991}:
\begin{equation}
\label{relation1}
\lim_{n \to \infty} 
\frac{s_{{\cal N}(n+1)+1}-s_{{\cal N}(n+1)}}
{s_{{\cal N}(n)+1}-s_{{\cal N}(n)}}
= \lambda \, .
\end{equation}
Moreover, 
although a general proof has not been given in the general case,
for many examples considered in \cite{SmFo1979} the following relation,
which is a little stronger than (\ref{relation1}), was obtained:
\begin{equation}
\lim_{n \to \infty} \frac{s_{{\cal N}(n+1)}-s}{s_{{\cal N}(n)}-s}
= \lambda < 1 \, ,
\end{equation}
which immediately implies the the extracted subsequence is linearly
convergent. To accelerate the convergence of the extracted 
sequence, various standard algorithms as e.g.~Aitken's $\Delta^2$ process 
can be applied.
}

%
%
%
%
%

%
%
\subsubsection{General $E$ Algorithm}
\label{ThetaAlgorithm}

A beautiful mathematical construction which deserves a particular
mention is that of the $E$ algorithm~\cite{Br1980E,Ha1979E}. 
Rather than a traditional sequence
transformation, the $E$ algorithm is a template acceleration algorithm built
in such a way to be exact for all sequences of the form
\begin{equation}
\label{E-kernel}
s_n = s + a_1 \, g_1 \left( n \right) + \ldots +  
a_k \, g_k \left( n \right)\,,
\end{equation}
where the $g_i$'s are given auxiliary sequences: with different choices for
the $g_i$'s, different acceleration algorithms are obtained. In particular,
many of the best known sequence transformations can be recovered as
particular cases of the $E$ algorithm. The purpose of the $E$ algorithm
is to successively eliminate the $g_i$ terms in order to recover the 
(anti-)limit $s$ from the $s_n$.

It is interesting to examine in some detail the derivation of the
$E$ algorithm, since the employed method is quite general, and can be used
by the reader to derive new sequence transformations adapted to specific
problems. A more comprehensive description can
be found for example in Refs.~\cite{Br1980E,BrRZ1991}.
The first step is to lay out the
elements of the approximation table for the
transformed sequence in the following standard arrangement
\begin{equation}
  \begin{array}{l@{\hspace{0.5cm}}l@{\hspace{0.5cm}}l%
      l@{\hspace{0.5cm}}l@{\hspace{0.5cm}}l@{\hspace{0.5cm}}l}
     E_{- 1}^{(0)} = 0 &  &  &  &  &  & \\
     & E_0^{(0)} = S_0 &  &  &  &  & \\
     E_{- 1}^{(1)} = 0 &  & E_1^{(0)} &  &  &  & \\
     & E_0^{(1)} = S_1 &  & E_2^{(0)} &  &  & \\
     E_{- 1}^{(2)} = 0 &  & E_1^{(1)} &  & E_3^{(0)} &  & \\
     & E_0^{(2)} = S_2 &  & E_2^{(1)} &  & \ddots & \\
     E_{- 1}^{(3)} = 0 &  & E_1^{(2)} &  & E_3^{(1)} &  & \\
     \vdots & E_0^{(3)} = S_3 &  & E_2^{(2)} &  & \ddots & \\
     \vdots & \vdots & E_1^{(3)} &  & E_3^{(2)} &  & \\
     \vdots & \vdots & \vdots & E_2^{(3)} &  & \ddots & \\
     \vdots & \vdots & \vdots & \vdots & E_3^{(3)} &  & \\
     \vdots & \vdots & \vdots & \vdots & \vdots & \ddots & \,.
   \end{array}
   \label{ExtrapTable}
\end{equation}
If we write the exactness requirement of Eq.~(\ref{E-kernel}) for the
quantities $s_n, \ldots, s_{n + k}$ and solve for the unknown $s \equiv
E_k^{(n)}$, we get for the elements $E_k^{(n)}$ of the $k$th column of the
approximation table the form
\begin{equation}
  E_k^{(n)} = \frac{\left|\begin{array}{ccc}
    s_n & \ldots & s_{n + k}\\
    g_1 \left( n \right) & \ldots & g_1 \left( n + k \right)\\
    \vdots &  & \vdots\\
    g_k \left( n \right) & \ldots & g_k \left( n + k \right)
  \end{array}\right|}{\left|\begin{array}{ccc}
    1 & \ldots & 1\\
    g_1 \left( n \right) & \ldots & g_1 \left( n + k \right)\\
    \vdots &  & \vdots\\
    g_k \left( n \right) & \ldots & g_k \left( n + k \right)
  \end{array}\right|} \label{E algorithm} .
\end{equation}
Then, making use of standard determinant relations it can be shown that the
actual computation of the determinants of Eq.~(\ref{E algorithm}) is not
really necessary (this finding is in complete analogy to the case of the
derivation of the $\varepsilon$-algorithm from the Shanks process); in turn,
introducing the auxiliary quantities
\begin{equation}
 g_{k, i}^{\left( n \right)} = \frac{\left|\begin{array}{ccc}
     g_i \left( n \right) & \ldots & g_i \left( n + k \right)\\
     g_1 \left( n \right) & \ldots & g_1 \left( n + k \right)\\
     \vdots &  & \vdots\\
     g_k \left( n \right) & \ldots & g_k \left( n + k \right)
   \end{array}\right|}{\left|\begin{array}{ccc}
     1 & \ldots & 1\\
     g_1 \left( n \right) & \ldots & g_1 \left( n + k \right)\\
     \vdots &  & \vdots\\
     g_k \left( n \right) & \ldots & g_k \left( n + k \right)
   \end{array}\right|}\,, 
\end{equation}
defined themselves in terms of ratios of determinants, a recursive evaluation
scheme for both the elements of the table and the $g_{k, i}^{\left( n \right)}$
can be derived, with the rules
\begin{subequations}
\begin{eqnarray}
  E_0^{\left( n \right)} & = & s_n \, , \, \quad n \in \mathbb{N}_0 \, , \\[2ex]
  g_{0, i}^{\left( n \right)} & = & g_i ( n ) \, , \quad 
    i \in \mathbb{N} \,, \\[2ex]
  E_k^{\left( n \right)} & = & E_{k - 1}^{\left( n \right)} 
    - \frac{E_{k - 1}^{\left( n + 1 \right)} 
    - E_{k - 1}^{\left( n \right)}}{g_{k - 1, k}^{\left( n + 1 \right)} 
    - g_{k - 1, k}^{\left( n \right)}} \, g_{k - 1, k}^{\left( n \right)} 
    \,, \\[2ex]
  g_{k, i}^{\left( n \right)} & = & g_{k - 1, i}^{\left( n \right)} 
    - \frac{g_{k - 1, i}^{\left( n + 1 \right)} 
    - g_{k - 1, i}^{\left( n \right)}}{g_{k - 1, k}^{\left( n + 1 \right)} 
    - g_{k - 1, k}^{\left( n \right)}} \,
    g_{k - 1, k}^{\left( n \right)} \, \quad 
    i \in \mathbb{N}, \, \, i \ge k + 1 \,.
\end{eqnarray}
\end{subequations}
Using this set of recursive rules, the approximation table can be immediately
computed by standard techniques once a 
prescription for the auxiliary sequences $g_i(n)$ of Eq.~(\ref{E-kernel})
is given. 

In fact, many extrapolation schemes can be derived as a special case of the
$E$-algorithm: among them we find (with 
$\left\{ x_n \right\}_{n=0}^\infty$ and 
$\left\{ y_n \right\}_{n=0}^\infty$ being
arbitrary sequences which specify some interpolation scheme):
\begin{enumerate}
\item The Richardson extrapolation process, corresponding to the choice 
     $g_i(n) = x^i_n$, leads to the Neville algorithm 
   [see Eq.~(3.1.3) of Ref.~\cite{PrFlTeVe1993} and 
    Eq.~(6.1-5) of Ref.~\cite{We1989}].
\item The $G$ transformation corresponds to the choice 
  $g_i \left( n \right) = x_{n + i - 1}$
  [see p.~95 of \cite{BrRZ1991} and Ref.~\cite{GrAt1968}].
\item The $\epsilon$ algorithm corresponds to
  the choice $g_i(n) = \Delta s_{n + i - 1}$.
\item Certain generalization of the $\epsilon$-algorithm correspond
  to the choice $g_i \left( n \right) = R^i \left( s_n
  \Delta x_n \right) / \Delta x_n$, being $R$ a difference operator
  generalizing $\Delta$ and defined by $R^{k + 1} \left( s_n \Delta x_n
  \right) = \Delta \left( R^k \left( s_n \Delta x_n \right) / \Delta x_n
  \right)$
  (see page 108 of \cite{BrRZ1991}).
\item The $\rho$-algorithm corresponds to 
  the choice $g_{2 i - 1}(n) = x_n^{- i} s_n$, 
  $g_{2 i} \left( n \right) = x_n^{- i}, i = 1, \ldots, k$.
\item The Germain--Bonne transformation~\cite{GB1978} entails to the choice 
  $g_i(n) = \left( \Delta s_n \right)^i$.
\item The $p$-process is given by 
  $g_1 ( n ) = x_n$ and 
  $g_i ( n ) = \Delta s_{n + i - 2}$ ($i \in \mathbb{N}$,  $i \geq 2$)
  (see Ref.~\cite{Br1977}).
\item Levin's generalized transformation corresponds to the choice $g_i
  \left( n \right) = x_n^{i - 1} \Delta s_n / y_n$
  (see page 113 of \cite{BrRZ1991}).
\end{enumerate}

{\color{black}
One of the fascinating properties of the $E$ algorithm is the fact that it 
can be used to accelerate the convergence of the sequences for which an 
asymptotic expansion of the error is known. That is, assuming that 
the error $r_n = s_n - s = a_1 \, g_1 \left( n \right) + 
a_2 \, g_2 \left( n \right) + \ldots $ has an asymptotic expansion with 
respect to the asymptotic sequence $\left(g_1, g_2, \ldots \right)$, i.e.
$g_{i+1}(n) = o \left( g_i(n) \right)$ for $i \in \mathbb{N}_0$, then 
theorems exist (see~pp. 68--69 of \cite{BrRZ1991} and \cite{Br1980E}) showing 
that the $E$ algorithm is indeed accelerative for the sequence. Moreover,
it can be shown for the columns of the extrapolation table 
(\ref{ExtrapTable}) that $E_k^{(n)}$ converges to $s$ faster than 
$E_{k-1}^{(n)}$: $E_k^{(n)} - s = o \left( E_{k-1}^{(n)} - s \right)$. 
The practical importance of this fact should not be underestimated,
since there are a number of methods known which may help to obtain
the asymptotic expansions for the error. Again, some of these methods
are discussed in detail in Refs.~\cite{Br1980E,BrRZ1991}.
}

However, the generality of the approach comes at a price,
in view of (i) the difficulty of deriving
general theorems about the convergence properties of such an acceleration
framework and (ii) an enhanced sensitivity to numerical problems, due to the
augmented complication of the iterative scheme (in fact, some of these problems
can be avoided in less general, more specialized and more computationally
efficient algorithms which may be derived for particular cases).

%
%
\subsection{Applications of Sequence Transformations to 
Convergence Acceleration}
\label{ConvAccelAppl}

%
%
\subsubsection{Simple Case: Series arising in plate contact problems}
\label{coneasy}

This review would be incomplete without the discussion of 
some actual applications of the algorithms discussed in 
Secs.~\ref{PadeApproximation} and 
\ref{NonlinearST}. The review~\cite{We1989}
contains a few very useful practical applications with 
numerical example. Also, we should mention that
more numerical examples have already been given, e.g.,
in Refs.~\cite{JeMoSoWe1999,JeAkMoSaSo2003,PeKi2002}.
Here, we discuss a few other examples 
in depth. In the current Section,
we start with some easy cases in 
Sec.~\ref{coneasy}. In some sense, the examples
discussed here complement Refs.~\cite{JeMoSoWe1999} and
\cite{JeAkMoSaSo2003} by considering a few cases where 
a direct application of the accelerators is possible in the 
case of nonalternating series.  
Specifically,  we first consider a very simple
example which occurs naturally in the context of 
so--called ``plate contact problems''. 
Such problems are well known in mechanics and in engineering applications and
concern the contact between the plates and unilateral supports. While, of 
course, any discussion on the physics of the ``plate contact problems'' 
is out of scope of the present report, we just mention that the mathematical
analysis of these problems may be traced back to the evaluation of 
special infinite series~\cite{DeKePaGl1984}, two of which 
have the following structure:   
\begin{align}
\label{RpSer} 
R_p(x) =& \; \sum_{k=0}^\infty \; \frac{x^{2k+1}}
   {{(2k+1)}^p} \,,
\\[2ex]
\label{TpSer} 
T_p(x,b) =& \; \sum_{k=0}^\infty \; \frac{1}
   {{(2k+1)}^p} \frac{ \cosh[(2k+1)x]}{\cosh[(2k+1)b]}  \,.
\end{align}
The practical use of these series, however, is very embarrassing
because of their slow convergence for $x \approx 1$ (for Eq.~(\ref{RpSer})) 
or $x \approx b$ (for Eq.~(\ref{TpSer})). In order to
overcome such a slow convergence, a number of different techniques
have been proposed during the last decade
\cite{Ma1997,BaGa1995,BoDe1992,Ga1991,DeLiDe1990}. 
In Ref.~\cite{AkSaJeBeSoMo2003,Be1999master}, for example, it was 
shown that the combined nonlinear--condensation transformation 
(CNCT) as discussed in subsection \ref{Sec:CNCT} of this report is able 
to efficiently accelerate the convergence
of the series (\ref{RpSer})--(\ref{TpSer}) in problematic parameter
regions. In order to illustrate this property,
we display in Table \ref{TableR2} the 
calculation of the function $R_p(x)$ for the parameter $p = 2$ and
the argument $x = 1$. For this set of parameters, the 
series (\ref{RpSer}) are summed by means of the ``straightforward''
term--by--term summation (second column) as well as of the 
CNCT method (third column). As seen from these calculations, 
the CNCT approach allows us to reproduce the ``exact'' value
of $R_2(1)$ with accuracy up to 19 decimal digits for a
transformation of order $n = 16$. In contrast, the partial sums 
$R^{(n)}_2(1)$ of 16th order may predict only the two first 
digits of $R_2(1)$. In analogy to Table~\ref{TableR2}
for $R_p(x)$, we also present Table \ref{TableT3} for
a typical evaluation of $T_p(x,b)$ in a region of slow
convergence of the series representation (\ref{TpSer}).

\begin{table}[htb]
\begin{center}
\begin{minipage}{12cm}
\begin{center}
\caption{Acceleration of the convergence
of $R_2(1)$ defined in Eq.~(\ref{RpSer}). In the first row, we 
indicate the quantity $n$ which is the order of the 
partial sum or alternatively the order of the transformation.
In the second row, the partial sums 
$R^{(n)}_p(x) = \sum_{k=0}^{n} x^{2k+1}/(2k+1)^p$ of the input series
are given while in the third row we indicate the CNCT
transforms (\ref{defcnc}).}
\label{TableR2}
\begin{tabular}{ccc}
\hline
\hline
\rule[-3mm]{0mm}{8mm}
n & $R^{(n)}_2(1)$ & ${\cal T}_{\rm CNC}(n)$ \\
\hline
$1  $ &  $ \underline{1}.111~111~111~111~111~111$
&        $ \underline{1.23}7~516~275~551~033~976$ \\
$2  $ &  $ \underline{1}.151~111~111~111~111~111$
&        $ \underline{1.233}~357~211~766~354~240$ \\
$3  $ &  $ \underline{1.}171~519~274~376~417~234$
&        $ \underline{1.233~70}1~899~765~011~829$ \\
$4  $ &  $ \underline{1.}183~864~953~388~762~913$
&        $ \underline{1.233~700}~970~001~046~762$ \\
$\dots$& $ \dots$ & $\dots$ \\
$13  $ & $ \underline{1.21}5~850~986~098~258~088$
&        $ \underline{1.233~700~550~136~169~82}0$ \\
$14  $ & $ \underline{1.2}17~040~046~740~350~834$
&        $ \underline{1.233~700~550~136~169~82}6$ \\
$15  $ & $ \underline{1.2}18~080~629~466~677~578$
&        $ \underline{1.233~700~550~136~169~827}$ \\
$16  $ & $ \underline{1.2}18~998~903~112~223~950$
&        $ \underline{1.233~700~550~136~169~827}$ \\
$17  $ & $ \underline{1.2}19~815~229~642~836~195$
&        $ \underline{1.233~700~550~136~169~827}$ \\
\hline 
\rule[-3mm]{0mm}{8mm}
exact   & $1.233~700~550~136~169~827$ & $1.233~700~550~136~169~827$ \\
\hline
\hline
\end{tabular}
\end{center}
\end{minipage}
\end{center}
\end{table}

\begin{table}[htb]
\begin{center}
\begin{minipage}{16cm}
\begin{center}
\caption{Same as Table~\ref{TableR2} for $T_3(1,1.00000001)$.}
\label{TableT3}
\begin{tabular}{ccc}
\hline
\hline
\rule[-3mm]{0mm}{8mm}
n & $T^{(n)}_3(1,1.00000001)$ & ${\cal T}_{\rm CNC}(n)$ \\
\hline
$0$ & $\underline{}0.999~999~992~384~058~448~444~917~150~320$ & 
  $\underline{1}.088~812~811~207~735~161~561~740~524~147$  \\
$1$ & $\underline{1.0}37~037~028~315~479~108~834~496~735~448$ & 
  $\underline{1.05}2~523~835~933~937~503~694~672~931~682$  \\
$2$ & $\underline{1.04}5~037~027~915~515~437~125~826~851~875$ & 
  $\underline{1.051~7}38~584~158~953~997~485~015~841~356$  \\
$3$ & $\underline{1.04}7~952~479~606~477~882~793~195~708~660$ & 
  $\underline{1.051~8}00~520~252~052~349~990~078~405~570$  \\
$4$ & $\underline{1.04}9~324~221~595~503~955~209~376~363~455$ & 
   $\underline{1.051~799}~867~127~602~653~337~384~365~118$  \\
$\dots$ & $\dots$ & $\dots$ \\
$22$ & $\underline{1.051}~681~688~751~729~032~310~253~771~189$ &   
  $\underline{1.051~799~780~317~229~791~448~360~99}2~429$  \\
$23$ & $\underline{1.051}~691~320~524~361~301~753~922~155~807$ &   
  $\underline{1.051~799~780~317~229~791~448~360~994}~831$  \\
$24$ & $\underline{1.051}~699~820~379~948~685~582~513~817~375$ & 
  $\underline{1.051~799~780~317~229~791~448~360~995}~124$  \\
$25$ & $\underline{1.051~7}07~358~954~780~387~793~955~705~424$ &   
  $\underline{1.051~799~780~317~229~791~448~360~995~16}0$  \\
$26$ & $\underline{1.051~7}14~075~905~484~661~391~959~259~950$ &   
  $\underline{1.051~799~780~317~229~791~448~360~995~16}4$  \\
$27$ & $\underline{1.051~7}20~086~420~586~089~799~171~715~372$ &   
  $\underline{1.051~799~780~317~229~791~448~360~995~165}$  \\
$28$ & $\underline{1.051~7}25~486~189~637~836~692~682~642~537$ &   
  $\underline{1.051~799~780~317~229~791~448~360~995~165}$  \\
\hline 
\rule[-3mm]{0mm}{8mm}
exact   & $1.051~799~780~317~229~791~448~360~995~165$ &
  $1.051~799~780~317~229~791~448~360~995~165$ \\
\hline
\hline
\end{tabular}
\end{center}
\end{minipage}
\end{center}
\end{table}

%
%
\subsubsection{Involved Case: Lerch Transcendent}
\label{concomp}

As a slightly more involved application of
convergence accelerators, we describe the basic strategy
for the evaluation of the Lerch transcendent~\cite{Ba1953vol1}
\begin{equation}
\label{lerch}
\Phi(x,s,v) = \sum_{k=0}^\infty \frac{x^k}{(k+v)^s}\,,  \qquad
|x|<1\,, \qquad v \neq 0,-1,\dots\, .
\end{equation}
This function naturally occurs in the evaluation 
of statistical distributions, which have application
in biophysics~\cite{JeAkMoSaSo2003}.
We assume $x$, $s$, and $v$ to be real rather than
complex.

For negative $x$, the series (\ref{lerch}) is alternating,
and its partial sums can be directly used as
input data for the delta transformation (\ref{dWenTr}).
I.e., for $x < 0$, we evaluate the transforms
\begin{equation}
\label{snLerch}
\delta^{(0)}_k(1, s_n) \,, \qquad
s_n =  \sum_{k=0}^n \frac{x^k}{(k+v)^s}\,.
\end{equation}
This approach corresponds to the
standard use of nonlinear sequence transformations as efficient
accelerators for alternating series~\cite{We1989}.

By contrast, for $x > 0$ the series (\ref{lerch}) is nonalternating,
and the direct use of the delta transformation is not recommended.
Additionally, one observes that the series (\ref{lerch}) is slowly
convergent for $x$ close to unity.
In this parameter region,
the use of the CNCT for positive $x$ and $x$ close to unity removes
the principal numerical difficulties.
Thus, for $x > 0$ close to unity, we evaluate the 
transforms ${\cal T}_{\rm CNC}(n)$ as defined
in Eq.~(\ref{defcnc}), on the basis of the input data
given by the partial sums $s_n$ as given in Eq.~(\ref{snLerch}).

For both positive and negative $|x| < 0.3 $, by contrast, it is 
computationally advantageous to directly sum the
terms in the defining power series (\ref{lerch}) term by term,
because of the extremely rapid ``direct'' convergence of
the power series in this domain.
For negative argument, the series is alternating, so one can 
use the $\delta$ transformation, whereas for positive argument, one
uses the CNC transformation. In order to illustrate the importance of the
correct choice of the acceleration algorithm, we display in Table
\ref{TableLerchPos} the results of the computations of the 
Lerch transcendent (\ref{lerch}) for the argument $x = 0.99$ and
the parameters $s = 1.1$ and $v = 0.1$. As seen from the Table, 
while both, the straightforward term--by--term summation and
the $\delta$ algorithm result in a slowly convergent
result, the 14th order of the CNC transformation 
allows us to reproduce the exact value of 
$\Phi(x = 0.99, s = 1.1 ,v = 0.1)$ with the accuracy of about
10$^{-14}$. In contrast, if one evaluates the 
Lerch transcendent for the negative argument $x = -0.99$ and
the same set of parameters $s$ and $v$, the $\delta$ transformation
leads to much faster convergence of the series (\ref{lerch}) if
compared to the CNCT method [cf. Table \ref{TableLerchNeg}]. 
These two simple examples, 
provides us with a first manifestation of the general wisdom that 
numerical acceleration algorithms have to be adapted to parameter
domains. 

\begin{figure}[t]
\begin{center}
\includegraphics[width=0.3\linewidth,angle=270]{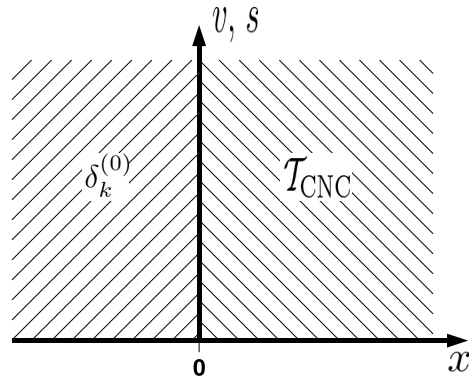}
\end{center}
\caption{\label{figL} Numerical algorithms used for the Lerch transcendent.}
\end{figure}

\begin{table}[htb]
\begin{center}
\begin{minipage}{14cm}
\begin{center}
\caption{Acceleration of the convergence of 
the Lerch transcendent $\Phi(x,s,v)$ defined in Eq.~(\ref{lerch}).
Results are presented for the set of parameters $x = 0.99$,
$s = 1.1$ and $v = 0.1$.  
In the first row, we indicate the quantity $n$ which is the order 
of the partial sum or alternatively the order of the transformation.
In the second row, the partial sums 
$s_n =  \sum_{k=0}^n x^k / (k+v)^s$ of the input series
are given while in the third and the fourth rows we indicate the 
transformation $\delta^{(0)}_n(1,s_0)$ and the CNCT transforms, 
respectively.}
\label{TableLerchPos}
\begin{tabular}{cccc}
\hline
\hline
\rule[-3mm]{0mm}{8mm}
n & $s_n$ & $\delta^{(0)}_n(1,s_0)$ & ${\cal T}_{\rm CNC}(n)$ \\
\hline
$1  $ &  $ \underline{1}3.480~716~950~334$
&        $ \underline{1}4.323~956~240~313$
&        $ \underline{16}.333~647~069~601$ \\
$2  $ &  $ \underline{1}3.914~057~335~485$
&        $ \underline{1}4.872~118~432~283$
&        $ \underline{16.27}4~855~206~560$ \\
$3  $ &  $ \underline{1}4.193~574~009~886$
&        $ \underline{1}5.328~655~340~349$
&        $ \underline{16.279~4}39~130~003$ \\
$\dots$& $ \dots$ & $\dots$ \\
$9  $ & $ \underline{1}4.965~099~286~381$
&        $ \underline{16}.183~960~654~046$
&        $ \underline{16.279~415~474}~265$ \\
$10  $ & $ \underline{1}5.036~154~906~635$
&        $ \underline{16.2}14~534~170~121$
&        $ \underline{16.279~415~474}~379$ \\
$11  $ & $ \underline{1}5.099~561~126~734$
&        $ \underline{16.2}35~397~170~154$
&        $ \underline{16.279~415~474~45}7$ \\
$12  $ & $ \underline{1}5.156~650~914~411$
&        $ \underline{16.2}49~604~756~607$
&        $ \underline{16.279~415~474~45}7$ \\
$13  $ & $ \underline{1}5.208~442~487~775$
&        $ \underline{16.2}59~256~744~711$
&        $ \underline{16.279~415~474~453}$ \\
$14  $ & $ \underline{1}5.255~730~569~885$
&        $ \underline{16.2}65~802~155~733$
&        $ \underline{16.279~415~474~453}$ \\
\hline 
\rule[-3mm]{0mm}{8mm}
exact   & $16.279~415~474~453$ & $16.279~415~474~453$ & $16.279~415~474~453$\\
\hline
\hline
\end{tabular}
\end{center}
\end{minipage}
\end{center}
\end{table}
\begin{table}[htb]
\begin{center}
\begin{minipage}{14cm}
\begin{center}
\caption{Same as Table \ref{TableLerchPos} for $x = -0.99$,
$s = 1.1$ and $v = 0.1$.}
\label{TableLerchNeg}
\begin{tabular}{cccc}
\hline
\hline
\rule[-3mm]{0mm}{8mm}
n & $s_n$ & $\delta^{(0)}_n(1,s_0)$ & ${\cal T}_{\rm CNC}(n)$ \\
\hline
$1  $ &  $ \underline{11}.697~791~285~548$
&        $ \underline{11.9}89~386~929~355$
&        $ \underline{ }9.312~621~614~169$ \\
$2  $ &  $ \underline{1}2.131~131~670~698$
&        $ \underline{11.96}5~797~543~541$
&        $ \underline{ }9.521~513~807~481$ \\
$3  $ &  $ \underline{11}.851~614~996~297$
&        $ \underline{11.967~0}38~897~324$
&        $ \underline{ }9.812~306~769~207$ \\
$\dots$& $ \dots$ & $\dots$ \\
$8   $ & $ \underline{1}2.009~956~315~957$
&        $ \underline{11.967~090~786~6}54$
&        $ \underline{1}0.403~057~902~502$ \\
$9   $ & $ \underline{11.9}29~460~886~859$
&        $ \underline{11.967~090~786~6}22$
&        $ \underline{1}0.479~117~052~492$ \\
$10  $ & $ \underline{1}2.000~516~507~113$
&        $ \underline{11.967~090~786~619}$
&        $ \underline{1}0.547~006~982~924$ \\
$11  $ & $ \underline{11.9}37~110~287~014$
&        $ \underline{11.967~090~786~619}$
&        $ \underline{1}0.608~315~040~497$ \\
\hline 
\rule[-3mm]{0mm}{8mm}
exact   & $11.967~090~786~619$ & $11.967~090~786~619$ & $11.967~090~786~619$\\
\hline
\hline
\end{tabular}
\end{center}
\end{minipage}
\end{center}
\end{table}
%
%

%
%
\subsection{Applications of Sequence Transformations to Resummations}
\label{Chap:DivSeqTran}

%
%
\subsubsection{Simple Case: Incomplete Gamma Function}
\label{diveasy}

We follow the same strategy as in Sec.~\ref{ConvAccelAppl}.
As a simple example of a resummation algorithm
(see Sec.~\ref{diveasy}), we discuss a possible
implementation of the incomplete Gamma Function
$\Gamma(0, x)$ for positive argument $x$,
based on the resummation of the divergent large-$x$
asymptotic expansion, where the
series is strictly alternating. Of course, the
upshot will be in demonstrating that the resummation algorithm
permits us to use the divergent large-$x$ asymptotic expansion
also at small positive $x$ with good effect.
A much more involved case, which demands an adaptive
algorithm, is the fully relativistic
hydrogen Green function, as discussed in Sec.~\ref{divcomp}.

We start our discussion on the application of nonlinear sequence 
transformations to the resummation of the divergent series from the
computationof the incomplete Gamma function given by 
\cite{AbSt1972}
\begin{equation}
   \label{IGDefGen}
   \Gamma(\alpha,z) \;=\; \int_z^\infty \; {\rm d}t
   \; e^{-t} \; t^{\alpha-1} \;. 
\end{equation}
For the sake of simplicity, below we restrict our analysis to
the special case of $\alpha=0$ and of positive, real argument:
$z \in \mathbbm{R}$, $z > 0$. For such $z$, the incomplete Gamma 
function reads as
\begin{equation}
   \label{IG_alpha_0}
   \Gamma(0,z) \;=\; \int_z^\infty \; {\rm d}t \; \frac{e^{-t}} {t} 
   \;=\; - {\rm Ei} (-z) \; ,
\end{equation}
where ${\rm Ei} (z)$ is the standard exponential integral.

Usually, the evaluation the incomplete Gamma function (\ref{IG_alpha_0}) is 
traced back to the computation of its asymptotic expansion. The
explicit form of such an expansion is however different for the cases of 
small and large arguments $z$. For example, while for large, positive
$z$ the incomplete Gamma function $\Gamma(0,z)$ has the 
asymptotic expansion
\begin{equation}
   \label{IGAsy}
   \Gamma(0,z) \; = \; \frac{e^{-z}}{z} \;
   \sum_{n=0}^\infty \; (-1)^n \; n! \; \left( \frac{1}{z} \right)^n \, ,
   \qquad z \to \infty\; , 
\end{equation}
a completely different series arises for $z \to 0$
\begin{equation}
   \label{IGLow}
   \Gamma(0,z) \;=\; \ln z - \gamma + z - \frac{z^2}{4}
   + \frac{z^3}{18} + {\rm O}(z^4)\, ,
   \qquad z \to 0 \; ,
\end{equation}
where $\gamma = 0.577216\dots$ is Euler's constant. It is clear that 
(\ref{IGLow}) has numerical advantages over (\ref{IGAsy}) for small $z$, 
whereas for large $z$, one would certainly refrain from using (\ref{IGLow}). 

As seen from Eqs.~(\ref{IGAsy}) and (\ref{IGLow}), asymptotic
expansions can be utilized to compute the Gamma function $\Gamma(0,z)$ in
the regions of the large ($z \to \infty$) and small ($z \to 0$) positive
arguments $z$. The problem arises, however, for the intermediate values 
of $z$. In order to resum the divergent series for $z \to \infty$ 
and, hence, to compute
the incomplete Gamma function for such (intermediate) $z$ region we may
apply the $\delta$ transformation. That
is, by using the partial sums of the alternating divergent 
series (\ref{IGAsy})
\begin{equation}
   \label{IGAsy_partial}
   s_n \; = \; \frac{e^{-z}}{z} \;
   \sum_{k=0}^n \; (-1)^k \; k! \; \left( \frac{1}{z} \right)^k \, ,
\end{equation}
as input data for Eq.~(\ref{dWenTr}), it is possible to recover 
the actual value of the incomplete Gamma function to high 
accuracy, even at relatively small $z$. In Table~\ref{TableGamma1},
for example, the resummation of the asymptotic expansion 
(\ref{IGAsy}), which is \textit{a priori} valid only for large $z$, is
performed for the case $z = 1$. As seen from this Table, while the
term--by--term summation of Eq.~(\ref{IGAsy}) fails to compute 
the Gamma function $\Gamma(0, z = 1)$, the 26th-order transformation
$\delta_{26}^{(n = 0)}(1,s_0)$ 
reproduces this values up to 16 decimal digits.   

\begin{table}[htb!]
\begin{center}
\begin{minipage}{12.0cm}
\caption{Resummation of the large-$z$ asymptotic series 
for the function $\Gamma(0, z) = \exp(-z) /z \, \sum^\infty_{k=0} k!/(-z)^k$ 
at argument $z=1$ [see Eq.~(\ref{IGAsy})].
In the first row, we indicate the quantity
$n$ which is the order of the 
partial sum or alternatively the order of the transformation.
In the second row, we list the partial sums 
$s_n = \exp(-z) /z  \sum^n_{k=0} k!/(-z)^k$ [see Eq.~(\ref{IGAsy_partial})],
whereas in the third row, the $\delta$ transforms
are displayed. These are defined in (\ref{dWenTr}).}
\label{TableGamma1}
\begin{center}
\begin{tabular}{crr}
\hline
\hline
\multicolumn{1}{c}{\rule[-3mm]{0mm}{8mm} $n$}      &       \multicolumn{1}{c}{$s_n$} &
\multicolumn{1}{c}{$\delta^{(0)}_n(1,s_0)$ } \\ 
\hline
 $  1    $    &     $
 0.000~000~000~000~000\times 10^0  $    &     $ \underline{ 
 0.2}45~252~960~780~961   $ \\

 $  2    $    &     $
 0.735~758~882~342~884\times 10^0 $     &   $ \underline{
 0.21}0~216~823~526~538   $ \\

 $  3   $    &     $
-1.471~517~764~685~769\times 10^0$    &     $ \underline{
 0.2}20~040~039~579~180 $ \\

 $  4    $    &     $
 7.357~588~823~428~846\times 10^0  $     &     $ \underline{}
 0.219~508~174~842~629$ \\
 $\dots$    &     $\dots$     &     $ \dots$ \\
$22$ &$
 3.955~539~479~532~130 \times 10^{20}$ & $\underline{  
 0.219~383~934~395~5}18$\\

$23$ &$
-9.114~871~523~102~565 \times 10^{21}$ & $\underline{  
 0.219~383~934~395~5}19$\\

$24$ &$
 2.191~353~397~822~361 \times 10^{23}$ & $\underline{  
 0.219~383~934~395~520}$\\

$25$ &$
-5.487~119~942~851~230 \times 10^{24}$ & $\underline{   
 0.219~383~934~395~520}$\\

$26$ &$
 1.428~755~174~056~189 \times 10^{26}$ & $\underline{  
 0.219~383~934~395~520}$\\
\hline 
\rule[-3mm]{0mm}{8mm}
exact & $0.219~383~934~395~520$ & $0.219~383~934~395~520$ \\
\hline
\hline
\end{tabular}
\end{center}
\end{minipage}
\end{center}
\end{table}

%
%
\subsubsection{Involved Case: Relativistic Hydrogen Green Function}
\label{divcomp} 

In Sec.~\ref{diveasy}, we have demonstrated how nonlinear sequence
transformations may be used for the resummation of the divergent series 
arising from the expansion of the incomplete Gamma function. Below 
another example for the application of the nonlinear sequence 
transformations will be discussed, which concerns the evaluation of the 
relativistic Green function. This function is known to be
an essential ingredient in  (relativistic) atomic physics and quantum 
mechanics, because it allows us to
perform a summation over a complete spectrum of 
intermediate, virtual atomic states. Such a
summation is required, for example, for describing 
quantum electrodynamic radiative 
corrections in atoms 
(see, e.g., Refs.~\cite{JeMoSo1999,JeMoSo2001pra,JeMo2004pra,JeMo2005p}),
for the computation of hyperpolarizabilities and for the 
evaluation of two-- (and multi--)photon processes
(see, e.g., Ref.~\cite{SuKoFr2005}). 

The fully relativistic 
Green function of an electron bound in a hydrogenlike atom
is given in terms of the Dirac Hamiltonian 
$H_{\rm D} = \vec{\alpha} \cdot \vec{p} + \beta \, m - Z\alpha/|\vec{x}|$,
where the $\vec{\alpha}$ and $\beta$ are standard 
$4 \times 4$-matrices~\cite{Ro1961} in the Dirac 
representation, $\vec{p}$ is the momentum operator, $Z$ is the
nuclear charge number and $\alpha$ is the fine-structure
constant. We use units in which $\hbar = c = \epsilon_0 = 1$ and 
write the Dirac--Coulomb Green function as
\begin{equation}
G(\vec{x}_2,\vec{x}_1,z) = 
\left< \vec{x}_2 \left| \frac{1}{H_{\rm D} - z} \right| \vec{x}_1 \right>\,.
\end{equation}
Here, $z$ is an energy argument which is assumed to be 
manifestly complex
rather than real values in some calculations. 
A closed-form representation of $G(\vec{x}_2,\vec{x}_1,z)$
is not known. However, it is possible and convenient to 
separate radial and angular variables, and to write the Green 
function in the following $4 \times 4$-matrix 
representation \cite{Ro1961,SwDr1991b} 
for the fully relativistic calculations, 
\begin{equation}
\label{DefinitionOfDCGF}
G(\vec{x}_2,\vec{x}_1,z) = \sum\limits_{\kappa \mu}
\left(\begin{array}{cc} 
G^{11}_\kappa(x_2,x_1,z) \, \,
\chi_\kappa^\mu(\hat{x}_1) \, 
\chi_\kappa^{\mu \dag}(\hat{x}_2) &
-{\rm i} \, G^{12}_\kappa(x_2,x_1,z) \, \,
\chi_\kappa^\mu(\hat{x}_1) \,
\chi_{-\kappa}^{\mu \dag}(\hat{x}_2) \\[2ex]
{\rm i}\,G^{21}_\kappa(x_2,x_1,z) \, \,
\chi_{-\kappa}^\mu(\hat{x}_1) \,
\chi_\kappa^{\mu \dag}(\hat{x}_2) &
G^{22}_\kappa(x_2,x_1,z) \,\,
\chi_{-\kappa}^\mu(\hat{x}_1) \,
\chi_{-\kappa}^{\mu \dag}(\hat{x}_2)
\end{array}\right)\,.
\end{equation}
Here, the radial variables are $x_i = |\vec{x}_i|$, and the unit
vectors are $\hat{x}_i = \vec{x}_i/|\vec{x}_i|$. The symbol
$\chi_\kappa^\mu(\hat{x})$ denotes the standard two-component
Dirac spinor, as given e.g. in Refs.~\cite{EiMe1995,VaMoKh1988}, 
\begin{align}
\chi_\kappa^\mu(\hat{x}) =& \sum_m C^{j\mu}_{l \mu-m \, \frac{1}{2} m}
Y_{l \mu-m}(\hat{x}) \, \chi_m \,, \quad
\chi_{\frac12} = \left( \begin{array}{c} 1 \\ 0 \end{array} \right) \,, \quad
\chi_{-\frac12} = \left( \begin{array}{c} 0 \\ 1 \end{array} \right) \,,
\nonumber\\[2ex]
l =& \left| \kappa + \frac12 \right| - \frac12 \,, \quad
j = |\kappa| - \frac12 \,, \quad
\kappa = (-1)^{j+l+1/2} \, \left(j + \frac12 \right) \,, \quad
\end{align}
and the $C^{j\mu}_{l \mu-m \, \frac{1}{2} m}$ are Clebsch--Gordan 
coefficients in the conventions of Ref.~\cite{VaMoKh1988}.
In Eq.~(\ref{DefinitionOfDCGF}), 
the expressions of the form $\chi_\kappa^\mu(\hat{x}_1) \, 
\chi_\kappa^{\mu \dag}(\hat{x}_2)$ are to be understood as
dyadic products, i.e.~as defining $2 \times 2$ matrices.
Of course, $\kappa$ has the meaning of a Dirac angular quantum number.
The quantities $l$ and $j$ are the orbital and the total 
angular momentum quantum numbers, respectively,
The components of the radial Green function are given in terms of the 
four quantities $G^{ab}_\kappa(x_1,x_2,z)$ with $a, b = 1, 2$,
and these functions have been historically problematic as far as 
their numerical evaluation is concerned. 

Indeed, as seen from Eq.~(\ref{DefinitionOfDCGF}), 
the evaluation of matrix elements involving the relativistic
hydrogen Green function entails both angular and radial
parts. In many cases, the angular part of the Green function can be evaluated
using standard, yet sometimes cumbersome angular momentum algebra, 
while the computation of the radial
functions $G^{ab}_\kappa(x_2, x_1, z)$ is known to be a 
very involved task. An explicit representation of 
these four components is given by [see, e.g., 
Eq.~(A.16) in \cite{Mo1974a} and
Eqs.~(101), (108) and (109) in~\cite{MoPlSo1998}]:
\begin{subequations}
\label{defDCGF}
\begin{eqnarray}
G^{11}_{\kappa}(x_2,x_1,z) &=& Q\,(1+z)\,
A_-(\lambda, \nu, 2 c x_2) \,
B_+(\lambda, \nu, 2 c x_1) \,,\\[2ex]
G^{12}_{\kappa}(x_2,x_1,z) &=& Q\,c\,
A_-(\lambda, \nu, 2 c x_2) \,
B_-(\lambda, \nu, 2 c x_2) \,,\\[2ex]
G^{21}_{\kappa}(x_2,x_1,z) &=& Q\,c\,
A_+(\lambda, \nu, 2 c x_2) \,
B_+(\lambda, \nu, 2 c x_2) \,,\\[2ex]
G^{22}_{\kappa}(x_2,x_1,z) &=& Q\,(1-z)\,
A_+(\lambda, \nu, 2 c x_2) \,
B_-(\lambda, \nu, 2 c x_2) \,.
\end{eqnarray}
\end{subequations}
Here, the overall prefactor is 
\begin{equation}
Q = \frac{1}{4\,c^2\,x_2\,x_1 \, \sqrt{x_2\,x_1}}\,
\frac{\Gamma(\lambda-\nu)}{\Gamma(1+2\,\lambda)}\,.
\end{equation}
The quantity $c$ is given in terms of the (complex) energy variable $z$ as
$c = \sqrt{1 - z^2}$, and up to order $(Z\alpha)^2$, the 
quantity $\lambda$ is equivalent to the modulus of the 
Dirac angular momentum quantum number $\kappa$ , 
\begin{equation}
\label{lambdadef}
\lambda = \sqrt{\kappa^2 - (Z\alpha)^2} = | \kappa | +
{\rm O}(Z\alpha)^2\,.
\end{equation}
Indeed, it is possible and sometimes even necessary to separate 
$\lambda$ into an integer part and a fractional remainder,
\begin{equation}
\label{DefinitionOfZeta}
\lambda = |\kappa| - \zeta\,, \qquad
\zeta = \frac{(Z\alpha)^2}{|\kappa| + \lambda}\,.
\end{equation}
We here identify $\zeta$ as the fractional part of $\lambda$.
Finally, $\nu$ is given in terms of the energy variable $z$ as
\begin{equation}
\label{DefinitionOfNu}
\nu = \frac{z}{\sqrt{1 - z^2}} \, (Z\alpha)\,.
\end{equation}
The quantities $A$ and $B$ in Eq.~(\ref{defDCGF}) are related 
to the two solutions, 
integrable at infinity and at the origin, respectively, 
of the corresponding radial Dirac equation,
and given in terms of Whittaker $M$ and $W$ functions.
The specific form is as follows:
\begin{subequations}
\begin{eqnarray}
A_\pm(\lambda, \nu, y) &=& 
\left( \lambda - \nu \right) \, M_{\nu - 1/2,\lambda}(y) \pm
\left( \kappa - \frac{Z\alpha}{\sqrt{1 - z^2}} \right) \, 
M_{\nu + 1/2,\lambda}(y) \,, \\[2ex]
B_\pm(\lambda, \nu, y) &=& 
\left( \kappa + \frac{Z\alpha}{\sqrt{1 - z^2}} \right) \,
W_{\nu - 1/2,\lambda}(y) \pm
W_{\nu + 1/2,\lambda}(y) \,.
\end{eqnarray}
\end{subequations}

As seen from Eq.~(\ref{defDCGF}), the computation of the radial Green
functions can be traced back to the evaluation of the special
Whittaker functions 
\begin{equation}
\label{WhichWM}
M_{\nu \pm 1/2,\lambda}(y) \,, \qquad
W_{\nu \pm 1/2,\lambda}(y) \,.
\end{equation}
For purely real and purely imaginary values of the energy variable $z$,
the quantity $c$ and thus the argument of the Whittaker functions remains 
purely real rather than complex. This situation is realized 
along special integration contours for the complex photon 
energy~\cite{Mo1974a,Mo1974b,JeMoSo1999,JeMo2004pra,JeMo2005p}, 
but not universally applicable. We will assume here that in many
practical applications, the complex argument of $y$ can be 
made small by choosing a suitable integration contour 
for the photon energy.

Here, we thus concentrate on numerical methods, some of which are based
on nonlinear sequence transformations, which permit the evaluation of the
$W_{\alpha_{\pm}, \lambda}(x)$ and $M_{\alpha_{\pm}, \lambda}(x)$ functions 
over a wide parameter range relevant for self-energy type 
calculation~\cite{Mo1974a,Mo1974b,JeMoSo1999,JeMo2004pra,JeMo2005p}.
This parameter range can be characterized by 
the first parameter $\nu \pm 1/2$ being complex, but small in absolute 
magnitude, and the second parameter $\lambda$ being real, but 
possibly large in absolute magnitude. Pathological cases 
in which the parameters 
are close to integers, also have to be treated. 

We rewrite the Whittaker functions $W$ and $M$ in terms of 
more fundamental hypergeometric $U$ and ${}_1F_1$ functions
The $M$ functions occurring in (\ref{WhichWM}) can be 
written as (see \cite{WhWa1944}, p.~337 and \cite{MaObSo1966}, p.~296)
\begin{eqnarray}
\label{Minto1F1}
M_{\nu \pm 1/2, \lambda}(y) &=& e^{-\frac{1}{2}\,y}\,
y^{\lambda + \frac{1}{2}}\,
{_1}F_1 \left(\frac{1}{2} \mp \frac{1}{2} + \lambda - \nu, 
2 \, \lambda + 1, y\right) 
\nonumber \\[2ex]
&=& e^{-\frac{1}{2}\,y}\,
y^{|\kappa| + \frac{1}{2} -\zeta}\,\,
{_1}F_1 \left(|\kappa| + 1 - \delta_\pm - \zeta - \nu,
2 \, |\kappa| + 1 - 2\,\zeta, y\right)\,,
\end{eqnarray}
whereas the $W$ functions can be rewritten in terms of the
hypergeometric functions $U$ (see~\cite{Sl1960}, p.~14)
\begin{eqnarray}
\label{WintoU}
W_{\nu \pm 1/2, \lambda}(y) &=& e^{-\frac{1}{2}\,y}\,
y^{\lambda + \frac{1}{2}}\,\,
U \left(\frac{1}{2}  \mp \frac{1}{2} + \lambda - \nu, 
2 \, \lambda + 1,y \right)
\nonumber \\[2ex]
&=& e^{-\frac{1}{2}\,y}\,
y^{|\kappa| + \frac{1}{2} -\zeta}\,
U \left(|\kappa| + 1 - \delta_\pm - \zeta - \nu,
2 \, |\kappa| + 1 - 2\,\zeta,y\right)\,.
\end{eqnarray}
We here define the variable $\delta_{\pm}$ as
\begin{equation}
\delta_{\pm} = \left\{  
\begin{array}{cc}
1 & \mbox{for the case $+$\,,} \\[2ex]
0 & \mbox{for the case $-$\,.}
\end{array} 
\right.
\end{equation}

The transformations (\ref{Minto1F1}) and (\ref{WintoU}) 
generate ${}_1 F_1$ and $U$ functions, which are given by
\begin{equation}
\label{FF1}
{}_1 F_1(a_{\pm},b_\lambda,y) \qquad \mbox{and} \qquad
U(a_{\pm},b_\lambda,y)\,.
\end{equation}
The parameters are 
\begin{equation}
a_{\pm} = |\kappa| + 1 - \delta_{\pm} - \zeta -\nu\,, \qquad
b_\lambda = 2\,|\kappa| + 1 - 2\,\zeta \,.
\end{equation}

For the $W$ function, it is possible to rewrite the $U$ function as
a ${}_2 F_0$ (see~\cite{WhWa1944}, p.~342,
\cite{MaObSo1966}, p.~317 and \cite{Sl1960}, p.~61)
\begin{eqnarray}
\label{Winto2F0}
W_{\nu \pm 1/2 ,\lambda}(y) &=&
e^{-\frac{1}{2}\,y} \, y^{\nu \pm 1/2} \,
{}_2 F_0\left(\frac{1}{2} \mp \frac12 + \lambda - \nu,
\frac{1}{2} \mp \frac12 - \lambda - \nu, -\frac{1}{y}\right) \nonumber\\
&=& e^{-\frac{1}{2}\,y} \, y^{\nu \pm 1/2} \,
{}_2 F_0\left(|\kappa| + 1 - \delta_\pm - \zeta - \nu,
-|\kappa| + 1 - \delta_\pm + \zeta - \nu, -\frac{1}{y}\right) \nonumber\\
& \sim &  e^{-\frac{1}{2}\,y} \, y^{\nu \pm 1/2} \,
\sum_{k=0}^{\infty}
\frac{\left( |\kappa| + 1 - \delta_\pm - \zeta - \nu  \right)_k \,
\left( -|\kappa| + 1 - \delta_\pm + \zeta - \nu \right)_k}%
{k!\,(-y)^k}\,.
\end{eqnarray}
This representation entails the 
Pochhammer symbol $(a)_k = \Gamma(a+k)/\Gamma(a)$ and 
constitutes the fundamental divergent asymptotic series of the $W$
function for large arguments (i.e., in inverse powers of $y$). 
Note, however, that because of the factorial growth of the 
numerators with $k$, the series eventually diverges no matter how large 
$y$ is. Nevertheless, we anticipate here that it can be used
as an input for the resummation algorithms introduced in 
Chap.~\ref{NonlinearST}.
We summarize that the transformation (\ref{Winto2F0}) 
has given rise to a 
${}_2 F_0$ function of arguments $a'_{\pm}$ and $b'_{\pm}$, where
\begin{equation}
\label{FF2}
{}_2 F_0(a'_{\pm},b'_{\pm},y)\,, \qquad
a'_{\pm} = a_{\pm}\,, \qquad
b'_{\pm} = - |\kappa| + 1 - \delta_{\pm} + \zeta - \nu\,. 
\end{equation}
The parameter $b'_\pm$ can assume values very close to negative
integers, if the absolute magnitude of $\nu$ and 
$\zeta$ is small. This entails a tremendous loss of numerical
significance in high-precision calculations, if the 
asymptotic form of the ${}_2 F_0$ function
(\ref{Winto2F0}) is used for large $|\kappa|$
(the problematic terms originate for 
a summation index $k = |\kappa| - 1 + \delta_\pm$).
In the numerical code, it is therefore crucial to separate all
parameters of the hypergeometric functions
into integer and fractional contributions.

%
%
\begin{figure}[htb]
\begin{center}
\includegraphics[width=0.3\linewidth,angle=270]{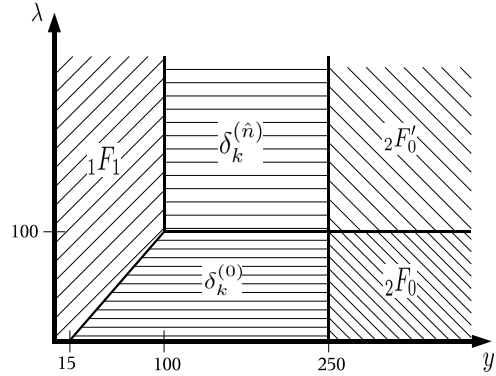}
\end{center}
\caption{\label{FigureW} Numerical evaluation of the
$W$ function.}
\end{figure}

A last representation useful for small $y$ is the decomposition
of the $W$ function in terms of two ${}_1 F_1$ functions,
\begin{eqnarray}
\label{Winto1F1}
& & W_{\nu \pm 1/2, \lambda}(y) =
  e^{-\frac{1}{2}\,y} \, y^{\lambda + \frac{1}{2}}
\, \left[
\frac{\Gamma(-2\,\lambda)}
  {\Gamma\left(\frac{1}{2} \mp \frac12 - \lambda - \nu \right)} \,
{}_1 F_1 \left(\frac{1}{2} \mp \frac12  + \lambda - \nu, 
  2\,\lambda + 1, y\right) \right.
\nonumber\\
& & \;\;\; + \left.
\frac{\Gamma(2\,\lambda)}%
{\Gamma\left(\frac{1}{2} \mp \frac12 + \lambda - \nu \right)} \,
{}_1 F_1 \left(\frac{1}{2} \mp \frac12 - \lambda - \nu, 
1 - 2\,\lambda, y\right)
\right]\, \nonumber\\
& & =
e^{-\frac{1}{2}\,y} \, y^{|\kappa| + \frac{1}{2} - \zeta}
\, \left[
\frac{\Gamma(-2\,|\kappa| + 2\,\zeta)}
  {\Gamma\left(-|\kappa| + 1 - \delta_\pm - \nu + \zeta\right)} \,
{}_1 F_1 \left( |\kappa| + 1 - \delta_\pm - \nu - \zeta, 
2\,|\kappa| + 1 - 2\,\zeta, y\right) \right.
\nonumber\\
& & \;\;\; + \left.
\frac{\Gamma(2\,|\kappa| - 2\,\zeta)}%
{\Gamma\left(|\kappa| + 1 - \delta_\pm - \nu - \zeta \right)} \,
{}_1 F_1 \left(-|\kappa| + 1 - \delta_\pm - \nu + \zeta, 
- 2\,|\kappa| + 1 + 2\,\zeta, x\right)
\right]\,.
\end{eqnarray}
One of the two ${}_1 F_1$ functions in Eq.~(\ref{Winto1F1})
has parameters $a_\pm$ and $b_\pm$, and the other is of the form
\begin{equation}
\label{FF3}
{}_1 F_1(a''_{\pm},b''_{\pm},y)\,, \qquad
a''_{\pm} = -|\kappa| + 1 - \delta_{\pm} -\nu + \zeta \,, \qquad
b''_{\pm} = - 2\,|\kappa| + 1 + 2\,\zeta\,.
\end{equation}
Again, numerical stability problems can result
(both parameters $a''_{\pm}$ and $b''_{\pm}$ can be close to negative 
integers), and these can also be solved by a separation of the 
arguments into integer and fractional contributions.

We have thus mapped the problem of the calculation of the 
radial Dirac--Coulomb wave functions given in
Eq.~(\ref{DefinitionOfDCGF}) onto the problem of 
evaluating the $M$ and $W$ functions listed in 
(\ref{WhichWM}), or of evaluating the ${}_1 F_1$ and $U$ function given in 
(\ref{FF1}), (\ref{FF2}) and~(\ref{FF3}), which is equivalent. 
The parameter range for $\nu$ required for the 
calculations reported in 
Refs.~\cite{JeMoSo1999,JeMoSo2001pra,JeMo2004pra,JeMo2005p} is approximately
given by $\nu \in (0, n)$ for the so-called low-energy part 
(soft virtual photons and by 
$\nu \in {\rm i}\, (0, Z\alpha)$ for the high-energy part
(hard virtual photons). 
The parameter range for $\nu$ thus is tiny in comparison 
to the parameter ranges for $\lambda$ and $y$, 
which both comprise the interval $(0, 10^7)$ in 
the high-precision 
calculations described in
Refs.~\cite{JeMoSo1999,JeMoSo2001pra,JeMo2004pra,JeMo2005p}.

%
%
\begin{figure}[htb]
\begin{center}
\includegraphics[width=0.3\linewidth]{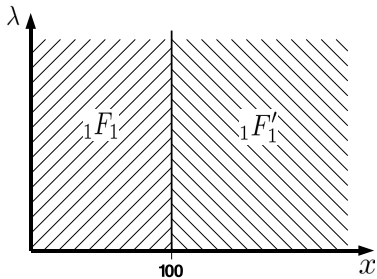}
\end{center}
\caption{\label{FigureM} Numerical evaluation of the
$M$ function.}
\end{figure}

For the $M$ and $W$ functions listed in Eq.~(\ref{DefinitionOfDCGF}),
we can thus devise a numerical algorithm
which in each case depends only on the value of
the second parameter $\lambda$ and of the argument $y$, 
not on the value of the first parameter $\nu$.
For the $W$ function, a rather sophisticated differentiated treatment
of the different parameters is required.  All the choices
are summarized in Fig.~\ref{FigureW}.
The explanation is a follows:
\begin{itemize}
\item Algorithm labeled ``${}_1 F_1$'':
We decompose the $W$ function 
into two hypergeometric ${}_1 F_1$ functions
as in Eq.~(\ref{Winto1F1}). The two ${}_1 F_1$ functions are
evaluated by virtue of their defining power series (in the argument $y$). 
Suitable truncation criteria for the power series
can be constructed according to Eqs.~(D.3) and 
D.10)~of Ref.~\cite{Mo1974b} for the first and the 
second ${}_1 F_1$ function on the 
right-hand side of Eq.~(\ref{Winto1F1}).
\item Algorithms labeled ``$\delta^{(0)}_k$'' and 
``$\delta^{(\hat{n})}_k$'': In an intermediate range of
arguments $\lambda$ for the $W$ function,
we use the divergent asymptotic series on the right-hand side
of Eq.~(\ref{Winto2F0}) as input for the 
sequence transformation (\ref{dWenTr}).
\begin{equation}
\label{DefinitionOfPartials}
s_n = \sum_{k=0}^{n} 
\frac{\left( |\kappa| + 1 - \delta_\pm - \zeta - \nu  \right)_k \,
\left( -|\kappa| + 1 - \delta_\pm + \zeta - \nu \right)_k}%
{k!\,(-y)^k}\,.
\end{equation}
The asymptotic series is resummed either via the nonlinear
sequence transformation 
\begin{equation}
{\delta}_{k}^{(0)} (1, s_0) \qquad \mbox{(algorithm ``$\delta^{(0)}_k$'')}
\end{equation}
or 
\begin{equation}
{\delta}_{k}^{(\hat{n})} (1, s_{\hat{n}}) \qquad 
\mbox{(algorithm ``$\delta^{(\hat{n})}_k$'')}
\end{equation}
where
\begin{equation}
\hat{n} = \Ent{{\rm Re}(|\kappa| - 1 + \delta_\pm - \zeta + \nu )}\,.
\end{equation}
Note that in this notation,
$k$ refers to the order of the transformation, which is 
subsequently increased until convergence is reached.
By contrast,
$\hat{n}$ refers to the element in the asymptotic series beyond which
the transformation is started, according to the
definition in Eq.~(\ref{dWenTr}).
The resummation of the asymptotic series
yields values for the $W$ function accurate to at least a relative
precision of at least $10^{-24}$, in the parameter regions
indicated in Fig.~\ref{FigureW}.
\item Algorithms ``${}_2 F_0$'' and ``${}_2 F'_0$'':
For large $y$, the asymptotic series (\ref{Winto2F0}) of the 
${}_2 F_0$ function is suitably
truncated, according to criteria found in the Eqs.~(D.6), (E.14)
and (E.15) of Ref.~\cite{Mo1974b}. For 
``${}_2 F_0$'', the asymptotic series
is summed from the first element $k=1$ in Eq.~(\ref{Winto2F0}).
For cases where {\em
both} $\lambda$ and $x$ are large, the asymptotic series is summed
in both upward and downward directions
from the element of maximum absolute
magnitude in both directions (algorithm  ``${}_2 F'_0$'').
\end{itemize}
The first decisive observation in the context of the current review
is that the nonlinear sequence transformation (\ref{dWenTr}) can
be used in order to bridge an intermediate region where a
suitable truncation of the asymptotic series (\ref{Winto2F0})
fails to produce results of sufficient accuracy, because the argument $y$ 
is not sufficiently large, and yet the power series representation
of the $W$ function in terms of two ${}_1 F_1$'s according to 
(\ref{Winto1F1}) also fails because of numerical losses due to 
cancellations among the two confluent hypergeometric functions.
The second decisive observation, in the context of the current review
is that a simple, brute-force implementation of this idea 
cannot succeed, and that considerable further sophistication
is required if large parameter ranges are to be covered. 
In particular, the separation into the ``$\delta^{(0)}_k$'' and
``$\delta^{(\hat{n})}_k$'' algorithms, counterintuitive at first glance,
constitutes one of the keys to a numerically effective evaluation
of the Dirac--Coulomb Green function (\ref{DefinitionOfDCGF}).

For the $M$ function in Eq.~(\ref{WhichWM}), we use an
evaluation scheme
illustrated in Fig.~\ref{FigureM}. For small arguments,
the $M$ function is evaluated by summing the power
series of the ${}_1 F_1$ function in Eq.~(\ref{Minto1F1}),
starting with the first element of this power series.
This region is denoted as ``${}_1 F_1$'' in Fig.~\ref{FigureW}.
The elements of the power series of the confluent
hypergeometric function are evaluated recursively
in analogy to Eq.~(D.5) in~\cite{Mo1974b},
and a termination criterion analogous to (D.6) in~\cite{Mo1974b}
is used.  
For large arguments, the power series 
is also used, but summation is started with the 
element of the power series with the largest absolute magnitude.
This maximum element is determined as outlined in Appendix E 
of~\cite{Mo1974b}, and summation in the 
upward and downward direction is terminated
by criteria analogous to (E.9) and (E.10) in~\cite{Mo1974b}.
The relevant region is denoted as ``${}_1 F'_1$'' in Fig.~\ref{FigureW}.

{\color{black}
In the current application just shown we have discussed the 
accelerated pointwise evaluation of special functions
by applying convergence acceleration and summation
processes to their defining expansions. In the same light,
it should be stressed that quite recently~\cite{Br2004},
the same strategy has been proposed for the pointwise evaluation
of Fourier series and other expansions (e.g., those into orthogonal
polynomials). Namely, it has been observed that the application
of the simple $\epsilon$ algorithm to Fourier series of 
functions with discontinuities leads to an acceleration
of the convergence of the Fourier series, and that 
the Gibbs ``overshooting'' phenomenon can be significantly reduced. 
A slight generalization of the problem using complex
variables turns out to be beneficial in the latter context. One defines 
the Fourier conjugate series ${\tilde S}(t)$ of a 
Fourier series $S(t)$ with a zero constant term by the following
relations,
\begin{align}
S(t) =& \; \sum_{k=1}^\infty a_k \, \cos(k\,t) + 
\sum_{k=1}^\infty b_k \, \sin(k\,t) \,, \\[2ex]
{\tilde S}(t) =& \; \sum_{k=1}^\infty a_k \, \sin(k\,t) - 
\sum_{k=1}^\infty b_k \, \cos(k\,t) \,, \\[2ex]
F(t) =& \; S(t) + {\rm i}\, {\tilde S}(t)  \; = \;
\sum_{k=1}^\infty (a_k - {\rm i}\, b_k) \, \exp({\rm i}\, k\,t) \,. 
\end{align}
In many cases~\cite{Br2004}, 
the series $S(t) = {\rm Re}\, F(t)$ can be numerically 
determined to good accuracy by applying the $\epsilon$ algorithm
to the complex series $F(t)$, and selecting the real part
of the calculated complex Pad\'{e} approximants to $F(t)$.
Note that the calculations described here for the 
relativistic hydrogen Green function also necessitate
the application of convergence accelerators ``in the complex
plane,'' as the energy variable $z$ and, consequently, the
parameter $\lambda$ defined in Eq.~(\ref{lambdadef}) 
can assume complex values. With these remarks,
we would like to close the discussion of the convergence 
accelerators and turn our attention to the summation of divergent 
series.}

%
%
\section{Borel Summation}
\label{Chap:DivBorel}

The contents of the current section on Borel summation
can be summarized as follows.
\begin{itemize}
\item In Sec.~\ref{QuickStartBorel}, a crude guidance for the 
reader is provided, and some basic properties of 
the Borel summation process are reviewed.
\item In Sec.~\ref{zerodim}, the performance 
of a number of resummation algorithms
is studied using zero-dimensional model theories as example.
\item In Secs.~\ref{even_oscillator_Borel-summation},~\ref{DivMATodd}
and~\ref{DivMATdwf}, we discuss both the mathematical
theorems governing the large-order behaviour of the 
perturbation theory of anharmonic oscillators, as well as 
the practical summation of the perturbative expansions,
in a number of cases of special interest.
\item In Sec.~\ref{RG}, some connections of both summation
processes as well as convergence accelerators to the 
renormalization group, and to the determination of 
critical exponents, is discussed.
\end{itemize}

%
%
\subsection{Quick Start}
\label{QuickStartBorel}

Why should there be another review article on Borel summation
in quantum mechanics and field theory? This question may 
well be asked in view of a number of excellent review 
articles on related subjects. Here is an incomplete list:
The development of large-order perturbation theory
with a special emphasis on the determination of critical exponents 
of second-order phase transitions has been 
authoritatively summarized in Refs.~\cite{LGZJ1980,ZJ1996}.
The development of generalized perturbative expansions 
with applications to quantum mechanical problems has been 
described in Ref.~\cite{ZJ1981npb},
and further works~\cite{ZJJe2004i,ZJJe2004ii,JeZJ2004plb}.
The generalized perturbative expansions, which are also known as resurgent
expansions, and their connection to perturbation series in quantum
chromodynamics has been explored in other works
(e.g., Refs.~\cite{Sh1994,Fo2000primer,Fo2001,Fo2002,St2002,St2004}).
The so-called Lipatov asymptotic form, which describes
factorially divergent contributions to perturbation series
due to the explosion of the number of Feynman diagrams in higher orders,
has been discussed in Ref.~\cite{Su2005rev}. Finally,
the significance of so-called renormalons has been elucidated in
Ref.~\cite{Be1999}: these are additional
factorially divergent contributions to perturbation series
due to a specific type of diagrams whose number does not 
explode in higher orders (but whose numerical coefficients do). 

The answer to the provocative question which was asked
in the beginning of the last paragraph is as follows:
We are trying here to accomplish three goals:
\begin{enumerate}
\item[(i)] Our first goal is to 
establish a connection between the physics literature 
on the subject, and several interesting theorems which have 
been proven on the mathematical side of the field, and to 
illustrate the concept of distributional Borel 
summability~\cite{CaGrMa1986,CaGrMa1993} by a number 
of concrete calculations and mathematical considerations.
\item[(ii)] Our second goal is to 
describe the current status, and possible further 
developments, of the field in the direction of an enhanced
perturbative calculation of critical exponents, where
a lot of effort has already been placed and where progress
eventually relies on a combined effort along a careful
technical analysis of the problem of parameterizing Feynman
amplitudes, on the sophisticated use of modern
high-performance computers, and on numerical algorithms.
\item[(iii)] Our third goal is to illustrate various 
connections of the divergent series to physical processes,
and to illustrate that a number of phenomena would
be less well understood today were it not for the fruitful
combination of results from mathematics and physics
in a field which has matured only recently.
\end{enumerate}
In Appendix~\ref{OtherFields},
we discuss very briefly
how the concept of a
resurgent expansion~\cite{CaNoPh1993}, which has already proven
to be very useful in quantum mechanics, could be useful to
justify certain remainder estimates for divergent perturbation series
in quantum chromodynamics.
Many of the sophistications discussed here can be
summarized under the unifying concept of the singularities
of the Borel transform in the ``Borel plane'' spanned by the
argument of the Borel transform.

Observe that the singularities in the Borel plane 
have got nothing to do with the so-called Bender--Wu 
cuts which are encountered when the coupling constant
of anharmonic problems is continued analytically into the 
complex plane~\cite{BeWu1969,BeWu1971,BeWu1973}. 
Namely, it has been shown that the energy levels 
of typical anharmonic oscillators (e.g., with a
$g\, x^4$ perturbation) behave as multi-valued 
functions when the coupling constant $g$ is 
continued analytically into the complex plane:
They have branch cuts. For the quartic perturbation,
the branch points and cuts are known (see Ref.~\cite{BeWu1969})
to occur near ${\rm arg}\,g \approx \pi/2$
and ${\rm arg}\,g \approx 3 \pi/2$. The branch cuts are 
``short'' in the sense that they only extend to angles of a
few degrees about the quoted complex arguments of $g$.
For a detailed discussion, we refer to 
Refs.~\cite{BeWu1969,BeWu1971,BeWu1973}, but a short 
detour is in order.

Let us suppose that we start with a particular real $g$ and follow
a particular energy eigenvalue
$E \equiv E(g)$ as a function of complex $g$,
as $g$ is varied along a path in the complex plane
and crosses a Bender--Wu branch cut, and let us suppose that 
$g$ is again varied on the other Riemann sheet toward
the real axis and to the starting value of $g$:
we end up with a different eigenvalue $E(g)$. 
In some sense, the Bender--Wu branch cuts offer ``loopholes''
in the complex plane that join the energy eigenvalues of the 
problem. Another curious observation is as follows:
suppose we vary $g$ scraping a branch cut, directly and 
exactly through a branch point, perpendicular to the 
scraped branch cut. Let us suppose, furthermore, that we
follow the evolution of the functions $E_{{\rm I} \to {\rm II}}(g)$
and $E_{{\rm II} \to {\rm I}}(g)$ that correspond to 
crossing from branch cut from sheet I to sheet II and vice 
versa, across the branch point: the surprising conclusion~\cite{BeWu1969} 
is that the two functions $E_{{\rm I} \to {\rm II}}(g)$
and $E_{{\rm II} \to {\rm I}}(g)$ will cross at the 
branch point, and this constitutes a level crossing
as a function of $g$. These and related observations have been crucial
for the development of the mathematical theory of the 
anharmonic oscillators and related problems, 
which is summarized here in the form of a number of 
theorems. In a further context, the investigation of complex 
coupling constants has inspired the development of dilation
analyticity~\cite{BaCo1971} (sometimes called 
``dilatation'' analyticity) which has proven to be a crucial
tool for the calculation of complex resonance energies which include 
an imaginary part (a decay width). Yet, the analytic
continuation of $g$ into the complex plane does not 
always provide additional insight in the case of the practical problem
of obtaining optimized resummed values of perturbation series
for a given, real, concrete value of $g$, based on a finite
number of known coefficients. Problems of that sort are
also investigated in this review.

In Sec.~\ref{Sec:QuickStart}, we have given a collection
of ``recipes'' as a guide to the selection of 
the convergence acceleration methods. 
An analogy of the ``recipe'' collection for the 
resummation of divergent series is much 
more difficult to provide, and we will resist the 
temptation to give one here. The reason is that while
the discussion in Sec.~\ref{seqtr} covers a number of 
algorithms, we here focus on only one: the Borel summation.
We trust that the numerous examples treated here will 
serve as an intuitive guidance for the interested reader,
when a new problem is encountered.

%
%
\subsection{Selected Methods}

%
%
\subsubsection{General Formulas}
\label{borel_general_formulas}

Since its introduction~\cite{Bo1899,Bo1928}, the
Borel method 
has been found to be a very powerful and elegant tool in quantum 
mechanics as well as in quantum field theory and, therefore, 
has been discussed in details in an 
enormous literature (see, e.g., pp.~233---238 of \cite{ShWa1994},
pp.~453---456 of \cite{Ka1993a}, 
pp.~880---887 of \cite{ZJ1996}, 
pp.~465---466 of \cite{ItZu1980}, 
pp.~373---376 of \cite{NeOr1988}, 
pp.~87---92 of \cite{Pa1988a}, 
pp.~54---56 of \cite{Ri1991}, 
and pp.~743---744 of \cite{Kl1995path},
as well as the seminal work~\cite{DuHa1999}). 
Let us start our discussion with the asymptotic, divergent 
perturbative expansion of some physical observable 
$f(g)$ in powers of a coupling parameter $g$.
We exchange the argument $z$ of the function defined in
Eq.~(\ref{DefP}) by $g$ and redefine the perturbative
coefficients according to $\alpha_n \to a_n$, as 
is customary in the literature,
\begin{equation}
\label{perturbative_expansion_Borel}
f(g) \sim \sum\limits_{n = 0}^{\infty} a_n \, g^n \,,
\qquad g \to 0\, .
\end{equation}
Moreover, we assume that the coefficients $a_n$ of this series
grow factorially as $n \to \infty$, and that $f(g)$ is analytic in a suitable 
sectorial region of the complex $g$-plane. Then the Borel transformed
series
\begin{equation}
\label{BorelTransP}
B_f(u) \; = \; \sum_{n=0}^{\infty} \, \frac{a_n}{n!} \, u^n
\; = \; \sum_{n=0}^{\infty} \, \frac{a_n}{\Gamma(n+1)} \, u^n
\end{equation}
converges in some circle around $u = 0$ by the Cauchy--Hadamard theorem
because by assumption $|a_n|/n! <A^n$ for some $A>0$.
That is, the series (\ref{BorelTransP}) possesses a nonzero
radius of convergence $\rho > 0$, and $B_f (u)$ is 
analytic in the circle of radius $\rho$ centered at $u = 0$.

We can now make use of the relationship
$n! \, = \, \int_{0}^{\infty} u^n {\rm e}^{-u} {\rm d} u$,
interchange summation and integration so that
$f(g)$ can be represented by the following Laplace integral (see
for example Theorem 122 on p.~182 of
\cite{Ha1949} or p.~141 of \cite{WhWa1944}),
\begin{equation}
f(g) \; = \;
\frac{1}{g} \, \int_{0}^{\infty} \, \exp (-u/g) \, B_f(u) \, {\rm d} u \, .
\label{LaplaceIntP}
\end{equation}
The case that the series (\ref{BorelTransP}) for $B_f(u)$
converges for all $u \in [0, \infty)$ actually cannot occur here since
we explicitly assumed above that the coefficients $a_n$ of the formal
power series (\ref{DefP}) for $f(g)$ grow factorially as $n \to
\infty$. The representation of $f(g)$ by the
Laplace integral (\ref{LaplaceIntP}) is valid provided that 
$B_f(u)$ admits an analytic continuation to a neighborhood of
the positive real semiaxis  such that $B_f(u)$ grows
at most exponentially as $u \to \infty$ (see for example p.~141 of
\cite{WhWa1944}), so that the Laplace--Borel (\ref{LaplaceIntP})
integral converges at the upper limit. 
Then, the Laplace integral (\ref{LaplaceIntP})
exists for some $g > 0$ and provides a finite expression for $f(g)$ 
even if the perturbation series (\ref{DefP}) diverges for every
$g \ne 0$. 

The Borel summation process can be easily generalized.
Let us now assume that a physical
quantity $F(g)$ admits an asymptotic expansion,
\begin{equation}
F(g) \sim \sum\limits_{n = 0}^{\infty} a_n \, g^n \,,
\qquad g \to 0 \,,
\qquad
\vert a_n \vert \; \le \; C \, \sigma^n \, \Gamma (kn + 1) \, ,
\end{equation}
for some rational number $k \geq 1$ and
for all integer $n$ with $C$ and $\sigma$ as suitable positive 
constants. Then we define the Borel transform of order $k$ as
\begin{equation}
B^{(k)}_F(u) \; = \; 
\sum_{n=0}^{\infty} \, \frac{a_n}{\Gamma (kn+1)} \, u^n \, .
\label{BorelTransP_k}
\end{equation}
By construction, this series converges in the circle of radius
$\rho \equiv \sigma^{-1}$ around $u = 0$. If
$B^{(k)}_F(u)$ can be analytically continued from its 
circle of convergence to the whole positive real semiaxis, and the
continuation is bounded by $e^{u^{1/k}}$ as $u \to +\infty$, then
$F(g)$ can be expressed by the Laplace integral (see
\cite{GrGrTu1971} or Example 3 on p.~45 and Problem 29(b) on p.~73 of
\cite{ReSi1978})
\begin{equation}
   F(g) \; = \; \frac{1}{k g} \,
   \int_{0}^{\infty} \, 
   B^{(k)}_F(u) \,
   \exp\left[-\left(\frac{u}{g}\right)^{1/k}\right] \, 
   \left(\frac{u}{g}\right)^{1/k-1} \, {\rm d} u \, .
   \label{LaplaceIntP_k}
\end{equation}
Under certain conditions, the Borel summation process can 
even be generalized in such a
way that perturbation expansions can be summed whose coefficients $a_n$
grow like
$\Gamma (kn+1) \exp(\alpha n^2/4)$, with $k$ and $\alpha$ being suitable
positive constants \cite{GrMa1984a,GrMa1984b}.

Let us now consider two simple examples, 
\begin{equation}
\label{defA}
{\cal A}(g) = \sum_{n=0}^\infty (-1)^n \, n! \, g^n
\end{equation}
and
\begin{equation}
\label{defN}
{\cal N}(g) = \sum_{n=0}^\infty n! \, g^n\,.
\end{equation}
The series ${\cal A}(g)$ actually constitutes 
the so-called Euler series (see Sec.~13.3 of Ref.~\cite{We1989}),
which has often been used as a paradigmatic example of 
a factorially divergent, asymptotic series in the literature.
The Borel transforms have a unit radius of convergence and read
\begin{equation}
B_{\cal A} (u) \; = \; \sum_{n=0}^{\infty} \, (-1)^n \, u^n
\; = \; \frac{1}{1+u} \, ,\qquad
B_{\cal N} (u) \; = \; \frac{1}{1-u} \,.
\label{BorelTransAN}
\end{equation}
By inserting the above formulas, which actually 
constitute analytic continuations for $u > 1$, into the
Laplace integral (\ref{LaplaceIntP}) we find the Borel sums
\begin{equation}
{\cal A} (g) \; = \; \frac{1}{g} \,
\int_{0}^{\infty} \, \frac{\exp(-u/g) \, {\rm d} u}{1 + u} \,,
\qquad
{\cal N} (g) \; = \; \frac{1}{g} \,
\int_{0}^{\infty} \, \frac{\exp(-u/g) \, {\rm d} u}{1 - u} \,,
\label{BorelSumAN}
\end{equation}
The Borel sum of ${\cal A} (g)$ is well defined and 
may be expressed as [see, e.g., Eq.~(3.02) on p.~40 of \cite{Ol1974}]
\begin{equation}
\label{AgSummed}
{\cal A} (g) \; = \; \frac{1}{g} \, \exp(1/g) \, E_1 (1/g) \,,
\end{equation}
where
\begin{equation}
E_1 (g) \; = \;
\int_{g}^{\infty} \, t^{-1} \, {\rm e}^{-t} \, {\rm d} t \, ,
\qquad \vert \arg (g) \vert \le \pi/2
\end{equation}
is an exponential integral.

In the case of ${\cal N} (g)$,
the analytic continuation (\ref{BorelTransAN}) has a pole
for $g=1$ so that the Laplace--Borel integral 
has a singularity for
every $g > 0$  on the integration path and does not define 
an analytic function of $g$. 
In view of these complications, it is not
immediately obvious in what sense the integral representation
(\ref{BorelSumAN}) can be associated with ${\cal N} (g)$ and  used for its
evaluation. One may choose a principal-value prescription 
or encircle the pole in either the positive or the negative 
sense. Indeed, we have [see, e.g.,  Eq.~(3.08) on p.~41 of \cite{Ol1974}]:
\begin{equation}
E_1 (-x \pm i 0) \; = \; - {\rm Ei} (x) \mp {\rm i}\, \pi \, ,
\quad x > 0 \, ,
\end{equation}
where
\begin{equation}
{\rm Ei} (x) \; = \; - {\rm PP} \,
\int_{-\infty}^{x} \, t^{-1} \, {\rm e}^{t} \, {\rm d} t
\end{equation}
is the exponential integral defined as a Cauchy principal value 
(principal part, $\mathrm{PP}$) for $x > 0$. The
notation $E_1 (-x \pm i 0)$ denotes the principal branch of $E_1 (-x)$
on the upper $(+)$ and lower $(-)$ side of the cut along
the negative real axis, respectively.
Thus, ${\cal N} (g)$ defined by the divergent series (\ref{defN}) can
for $g > 0$ be expressed as follows:
\begin{equation}
\label{NB}
{\cal N} (g) \; = \; \frac{1}{g} \, \exp (-1/g) \,
\left[ {\rm Ei} (1/g) + \left\{ \begin{array}{cc}
0 & \mbox{(principal value prescription)} \\
{\rm i} \, \pi & \mbox{(lower contour)}\\
-{\rm i} \, \pi & \mbox{(upper contour)}\end{array}
\right. \right] \,.
\end{equation}
We will later see that all these possibilities are actually 
realized in physically relevant examples. 
The sign of the imaginary part, if it exists, can
mostly be fixed on the basis on an immediate consideration
(e.g., a resonance energy eigenvalue has to describe 
a decaying, not an exponentially growing state as time
increases). However, there are also situations, as already mentioned in 
Sec.~\ref{borel_general_formulas}, 
where the presence of a singularity along the 
integration path of the Laplace--Borel integral
indicates the existence of additional contributions that cannot 
be obtained on the basis of perturbation theory alone.

%
%
\subsubsection{Mathematical Foundations of Borel Summability}
\label{borelgoes}

One of the purposes of this review is to 
form a bridge between mathematics and physics,
by discussing exact available mathematical results
regarding the Borel summability of specific perturbation series.
We will thus need, in the current subsection, concepts recalled
in Appendix~\ref{appbase}, in particular the notions of the 
resolvent, the spectrum and the Green function 
defined in Appendix~\ref{appbaseres} and that of an 
analytic continuation of the resolvent to the second sheet
of its Riemann surface, which defines a resonance and is
described in Appendix~\ref{appbaseana}.

Let us also recall that for the convergent case,
the concept of holomorphy establishes a one-to-one
correspondence between convergent power series and functions of
complex variables: this is the Weierstrass notion of analytic functions. 
If convergent power series are replaced by divergent ones, this
one-to-one correspondence is in general impossible, because infinitely
many functions of a complex variable singular at a point (without loss the
origin) may admit the same divergent power series expansion.  Consider
the  example ${\cal A}(g)$ given by (\ref{AgSummed}).  We know
that its series expansion as $|g|\to 0$, $|{\rm arg} \, g|<\pi$ is 
$\sum_{n=0}^{\infty} n! \, (-g)^n$. However, any function 
${\cal A}(g)+e^{-g^{-1/q}}$, with $q\geq 0$, admits the same
divergent expansion as $|g|\to 0$ in the same directions.
Additional requirements, such as uniform remainder estimates, are
thus required to set up one-to-one correspondences between 
divergent power series and  classes of analytic functions. In any
physical problem giving rise to a divergent perturbation expansion,
the most important thing to check in order to assign the correct
meaning to the expansion is that this series and the
 solution fulfill the appropriate additional requirements.

The fundamental
result in this connection is a classical theorem of 
Carleman~\cite{Ca1926}. Before we state the theorem,
we need to recall some basic definitions and in particular,
the fact that the condition ${\rm Re}\, g^{-1} > R^{-1}$ 
for a complex coupling constant $g$ defines the inner region
of a circle of radius $R$ centered at ${\rm Re}\,g = R/2$,
${\rm Im}\,g = 0$ in the complex $g$-plane. 

\begin{definition}
\label{asin}
Let $R > 0$ and let $f(g)$ be a function analytic in the circle
$C_R  \equiv  \{g\in\C: {\rm Re}\,g^{-1} > R^{-1}\}$ of radius $R$. 
We say that a power series 
$\sum_{n=0}^{\infty} a_n \, g^n$ is asymptotic to $f$, or
that $f$ admits a $\sum_{n=0}^{\infty}a_n \, g^n$ as an asymptotic
expansion (to all orders) as $g\to 0$ in $C_R$ if, for each fixed $N$:
\begin{equation}
\label{asin1}
\left|f(g)- \sum_{n=0}^{N} a_n \, g^n \right| = 
O(g^{N+1}), \qquad g\to 0, \:\: g \in C_R\,.
\end{equation}
\end{definition}

It is easy to check that a function has at most one
asymptotic expansion as $g\to 0$ in $C_R$.
On the contrary, infinitely many functions may admit the same
asymptotic expansion in a prescribed region.  For example, if the function
$f(g)$ satisfies Definition \ref{asin} above, then any function of the
form $f(g)+A \, e^{-1/g}$, $A\in\C$, admits as well
$\sum_{n=0}^{\infty}a_n \, g^n$ as an asymptotic expansion as $g\to 0$ in
$C_R$ because $A \, \exp(-1/g)$ has an identically zero asymptotic
expansion therein.
In order to establish a one-to-one correspondence between a function and
a (possibly divergent) power series expansion a stronger condition than
the one given in Definition \ref{asin} is needed, which is as follows.

\begin{definition}
\label{asin2}
Let $C_R$ be as above. We say that a function $f(g)$ analytic in $C_R$
obeys a strong asymptotic condition and admits 
$\sum_{n=0}^{\infty}a_ng^n$ as a strong asymptotic series in $C_R$ if there
are $A,B>0$ such that
\begin{equation}
\label{asin3}
\left| f(g)- \sum_{n=0}^{N-1}a_n \, g^n \right| \leq 
A \, B^N \, N! \, |g|^N
\,,\qquad \forall\;g\in C_R\,.
\end{equation}
\end{definition}
 
We can finally state Carleman's theorem:
\begin{theorem}(Carleman)
\label{Carleman}
If $\sum_{n=0}^{\infty} a_n \, g^n$ is a strong asymptotic series 
for two analytic functions $f$ and $g$, then $f\equiv g$. 
Moreover, there are $A$ and $B>0$ such that 
\begin{equation}
\label{asin4}
|a_n|\leq A \, B^n \, n!\,, \qquad \forall\; n \in \N_0\,.
\end{equation}
{\it Proof.} See, e.g., Carleman's book~\cite{Ca1926}. 
\end{theorem}

The following remarks are in order.
Carleman proves also the necessity of the condition (\ref{asin3}),
in the sense that no weaker condition can guarantee the result.
Functions admitting an asymptotic
expansion with the above properties are thus determined by
the series even if divergent and they are therefore called {\it
quasi-analytic}.  
Observe that the convergence of a power series yields by definition a
constructive approximation method for its sum. By the same definition,
this cannot be extended to divergent series even when the conditions of
Carleman's theorem are satisfied. 

Therefore the notion of a regular summation method is
introduced: given a pair $(f(g), 
\sum_{n=0}^{\infty} a_n \, g^n)$
fulfilling the conditions of Carleman's theorem, a regular
summation method is a constructive procedure which makes it possible
to recover $f(g)$ out of the divergent power series and yields the
ordinary sum when the series is convergent. In other words, when
applicable, a summation method makes it possible to associate a 
unique analytic function to a given divergent series.  In
general the validity of a regular summation method requires much
stronger conditions on 
$f(g)$ and $\sum_{n=0}^{\infty}a_ng^n$ than those of  Carleman's
theorem. An example is the Stieltjes method, equivalent to
the convergence of the Pad\'e approximants, which requires very
restrictive conditions both on $f$ and  on 
$\sum_{n=0}^{\infty} a_n \, g^n$ as recalled below.
A classical result of Nevanlinna~\cite{Ne1919}, however 
(see also Sokal, Ref.~\cite{So1980}), 
singles out the Borel method as described above 
as the one which
requires just the conditions of Carleman's theorem. We have indeed
\begin{theorem}{(Nevanlinna)}
\label{N}
Under the conditions of Carleman's theorem, the divergent series 
$\sum_{n=0}^{\infty}a_n \, g^n$ is Borel summable to $f(g)$ for $g$
real, $0\leq g <R$. This means the following:
(i) The Borel transform $B_f(u) = \sum_{n=0}^{\infty} \frac{a_n}{n!} u^n$, 
a priori holomorphic within a circle of convergence of radius
$B^{-1}$, i.e.~for $|u| < B^{-1}$ where 
$B$ is given in Eq.~(\ref{asin4}), admits an 
analytic continuation to the strip
$\{u\in\C\,:\,|\Im{u}| < T\,,{\rm Re}\,u>0 \}$.
(ii) The Laplace integral 
$1/g \int_0^{\infty} B(u) \,{\rm e}^{-u/g}\,{\rm d}u$
converges absolutely for $0\leq g < T$ and for these values of $g$
one has
\begin{equation}
f(g) = \frac{1}{g} \, \int_0^{\infty} B_f(u) \, {\rm e}^{-u/g}\,{\rm d}u\,.
\end{equation}
{\it Proof.} For a modern presentation, see~\cite{So1980}.
\end{theorem}

In many instances, as we shall explicitly see later on in the
examples from quantum physics, it is actually possible to verify a
stronger statement for Borel summability, valid for functions
$f(g)$ holomorphic in the sector 
$|{\rm arg} \, g| < \frac{\pi}{2}+\epsilon$ with $\epsilon>0$. Namely, 
we have the following theorem.
 
\begin{theorem}{(Watson-Nevanlinna)}
\label{WN}  
Let the pair $(f(g), \sum_{n=0}^{\infty} \frac{a_n}{n!} \, g^n)$
fulfill the following conditions:
(i) $f(g)$ is holomorphic in the open sector
$\Omega_{\pi/2+\epsilon}  \equiv 
\{g\in \C\setminus\{0\}\,:\,|{\rm arg}\, g| < \frac{\pi}{2}+\epsilon\}$,
with $\epsilon > 0$, and continuous up to the boundary.
(ii) $f(g)$ admits
$\sum_{n=0}^{\infty} a_n \, g^n$ as a strong asymptotic expansion 
to all orders as $|g|\to 0$ in
$\Omega_{\pi/2+\epsilon}$, i.e.  there are $A>0$, $B>0$ such that,
for $N \in \mathbb{N}$
\begin{equation}
\label{MF3}
\left| f(g) - \sum_{n=0}^{N-1} a_n\,g^n\right|
\leq 
A \, B^N \, N!\, |g|^N, \qquad 
|g|\to 0, \quad g\in \Omega_{\pi/2+\epsilon}
\end{equation}
Then the series $\sum_{n=0}^{\infty}a_ng^n$ is Borel summable to
$f(g)$ for all complex $g$ in the closure of the sector 
$\Omega^B_{\epsilon}  \equiv  \{g\in\C:\, 0 < |g| < B^{-1}\; ;
|{\rm arg}\, g|<\epsilon\}$.

{\it Proof.} This classic result is proven for instance in the book
of Hardy~\cite{Ha1949}. Another proof can be found 
in Ch.~XII.3 of Ref.~\cite{ReSi1978}.
\end{theorem}

As before, all these results can be extended to series diverging
as $n!^{\alpha}$ for any $\alpha >1$ provided the function $f(g)$ is
analytic in a wider sector, the rule of thumb being: a sector of
opening $\pi$ both in analyticity and in the validity of the
remainder estimate is necessary to absorb any
$n!$. Thus, an analyticity sector of opening at least $2\pi$ is necessary
to absorb a divergence as $n!^{2}$, and so on. Obviously,
the analyticity
is intended on a Riemann surface sector when the opening angle exceeds
$2\pi$. We limit ourselves to state the analogue of the
Watson-Nevanlinna criterion for the Borel summability of order
$k>1$, in the sense of Eq.~(\ref{LaplaceIntP_k}),
remembering that $k=1$ corresponding to ordinary Borel summability.

\begin{theorem}{(Nevanlinna-Leroy)}
\label{NL}
Consider once again the pair $(f(g), 
\sum_{n=0}^{\infty}\frac{a_n}{n!}g^n)$, this time under the following
conditions: (i) For some rational number 
$k > 0$, $f(g)$ is analytic in the circle
$C_R \equiv \{ g\in \C: {\rm Re} \, g^{-1/k} > R^{-1}\}$.
(ii) The function $f(g)$ admits $\sum_{n=0}^{\infty}a_n \, g^n$ as a 
strong asymptotic expansion to all orders as $|g|\to 0$ in
$C_R$, i.e.~there are $A > 0$, $B > 0$ so that,
for $N \in \N$ and $|g|\to 0, g\in C_R$:
\begin{equation}
\label{MF4}
\left| f(g) - \sum_{n=0}^{N-1} a_n \, g^n \right|
\leq A \, B^N \, \Gamma(k \, N+1)\,|g|^N \,.
\end{equation}
Then the series $\sum_{n=0}^{\infty} a_n\,g^n$ is Borel-Leroy summable
of order $k$ to $f(g)$ for $g$ real, 
$0\leq g <R$. This means that the Borel-Leroy
transform of order $k$,
\begin{equation}
B^{(k)}_f(u) = \sum_{n=0}^{\infty}
\frac{a_n}{\Gamma(k \, n+1)} \, u^n \,,
\end{equation}
a priori holomorphic for $|u|<B^{-1}$, admits an analytic continuation to
the strip  $\{u\in\C\,:\,|{\rm Im} \, u| < R\,,\,{\rm Re}\, u > 0\}$ 
and, for $0\leq u <R$, one has
\begin{equation}
f(g) = \frac{1}{k\,g} \,
\int_{0}^{\infty} \, B^{(k)}_f(u) \, 
\exp\left[ - \left(\frac{u}{g}\right)^{1/k} \right] \, 
\left(\frac{u}{g}\right)^{1/k-1} \, {\rm d} u \,,
\end{equation}
where the integral in the right side converges absolutely.
\end{theorem}

As above, in many examples where the perturbation series diverges as
fast as $(k\,n)!$ with $k>1$, it is actually possible 
to verify more, namely the validity of the 
generalized Watson-Nevanlinna criterion, where the 
analyticity requirement is extended from a 
disk $C_R$ to an open Riemann surface sector.

\begin{theorem}{(Watson-Nevanlinna-Leroy)}
\label{WNL}
Consider once more the pair $(f(g), \sum_{n=0}^{\infty}\frac{a_n}{n!}g^n)$, 
this time under the following conditions:
(i) For some rational number $k > 0$, 
$f(g)$ is analytic in the open Riemann surface sector
$\Omega_{k\pi/2+\epsilon} = 
\{g\in\C\setminus\{0\}\,:|{\rm arg} \, g|<\frac{k\pi}{2}+\epsilon\}$.
(ii) The function $f(g)$ admits
$\sum_{n=0}^{\infty} a_n \, g^n$ as a strong asymptotic expansion 
to all orders as $|g|\to 0$ in $\Omega_{k\pi/2+\epsilon}$, 
i.e.~the condition (\ref{MF4}) holds for  $|g|\to 0$, 
$g \in \Omega_{k\pi/2+\epsilon}$.
Then the series $\sum_{n=0}^{\infty} a_n \, g^n$ is Borel-Leroy summable
of order $k$ to $f(g)$ for all complex $g$ in the sector 
$\Omega^B_{\epsilon} = \{g\in\C \; : \; 
0 < |g| < B^{-1},\,\,|{\rm arg} \, g| < \epsilon\}$. 
This means that for $0< |g| <B^{-1}$, $|{\rm arg} \, g| < \epsilon$ one has
\begin{equation}
   f(g) = \frac{1}{k\,g} \,
   \int_{0}^{\infty} \, B^{(k)}_f(u) \, 
   \exp\left[ - \left(\frac{u}{g}\right)^{1/k} \right] \, 
   \left(\frac{u}{g}\right)^{1/k-1} \, {\rm d} u \,,
\end{equation}
where the integral in the right side converges absolutely. 
\end{theorem}

We include the following remarks.
For the proof of both theorems, see e.g.~\cite{ReSi1978,So1980}. 
Actually, the Theorems \ref{NL} and \ref{WNL}
follow from Theorems \ref{N} and \ref{WN} 
via the change of variable $g \mapsto g^{1/k}$, respectively.  
As already recalled, the Borel summation method is
regular, i.e.~it yields the ordinary sum when applied to a
convergent series. Actually it yields more. Given a  power series
$\sum_{n=0}^{\infty}{a_n}g^n$ with radius of convergence $r$, the
Borel transform 
$B_f(u)= \sum_{n=0}^{\infty}\frac{a_n}{n!} \, u^n$ is an entire
function, and the Borel -Laplace integral 
$1/g \int_0^{\infty}B_f(u) {\rm e}^{-u/g}\,du$
converges (uniformly) to the sum $f(g)$ in the Borel polygon. The Borel
polygon (see e.g. \cite{Ca1926} for that notion) always contains the
convergence disk, but coincides with it only if its boundary is  a
natural boundary for $f$, namely if $f$ cannot be analytically continued
beyond its circle of convergence. In other words, the Borel method
applied to a convergent series gives the analytic continuation of the sum
in a region beyond the circle of convergence. This is actually the
motivation of its introduction by \'Emile Borel in 1899 in Ref.~\cite{Bo1899}. 

%
%
\subsubsection{Borel-Pad{\'e} Method}
\label{subsubsection_borel_pade}

In Sections \ref{borel_general_formulas} and \ref{borelgoes} we have 
discussed the general formulas and the mathematical background of the 
Borel resummation approach. Within this approach, the Borel transform 
(\ref{BorelTransP}) of a power series usually has a finite radius of 
convergence about the origin $g$ = 0. For the evaluation of the Borel 
integral, therefore, $B_f(u)$ has to be continued analytically beyond the 
radius of convergence. Such a continuation, however, requires in principle 
the explicit knowledge of all coefficients of the power series; this
situation is rather uncommon in physical applications where only a 
few series coefficients are typically known. Therefore, a Borel method can 
be a useful numerical tool only if a reasonably accurate approximation to 
the analytic continuation of the Borel transformed series can be 
constructed from a finite number of expansion coefficients. A number of 
approaches have been proposed, therefore, for the approximate analytic 
continuation of Borel transformed perturbation series. For instance, 
in Ref.~\cite{GrGrSi1970} it has been argued that the analytical 
continuation can be achieved by evaluating Pad\'e approximants. The first 
$n + 1$ terms of the Borel transformed series (\ref{BorelTransP}) can be 
used to construct a diagonal or off--diagonal Pad\'{e} approximant
\begin{equation}
   \label{Borel_pade_approximant}
   {\mathcal P}_n(u) = \bigg[ \Ent{n/2} \bigg/
   \Ent{(n+1)/2} \bigg]_{B_f}\!\!\!\left(u\right)\,,
\end{equation}
where the symbol $\Ent{x}$ has been defined in Sec.~\ref{Sec:SpecSymSpecFun}.
We then evaluate the (modified) Borel integral 
\begin{equation}
\label{contours}
{\mathcal T}\!{\mathcal P}_n(g) = \frac{1}{g} \int_{C_{j}} {\rm d}u \, 
\exp(-u/g)\,{\mathcal P}_n(u) \,,
\end{equation}
where the integration contours $C_{j}$ with $j = -1, 0, +1$ have been 
defined in Refs.~\cite{FrGrSi1985,Je2000prd,Je2001pra}.
These contours are just straight lines tilted in the complex
plane so that the singularities of the Pad\'e approximants
are appropriately encircled when Jordan's Lemma is applied, 
to give the pole contributions that finally yield numerical
approximations to the imaginary part of the Borel sum.
Of course,
the result obtained along $C_{-1}$ is the complex conjugate of the results 
along
$C_{+1}$. The arithmetic mean of the results obtained by the integration 
along these contours is attributed then to the contour $C_{0}$ 
which involves no imaginary part and can alternatively
be obtained by averaging over $C_{-1}$ and $C_{+1}$. 
The limit of the sequence 
$\{ {\mathcal T}\!{\mathcal P}_n(g) \}_{n = 0}^{\infty}$, 
if it exists, is assumed to represent 
the complete, physically relevant solution:
\begin{eqnarray}
   \label{pade_borel-limit}
   f(g) \, = \, \lim_{n \to \infty} {\mathcal T}\!{\mathcal P}_n(g) \, .
\end{eqnarray}
In Refs.~\cite{FrGrSi1985,Je2000prd,Je2001pra},
it has been shown that integrations along 
the contours $C_j$ have been 
shown to provide the physically
correct analytic continuation of resummed perturbation series 
for those cases where the
evaluation of the standard Laplace--Borel integral is impossible 
due to multiple branch
cuts or due to spurious singularities in view of the finite order of the  
Pad\'{e} approximant (\ref{Borel_pade_approximant}).
This is confirmed in the current review via the example of the 
cubic anharmonic oscillator treated in Sec.~\ref{DivMATodd}.

%
%
\subsubsection{Conformal Mapping Method}
\label{subsubsection_borel_pade-conformal}

Apart from the Pad{\'e} approximants discussed in the previous Section, 
the analytic continuation of the Borel transform can be achieved by 
various other means. Examples are order dependent 
mappings~\cite{SeZJ1979,LGZJ1983} and two-point Pad\'e
approximants \cite{Gr1979} or Pad\'e-type approximants \cite{Ma1987a},
or the first confluent form of the $\epsilon$ algorithm~\cite{Ma1984}.
Another possibility consists in a 
conformal mapping. In Ref.~\cite{JeSo2001}, it has been found that a 
combination of a analytic continuation of the Borel plane via conformal 
mapping and a subsequent numerical approximation by Pad\'{e} approximants 
leads to significant acceleration of the convergence of (slowly 
convergent) series. Such a combination is primarily useful in those cases 
where only the leading large-order asymptotics of the perturbative 
coefficients $a_n$ in Eq.~(\ref{DefP}) are known to sufficient accuracy, 
and the subleading asymptotics have not yet been determined. For such 
asymptotic series, moreover, we will assume below that the coefficients 
$a_n$ display an alternating sign pattern. This implies the existence of a 
single pole along the negative real axis corresponding to the leading 
large-order growth of the perturbative coefficients, which we assume for 
simplicity to be at $u$ = - 1. For Borel transforms which have only a 
single cut in the complex plane which extends from $u$ = -1 to $u = - 
\infty$, one may employ the following conformal mapping:
\begin{equation}
   \label{conformal_u_z}
   z(u) = \frac{\sqrt{1 + u} - 1}{\sqrt{1 + u} + 1} \, ,
\end{equation}
where $z$ is called the conformal variable. The conformal mapping 
(\ref{conformal_u_z}) maps the complex $u$-plane with a cut along 
$(-1,-\infty)$ unto the unit circle in the complex $z$-plane.

From Eq.~(\ref{conformal_u_z}) we can express the Borel integration 
variable $u$ as a function $z$:
\begin{equation}
   \label{conformal_z_u}
   u(z) = \frac{4z}{(z - 1)^2} \, .
\end{equation}
By expanding now the right--hand side of this relation in powers of $z$, we can
rewrite the $m$th partial sum of the Borel transform (\ref{BorelTransP}) as:
\begin{equation}
\label{BorelTransP_z}
B^m_f (u(z)) \, = \, \sum\limits_{n = 0}^m C_n z^n + O(z^{m+1}) \, ,
\end{equation}
where the coefficients $C_n$ as a function of $a_n$ are uniquely
determined.
Making use of Eq.~(\ref{BorelTransP_z}),
we finally define the (partial sum of the)
Borel transform expressed as a function of the conformal variable $z$:
\begin{equation}
   \label{BorelTransP_z_conformal}
   B^m_f (z) \, = \, \sum\limits_{n = 0}^m C_n z^n \,
   ,
\end{equation}
and apply this transform in order to construct the low--diagonal Pad{\'e} 
approximant:
\begin{equation}
   \label{Borel_pade_approximant_z}
   {\mathcal P}_m(z) = \bigg[ \Ent{m/2} \bigg/
   \Ent{(m+1)/2}
   \bigg]_{B^m_f}\!\!\!\left(z\right)\, .
\end{equation}
Similar as before, as a last step we have to evaluate the Laplace--Borel 
integrals along the integration contours $C_j$ [see Eq.~(\ref{contours})]:
\begin{equation}
\label{contours_conf}
{\mathcal T}^C\!{\mathcal P}_m(g) = \frac{1}{g} \, \int_{C_{j}} {\rm d}u \, 
\exp(-u/g)\,{\mathcal P}_m(z(u)) \, ,
\end{equation}
at increasing $m$ and to examine the apparent convergence of the transforms
${\mathcal T}\!{\mathcal P}_m(g)$. 
The application of such a Borel--Pad{\'e} method combined 
with a conformal mapping of the
Borel plane will be discussed in detail in the following 
Section \ref{zerodim} for the calculation of series originating from
zero-dimensional model theories.

%
%
\subsection{Zero--Dimensional Theory}
\label{zerodim}

Until now we have discussed the basic relations as well as the 
mathematical background of the Borel summation and related methods. Now we 
are ready to use these methods for the analysis of the problems which 
arise in the different applications of quantum mechanics and 
quantum field theory. In this subsection, for example, we will show how the 
Borel summation method may help us in computing the partition function 
$Z(g)$ of the zero-dimensional $\phi^4$--theory:
\begin{equation}
\label{Zdef}
Z(g) = \int\limits_{-\infty}^{\infty}
\frac{{\rm d} \Phi}{\sqrt{2 \pi}} \, \exp\left[ - \frac{1}{2} \, \Phi^2 \, 
- g \, \Phi^4 \right] \, ,
\end{equation}
where we assume $g > 0$. In fact, this function has been used as a 
paradigmatic example for the divergence of perturbative expansions in 
quantum field theory (see~\cite{ZJ1996,ItZu1980}). In particular, it was 
shown that the coefficients in the expansion of this integral in powers of 
$g$ are equal to the number of vacuum polarization diagrams in a $\phi^4$ 
theory. The series (\ref{Zdef}) is an example for the ``ordinary'' Borel 
summability introduced in Sec.~\ref{borelgoes}.
  
By expanding (\ref{Zdef}) in powers of the coupling parameter $g$ for 
$g \to 0$ we may obtain the strictly alternating series:
\begin{equation}
   \label{series_Z}
   Z(g) \, = \, \sum_{N=0}^{\infty} C_N (-g)^N \, = \,\sum_{N=0}^{\infty}
   \frac{4^{N} \, \Gamma(2 N + 1/2)}
   {\sqrt{\pi} \, \Gamma(N + 1)} \, (-g)^N\, .
\end{equation}
By using the known relations for the asymptotics Gamma function for large 
$N$ \cite{AbSt1972}, it is easy to find the large--order asymptotics of 
the perturbative coefficients $C_N$:
\begin{equation}
\label{asympgamma}
  C_N \, = \, \frac{4^{N}}{\sqrt{\pi}} \, \frac{\Gamma(2 N + 1/2)}{\Gamma(N + 1)} \sim
  \frac{16^N}{\sqrt{2} \pi} \, \Gamma(N) \, \left[ 1 - 
  \frac{1}{16 N} + \frac{1}{512 N^2}  + O\left(\frac{1}{N^3}\right) \right] \,
\end{equation}
The factorial growth of the (absolute value of the) coefficients $C_N$ 
leads to a divergence of the perturbative expansion in 
Eq.~(\ref{series_Z}), even for small coupling $g$. Therefore, we have to 
apply special resummation techniques in order to recover the partition 
function $Z(g)$ from the divergent series (\ref{asympgamma}). In our test 
calculations, we have applied four resummation techniques: (i) the 
$\delta$ transformation defined in Eq.~(\ref{dWenTr}),
(ii) the COMPOS algorithm~\cite{BrRZ1991} 
discussed in Section \ref{Sec:QuickStart},
(iii) the Borel--Pad{\'e} method as discussed in 
Eqs.~(\ref{Borel_pade_approximant})--(\ref{pade_borel-limit}) as well as 
(iv) the Borel--Pad{\'e} method combined with a conformal mapping of the 
Borel plane [see Eqs.~(\ref{conformal_u_z})--(\ref{contours_conf})]. The 
results obtained from these methods for a coupling parameter $g = 0.1$ 
are compared in Table~\ref{Table8}. As seen from this Table, the 
Borel--Pad\'e approximation is clearly less powerful in the summation of 
the divergent perturbative expansion~(\ref{series_Z}) than the nonlinear 
sequence transformations, and for large coupling $g \approx 10$, the Pad\'e 
method fails. However, if the Borel--Pad\'e method is combined with 
the conformal mapping of the Borel plane, it may significantly accelerate 
the summation of the divergent series. For instance, as seen from the 
fifth column of the Table~\ref{Table8}, such a combined method method 
appears to be as powerful as the nonlinear sequence transformations for the 
summation of the series (\ref{series_Z}) taken at the relatively 
large coupling of $g = 0.1$.

%
%
\begin{table}[t]
\begin{center}
\begin{minipage}{15.5cm}
\begin{center}
\caption{Evaluation of the perturbation series for $Z(g)$ 
[zero-dimensional $\phi^4$-theory, see Eqs.~(\ref{Zdef})
and~(\ref{series_Z})] for $g = 0.1$. In the first column, we indicate the
quantity $n$ which is the order of the partial sum or alternatively the
order of the transformation. In the second column,  
we list the Weniger transforms $\delta_n^{(0)}(1,s_0)$, whereas in the
third column the results of the COMPOS algorithm are displayed (the
composition starts with the fourth element of the series).
Apart from these nonlinear sequence  
transformations, the Borel--Pad{\'e} method as well as the Borel--Pad{\'e}
method combined with a conformal mapping of the Borel plane
${\mathcal T}^CZ_M(g = 0.1)$ are also used
for the resummation of the divergent series. The results of such
resummations are presented in the fourth and fifth columns,
respectively.}
\label{Table8}
\begin{tabular}{lrrrr}
\hline
\hline
\multicolumn{1}{c}{$n$}
& \multicolumn{1}{c}{${\delta}_{n}^{(0)} \bigl(1, s_0 \bigr)$}   
& \multicolumn{1}{c}{COMPOS}
& \multicolumn{1}{c}{${\mathcal T}Z_M(g = 0.1)$}
& \multicolumn{1}{c}{${\mathcal T}^CZ_M(g = 0.1)$}
\\
\hline
$ 1 $&
  $\underline{0.8}90~909~091$&
  $\underline{}$&
  $\underline{0.8}01~186~280$&
  $\underline{0.85}9~380~049$\\
$ 2 $&
  $\underline{0.85}0~511~945$&
  $\underline{}$&
  $\underline{0.8}70~363~571$&
  $\underline{0.8}65~756~316$\\
$ 3 $&
  $\underline{0.85}6~782~997$&
  $\underline{}$&
  $\underline{0.85}1~692~619$&
  $\underline{0.8}61~937~975$\\
$ 4 $&
  $\underline{0.857}~694~858$&
  $\underline{0.8}77~198~421$&
  $\underline{0.85}9~335~944$&
  $\underline{0.85}8~100~169$\\
$\dots$ &
  $\dots$ &
  $\dots$ &
  $\dots$ \\
$ 17 $&
  $\underline{0.857~608~585}$&
  $\underline{0.857~608~585}$&
  $\underline{0.857~608}~475$&
  $\underline{0.857~608~58}7$\\
$ 18 $&
  $\underline{0.857~608~585}$&
  $\underline{0.857~608~585}$&
  $\underline{0.857~608}~638$&
  $\underline{0.857~608~58}6$\\
$ 19 $&
  $\underline{0.857~608~585}$&
  $\underline{0.857~608~58}2$&
  $\underline{0.857~608}~554$&
  $\underline{0.857~608~585}$\\
$ 20 $&
  $\underline{0.857~608~585}$&
  $\underline{0.857~608~58}1$&
  $\underline{0.857~608}~600$&
  $\underline{0.857~608~585}$\\
$ 21 $&
  $\underline{0.857~608~585}$&
  $\underline{0.857~608~585}$&
  $\underline{0.857~608~5}76$&
  $\underline{0.857~608~585}$\\
$ 22 $&
  $\underline{0.857~608~585}$&
  $\underline{0.857~608~585}$&
  $\underline{0.857~608~5}90$&
  $\underline{0.857~608~585}$\\
$ 23 $&
  $\underline{0.857~608~585}$&
  $\underline{0.857~608~585}$&
  $\underline{0.857~608~58}2$&
  $\underline{0.857~608~585}$\\
\hline
exact&
  $0.857~608~585$&
  $0.857~608~585$&
  $0.857~608~585$&
  $0.857~608~585$\\
\hline
\hline
\end{tabular}
\end{center}
\end{minipage}
\end{center}
\end{table}
%
%
%

%
%
\subsection{Even Anharmonic Oscillators}
\label{even_oscillator_Borel-summation}

{\em Mathematical discussion of even anharmonic oscillators.}
The quantum mechanics of anharmonic oscillators in
the configuration space $\mathbbm{R}^N$ is 
described by the Schr\"odinger equation $H_{\ell}\psi=E\psi$ in the
Hilbert space $L^2(\R^N)$. Here, $H_{\ell}$ is the Schr\"odinger operator
whose action (on the appropriate domain) is defined by 
\begin{equation}
\label{Hclass}
H_{\ell} =
\frac12 \sum_{i=1}^N p^2_i +
\frac12 \sum_{i=1}^N \omega_i^2 \, q^2_i +
g \, V_\ell(q_1,\ldots,q_N)\,.
\end{equation}
We here assume that the order $\ell$ of the expansion
about the minimum of the harmonic potential is exclusively selected.
We obtain by definition an anharmonic
oscillator or order $\ell$ in $N$ degrees of freedom.
We can write (\ref{Hclass}) as
\begin{eqnarray}
\label{I6}
H_{\ell} = H_0 + g \, V_{\ell}(q_1,\ldots,q_N) \,,
\end{eqnarray}
where 
\begin{equation}
\label{HVL}
H_0 =
\frac12 \sum_{i=1}^N p^2_i +
\frac12\sum_{i=1}^N \omega_i^2 \, q^2_i
\end{equation}
is the Schr\"odinger operator of the $N$-dimensional harmonic oscillator.
Here, as usual,
\begin{equation}
p_i \equiv -{\rm i}\, \frac{\partial}{\partial q_i}
\end{equation}
is the canonical quantization of the $i$th momentum, and we still denote 
by $q_i$ the multiplication operator quantizing the $i$th coordinate
$q_i$ (the reduced Planck constant is $\hbar = 1$ throughout this review). 
We recall that the eigenvalues of $H_0$ are $E_{n_1,\ldots,n_N} =
(n_1 + 1/2) \, \omega_1 + \ldots + (n_N + 1/2) \, \omega_N$.

The even anharmonic oscillators in dimension $N$ (or,
equivalently, in $N$ degrees of freedom) are by definition the
anharmonic oscillators such that the potential $V_\ell$ in the classical
Hamiltonian (\ref{Hclass}) is a polynomial in $N$ variables of even degree
$2s$, $s=2,3,\ldots$. The basic property is that, if $V_\ell$ is bounded
from
below, then it tends to $+\infty$ as $\|q\|\to\infty$ for $g\geq 0$.
Hence any constant energy surface is compact, and any classical motion
is localized. Then (see e.g. Theorem
XIII.67  of~\cite{ReSi1978}) the operator
$H_\ell$ is  self-adjoint in the Hilbert space
$L^2(\R^n)$  and its spectrum
$\sigma(H_\ell)$  is discrete. Namely, it
consists entirely of isolated eigenvalues with finite multiplicity which
accumulate at
$+\infty$. Since we may assume without loss $V_\ell\geq 0$, we can
conclude that the eigenvalues of $H_\ell$ (repeated according to their
multiplicities) form a positive sequence $0<E_0(g)<E_1(g)\leq E_2(g)
\ldots\to +\infty$, and the corresponding eigenvectors form a complete
orthonormal basis. The ground state $E_0(g)$ is always simple
(i.e., non-degenerate).

The quantity we look at in this context is the standard bound state
perturbation theory, or
\RSPT, for any eigenvalue $E_k(g)$ around any unperturbed eigenvalue
$E_k(0) \equiv E_k$ of the harmonic oscillator.  The first relevant result is
that, no matter how the anharmonic terms are selected to form the
perturbation,  the \RSPT\ always generates a divergent expansion. The
reason is that the perturbation is not small with respect to the
unperturbed operator: no matter how small we can take $g$, the perturbing
potential
$gV_\ell$ will always overtake the harmonic one for $\|q\|$ large enough.
For the sake of simplicity, from now on we consider only the one-dimensional
case. Here the unperturbed eigenvalues are all simple, and thus we avoid the
discussion of degenerate perturbation theory. 
In higher dimension this discussion is
unavoidable. The results are however the 
same. For details the reader is referred to
\cite{HuPi1983}. The first result is:
 
\begin{theorem}
\label{3.1}
Let $E$ be any eigenvalue of $H_{0}$, and let
$V_{\ell=2s}$ be an even perturbation. It follows:
\begin{enumerate}
\item
For $g>0$ sufficiently small there exists one and only one eigenvalue
$E(g)$ of $H_{\ell}(g)$ near $E$, $\lim_{g\to 0}E(g)=E$. 
\item The \RSPT\ for $E(g)$ near
$E$ exists to all orders, and represents a full asymptotic expansion
for $E(g)$ as $g\to 0$; namely
\begin{equation}
E(g)-\sum_{n=0}^M a_n \, g^n = \mathrm{O}(g^{M+1}), \quad 
\forall M \in \mathbbm{N}\,.
\end{equation}
Here, $a_n$ are the coefficients of the perturbations
series ($n \in \mathbbm{N}_0$), and $a_0=E$ is the unperturbed eigenvalue.
\item 
The series is divergent as $a_n \sim [(s-1)n]!$ for 
$n \to \infty$. More precisely, there are constants $0<A<B$ such that
\begin{equation}
\label{bounds}
A^n \, \Gamma((s-1)n +1)< a_n < B^n \Gamma ((s-1)n + 1)
\quad \forall n \in \mathbbm{N}_0 
\end{equation}
\end{enumerate}
\end{theorem}

For example, if $s=2$, i.e. $\ell=4$ the series diverges as $n!$,
and if $s=4$ it diverges as $(3n)!$, 
independently of the space dimension.  The statement of the
theorem summarizes a number of contributions. The first proof of
the divergence, in one dimension, 
goes back to Ref.~\cite{BeWu1969}; again in
one dimension, and for nondegenerate unperturbed eigenvalues the bounds
(\ref{bounds}) were proven in Ref.~\cite{Si1982} and in the 
general degenerate case in Ref.~\cite{HuPi1983}. 

The question now arises of the summability of the above
divergent expansions. There is a first sharp distinction between the
two following alternative cases:
\begin{enumerate}
\item The eigenvalues of $H_0$ are stable as eigenvalues of the operator
family $H_{\ell}(g)$ for $g$ small. This means the following: let $E$ be an
eigenvalue of $H_0$, and $\Gamma$ any circumference of center $E$
and  radius $r$ independent of $g$ and smaller than the isolation
distance of $E$ (i.e., the distance between $E$ and the closest different 
eigenvalue).  Then $H_{\ell}(g)$ has exactly one eigenvalue $E(g)$
inside $\Gamma$ for $g\geq 0$ suitably small. The stability implies 
obviously the right continuity of the perturbed eigenvalues at $g=0$: 
$\lim_{g\to 0_+}E(g)=E$. 
\item The eigenvalues of $H_0$ are unstable as eigenvalues of
$H_{\ell}(g)$ for $g$ small. In that case, the eigenvalues of $H_{\ell}(g)$
either disappear into the continuous spectrum, or those inside
$\Gamma$  are at least $2$ for any $g>0$.  Notice that
instability does not a priori prevent continuity of the perturbed
eigenvalues as $g\to 0_+$.
\end{enumerate}
In the stable case the \RSPT\ is Borel summable to the eigenvalues of
order $s$ for all finite $N$. More precisely:

\begin{theorem}
\label{3.2}
Let $V_{\ell=2s}$ be an even perturbation. It follows:
\begin{enumerate}
\item
Any eigenvalue $E$ of $H_0$ is stable for $g\geq 0$ small enough.
\item Each eigenvalue $E(g)$ of $H_\ell(g)$ extends to an analytic function of
$g$ in the sector $|{\rm arg} \, g| < \frac{\pi \, (s+1)}{2}$ in an
$s$-sheeted Riemann surface.
\item For any $\epsilon>0$, there exists some $ B_\epsilon(E)$ so that
$E(g)$ and its Rayleigh-\Sc\
perturbation expansion fulfill the conditions of the Watson-Nevanlinna
criterion (Theorem \ref{WN}) in any sector 
$\{ g\in\C\,:\, |g| < B_\epsilon(E)\,,\,
|{\rm arg} \, g| < \frac{\pi \, (s+1)}{2} - \epsilon \}$. 
As a consequence, by Theorem \ref{WN}:
\item The divergent Rayleigh-\Sc\
perturbation expansion for $E(g)$ is Borel summable to $E(g)$ for
$0 \leq |g| < B_\epsilon(E)$ and 
$|{\rm arg} \, g|<\frac{\pi}{2}-\epsilon$. 
\end{enumerate}

\end{theorem}
{\it Proof.} Assertion 1 was proven by 
Simon~in Ref.~\cite{Si1970}, as well as Assertion 2 for
$N=1$ (providing a rigorous justification of a 
discovery of Bender and Wu). 
Assertion 3 (and hence 4) was proven by 
Graffi and Grecchi in Ref.~\cite{GrGr1978}
for the ground state and all $N$,
and by Hunziker-Pillet~\cite{HuPi1983} for all states
and all $N$. 

Here, $N$ denotes, as before, the dimension of configuration
space.

Observe that the condition $|{\rm arg} \, g|<\frac{\pi}{2}$
in Assertion 4 effectively ensures that ${\rm Re}\,g > 0$ and thus
the stability of perturbation theory.
If $N=1$, so that $H_{\ell}$ reduces to the one-dimensional operator 
$H_{\ell = 2 s}=p^2+q^2+g \, q^{2s}$, and much
sharper results are available.  Namely:

\begin{theorem}
\label{3.3}
Let $N=1$ and $H_{2 s} =p^2+q^2+g \, q^{2s}$. It follows:
\begin{enumerate}
\item For any $s\geq 2$ any eigenvalue $E(g)$ is analytic in the cut plane
$|{\rm arg} \, g|<\pi$.
\item For $s=2,3$, the $[N/N]$ and $[N/N-1]$ Pad\'e approximants of the
\RSPT\ converge to $E(g)$, uniformly on compacts in the cut plane.
\item For $s>3$, the Pad\'e approximants do not converge to $E(g)$.
\end{enumerate}
\end{theorem}

{\it Proof.} Assertion 1 was proven by Loeffel and Martin \cite{LoMa1970}.
Assertion 2 was proven 
by Loeffel, Martin, Simon and Wightman~\cite{LoMaSiWi1969} 
and represents to the
present day the only non-exactly solvable example in which the
convergence of the Pad\'e approximants has been proven. Assertion 3 was
proven by Graffi and Grecchi~\cite{GrGr1975}. 

The proof relies on the verification that, for all $s$,  
the Stieltjes moment problem (see Sec.~\ref{FunStiSer}) generated
by the perturbation
coefficients $(-1)^n \, a_n$ is solvable, i.e. there is a (positive) measure
$\Phi$ on $[0,+\infty)$ such that
\begin{equation}
\label{3.2.2}
a_n = \int_0^{\infty} x^n \, {\rm d}\Phi(x), \qquad n \in \mathbb{N}_0 \, .
\end{equation}
If this is true, all functions
\begin{equation}
\label{3.2.3}
F_{\Phi}(g) = \int_0^{\infty} \frac{{\rm d}\Phi(x)}{1+gx}\,,
\end{equation}
which are analytic in the cut plane $|{\rm arg} \, g|<\pi$, 
admit the \RSPT\
expansion as Taylor expansion at $g=0$ (notice that the example 
${\cal N}(g)$ of Section \ref{borel_general_formulas} 
is a particular case of (\ref{3.2.3}) with 
${\rm d}\Phi(x) = {\rm e}^{-x}\,{\rm d}x$). 
When $s = 2, 3$, the solution
$\Phi(x)$ of the Stieltjes  moment problem (\ref{3.2.2}) is unique. In
that case it is known that 
$F_{\Phi}(g)$ can be expanded in a convergent Stieltjes type continued
fraction
\begin{equation}
F_{\Phi}(g)= 
\frac{b_0}{\displaystyle a_0+ \frac{b_1 \, g}
  {\displaystyle a_1+{\frac{b_2 \, g}
   {\displaystyle a_2+ \frac{b_3 \, g}
    {a_3+\ldots}}}}}
\end{equation}
where  $a_k>0$ and $b_k>0$ for all $k$. The fraction converges in the cut
plane, and the coefficients
$a_k$,
$b_k$ of the continued fraction can be expressed in terms of the
coefficients $a_k$ of the series. Now the even and odd approximants of
the continued fraction (i.e. its even and odd truncations) are just the
$[N/N-1]$ and $[N/N]$ Pad\'e approximants to the perturbation
series in the sense of Sec.~\ref{PadeApproximation}.
By contrast, 
if $s>3$, the moment problem has infinitely many solutions; hence the
continued fraction diverges and so do the Pad\'e approximants.

{\em Practical resummations of energy eigenvalues 
of even anharmonic oscillators.}
We attempt to verify the assertions of Theorems 
(\ref{3.1}), (\ref{3.2}) and (\ref{3.3}) by concrete calculations, 
using the quartic anharmonic oscillator in the normalization
\begin{equation}
\label{even_anharmonic_oscillator_x4}
H(g) = \frac{1}{2} p^2 + \frac{q^2}{2} + g q^4 
\end{equation}
as an example. The Rayleigh--Schr\"odinger 
perturbation expansions of its eigenvalues
\begin{equation}
\label{RS_perturbation_expansion}
E_{N}(g) = N + \frac12 + \sum\limits_{k = 1}^{\infty} A_{N}^{k} g^k \, ,
\end{equation}
leads to well-known perturbation coefficients $A_{N}^{k}$.
E.g., for the ground--state energy $E_{N = 0}(g)$, the first six perturbation
coefficients are given by:
\begin{eqnarray}
   \label{perturbation_coefficuients_quartic_oscillator}
   A_{0}^{1} &=&  \frac{3}{4}  \,, \quad 
   A_{0}^{2}  =  -\frac{21}{8}  \,, \quad 
   A_{0}^{3}  =   \frac{333}{16} \,, \nonumber \\[1ex]
   A_{0}^{4} &=& -\frac{30885}{128} \,, \quad
   A_{0}^{5}  =   \frac{916731}{256} \,, \quad
   A_{0}^{6}  =  -\frac{65518401}{1024}
   \, . 
\end{eqnarray}
These coefficients grow factorially and, hence, the perturbation series 
(\ref{RS_perturbation_expansion}) for the ground state $N$ = 0 of the quartic
anharmonic oscillator is divergent. For instance, as seen 
from the second column of the Table \ref{table_quartic_oscillator},
we can verify that a straightforward term--by--term summation of the series 
(\ref{RS_perturbation_expansion}) at 
strong coupling $g$ = 1.0 fails to reproduce 
the exact ground-state energy. In order to find this energy, therefore, more
powerful resummation techniques have to be used. As it has been proven
in Theorem \ref{3.3}, for example, one may
resum the Rayleigh--Schr\"odinger perturbation series by means
of Pad\'{e} approximants. That is, by making use of the first
$\Ent{(n+1)/2}$ terms of the series
(\ref{RS_perturbation_expansion}), we may construct
the Pad\'{e} approximants
$\bigg[ \Ent{n/2} \bigg/ \Ent{(n+1)/2} \bigg]_{E_{\rm 0}}\!\!\!\left(g\right)$ 
(we here choose
the lower-diagonal ones, as the higher degree of the denominator 
polynomial usually leads to a little more favourable numerical results).
As seen from the third column of the Table \ref{table_quartic_oscillator},
the application of the Pad\'e approximant leads to
rather slow convergence. For instance, for $n = 54$,
the Pad\'e approximant reproduces only four significant digits
of the ground state energy of the quartic potential. 

In fact, a much
faster convergence can be achieved by combining the Pad\'e approximant
with the Borel summation method. Since the concept of this
combined Borel--Pad\'e method has already been discussed in 
Sec.~\ref{subsubsection_borel_pade}, here we just recall its basic
formulas. That is, the Borel transform of the perturbation series
(\ref{RS_perturbation_expansion}) is defined by
\begin{eqnarray}
\label{fBorel_quartic}
B(u)_{E_0} & = & \sum_{k=0}^{\infty} \frac{A_{N=0}^k}{k!} \, u^k\, ,
\end{eqnarray}
and later used to construct the Pad\'{e} approximants 
\begin{equation}
\label{pade_from_borel_quartic}
{\mathcal P}_n(u) = \bigg[ \Ent{n/2} \bigg/
\Ent{(n+1)/2}
\bigg]_{B_{E_0}}\!\!\!\left(u\right)\,,
\end{equation}
where $\Ent{x}$ is defined in 
Sec.~\ref{Sec:SpecSymSpecFun}. The resummation is accomplished then by
evaluation of the (modified)
Borel integral along the integration contour $C_{0}$ 
introduced in~\cite{Je2000prd}:
\begin{equation}
  \label{contours_quartic}
  {\mathcal T}\!{\mathcal P}_n(g) = \frac{1}{g} \, \int_{C_{0}} {\rm d}u \, 
  \exp(-u/g)\,{\mathcal P}_n(u) \,.
\end{equation}
For this contour, the final result involves no imaginary part. 

\begin{table}[t]
\caption{\label{table_quartic_oscillator} Resummation of the 
asymptotic series (\ref{RS_perturbation_expansion}) taken at $g$ = 1.0 
for the ground--state energy of the quartic anharmonic 
oscillator (\ref{even_anharmonic_oscillator_x4}).
In the first row, we indicate the quantity $n$ which is the order of the 
perturbation expansion. In the second row, we list the partial sums 
$s_n = \sum^n_{k=0} A_{N=0}^{k} g^k $. In the third row, we list the Pad\'{e}
approximants of the perturbation series,
whereas in the third row, the Borel--Pad\'{e} transformation 
is presented [cf. Eqs.~(\ref{fBorel_quartic})--(\ref{contours_quartic})].}
\vspace*{0.3cm}
\begin{center}
\begin{tabular}{crrr}
\hline
\hline
\multicolumn{1}{c}{\rule[-3mm]{0mm}{8mm} n}      &       
\multicolumn{1}{c}{$\sum^n_{k=0} A_{0}^{k} g^k$} &
\multicolumn{1}{c}{ $\bigg[ \Ent{n/2} \bigg/
   \Ent{(n+1)/2}
   \bigg]_{E_{0}}\!\!\!\left(g\right) $ } &
\multicolumn{1}{c}{${\mathcal T}\!{\mathcal P}_n(g)$}
\\ \hline
$  1    $    & 
$ 1.250~000~000 \times 10^{0} $ &
$-1.000~000~000 \times 10^{0} $ &   
$ \underline{}
  1.657~119~541 \times 10^{-1}   $ \\

$  2    $    &     
$-1.375~000~000 \times 10^{0}  $ &
$ 6.666~666~667 \times 10^{-1} $ &     
$ \underline{}
  7.197~424~642  \times 10^{-1}  $ \\

$  3    $    &     
$ 1.943~750~000 \times 10^{+1} $ &
$ 1.542~372~881 \times 10^{0}  $ &     
$ \underline{}
  9.643~165~557 \times 10^{-1} $ \\

$  4    $    &    
$-2.218~515~625 \times 10^{+2} $ &
$ 7.338~453~886 \times 10^{-1} $ &     
$ \underline{}
  7.889~653~148  \times 10^{-1}$ \\

$  5    $    &     
$ 3.359~128~906  \times 10^{+3} $  &
$ 9.925~839~156  \times 10^{-1} $ &     
$ \underline{
  8}.122~744~488 \times 10^{-1}$ \\

 $ ...    $    &     $
...$     
&
...
&     $ \underline{}
... $ \\

$ 49    $    &    
$ 8.809~956~028  \times 10^{+84} $ &
$ \underline{
  8.03}8~179~244  \times 10^{-1} $ &
$ \underline{
  8.037~706~51}4 \times 10^{-1}$ \\

$ 50    $    &     
$-1.309~231~884 \times 10^{+87} $ &
$ \underline{
 8.037}~397~565  \times 10^{-1} $ &     
$ \underline{
 8.037~706~51}1 \times 10^{-1}$ \\

$ 51    $    &     
$ 1.984~872~788 \times 10^{+89} $ &
$ \underline{
8.03}8~080~847  \times 10^{-1} $ &
$ \underline{
  8.037~706~51}3 \times 10^{-1}$ \\

$ 52    $    &     
$-3.068~686~346 \times 10^{+91} $ &
$ \underline{
  8.037}~459~721   \times 10^{-1} $ &     
$ \underline{
  8.037~706~512} \times 10^{-1}$ \\

$ 53    $    &     
$ 4.836~297~199 \times 10^{+93}$ &
$ \underline{
8.03}8~004~311  \times 10^{-1}$ 
&     
$ \underline{
8.037~706~512} \times 10^{-1}$ \\

 $ 54    $    &     $
-7.767~069~326 \times 10^{+95}$     
&
$ \underline{
8.037}~508~557  \times 10^{-1}$
&     $ \underline{
8.037~706~512} \times 10^{-1}$ \\

\hline 
\rule[-3mm]{0mm}{8mm}
exact   & $8.037~706~512 \times 10^{-1}$ &
  $8.037~706~512 \times 10^{-1}$ & $8.037~706~512 \times 10^{-1}$\\
\hline
\hline
\end{tabular}
\end{center}
\end{table}

By making use of Eqs.~(\ref{fBorel_quartic})--(\ref{contours_quartic}), we are
now able to carry out a resummation of the perturbation series
(\ref{RS_perturbation_expansion}) for the ground state of the quartic harmonic
oscillator. The results of such a resummation, performed for the coupling 
parameter $g$ = 1.0 are presented in the third column of the Table 
\ref{table_quartic_oscillator}. As seen from this Table, the 
transform ${\mathcal T}\!{\mathcal P}_n(g)$ of 53rd order allows 
us to reproduce the exact ground--state energy up to 10 decimal digits 
demonstrating that with the Borel--Pad\'{e} resummation method one can use 
the perturbation series at large coupling parameters $g$. 

\begin{figure}[tb!]
\begin{center}
\begin{minipage}{12.0cm}
\begin{center}
\includegraphics[width=0.8\linewidth]{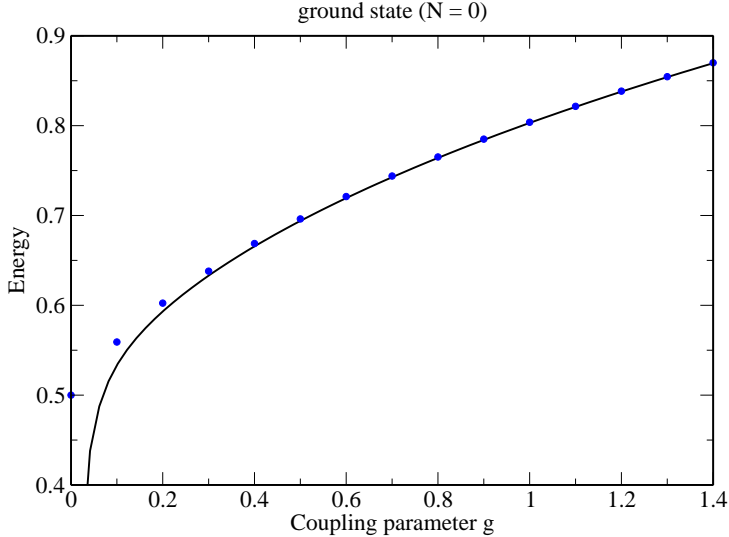}
\end{center}
\caption{\label{PM_scaling} 
Exact values (filled circles) for the ground state 
energy of the quartic anharmonic oscillator as a function of $g$, 
together with the partial sum of the strong--coupling expansion 
(solid line) as defined by the 
first three nonvanishing terms in powers of $ g^{-2k/3}$
[see Eq.~(\ref{strong_coupling_expansion})].}
\end{minipage}
\end{center}
\end{figure}

As mentioned above, calculations presented in Table 
\ref{table_quartic_oscillator} have been carried out for the coupling 
parameter $g = 1.0$. While this parameter already corresponds to the 
large--coupling regime, one often needs to study the energy spectrum 
of the anharmonic oscillator at even larger couplings. In place of a
Borel resummation of the perturbation expansion
(\ref{RS_perturbation_expansion}), for such (very) strong coupling $g$ it is 
usually convenient to perform the so--called Symanzik scaling 
$q \mapsto g^{-1/6} q$ in the 
Eq.~(\ref{even_anharmonic_oscillator_x4}) and rewrite 
the quartic potential into 
another one with the same eigenvalues but a fundamentally different structure
\begin{eqnarray}
\label{scaled_hamiltonian}
H(g) = g^{1/3} 
\left[ H_{\rm L} +  \frac{q^2}{2} g^{-2/3}
\right] \, ,
\end{eqnarray}
where the large-coupling Hamiltonian $H_{\rm L}$ does not depend on $g$:
\begin{eqnarray}
\label{HL}
H_{\rm L} = -\frac{1}{2} \left( \frac{d}{dq} \right)^2 + q^4 \, .
\end{eqnarray}
As seen from these Equations, the eigenvalues of the quartic Hamiltonian
for the $g \rightarrow \infty$ are determined in leading order by the 
eigenvalues $E_{N}^{(\rm L)}$ of the ``unperturbed'' Hamiltonian (\ref{HL}):
\begin{equation}
\label{leading_asymptotics}
E_{N}(g) \approx g^{1/3} E_{N}^{(\rm L)} \, .
\end{equation}
An even more accurate estimate of the eigenvalues $E_{N}(g)$ for 
large $g$ can be obtained if apart from the leading term, 
higher--order terms are included in
asymptotics (\ref{leading_asymptotics}). In order to compute these 
higher--order terms we may apply 
standard Rayleigh--Schr\"odinger perturbation theory
to Eq.~(\ref{scaled_hamiltonian}) and use the fact that the
``perturbation'' $V(g) = q^2 g^{-2/3}/2$ is 
Kato--bounded with respect to the unperturbed Hamiltonian $H_{\rm L}$ for 
large $g$. Within such an approach, the convergent 
strong--coupling perturbation expansion can be
written for each energy $E_{\rm N}(g)$:
\begin{equation}
\label{strong_coupling_expansion}
E_{N}(g) = g^{1/3} \sum\limits_{k = 0}^{\infty} L_N^k \, g^{-2k/3} \,.
\end{equation}
Here, we have $L_{N}^0 \equiv E^{\rm (L)}_{N}$, and the higher perturbation
coefficients $L^{k}_{N > 0}$ can be calculated to high numerical
accuracy in the basis of the wavefunctions
of the unperturbed Hamiltonian (\ref{HL}). As seen from 
Fig.~\ref{PM_scaling}, the asymptotic expansion 
(\ref{strong_coupling_expansion}) as defined by first three 
nonvanishing terms $L_0^0$, $L_0^1$ and $L_0^3$ is able to
reproduce with a good accuracy the exact energy of the ground state of quartic
Hamiltonian (\ref{even_anharmonic_oscillator_x4}) starting with 
relatively moderate coupling of $g = 0.4$. 

%
%

%
%
\subsection{Odd Anharmonic Oscillators}
\label{DivMATodd}

{\em Mathematical Discussion of odd anharmonic oscillators.}
The odd anharmonic oscillators in dimension $N$ (or,
equivalently, in $N$ degrees of freedom) are by definition the
anharmonic oscillators such that the potential $V_\ell$ in the
Hamiltonian (\ref{Hclass}) is a polynomial in $N$ variables of odd degree
$2s+1$, $s=1,2,\ldots$. The most famous example is probably the potential
$V \equiv q_1^2 \, q_2$, yielding the two-dimensional 
H\'enon-Heiles Hamiltonian
\begin{equation}
\label{3.4.1}
{\cal H} = \frac12 \, \left(p_1^2 + p_2^2\right) + 
\omega_1^2 \, q_1^2 + \omega_2^2\, q_2^2 + g \, q_1^2 \, q_2\,,
\end{equation}
which represents one of the fundamental models for the onset of chaos
in classical Hamiltonian mechanics.

The basic difference to the even case is that here constant energy
surfaces are in general not compact. Then the classical motions reach
infinity when the initial conditions belong to open regions in phase
space, and actually do it in a finite time. The quantum aspect of
this mechanical property is the lack of self-adjointness of the
corresponding \Sc\ operator, and consequently the lack of uniqueness of 
quantum dynamics. 
To better understand this point, consider the simplest
one-dimensional case, the cubic anharmonic oscillator, whose
Hamiltonian is (in an appropriate normalization):
\begin{equation}
\label{3.4.2} 
H_{\ell = 3} = p^2 + q^2 + g \, q^3 \,.
\end{equation}
The potential $V_{\ell=3}(q) = q^2+g \, q^3$ has a minimum at 
$q = 0$ with value $V_3(0) = 0$ and a
maximum at $q=-2/3g$ with value $V_3(-2/3g) = \frac{4}{27g^2}$. Therefore,
all classical trajectories 
with initial conditions fulfilling
$E > \frac{4}{27g^2}$ reach $-\infty$ in a
finite time (with at most one inversion of velocity) because the time
required is well known to be
\begin{equation}
t = \int_{q_0}^{-\infty}\frac{dq}{\sqrt{E-q^2-g \, q^3}}
\end{equation}
and this integral is convergent. Under these conditions (the classically
incomplete case: see e.g.~\cite{ReSi1978}) the corresponding \Sc\ operator  
$p^2+q^2+g \, q^3$ is not essentially self-adjoint because there are
infinitely many different ways in which it can be
realized as a self-adjoint operator in the Hilbert space $L^2(\R)$.
On the other hand the \RSPT\ exists.
Actually the perturbation theory of the cubic anharmonic oscillator
represents the very first problem considered by Heisenberg with matrix
mechanics: see e.g. Ref.~\cite{vW1968},
and the first obvious question is
to isolate the quantity to associate with it. 

We consider this problem only in one dimension, and for a general
interaction of the type $q^{2s+1}$,  i.e.~we consider the \Sc\ operators
\begin{equation}
\label{3.4.3} 
H_{2s + 1}(g) \equiv  p^2+q^2+g \, q^{2s+1}, \quad s=1,2,3\ldots,
\end{equation}
The lack of self-adjointness for
$g\neq 0$ entails a fortiori the instability of the harmonic oscillator
eigenvalues. The asymptotic power series 
$E(g) \sim \sum_{n=0}^{\infty} a_n \, g^n$  generated by
\RSPT\ near any unperturbed eigenvalue $E=a_0$ has indeed the
following properties:
\begin{enumerate}  
\item It contains only even powers; namely, $a_{2n+1}=0$. This
is because the unperturbed operator is even under the
parity operation $(P\psi)(q)=\psi(-q)$, and the perturbation is
odd.
\item The nonzero coefficients $a_{2n}$ have constant sign,
the perturbation series is nonalternating. 
\item The even coefficients behave as 
$a_{2n} \sim (sn)!$ for $n \to \infty$.
\end{enumerate}
The proof is given in Ref.~\cite{CaGrMa1980}.
If the series has only even powers of constant sign, the series 
$\sum_{n=0}^{\infty}a_n \, ({\rm i}\, g)^n$ has alternating signs. 
It is thus
conceivable that, in examining the summability properties of \RSPT, the
natural starting point should be the examination of the operator
$H_{2s + 1}(g)$ for $g$ complex instead of $g$ real.
One has indeed 

\begin{theorem}
\label{3.4} 
Let $H_{2s + 1}(g)$ be the Hamiltonian of an odd anharmonic 
oscillator as discussed. It follows:
\begin{enumerate}
\item If $|{\rm arg} \, \beta|<\frac{\pi}{2}$, the operator 
$H_{2s + 1}({\rm i}\, \beta)$
can be realized in a unique way as a closed operator in $L^2(\R)$ with
discrete spectrum; moreover $(H_{2s + 1})^{\ast}({\rm i}\,\beta) =
H_{2s + 1}(-{\rm i} \, \overline\beta)$, where $\overline\beta$ denotes
the complex conjugate of $\beta$. 
\item The eigenvalues of the unperturbed operator $H_0$ are stable as
eigenvalues $E({\rm i}\,\beta)$ of the argument ${\rm i}\,\beta$ for 
$|\beta|$ small and $|{\rm arg} \, \beta| < \frac{\pi}{2}$; 
that is, if $E_0$ is any eigenvalue of
the one-dimensional harmonic oscillator (and hence simple) there is
$B(E_0)>0$ such that $H({\rm i}\,\beta)$ has exactly one eigenvalue 
$E({\rm i}\,\beta)$ near $E_0$ for $|\beta| < B(E_0)$  and 
$|{\rm arg} \, \beta| < \frac{\pi}{2}$.
\item Any eigenvalue $E({\rm i}\,\beta)$ is a holomorphic function 
of $\beta$ in the region
$\{\beta:\,\,0<|\beta|<B(E_0)\,;\,|{\rm arg} \, \beta|<\frac{\pi}{2}\}$;
the function $E({\rm i}\,\beta)$ admits an analytic continuation to any sector 
$S_\delta=\{\beta\,:\,0<|\beta|<B_\delta(E_0)\,;\,
-\frac{\pi \, (2s-1)}{8} + \delta < {\rm arg} \, \beta <
\frac{\pi \, (2s+7)}{8}-\delta\}$, 
$\forall\,\delta >0$ on a $(2s+3)$-sheeted Riemann surface. 
\item The function $E({\rm i}\,\beta)$ and the 
Rayleigh-\Sc\ perturbation series
fulfill the conditions of the Watson-Nevanlinna Theorem \ref{WN} 
in any sector 
$\{\beta\,:\, 0 \leq |\beta| < B_\delta(E_0) \,,\,
-\frac{\pi (2s-1)}{8} + \delta< {\rm arg} \, \beta < 
\frac{\pi (2s+7)}{8} - \delta\}$, $\forall \delta>0$.
\item The Rayleigh-\Sc\ perturbation series is Borel summable to
$E({\rm i}\,\beta)$ in any sector 
$\{\beta\,:\,0\leq |\beta| < B_\delta(E_0),\,\,
|{\rm arg} \, \beta| < \frac{\pi}{2} - \delta\}$, $\forall \delta>0$ or
equivalently, to $E(g)$ in any sector
$\{|g|<B_\delta(E_0)\,;\,\delta <
{\rm arg} \, g < \pi-\delta\}$, if we set $g={\rm i}\,\beta$. 
\end{enumerate}
\end{theorem}
{\it Proof.} These results are proven in~\cite{CaGrMa1986,CaGrMa1993}. 

The above results clarify the meaning of the perturbation series as long as
the summation is performed along nonreal directions; if the Borel sum is
continued to $g$ real the result is a complex function. The question
thus arises of establishing  the relation between this complex-valued
function and the real-valued perturbation series, and of finding the
meaning of the imaginary part. The physical intuition
is that this complex valued function should represent a resonance of
the problem, even though there is no continuous spectrum involved: the
real part should be the position, and the imaginary part the width,
because it  should be proportional to the penetration of the barrier.
The  distributional Borel summability puts on a rigorous basis this
intuition. Namely we have:

\begin{theorem}
\label{3.5}
Let $E(g)$ be the analytic continuation to $\Im g=0$ of the
function $E(g)$, which exists by the above Theorem. Then:
\begin{enumerate}
\item
$E(g)$ and the perturbation series $\sum_{n=0}^{\infty}a_n \, g^n$
fulfill the conditions of the distributional Nevanlinna-Leroy 
criterion (see Theorem~\ref{T.5.1} and~Definition~\ref{defDNL}
in Appendix B)
of order $\frac{2s-1}{2}$. 
\item The divergent perturbation series $\sum_{n=0}^{\infty} a_n \, g^n$
is Borel summable in the distributional sense of order $s$ to $E(g)$
for $|g|$ suitably small. Moreover, $\Re E(g)$ is even in $g$ and
$\Im E(g)$ is odd.
\end{enumerate}
\end{theorem}
{\it Proof.} See Ref.~\cite{Ca2000}.

A few remarks are in order.
(i)~Equation~(1.6) of Ref.~\cite{Ca2000} clarifies 
the role of the standard WKB analysis to the barrier penetration
and the imaginary part $\Im E(g)$, in the context of 
odd anharmonic oscillators.
(ii)~The analogue of Theorem \ref{3.5} for the 
H\'{e}non-Heiles Hamiltonian (3.4.1) and 
for the ground state energy level of the harmonic oscillator 
has been recently proven in Ref.~\cite{Ca2006}.
(iii)~Consider the odd oscillators for purely imaginary values of the
coupling $g$, i.e.~the operators $H({\rm i}\,\beta)$, 
$\beta\in\R$. Let $P$ be
the standard parity operation: $(P\psi)(x)=\psi(-x)$, and $T$ the
time-inversion operation, equivalent, as we know, to complex
conjugation:  $(T\psi)(x)=\overline{\psi}(x)$. $P$ and $T$ are
unitary operations, and so is their product $PT=TP$. Now it is
immediate to check that 
\begin{equation}
[H({\rm i}\,\beta),PT]=[H({\rm i}\,\beta),TP]=0 \,,
\end{equation}
i.e.~the odd anharmonic oscillators are examples of $PT$-symmetric
operators. It has been conjectured around 1985 by Bessis and
Zinn-Justin that the spectrum of the odd $PT$-symmetric anharmonic
oscillators should consist of real eigenvalues for all values of
$\beta\in\R$, even if the operator
$H(\beta)$ is obviously not self-adjoint and not even normal because it
is easily verified that
$[H({\rm i}\,\beta),H({\rm i}\,\beta)^{\ast}]= 
[H({\rm i}\,\beta),H(-{\rm i}\,\beta)]\neq 0$. Note that by
the Borel summability the reality of the spectrum
is valid for $\beta$ small
(depending on the eigenvalue index $k$) because the eigenvalues of the
harmonic oscillators are stable for $\beta$ small and the perturbation
series  for each eigenvalue is real when $g={\rm i}\, \beta$,
$\beta\in\R$. Notice that in this general setting, the non-normality 
makes also the problem of the existence of eigenvalues
a nontrivial problem in situations where perturbation theory cannot be
applied, such as the one obtained for $\beta$ large. 
Later Bender (Ref.~\cite{Be2004}) conjectured that
$PT$-symmetric Hamiltonians should always have real spectrum, so that
$PT$-symmetry could be considered a condition equivalent to
self-adjointness in ensuring the reality of the spectrum as long as 
\Sc\ operators are concerned. This latter conjecture has been 
disproven~\cite{CaGrSj2005},
but the original Bessis-Zinn-Justin conjecture has been recently proven
by Shin~\cite{Sh2002cmp} 
(see also~Refs.~\cite{DoDuTa2001i,DoDuTa2001ii,DoDuTa2005}).

\begin{theorem}
\label{3.6}
Let $H_{2s+1}(g)$ be the Hamiltonian of the one-dimensional
odd anharmonic oscillator. 
For any $s\in\N$ and any $g={\rm i}\,\beta$, 
$\beta\in\R$, the spectrum of the operator
$H_{2s+1}(g)$ consists of countably many simple positive eigenvalues
$E_N({\rm i}\,\beta)$.
\end{theorem}
{\it Proof.} See Shin~\cite{Sh2002cmp}. 

Finally, we note that similar results as for anharmonic oscillators
hold for the resonances of the Stark effect, see Appendix~\ref{appdiststar}.

{\em Practical resummations and
complex scaling of eigenvalues for the cubic potential.}
As for the even anharmonic oscillators, we now supplement the 
mathematically inspired analysis by some concrete calculations.

Before starting the numerical 
analysis of the cubic potential, however, we may mention 
that the problem of ``recovering'' the complex resonance energies
from a nonalternating perturbation series 
has been discussed for the Stark effect
in~\cite{FrGrSi1985,Je2001pra} and for the quantum 
electrodynamics (QED) effective action in~\cite{Je2000prd}. 
For the Stark effect, for example, the situation is as follows:
the problem is spectrally unstable: as soon as the electric field is
turned on, all eigenvalues of the hydrogen atom disappear in the 
continuous spectrum (a phenomenon first proven by Titchmarsh~\cite{Ti1957}).
Thus the perturbation theory is always divergent. The physical
reason is that the linear potential of the electric field generates a tunneling
phenomenon. As first pointed out by Oppenheimer~\cite{Op1928}, the hydrogen
eigenvalues are thus expected to turn into resonances: their width (inverse
life-time) should be proportional to the probability 
of penetrating the barrier. 
In the 1970s and 1980s, this picture has been set on
rigorous mathematical grounds: moreover,
the perturbation theory of the problem
has fully been understood. Not only it can be uniquely associated with the 
resonances via ordinary and distributional Borel summability, but it defines
very effective approximation schemes for computing both the location and
the width of the resonances.

Here, we can rely on the literature for the Stark effect 
and for the QED effective action and treat the conceptually 
simpler case of the cubic anharmonic oscillator as an example.
Calculations for such a potential
will be performed in order to illustrate how to derive the complex
resonance eigenvalues 
(including the width) from a real perturbation series using Borel 
summation in complex directions of the parameters, in the sense of
distributional Borel summability. 
We discuss the problem of determination of the complex
resonance energies for the case of the cubic oscillator,
in the normalization:
\begin{equation}
\label{cubic_anharmonic_oscillator}
H(g) = \frac{p^2}{2} + \frac{q^2}{2} + \sqrt{g} \, q^3 \, . 
\end{equation}
The energy spectrum of the anharmonic 
oscillator can be analyzed numerically, e.g.~via 
a diagonalization of the 
Hermitian matrix within a (large) basis of harmonic oscillator 
wavefunctions. One has to take into account, though, that
the bound states of the (odd) harmonic oscillator become 
resonances, labeled by an index $N$ according to the 
above Theorems~\ref{3.4} and~\ref{3.5}:
\begin{equation}
\label{energy_cubic_oscillator}
E_{N}(g) = {\rm Re} \, E_{N}(g) - {\rm i} \, \frac{\Gamma_{N}(g)}{2} \, ,
\end{equation}
where $\Gamma_{N}(g)$ is the width. An appropriate ansatz for studying the 
energy spectra of the odd anharmonic oscillators has been proposed in 
Refs.~\cite{BaCo1971} and~\cite{YaEtAl1978} and is based on the 
complex (or so-called dilation) transformation of the coordinate $q$:
\begin{equation}
\label{complex-scaling_q}
q \mapsto q \, {\rm e}^{{\rm i} \, \theta} \, ,
\end{equation}
where $\theta$ may be real or complex. By inserting this transformed 
coordinate into Eq.~(\ref{cubic_anharmonic_oscillator}), we may obtain 
the complex-scaled Hamiltonian for the cubic oscillator:
\begin{eqnarray}
\label{cubic_anharmonic_oscillator_scaled}
H(g, \theta) &=& - \frac{1}{2} \frac{d^2}{dq^2} {\rm e}^{-2 {\rm i} \theta} 
+ \frac{q^2}{2} {\rm e}^{2 {\rm i} \theta} 
+ \sqrt{g} \, q^3 {\rm e}^{3 {\rm i} \theta} \nonumber \\
&=& {\rm e}^{-2 {\rm i} \theta}
\left( - \frac{1}{2} \frac{d^2}{dq^2} 
+ \frac{q^2}{2} {\rm e}^{4 {\rm i} \theta} 
+ \sqrt{g} \, q^3 {\rm e}^{5 {\rm i} \theta} \right) \, .
\end{eqnarray}
As discussed in Ref.~\cite{YaEtAl1978}, 
by finding the eigenvalues of this dilationally
transformed Hamiltonian one may find that when $\theta$
becomes large enough, one of the rotated branches passes the position of a resonance
pole, and a new eigenvalue of the Hamiltonian 
(\ref{cubic_anharmonic_oscillator_scaled}) appears at the position 
of the resonance pole. This resonance eigenvalue, once uncovered, 
remains fixed in the complex plane
and is independent of further increases in $\theta$.
Hence, the evaluation of the eigenvalues of the Hamiltonian 
(\ref{cubic_anharmonic_oscillator_scaled}) allows us to determine the (complex)
energies of the original cubic Hamiltonian (\ref{cubic_anharmonic_oscillator}).

Now we are ready to calculate the energy values of the complex--transformed 
(\ref{cubic_anharmonic_oscillator_scaled}) and, hence, of the original
cubic Hamiltonian (\ref{cubic_anharmonic_oscillator}). Again, these 
calculations can be performed via matrix diagonalization in the basis 
of harmonic oscillator wavefunctions. The result of such diagonalization
procedure for a coupling parameter $g = 0.01$ is presented in the 
last line of the Table \ref{table_cubic} and is found in perfect
agreement with previous calculations \cite{AlCa2000}. 

We not study the resummation of the perturbation series 
within the complex plane. In order to discuss this resummation
procedure let us again recall that within the standard 
Rayleigh--Schr\"odinger perturbation approach, the 
eigenvalues of the cubic potential are given by the series
\begin{equation}
\label{RS_perturbation_expansion_cubic}
E_{N} = N + \frac12 + \sum\limits_{k = 1}^{\infty} A_{N}^{k} g^k \, ,
\end{equation}
where the first six perturbation coefficients for the ground 
state $N$ = 0 are:
\begin{eqnarray}
\label{perturbation_coefficuients_cubic_oscillator}
A_{0}^{1} &=& -\frac{11}{8}  \,, \quad 
A_{0}^{2}  =  -\frac{456}{32}  \,, \quad 
A_{0}^{3}  =  -\frac{39709}{128} \,, \nonumber \\[1ex]
A_{0}^{4} &=& -\frac{19250805}{2048} \,, \quad
A_{0}^{5}  =  -\frac{2944491879}{8192} \,, \quad
A_{0}^{6}  =  -\frac{1075012067865}{65536} \, .
\end{eqnarray}
Similar to the case of the quartic oscillator, this series is
divergent and, hence, its term--by--term summation  
cannot reproduce
the energy values of the cubic potential. Apart from such a
non--summability, moreover, another problem should be mentioned. 
Namely, no complex part (i.e. width) of the eigenvalues can be
obtained by straightforward summation of 
Eq.~(\ref{RS_perturbation_expansion_cubic}). We may calculate
these widths, however, if we apply the Borel--Pad\'{e} resummation 
method with an integration contour deformed into the complex plane
as discussed in 
Eqs.~(\ref{fBorel_quartic})--(\ref{contours_quartic}). For such
an integration, however, we have to choose the proper contour
in the complex plane in order to reproduce the correct sign for
the imaginary part of the energy. In our calculations for 
the cubic potential, for example, the contour $C_{-1}$ has 
been chosen because it leads to the correct negative 
sign of the imaginary part of the resonance energy eigenvalue.
Notice that the distributional Borel summability also does
not fix the sign of imaginary part on the basis of the purely
real perturbation series (see Theorem~\ref{3.5}), and the choice
of the contour has to be inserted based on an ``external,''
physical consideration.
As seen from the third column of the Table \ref{table_cubic},
the Borel--Pad\'{e} method allows us
to reproduce the complex resonance energy of the ground state
of cubic potential for a coupling parameter $g = 0.01$.
 
\begin{table}[t]
\caption{\label{table_cubic} Resummation of the 
asymptotic series (\ref{RS_perturbation_expansion}) taken at $g = 0.01$ 
for the ground-state
energy of the cubic anharmonic oscillator (\ref{cubic_anharmonic_oscillator}).
In the first row, we indicate the quantity $n$ which is the order of the 
perturbation expansion. In the second row, we list the partial sums 
$s_n = \sum^n_{k=0} A_{0}^{k} g^k $ of the purely real perturbation 
theory, whereas in the fourth row, the Borel--Pad\'{e} transformation 
is presented [cf. Eqs.~(\ref{fBorel_quartic})--(\ref{contours_quartic})].}

\begin{center}
\begin{tabular}{crr}
\hline
\hline
\multicolumn{1}{c}{\rule[-3mm]{0mm}{8mm} n}      &       
\multicolumn{1}{c}{$ \sum^n_{k=0} A_{0}^{k} g^k $} &
\multicolumn{1}{c}{${\mathcal T}\!{\mathcal P}_n(g)$}
\\ 
\hline
 $  1    $    &     $
 \underline{0.48}6~250~000$ & 
                    $ 
 \underline{0.48}6~949~910 - 
 \underline{0.000~00}0~000 \, {\rm i}$ \\

 $  2    $    &     $
 \underline{0.484}~796~875$ & 
                    $ 
 \underline{0.484}~497~987 - 
 \underline{0.000~00}0~093 \, {\rm i}$ \\
 $  3    $    &     $ 
 \underline{0.484}~486~648$ & 
                    $ 
 \underline{0.484~3}41~583 - 
 \underline{0.000~00}4~008 \, {\rm i}$ \\
 $  4    $    &     $ 
 \underline{0.484~3}92~650$ &     
                    $ 
 \underline{0.484~315}~106 - 
 \underline{0.000~0}10~069 \, {\rm i}$ \\
 $  5    $    &     $ 
 \underline{0.484~3}56~707$ &     
                    $ 
 \underline{0.484~315}~279 - 
 \underline{0.000~00}9~966 \, {\rm i}$ \\
 $  ...    $    &     $ \underline{}
 ...$     &     $       \underline{}
 ...$ \\
 $  10    $    &    $ 
 \underline{0.484~3}20~462$ &     
                    $ 
 \underline{0.484~31}6~006 - 
 \underline{0.000~008~0}57 \, {\rm i}$ \\
 $  11    $    &    $ 
 \underline{0.484~31}8~395$ &     
                    $ 
 \underline{0.484~31}6~001 - 
 \underline{0.000~008~0}52 \, {\rm i}$ \\
 $  12    $    &    $ 
 \underline{0.484~31}6~573$ &     
                    $ 
 \underline{0.484~315~99}9 - 
 \underline{0.000~008~0}58 \, {\rm i}$ \\
 $  13    $    &    $ 
 \underline{0.484~31}4~836$ &     
                    $ 
 \underline{0.484~315~99}8 - 
 \underline{0.000~008~0}58 \, {\rm i}$ \\
 $  14    $    &    $ 
 \underline{0.484~31}3~053$ &     
                    $ 
 \underline{0.484~315~99}9 - 
 \underline{0.000~008~0}59 \, {\rm i}$ \\
 $  15    $    &    $ 
 \underline{0.484~31}1~093 $ &     
                    $ 
 \underline{0.484~315~997} - 
 \underline{0.000~008~060} \, {\rm i}$ \\
 $  16    $    &    $ 
 \underline{0.484~3}08~792$ &     
                    $
 \underline{0.484~315~997} - 
 \underline{0.000~008~060} \, {\rm i}$ \\
\hline 
\rule[-3mm]{0mm}{8mm}
exact   & $0.484~315~997$ &
  $0.484~315~997 - 0.000~008~060 \, {\rm i}$ \\
\hline
\hline
\end{tabular}
\end{center}
\end{table}

%
%
\subsection{Double--Well and Fokker--Planck Potential}
\label{DivMATdwf}

{\em Mathematical discussion of double-well like potentials.}
Let us restrict ourselves
to the discussion of the one-di\-mensional case. 
The simplest example is
represented by the symmetric double-well Hamiltonian,
which reads (in an appropriate normalization)
\begin{equation}
\label{I4bis}
H^{\rm DW}(g)=p^2+q^2(1 - g \, q)^2
\end{equation}
The translation $q\mapsto q+1/2g$ maps the Hamiltonian to the 
form $p^2 + g^2(q+1/2g)^2(q-1/2g)^2$ which is manifestly an
even function with two symmetric quadratic minima at $q=\pm 1/2g$.
The Hamiltonian (\ref{I4bis}) has minima at $q=0$ and $q=1/g$ and
the second one disappears as $g\to 0$.

Within even anharmonic oscillators examples exhibiting
spectral instability have a special status and a typical examples 
is just the double-well oscillator. 
As already mentioned, the basic phenomenon in this context is the
spectral instability, or, equivalently,  asymptotic degeneracy of the
eigenvalues. For the symmetric double well, the well known physical
picture goes as follows (see e.g.~\cite{LaLi1958}): in classical
mechanics the two wells are identical, and therefore generate the same
motions, in the sense that the phase curves in the two separate wells
are the same. Quantum mechanics just adds to this picture the tunneling
phenomenon.  Therefore we may expect to find, for $g$ small,
a doublet of eigenvalues of the double well operator $H^{\rm DW}(g)$
close to each unperturbed eigenvalue $E$,
and therefore $E$ is unstable. The first mathematical proof of this
picture is due to Harrell~\cite{Ha1978} who showed that 
as $|g|\to 0$, there are exactly two eigenvalues 
$E_{\pm}(g)$ of $H^{\rm DW}(g)$ near 
any unperturbed  eigenvalue $E$, and that the 
Rayleigh-Schr\"odinger perturbation expansion is an asymptotic
expansion for both eigenvalues of the doublet.

As we have seen, to understand the meaning of perturbation theory,
it is in general
necessary to understand the analyticity property of the eigenvalues for
complex values of the coupling constant $g$. In the case of 
only a single well, 
this is possible thanks to the stability of the eigenvalues even for
complex values of the coupling constant. It is remarkable that also in
the unstable case of the double-well,
one can identify and construct a pair of operators
$H_{\pm}(g)$, the so-called single well operators which admit
$E_{\pm}(g)$ as stable eigenvalues as $g\to 0$ in a suitable
complex region. The operators $H_+(g)$ and $H_-(g)$ are obtained by projecting
orthogonally $H(g)$ onto the subspace of the even and odd functions with
respect to the top of the barrier at $x=1/2g$, respectively. 
This represents the mathematical justification of the
physical intuition that the two perturbed quantum states ``live'' in
two separate but equal wells, and makes it possible to perform the 
stability analysis as already mentioned. The result is

\begin{theorem}
\label{DWB}
Consider the double well operator $H^{\rm DW}(g)$.
\begin{enumerate} 
\item $\exists R>0$ so that both eigenvalues $E_{\pm}(g)$ are analytic in
the Nevanlinna disk $C_R \equiv \{g:{\rm Re} \, g^{-2} > R^{-1}\}$.
\item
The  eigenvalues $E_{\pm}(g)$ are stable near the eigenvalue
$E$ of the unperturbed problem 
as $|g|\to 0$, $g\in C_R$, if considered as eigenvalues of the
operators $H_{\pm}(g)$, respectively.
\end{enumerate}
\end{theorem}
{\it Proof.} See~\cite{CaGrMa1988,CaGrMa1996,CaGrMa1993}.

The summability statement for the perturbation theory can hold only in
an indirect way, identified in \cite{CaGrMa1996}. Two further operators have
to be associated with the differential expression $p^2+q^2(1-g\,q)^2$, 
namely the resonance operators $T_{\pm 1}(g)$. 
The operator $T_{+1}(g)$ is
defined by the the differential expression $p^2+q^2(1-q\,q)^2$ by
modifying the $L^2$ boundary condition at $+\infty$: the functions in
its domain are required to be entire analytic and vanish as 
$|q| \to +\infty$ within the 
sector $S_1 \equiv \{ q\in\C: \frac{\pi}{6} < {\rm arg} \, q <\frac{\pi}{2}\}$.
Analogously, the operator $T_{-1}(g)$ is
defined by the vanishing boundary condition in the symmetrical sector 
$S_{-1} \equiv \{ x\in\C: -\frac{\pi}{2} < {\rm arg}\,x <
-\frac{\pi}{6} \}$. These considerations allow us 
to express the matrix elements of the resolvent of 
$H^{\rm DW}(g)$ as linear combinations of those of $T_{\pm 1}$. 

We now define various projection operators.
Let 
\begin{equation}
P_\pm (g) \, \psi(x) = \frac12\, \left[\psi(x)\pm \psi(g^{-1}-x)\right] \,,
\qquad \psi\in L^2(\mathbbm{R}) \,,
\end{equation}
represent the projections onto the subspaces of the even (respectively odd) 
functions with respect to the barrier point $x=1/2g$.
Then, $H^{\rm DW}_\pm (g) = P_\pm (g) \, H^{\rm DW}(g) \, P_\pm (g)$. 
We also define the following projectors,
\begin{subequations}
\begin{align}
\label{4}
Q(g) =& \; 2 \pi {\rm i} - \oint_{|z-E|=r} 
\frac{1}{z -  H^{\rm DW} (g)} {\rm d}z\,,
\\[2ex]
Q_\pm(g) =& \; 2 \pi {\rm i} - \oint_{|z-E|=r} 
\frac{1}{z -  H^{\rm DW}_\pm (g)} {\rm d}z\,.
\end{align}
\end{subequations}
These project out those eigenfunctions of 
$H^{\rm DW} (g)$ and $H^{\rm DW}_\pm (g)$ whose energies $E_n$
fulfill $| E_n - E | < r$. 
Let $\psi_0$ be the unperturbed eigenvector
corresponding to $E$. With the convention 
$\psi_\pm = P_\pm(g) \, \psi_0$, we now define the expressions
\begin{subequations}
\begin{align}
N_\pm(g) =& \; 
\left< \psi_0 \left| H^{\rm DW}_\pm (g) \, Q_\pm(g) \right| \psi_0 \right> 
\; = \; 
\left< \psi_\pm \left| H^{\rm DW}(g) \, Q(g) \right| \psi_\pm \right> \,,
\\[2ex]
D_\pm(g) =& \; \left< \psi_0 \left| Q_\pm(g) \right| \psi_0 \right> 
\; = \; \left< \psi_\pm \left| Q(g) \right| \psi_\pm \right> \,.
\end{align}
\end{subequations}
By standard Rayleigh-Schr\"odinger perturbation theory,
we can now write any double well eigenvalue $E_\pm(g)$ under the form
\begin{equation}
\label{1}
E_\pm(g) = \frac{N_\pm(g)}{D_\pm(g)} = 
\frac{\left< \psi_\pm \left| H^{\rm DW}(g) \, Q(g) \right| \psi_\pm \right>}
{\left< \psi_\pm \left| Q(g) \right| \psi_\pm \right>} \,.
\end{equation}
If we restrict our attention now to a specific 
manifold of states with a definite parity, then we may write
the energy eigenvalue of the symmetric double-well as 
\begin{equation}
E(g) = \frac{N(g)}{D(g)}\,.
\end{equation}
Following Ref.~\cite{CaGrMa1996},
it is possible to prove a representation
formula of the type
\begin{equation}
\label{ng}
N(g)=\frac12[\Phi(g)+\overline{\Phi}(g)]
\end{equation}
and analogously for $D(g)$, $g\in C_R$. Then one proves:

\begin{theorem} 
\label{bddb} 
With the above definitions, we have the assertions:
\begin{enumerate}
\item The function $\Phi(g)$ admits a power series 
expansion in $g$; $\Phi(g)$ and its expansion
fulfill the conditions of Theorem \ref{T.5.1} (see Appendix B).  
More precisely:
\item
The power series expansion is Borel summable in the distributional sense
(see Definition \ref{DB}, Theorem \ref{T.5.1} and the remark after it)
i.e.: $\Phi(g)$ is the upper sum, $\overline{\Phi}(g)$ is the lower sum,
$N(g)$ is the distributional Borel sum, and
$\Phi(g)-\overline{\Phi}({g})$ is the discontinuity. 
\item An analogous statement holds for $D(g)$ and its perturbation expansion. 
\end{enumerate}
\end{theorem}
{\it Proof.} See \cite{CaGrMa1996}.

Let us add a couple of remarks concerning the effect that
the asymptotic degeneracy phenomenon and the related tunneling are
fragile, i.e. they critically depend on the exact symmetry. Indeed
these phenomena disappear when the the symmetry is broken even by an
infinitesimal amount. The following two observations can be made.
(i) Consider for example the \Sc\ operator
$H_{g,\epsilon} =  p^2+q^2-2g(1+\epsilon)q^3+g^2\,q^4$, $\epsilon>0$
which corresponds to shifting the second minimum up by $\epsilon$ from $0$.
Then the unperturbed eigenvalues are stable and the perturbation
expansion near any unperturbed eigenvalue is Borel summable (for the
proof, see~\cite{GrGr1984}). 
(ii) Suppose we perform an arbitrarily
small variation of the profile of the potential $q^2(1-gq)^2$ away from
the minima. Then the constant $A$ changes by a finite amount, so
that the tunneling probability changes by several orders of magnitudes
(see for instance \cite{Si1982,JLMaSc1981}).
The latter observation has been termed the ``flea over the elephant''
by Simon in Ref.~\cite{Si1985}.

A remarkable example by Herbst and Simon \cite{HeSi1978plb}
shows that this phenomenon takes
place even within perturbation theory of stable, but asymmetric 
multi-well anharmonic oscillators. Consider indeed the following \Sc\ operator
\begin{equation}
\label{3.2.4}
H_{\rm FP}(g) = p^2 + q^2 \, (1-g \, q)^2 + 2 g \, q - 1 \,,
\end{equation}
where the symmetry of the minima is broken by the linear term. 
This symmetry breaking makes the unperturbed eigenvalues stable; however
the \RSPT\ for the ground state $E(g)$ is identically zero, while $E(g)$
itself is not zero. 

This problem finds a natural solution 
here within the context of resurgent expansions and
has also been treated in~\cite{CaGrMa1996}, 
where a general Schr\"odinger operator is examined of the type
\begin{equation}
\label{dbg}
H(g,j) = p^2+q^2(1-gq)^2-j(gq-1/2)\,,
\end{equation}
which reduces to the symmetrical double well for $j=0$ and to the
Fokker--Planck Hamiltonian~\cite{HeSi1978plb,JeZJ2004plb} for $j=-2$. 
The precise mathematical statement of the
solution is too technical to be reproduced here. Roughly speaking,
however, the results are as follows:

(i) If $j\neq 2p$, $p=0,\pm1,\ldots$, the same results as Theorem \ref{bddb}
hold. More precisely, the eigenvalues of $H(g,j)$ split
into two disjoint subsets $\{E_N(g,j)\}$ and $\{E^{\prime}_N(g,j)\}$.
The eigenvalues $\{E_N(g,j)\}$ are stable near the eigenvalues $2N+1+j/2$
of $H(g,j)$, the harmonic oscillator up to the shift $j/2$; The
eigenvalues $\{E^{\prime}_N(g,j)\}$ converge to $2N+1-j/2$ and therefore
have nothing to do with the limiting problem. They are instead stable
with respect to another limiting problem at $g=0$, namely the harmonic
oscillator up to the shift $-j/2$. The eigenvalues $E_N(g,j)$ admit a
Borel summable perturbation expansion, while the perturbation expansion
for any single eigenvalue $E^{\prime}_N(g,j)$ is identically zero.
(ii) If $j=2p$, $p=\pm1,\pm2\ldots$ an eigenvalue of the first limiting
problem may  coincide with an eigenvalue of the second one. Therefore a 
$2\times 2$ matrix has to be introduced, with elements still defined in
terms of matrix elements of the resolvents of the corresponding single
well operators. The two perturbed eigenvalues are the eigenvalues of this
matrix. The coefficients of the matrix have expression analogous to
(\ref{ng}), and are likewise Borel summable in the distributional sense.

{\em Practical resummations of double-well like potentials.}
We start from the case of the symmetric 
double--well potential, this time in the normalization:
\begin{equation}
\label{dw_potential_general}
H_{\rm dw}(g) = \frac{1}{2} \, p^2 \, + \,
\frac{1}{2} q^2 \left( 1 - \sqrt{g} \, q \right)^2 \, =
\frac{p^2}{2} \, + \, \frac{q^2}{2} 
\, - \, \sqrt{g} \, q^3 \, + \, g \frac{q^4}{2}
, 
\end{equation}
where, again, $g$ is a real, positive coupling parameter. Obviously, if 
$g \, = \, 0$, then Eq.~(\ref{dw_potential_general}) 
represents the Hamiltonian of the quantum 
harmonic oscillator whose eigenvalues are
given by the well known formula $E_N \, = \, N + 1/2$, where 
$N \in \mathbb{N}_0$ is the principal quantum number. For 
nonvanishing coupling, by contrast, no analytic expression for the 
energies $E_N(g \ne 0)$ can be derived and, hence, approximate methods 
have to be applied for solving Eq.~(\ref{dw_potential_general}). First, 
we again start from the Rayleigh--Schr\"odinger perturbation theory which allows
us to write the perturbation series (\ref{RS_perturbation_expansion}) series 
for the energy values $E_N(g)$. For the ground state $N = 0$, the first six   
coefficients $A_{N = 0}^{k}$ of these series are given by 
(see, e.g.,~\cite{JeHomeHD}):
\begin{eqnarray}
\label{perturbation_coefficuients_dw}
A_{0}^{1} &=& -1  \,, \quad 
A_{0}^{2}  =  -\frac{9}{2}  \,, \quad 
A_{0}^{3}  =  -\frac{89}{2} \,, \nonumber \\[1ex]
A_{0}^{4} &=& -\frac{5013}{8} \,, \quad
A_{0}^{5}  =  -\frac{88251}{8} \,, \quad
A_{0}^{6}  =  -\frac{3662169}{16} \, . 
\end{eqnarray}
As in the case of the cubic anharmonic oscillator 
(\ref{perturbation_coefficuients_cubic_oscillator}), these coefficients grow
factorially and have the same sign. The perturbation series 
(\ref{RS_perturbation_expansion}) with these coefficients are, hence, divergent 
and not Borel--summable. As mentioned already in the discussion above, however, 
apart from the non-summability of the
series (\ref{RS_perturbation_expansion})
another and even more critical argument against the 
application of the Rayleigh--Schr\"odinger perturbation 
theory to the double--well potential has been 
mentioned~\cite{ZJ1981jmp,ZJ1981npb,ZJ1983npb,ZJ1984jmp,%
JeZJ2004plb,ZJJe2004i,ZJJe2004ii}. 
Namely, while the perturbation series 
(\ref{RS_perturbation_expansion}), if being resummed, 
would provide only one energy 
level for every quantum number $N$, two eigenstates are known to 
appear in the case of nonvanishing coupling $g > 0$. Such a splitting of the
level of the unperturbed Hamiltonian into two levels (the instability
of the eigenvalues from a mathematical point of view) has been
explained by Landau and Lifshitz \cite{LaLi1958} within a semiclassical WKB
approach and has been attributed to the effect of quantum tunneling through the
potential barrier between the two wells. However, while the semiclassical
approach has provided a qualitative and perhaps semi-quantitative 
understanding of the energy spectrum, it can hardly be extended for a 
full and accurate quantitative analysis of the double--well 
(as well as multi--well) potentials. 

\begin{table}[t]
\begin{center}
\begin{minipage}{14cm}
\begin{center}
\caption{\label{table_splitting_dw_1} 
Energy splitting between the "ground" states
$E_{0, \varepsilon = \pm 1}(g)$ of the double--well 
Hamiltonian at the coupling parameter $g$ = 0.005. 
In the first row, we indicate the 
quantity $n$ which is the order of the expansion (\ref{energy_splitting_dw}). 
In the second row, we list the partial sums 
(\ref{energy_splitting_dw}),
whereas in the third row, the Borel--Pad\'{e} transformation 
is presented [cf. Eqs.~(\ref{fBorel_quartic})--(\ref{contours_quartic})].}
\vspace*{0.3cm}
\begin{tabular}{crr}
\hline
\hline
\multicolumn{1}{c}{\rule[-3mm]{0mm}{8mm} $n$} &       
\multicolumn{1}{c}{$ 2 \, 
\frac{{\rm e}^{-1/6g}}{\sqrt{\pi g}} \sum^n_{l=0} e_{0, 10l} \, g^{l}$} &
\multicolumn{1}{c}{${\mathcal T}\!{\mathcal P}_n(g)$}
\\ \hline

 $  0    $    &     $ \underline{ 
 5}.327~056~794 \times 10^{-14}  $    &     $ \underline{ }
                    $ \\

 $  1    $    &     $\underline{ 
 5.16}9~464~698 \times 10^{-14} $     &   $ \underline{
 5.1}78~046~952 \times 10^{-14}   $ \\

 $  2    $    &     $ \underline{
 5.166}~551~926  \times 10^{-14}  $     &     $ \underline{
 5.166}~468~039  \times 10^{-14} $ \\

 $  3    $    &     $ \underline{
 5.166~3}79~090  \times 10^{-14}$     &     $ \underline{
 5.166~36}7~389  \times 10^{-14}$ \\

 $  4    $    &     $ \underline{
 5.166~36}4~871  \times 10^{-14}$     &     $ \underline{
 5.166~363}~195  \times 10^{-14}$ \\

 $  5    $    &     $ \underline{
 5.166~363}~450  \times 10^{-14}$     &     $ \underline{
 5.166~363~2}50  \times 10^{-14}$ \\

 $  6    $    &     $ \underline{
 5.166~363~2}86  \times 10^{-14}$     &     $ \underline{
 5.166~363~26}2  \times 10^{-14}$ \\

 $  7    $    &     $ \underline{
 5.166~363~26}5  \times 10^{-14}$     &     $ \underline{
 5.166~363~261}  \times 10^{-14}$ \\

 $  8    $    &     $ \underline{
 5.166~363~26}2  \times 10^{-14}$     &     $ \underline{
 5.166~363~261}  \times 10^{-14}$ \\

 $  9    $    &     $ \underline{
 5.166~363~261}  \times 10^{-14}$     &     $ \underline{
 5.166~363~261}  \times 10^{-14}$ \\

\hline 
\rule[-3mm]{0mm}{8mm}
exact   & $5.166~363~261 \times 10^{-14}$ &
  $5.166~363~261 \times 10^{-14}$ \\
\hline
\hline
\end{tabular}
\end{center}
\end{minipage}
\end{center}
\end{table}

In this direction, an alternative route for studying the
eigenvalues of the double--well--like potentials has been proposed 
and described in detail
in Refs.~\cite{ZJ1981jmp,ZJ1981npb,ZJ1983npb,ZJ1984jmp,%
JeZJ2004plb,ZJJe2004i,ZJJe2004ii} and is 
based on the Feynman path integral approach. 
We restrict ourselves here to a rather short account of the basic formulas. 
In particular, it has been conjectured that the eigenvalues of the Hamiltonian 
(\ref{dw_potential_general}) can be found by solving the generalized 
Bohr--Sommerfeld quantization condition
\begin{eqnarray}
\label{qcond_dw}
\frac{1}{\sqrt{2 \pi}}\Gamma\left(\frac{1}{2} - B_{\rm dw}(E, g) \right) \,
\left( - \frac{2}{g} \right)^{B_{\rm dw}(E, g)} \, 
{\rm exp}\left[- \frac{A_{\rm dw}(E, g)}{2} \right] \, = \, \varepsilon i \, ,
\end{eqnarray}
where $\varepsilon = \pm 1$ is the parity, 
and the functions $B_{\rm dw}(E, g)$ and 
$A_{\rm dw}(E, g)$ are determined by the perturbative expansion and the 
perturbative expansion about the instantons, respectively. 
The evaluation of these functions in terms of series in variables 
$E$ and $g$ has been
described in detail elsewhere \cite{ZJJe2004i,ZJJe2004ii,ZJ1984jmp} 
for quite general classes of potentials. 
For the particular case of the double--well 
potential, for example, the function $B_{\rm dw}(E, g)$ has the following 
expansion:
\begin{eqnarray}
\label{B_expansion_dw}
B_{\rm dw}(E, g) \, = \, E \, + \, g \left( 3E^2 + \frac{1}{4} \right) \, + 
g^2 \left(35E^3 + \frac{25}{4}E\right)
\, + \, \mathrm{O}(g^3) \, , 
\end{eqnarray}
and defines the perturbation expansion (\ref{RS_perturbation_expansion}) which 
can be easily found by inverting the equation $B_{\rm dw}(E, g) \, = \, N$.
The instanton contributions to the eigenvalues of the Fokker--Planck Hamiltonian
are described by the function
\begin{eqnarray}
   \label{A_expansion_dw}
   A_{\rm dw}(E, g) \, = \, \frac{1}{3g} \, + \, 
   g \left( 17E^2 +\frac{19}{12}\right) \, + \, 
   g^2 \left(227E^3 + \frac{187}{4}E\right) \, + \, 
   \mathrm{O}(g^3) \, .
\end{eqnarray}

By making use of the generalized Bohr--Sommerfeld quantization condition 
(\ref{qcond_dw}) together 
with the expansions (\ref{B_expansion_dw})--(\ref{A_expansion_dw}) of the 
$A_{\rm dw}$ and $B_{\rm dw}$ functions,
we may finally derive the eigenvalues of the double--well 
Hamiltonian. That is, by expanding (\ref{qcond_dw}) in powers of the two small 
parameters $\exp(-1/6g)$ and $g$, the eigenvalues of the Hamiltonian 
(\ref{dw_potential_general}) we may find a 
so--called generalized perturbation expansion (resurgent expansion)
\begin{eqnarray}
   \label{resurgent_expansion_dw}
   E_{N, \varepsilon}(g) = \sum_{l = 0}^\infty A^l_{N} \, g^l +
   \sum^{\infty}_{n=1} \left( \frac{2}{g}\right)^{Nn}
   \left(-\varepsilon \, \frac{{\rm e}^{-1/6g}}{\sqrt{\pi g}} \right)^n \,
   \sum^{n-1}_{k=0} \left\{ \ln\left(-\frac{2}{g}\right) \right\}^k \,
   \sum^{\infty}_{l=0} e_{N, nkl} \, g^{l} \, ,
\end{eqnarray}
which apart of the perturbation series also accounts for the instanton 
contributions.
As seen from this expansion, apart from the principal quantum number $N$,
eigenvalues of the double--well potential are characterized also by the parity
$\varepsilon \, = \, \pm 1$. That is, each unperturbed level with the quantum
number $N$ splits up into two states of opposite parity. Such a splitting is
described by the second term in the right--hand side of the 
Eq.~(\ref{resurgent_expansion_dw}) and is attributed to the instanton
contributions. The order of the instanton contribution is defined by the 
index $n$, with $n = 1$ being
the one--instanton contribution, etc. Despite of its name, the leading, 
one--instanton term 
involves a summation over all possible $n$-instanton configurations but
neglects instanton interactions \cite{ZJJe2004i}. As seen from 
Eq.~(\ref{resurgent_expansion_dw}), the evaluation
of the energies within such a (first-order) approximation, 
also referred as a ``dilute instanton gas'' approximation, requires the 
knowledge of the $e_{N, nkl}$ coefficients. Since these coefficients are 
available for download~\cite{JeHomeHD}, we recall here just the six leading 
ones for the ground state $N = 0$ of the double-well potential:
\begin{eqnarray}
e_{0, 1 0 0} &=& 1 \,, \quad 
e_{0, 1 0 1} = -\frac{71}{12} \,, \quad 
e_{0, 1 0 2} = -\frac{6299}{288} \,, \quad 
e_{0, 1 0 3} = -\frac{2691107}{10368} \,, \\
e_{0, 1 0 4} &=& -\frac{2125346615}{497664} \,, \quad
e_{0, 1 0 5} = -\frac{509978166739}{5971968} \,, \quad 
e_{0, 1 0 6} = -\frac{846134074443319}{429981696} \,. \nonumber
\end{eqnarray}
By inserting these coefficients in Eq.~(\ref{resurgent_expansion_dw}), we are 
able to perform a numerical check of the validity of the one-instanton 
expansion for the energy splitting 
\begin{equation}
\label{energy_splitting_dw}
\Delta E_0 \, = \, E_{0, -1} - E_{0, 1} \, \sim \, 
\left(2 \, \frac{{\rm e}^{-1/6g}}{\sqrt{\pi g}} \right) \,
\sum^{\infty}_{l=0} e_{0, 10l} \, g^{l}
\end{equation}
of the ground-state level $N$ = 0 due to the quantum tunneling. As seen 
from the second column of the Tables \ref{table_splitting_dw_1} and 
\ref{table_splitting_dw_2}, however, 
a straightforward, term-by-term summation of the series 
(\ref{energy_splitting_dw}) may well reproduce 
the energy splitting $\Delta E_0$ for
the case of the small coupling $g = 0.005$
but leads to the divergence of the result
for strong coupling of $g = 0.03$. 
In order to overcome such a divergence we again 
employ the Borel--Pad\'{e} resummation method as given by 
Eqs.~(\ref{fBorel_quartic})--(\ref{contours_quartic}) in which 
as input data we
used coefficients $e_{0, 1 0 l}$. However, while the Borel--Pad\'{e} method
provides a good convergence of the 
series (\ref{energy_splitting_dw}), the result
of resummation shows significant 
discrepancy with the ``true'' energy splitting as
computed by the diagonalization of the 
Hamiltonian (\ref{dw_potential_general}) in 
the basis of the harmonic oscillator wavefunctions. Such a discrepancy indicates
the importance of the higher--instanton terms which take into account
the instanton interactions for the case of strong coupling. 
The evaluation of the 
higher--order corrections ($n \ge$ 2) to the 
energies of the double--well potential
is, however, a very difficult task since it requires a double resummation of the 
resurgent expansion, in powers of both $g$ and $\exp(-1/6g)$. 
In this contribution, 
one may address the question whether it is possible to 
resum the divergent series that gives rise to the 
$A_{\rm dw}$ and $B_{\rm dw}$ functions directly and look for solutions of the 
quantization condition (\ref{qcond_dw}) without recourse to the resurgent 
expansion. As it was discussed recently \cite{SuLuZJJe2006}, 
this appears to be the only 
feasible way to evaluate the multi-instanton expansion 
(in power of $n$) because 
the quantization condition incorporates all instanton orders.
We briefly summarize the results of this investigation here,
which has been carried out for a potential closely related
to the symmetric double-well.

\begin{table}[t]
\caption{\label{table_splitting_dw_2} 
Energy splitting between the "ground" states
$E_{0, \varepsilon = \pm 1}(g)$ of the double--well 
Hamiltonian at the coupling parameter $g = 0.03$. 
In the first row, we indicate the 
quantity $n$ which is the order of the expansion (\ref{energy_splitting_dw}). 
In the second row, we list the partial sums 
(\ref{energy_splitting_dw}),
whereas in the third row, the Borel--Pad\'{e} transformation 
is presented [cf. Eqs.~(\ref{fBorel_quartic})--(\ref{contours_quartic})].}
\vspace*{0.3cm}
\begin{center}
\begin{tabular}{crr}
\hline
\hline
\multicolumn{1}{c}{\rule[-3mm]{0mm}{8mm} $n$} &       
\multicolumn{1}{c}{$ 2 \, \frac{{\rm e}^{-1/6g}}{\sqrt{\pi g}} \sum^n_{l=0} e_{0, 10l} \, g^{l}$} &
\multicolumn{1}{c}{${\mathcal T}\!{\mathcal P}_n(g)$}

\\ \hline

 $  0    $    &     $ \underline{} 
 2.518~531~055  \times 10^{-2}  $    &     $ \underline{ }
                    $ \\

 $  1    $    &     $\underline{ }
 2.071~491~792  \times 10^{-2} $     &   $ \underline{}
 2.178~799~750  \times 10^{-2}   $ \\

 $  2    $    &     $ \underline{}
 2.021~916~083  \times 10^{-2}  $     &     $ \underline{}
 2.011~031~238  \times 10^{-2} $ \\

 $  3    $    &     $ \underline{}
 2.004~265~988   \times 10^{-2}$     &     $ \underline{
 1.98}9~410~405   \times 10^{-2}$ \\

 $  4    $    &     $ \underline{}
 1.995~553~827   \times 10^{-2}$     &     $ \underline{
 1.983~5}47~787  \times 10^{-2}$ \\

 $  ...    $    &     $ \underline{}
 ...$     &     $ \underline{}
 ...$ \\

 $  33    $    &     $ \underline{}
-1.487~012~669        \times 10^{2}$     &     $ \underline{
 1.98}0~196~846  \times 10^{-2}$ \\

 $  34    $    &     $ \underline{}
-4.525~128~589  \times 10^{2}$     &     $ \underline{
 1.98}0~196~848  \times 10^{-2}$ \\

 $  35    $    &     $ \underline{}
-1.418~424~114   \times 10^{3}$     &     $ \underline{
 1.98}0~196~848  \times 10^{-2}$ \\

 $  36    $    &     $ \underline{}
-4.575~754~388   \times 10^{3}$     &     $ \underline{
 1.98}0~196~848 \times 10^{-2}$ \\

\hline 
\rule[-3mm]{0mm}{8mm}
exact   & $1.983~530~115 \times 10^{-2}$ &
  $1.983~530~115 \times 10^{-2}$ \\
\hline
\hline

\end{tabular}
\end{center}
\end{table}

We discuss the main ideas of such a ``direct summation'' 
approach for the case of the so--called Fokker--Planck potential:
\begin{eqnarray}
\label{fp_potential_general}
H_{\rm FP}(g) & = & \frac{1}{2} p^2 \, + \,
\frac{1}{2} q^2 \left( 1 - \sqrt{g}q \right)^2 \, + \, \sqrt{g} q \, - \, 
\frac{1}{2} \nonumber \\ 
& = & \frac{1}{2} p^2 \, + \, 
\frac{1}{2} q^2 \, - \, \frac{1}{2} \, + \, \sqrt{g}q \, - \,
\sqrt{g}q^3 \, + \, \frac{g}{2} q^4 \, ,     
\end{eqnarray}
which can be considered as a modification of the double--well potential
(\ref{dw_potential_general}).  In contrast to the double--well case, however, 
the Fokker--Planck potential contains a linear symmetry--breaking term. 
Because of this term, the ground state of the 
potential (\ref{fp_potential_general}) 
is located in one 
of the wells and, hence, does not develop any degeneracy due to parity
(see also the mathematically inspired discussion earlier
in this Section of the review).  
Excited states, however, can still be degenerate for $g \to 0$ in view of the 
(perturbatively broken) parity $\varepsilon \, = \, \pm 1$ of the quantum 
eigenstates. Again, in order to analyze such a degeneracy we may start from the
generalized Bohr--Sommerfeld quantization condition which for the case of
Fokker--Planck Hamiltonian reads as:
\begin{eqnarray}
   \label{quantization_condition_fp}
   \frac{1}{\Gamma(-B_{\rm FP}(E, g)) \Gamma(1-B_{\rm FP}(E, g))} \, + \,
   \left(-\frac{2}{g}\right)^{2B_{\rm FP}(E, g)} 
   \frac{{\rm e}^{-A_{\rm FP}(E, g)}}{2 \pi} \, = \, 0 \, ,
\end{eqnarray}
where the series for the functions $B_{\rm FP}(E, g)$ and 
$A_{\rm FP}(E, g)$ are:
\begin{eqnarray}
   \label{B_expansion_fp}
   B_{\rm FP}(E, g) \, = \, E \, + \, 3E^2g \, + 
   \left(35E^3 + \frac{5}{2}E\right)g^2
   \, + \, 
   \mathrm{O}(g^3) \, , 
\end{eqnarray}
and
\begin{eqnarray}
   \label{A_expansion_fp}
   A_{\rm FP}(E, g) \, = \, \frac{1}{3g} \, + \, 
   \left( 17E^2 +\frac{5}{6}\right)g \, + \, 
   \left(227E^3 + \frac{55}{2}E\right)g^2 \, + \, 
   \mathrm{O}(g^3) \, . 
\end{eqnarray}
As mentioned above, these expressions may be used directly in order to
investigate the energy spectrum of the Fokker--Planck potential. That is,
if one defines the left--hand side of the quantization condition 
(\ref{quantization_condition_fp}) as a function of two variables $E$ and $g$:  
\begin{align}
\label{Q_function}
Q(E, g) =& \, 
\frac{1}{\Gamma(-B_{\rm FP}(E, g)) \, \Gamma(1-B_{\rm FP}(E, g))} 
+ \left(-\frac{2}{g}\right)^{2B_{\rm FP}(E, g)} 
\frac{{\rm e}^{-A_{\rm FP}(E, g)}}{2 \pi} \,.
\end{align}
The zeros of this function at some fixed $g$ determine the energy spectrum of 
the (\ref{fp_potential_general}):
\begin{equation}
\label{Q_zeros}
Q(E_{N}(g), g) \, = \, 0, \qquad N \in \mathbb{N}_0  \, .
\end{equation}
A numerical analysis of the 
function $Q(E, g)$ can be used, therefore, in order to 
find the eigenvalues of the Fokker--Planck Hamiltonian.

As seen from Eq.~(\ref{Q_function}), an analysis of the function
$Q(E, g)$ can be traced back to 
the evaluation of the functions $A_{\rm FP}(E, g)$ 
and $B_{\rm FP}(E, g)$ which constitute the series in \textit{both} 
variables $E$ and $g$ 
[cf. Eqs.~(\ref{B_expansion_fp})--(\ref{A_expansion_fp})]. 
In order to perform these summations, we may rewrite the functions 
$A_{\rm FP}(E, g)$ and $B_{\rm FP}(E, g)$ as a (formal) power series in terms 
of some variable $x$, taken at $x$ = 1:
\begin{eqnarray}
   \label{a_x_series}
   B_{\rm FP}(E, g) \, = \, \sum\limits_{k = 0}^{n} 
   b^{(k)}_{\rm FP}(E, g) x^k |_{x=1}\, ,               
\end{eqnarray}
\begin{eqnarray}
   \label{b_x_series}
   A_{\rm FP}(E, g) \, = \, \frac{1}{3g} + \sum\limits_{k = 1}^{n} 
   a^{(k)}_{\rm FP}(E, g) x^k |_{x=1}\, ,               
\end{eqnarray}
where the coefficients $a^{(k)}_{\rm FP}(E, g)$ and $b^{(k)}_{\rm FP}(E, g)$ 
are defined uniquely from (\ref{B_expansion_fp}) and (\ref{A_expansion_fp}): 
$b^{(0)}_{\rm FP}(E, g) \, = \, E$, $b^{(1)}_{\rm FP}(E, g) \, = \, 3E^2g$,
$a^{(1)}_{\rm FP}(E, g) \, = \, \left( 17 E^2 + 5/6 \right) g$, etc. Of 
course, the power series (\ref{a_x_series}) and (\ref{b_x_series}) are
mathematically equivalent to the expansions (\ref{B_expansion_fp}) and 
(\ref{A_expansion_fp}) but more convenient for further numerical computations 
as discussed below.   

Now we are ready to analyze 
the energy spectrum of the Fokker--Planck Hamiltonian 
[cf.~Eq.~(\ref{Q_zeros})]. 
As mentioned above, to perform such an analysis for any particular $g$ we have 
to (i) resum the series for the functions $A_{\rm FP}(E, g)$ and 
$B_{\rm FP}(E, g)$ and (ii) to insert the resulting generalized Borel sums 
$\overline{A}_{\rm FP}(E, g)$ and $\overline{B}_{\rm FP}(E, g)$ into 
Eq.~(\ref{Q_function});
these can be either the upper or the lower 
distributional Borel sums, but must be chosen consistently. 
We may interpret then the $Q(E, g)$ as a function of 
$E$ (at fixed $g$) and (iii) numerically determine the zeros of this 
function which correspond to the energy values $E^{(M)}_{\rm FP}$ of the
Fokker--Planck Hamiltonian [cf. Eq.~(\ref{Q_zeros})]. 
As an illustration of the
calculations, we have computed the function $Q(E, g)$ for the coupling 
$g = 0.1$ and the energy $E \, = \, 5.199~138~696 \times 10^{-3}$.
For this set of parameters, the 
series (\ref{a_x_series})--(\ref{b_x_series}) 
are divergent and we shall again use the Borel--Pad\'{e} resummation 
method in order to obtain the functions $A_{\rm FP}(E, g)$ and 
$B_{\rm FP}(E, g)$. The resulting distributional Borel sums
$\overline{A}_{\rm FP}(E, g)$ and $\overline{B}_{\rm FP}(E, g)$ are presented 
in the second and third columns of Table~\ref{table_splitting_dw_3}. 
By inserting these generalized sums into Eq.~(\ref{Q_function}) the function
$Q(E, g)$ has been calculated and is displayed in the forth column. As seen from
the Table, with the increasing the order $n$ of the series 
(\ref{a_x_series})--(\ref{b_x_series}), the function $Q(E, g)$ approaches 
zero, indicating that the energy $E \, = \, 5.199~138~696 \times 10^{-3}$
is an eigenvalue of the Fokker--Planck Hamiltonian for a coupling 
$g = 0.1$.

While the above calculations have been performed mainly
for illustration purposes, the ``direct summation'' approach as given by
Eqs.~(\ref{Q_function})--(\ref{b_x_series}) has been recently employed
for the detailed calculations of the ground as well as the excited state
energies of the Fokker--Planck potential for a wide range of coupling
parameters~\cite{SuLuZJJe2006}. 

\begin{table}[t]
\begin{center}
\begin{minipage}{14cm}
\begin{center}
\caption{\label{table_splitting_dw_3} The function 
$Q(E, g)$ [cf. Eq.~(\ref{Q_function})] for the Fokker--Planck Hamiltonian 
calculated at the coupling $g$ = 0.1 and the energy 
$E \, = \, 5.199~138~696 \times 10^{-3}$. 
In addition to this function (fourth column),
the generalized Borel transforms $\overline{B}_{\rm FP}(E, g)$ and 
$\overline{A}_{\rm FP}(E, g)$ of the series
(\ref{a_x_series}) and (\ref{b_x_series}) are displayed
as a function of the order $n$ (first column).
Indeed, the exact values for the functions $ \overline{A}_{\rm FP}(E, g)$
and $ \overline{B}_{\rm FP}(E, g) $ are not known.}
\vspace*{0.3cm}
\begin{tabular}{crrr}
\hline
\hline
\multicolumn{1}{c}{\rule[-3mm]{0mm}{8mm} $n$}      &       
\multicolumn{1}{c}{$ \overline{A}_{\rm FP}(E, g) $} &
\multicolumn{1}{c}{$ \overline{B}_{\rm FP}(E, g) $} &
\multicolumn{1}{c}{$ Q(E, g) $}

\\ \hline

 $  1    $    &   
 $3.421~230 + 0.000~000  \, {\rm i}$ &  
 $0.005~207 + 0.000~000  \, {\rm i}$ &  
 $\underline{0.000}~186 + 
 \underline{0.000}~175  \, {\rm i}$ \\

 $  2    $    &  
 $3.418~180 + 0.000~000  \, {\rm i}$ &  
 $0.005~198 + 0.000~000  \, {\rm i}$ &    
 $\underline{0.000}~211 + 
  \underline{0.000}~175  \, {\rm i}$ \\

 $  3    $    &     
 $3.420~640 + 0.000~000  \, {\rm i}$ &  
 $0.005~372 + 0.000~011  \, {\rm i}$ &
 $\underline{0.000~0}31 + 
 \underline{0.000}~171  \, {\rm i}$ \\

 $  4    $    &     
 $3.437~970 +  0.015~857 \, {\rm i}$ &  
 $0.005~366 +  0.000~105 \, {\rm i}$ &
 $-\underline{0.000~0}57 +  
  \underline{0.000~00}6  \, {\rm i}$ \\
 
 $  5    $    &     
 $3.428~560 +  0.019~738 \, {\rm i}$ &  
 $0.005~366 +  0.000~105 \, {\rm i}$ &
 $-\underline{0.000~00}6 -  
  \underline{0.000~0}26  \, {\rm i}$ \\

 $  ...    $    &    
 $  ...    $    &     
 $  ...    $    &     
 $  ...    $ \\
 
 $  11    $    &     
 $3.430~080 +  0.017~575  \, {\rm i}$ &  
 $0.005~355 +  0.000~092  \, {\rm i}$ &
 $-\underline{0.000~00}4 -  
  \underline{0.000~00}2  \, {\rm i}$ \\
 
 $  12    $    &     
 $3.429~310 +  0.017~691  \, {\rm i}$ &  
 $0.005~354 +  0.000~090  \, {\rm i}$ & 
 $\underline{0.000~000} +  
  \underline{0.000~000}  \, {\rm i}$ \\

 $  13    $    &     
 $3.429~560 +  0.017~320  \, {\rm i}$ &  
 $0.005~354 +  0.000~090  \, {\rm i}$ & 
 $\underline{0.000~000} +  
  \underline{0.000~00}1  \, {\rm i}$ \\

 $  14    $    &    
 $3.429~530 +  0.017~312  \, {\rm i}$ &  
 $0.005~354 +  0.000~090  \, {\rm i}$ & 
 $\underline{0.000~000} +  
  \underline{0.000~00}1  \, {\rm i}$ \\

 $  15    $    &     
 $3.429~560 +  0.017~321  \, {\rm i}$ &  
 $0.005~354 +  0.000~090  \, {\rm i}$ & 
 $\underline{0.000~000} +  
  \underline{0.000~000}  \, {\rm i}$ \\

 $  16    $    &     
 $3.429~570 +  0.017~540  \, {\rm i}$ &  
 $0.005~354 +  0.000~090  \, {\rm i}$ & 
 $\underline{0.000~000} +  
  \underline{0.000~000}  \, {\rm i}$ \\

 $  17    $    &     
 $3.429~470 +  0.017~440  \, {\rm i}$ &  
 $0.005~354 +  0.000~090  \, {\rm i}$ & 
 $\underline{0.000~000} +  
  \underline{0.000~000}  \, {\rm i}$ \\

 $  18    $    &     
 $3.429~470 +  0.017~450  \, {\rm i}$ &  
 $0.005~354 +  0.000~090  \, {\rm i}$ & 
 $\underline{0.000~000} +  
  \underline{0.000~000}  \, {\rm i}$ \\
 
\hline 
\rule[-3mm]{0mm}{8mm}
exact   &  &  &  $0.000~000 + 0.000~000 \, i$ \\
\hline
\hline
\end{tabular}
\end{center}
\end{minipage}
\end{center}
\end{table}

%
%
\subsection{Renormalization Group and Critical Exponents}
\label{RG}

\subsubsection{Orientation}

Universal properties of critical phenomena have found a natural explanation
thanks to the theory of renormalization group (RG, 
\cite{Wi1975},~\cite{Wi1983}). 
This powerful conceptual tool provides a framework to understand
the surprising behaviour that physical systems exhibit in the vicinity 
of second-order phase transitions (like universality, and 
the existence of a scaling behaviour parameterized by 
critical exponents). Computations of critical quantities are intimately
related to the techniques described in the present review: for instance,
the so-called approach at fixed dimension $d=3$ combines the 
technical difficulties and sophistications of Feynman diagram 
calculations with the mathematical intricacies of the divergent
character of the perturbative expansions; on the other hand a different method,
based on so-called high-temperature series, leads to truncated convergent 
expansions whose analysis is connected to the numerical methods discussed 
in Sec.~\ref{seqtr} of this review.

However, before we analyze these two approaches in a more detailed
manner, let us first recall some basic facts. Although this may perhaps 
come as a surprise, due to a number of technical reasons the extraction
of high-precision estimates from the RG is a hard task, requiring a lot 
of effort and computational resources. As a matter of fact, even if much
effort has been put in recent years towards a precise determination of
critical quantities, only very recently have theoretical results reached
an accuracy comparable to that of experiment; in addition, not always do
the different methods necessarily agree among themselves, or with the
experimental results.

For example, the most precise theoretical estimate nowadays available
(obtained in \cite{CaHaPeVi2006} from a sophisticated combination of 
Monte-Carlo and high-temperature techniques) predicts for the critical 
exponent $\alpha$ in systems described by an $O(2)$-symmetric
$\left(\phi^2\right)^2$-theory a value of $\alpha^{N=2}_{\rm th}=-0.0151(3)$;
this figures are incompatible with the currently accepted experimental result
of $\alpha^{N=2}_{\rm exp} = -0.01285(38)$
(obtained on the Space Shuttle by the microgravity experiment 
MISTE~\cite{LiEtAl1996,MISTE}), and a convincing explanation for such a 
discrepancy still has to be found. On the other hand, the theoretical 
predictions produced by the class of the so-called
field-theoretical perturbative methods appear to be in general agreement
with the experiment, but usually at the price of a much larger uncertainty
(giving for the quantity $\alpha^{N=2}$ mentioned above estimates like
$\alpha^{N=2}_{\rm th'}=-0.011(4)$ of Ref.~\cite{GuZJ1998bx}, where the 
theoretical error has been estimated conservatively, or the less conservative 
$\alpha^{N=2}_{\rm th''}=-0.0129(6)$ of Ref.~\cite{Kl1999}).

In any case much more precise experiments which could help to solve these
open questions are planned for the future, even if the recent
changes in NASA's priorities have unfortunately lead to the cancellation 
of an entire family of devices which were almost ready to fly
(see, e.g., Ref.~\cite{LaEtAl2004} for their detailed description).

In general, many computational tools have been developed to formulate numerical
predictions for universal physical quantities like critical exponents:
among them we find lattice-based methods (such as Monte-Carlo
methods \cite{PaSwWaWi1984,BaFeMMMu1996,CaEtAl2001,CaEtAl2002,CaEtAl2006}
and high-temperature series \cite{FeMoWo1971,BuCo1997}),
perturbative field-theoretical methods (like the
$\epsilon$-expansion \cite{WiFi1972,Wi1972vs,
WiKo1974,GoLaTkCh1983,KlSF2001jpa,KlNeSFChLa1991,Fi1998,ZJ2001}, 
the approach at fixed dimension $d=3$ 
(see Refs.~\cite{BaNiGrMe1976,NiMeBa1977,BaNiMe1978})
or the minimal renormalization without $\epsilon$-expansion \cite{ChDo2004}),
and non-perturbative field-theoretical methods 
(based on the so-called ``exact'' RGs,
see, e.g., Refs.~\cite{BaBe2001,CaDeMoVi2003,WeHo1973,vGWe2001}).
Almost all of these techniques are based on the construction 
of more and more precise approximations for the desired quantities, 
which naturally lead to the evaluation of a truncated series.

In Refs.~\cite{WiFi1972,Wi1972vs,WiKo1974}, the so-called
$\epsilon$-expansion led to the first convincing a priori prediction
for critical exponents, thus validating the RG approach.
A significant breakthrough toward the precise computation of RG quantities
has been achieved soon after by another perturbative field-theoretical method, 
the approach at fixed dimension $d=3$, 
originally suggested in \cite{ParisiTwoLoops,Pa1980}.
In a seminal series of articles~\cite{BaNiGrMe1976,NiMeBa1977,BaNiMe1978},
the calculation of the RG $\beta$-function for the 
$O(N)$-symmetric $\left(\phi^2\right)^2$-theory in $d=3$ advanced
from the $2$-loop level (see~\cite{ParisiTwoLoops,BrLGZJ1973}) 
to the level of $6$ loops.
The treatment of Refs.~\cite{BaNiGrMe1976,NiMeBa1977,BaNiMe1978}
(consisting in the numerical 
computation of $\approx 1000$ integrals in dimensions up to $6$ with $8$-$10$ 
significant digits) 
opened the possibility of formulating quite precise 
predictions for a large number of physical quantities, 
which as of today remain among the best ones available.
Quite recently, other methods (notably Monte-Carlo estimates, eventually
supplemented by the output of some {\it ad hoc} high-temperature series)
have reached a comparable or even better accuracy (see for example
\cite{CaHaPeVi2006}).

It would be very difficult (and out of the scope of the current
review) even only
to mention all the existing methods to compute RG-related quantities.
For example, the complete description 
of the techniques employed to obtain critical quantities by means of
just one single method (the $5$-loop evaluation of the $\epsilon$-expansion)
is of the size of a book~\cite{KlSF2001}. Even more so, 
a whole series of books \cite{DombAndGreen} has been devoted over the years 
to keep track of all progress in the field. 

As mentioned before, in the present work we will try to give an overview 
of two methods in particular,
(i) high-temperature series (see Sec.~\ref{hightemp}) and
(ii) the approach at fixed dimension $d=3$ (see Sec.~\ref{fixeddim}).
The choice of focusing on these two approaches to critical systems
is justified by the following considerations.

Both methods deliver their results in form of series; however, reflecting the 
very different nature of the two computational schemes the estimates derived 
from HT series usually possess a finite radius of convergence, while the
approach at finite dimension produces descriptions of the critical quantities 
in terms of divergent series; in addition, the number of accessible terms
of the series is very different in the two cases (typically $\sim 25$ for the
HT expansions, and only $6$-$7$ terms for the scheme at fixed dimension).

From a practical point of view, however, many analogies between the two
methods can also be found: in both cases the computation of the
coefficients of the series requires the evaluation of a factorially-growing
number of different contributions in higher 
order; in both cases adequate prescriptions 
(an acceleration scheme for the first method, and a resummation for the second)
have to be supplemented if one wishes to be able to extract meaningful physical
results from the obtained series. Quite surprisingly, the final precision
provided by the many terms of HT series is usually comparable to that obtained
from the few terms of the scheme in fixed dimension.

Indeed, the two methods show an intriguing mixture of similarities and 
differences, both being good examples of situations where a sophisticated
numerical analysis of some series is helpful or mandatory in obtaining refined
physical results; for this reason we think that a detailed inspection and 
comparison of such methods should be considered particularly interesting
in the framework of the present review.

Finally, in subsection \ref{rgremarks} we present a general outlook of the 
perspectives to extend the results obtained from both the HT series expansions
and the techniques at fixed dimension.

%
%
\subsubsection{High--Temperature Series}
\label{hightemp}

The general idea of high-temperature expansions may be understood as
follows. Let us assume that we are dealing with 
a statistical-mechanics system which undergoes some 
(second-order) phase transition and is described by some suitable
partition function $Z$, which depends on the inverse temperature
$\beta=1/(k_B T)$. An approximate description of the system valid only 
at temperatures $T\gg T_{\rm critical}$ may then be derived if we expand the
partition function in terms of $\beta\sim 1/T$. (Of course, a 
complementary ``low temperature''-expansion valid when $T\ll T_{\rm critical}$
may also be derived; the two expansions are related, and a way of exploiting
such connection is described in \cite{SyEsHeHi1966}).

An interesting and non-trivial property of such an expansion is the possibility
of writing it in purely diagrammatic terms; even if the number of diagrams 
which must be 
generated and evaluated grows very quickly with the perturbation order,
only sophisticated graph-generation and combinatorial techniques are required 
to complete the evaluation of the series. This feature should be contrasted
with the method at fixed dimension explained in the following subsection, where
not only the number of diagrams, but also the computational complexity of
the integral associated to each diagram grows with the order in perturbation
theory.

Many evaluation schemes for HT series have been presented in the literature
(for an introduction to this subject the reader is referred to 
\cite{ItDr1989}); a very successful specialization of the 
technique is given by the so-called linked-cluster expansion (LCE) as 
described in \cite{Wo1974}. We usually find the LCE applied
to a class of physical systems described by the following partition function,
which is a generalization of the Ising model:
\begin{equation}
   Z=\prod_i\left[\int {\rm d}\phi_i\;f\!\!\left(\phi_i\right)
   {\rm e}^{h_i\phi_i}\right]
   {\rm e}^{\beta\sum_{\left<ij\right>}\phi_i\phi_j}\,.
\label{LCEPartFunction}
\end{equation}
Here, $\phi_i$ is a scalar field, while $f$ is a non-negative distribution 
function which decreases faster than ${\rm e}^{-\phi^2}$ as $\phi\rightarrow\infty$
and is normalized by the relation $\int {\rm d}\phi\;f\!\left(\phi\right)=1$; 
the sum runs over all pairs of nearest neighbors on a lattice which is usually
taken to be a generalization of quadratic lattices to arbitrary dimensions.

An interesting feature of this way of setting up the problem is the possibility
of encapsulating the physical content of many different systems into the choice
of the distribution function $f$ which enters Eq.~(\ref{LCEPartFunction});
for example, the choice
\begin{equation}
f\!\!\left(\phi\right)=
\sum_{m=-S}^S\delta\!\!\left(\phi+2m\sqrt{\frac{3}{4S(S+1)}}\right)
\end{equation}
reproduces the usual spin-$S$ Ising model, while many other models which are
well-known in statistical mechanics (like Blume-Capel, Klauder, 
double Gaussian and range models) can be recovered by different choices.
In addition, the computation of the LCE can be set up in such a way that
its final results are directly expressed in terms of the symbolic expressions
for the cumulant moments of $f$. These quantities $\mu_{2n}$ are defined
by the relation
\begin{equation}
\sum_n\frac{\mu_{2n}}{\left(2n\right)!}h^{2n}=
\ln\int {\rm d}\phi\;f\!\left(\phi\right){\rm e}^{h\phi}\,.
\end{equation}
Thus, once the result is obtained as a function of the $\mu_{2n}$,
it can be specialized to any of the possible models just by replacing the 
symbolic moments with the actual values they assume when a specific choice of
$f$ has been made.

The physically interesting quantities to be computed are as usual the
connected $q$-point functions at zero magnetic field defined by
\begin{equation}
G_q(x_{i_1},...,x_{i_q}) =
\left. \frac{\partial^q \ln Z[h]}{
\partial h_{i_1} \, \dots  \, \partial h_{i_q}} \right|_{h=0}
\end{equation}
with their moments
\begin{eqnarray}
\chi_q &\equiv& \sum_{x_2,...,x_q} G_q(x_1,...,x_q), \nonumber \\
M_2^{(q)} &\equiv& \sum_{x_2,...,x_q} (x_1-x_2)^2 G_q(x_1,...,x_q).
\label{m2q}
\end{eqnarray}

The sophisticated graph-theoretical 
methods which allow the high-temperature series to be actually generated
are described, e.g., in Ref.~\cite{Ca2001}. 
We just mention in passing that, once the form of the lattice and the
expansion variables have been chosen, the core of the method is the 
possibility of rewriting the truncated expansion of the sum on the 
right-hand side of Eq.~(\ref{LCEPartFunction}) 
as the sum of many contributions, each one associated
to a particular graph; the new summation runs over a set of graphs satisfying
some topological constraint, and it is possible to formulate graphical rules
to describe the coefficient attached to each graph of the expansion.
In addition, many different expansions may be generated, due to the fact that 
the contributions to the original expansion can in turn be rearranged in terms
of new contributions associated to a different set of graphs.
In the end, the method consists in the (recursive) generation of all the 
graphs satisfying the constraints dictated by the chosen expansion, together
with the numbers specifying their associated contributions.

Here, we would like to discuss in some more detail how the
obtained data can be interpreted. A typical HT series reads 
(see \cite{Ca2001})
\begin{eqnarray}
\chi_2 &=& 
  1 + 8\,v + 56\,{v^2} + 392\,{v^3} + 2648\,{v^4} + 17864\,{v^5} + 
  118760\,{v^6} + 789032\,{v^7} 
\nonumber \\ &&\; +\, 
  5201048\,{v^8} + 34268104\,{v^9} + 
  224679864\,{v^{10}} + 1472595144\,{v^{11}} + 9619740648\,{v^{12}} 
\nonumber \\ &&\; +\,
  62823141192\,{v^{13}} + 409297617672\,{v^{14}} + 
  2665987056200\,{v^{15}} + 
  17333875251192\,{v^{16}}
\nonumber \\ &&\; +\,
  112680746646856\,{v^{17}} + 
  731466943653464\,{v^{18}} + 4747546469665832\,{v^{19}} 
\nonumber \\ &&\; +\,
  30779106675700312\,{v^{20}} + 199518218638233896\,{v^{21}} + 
  1292141318087690824\,{v^{22}}
 \nonumber \\ &&\; +\,
  8367300424426139624\,{v^{23}} + 
  54141252229349325768\,{v^{24}} 
 \nonumber \\ &&\; +\,
  350288350314921653160\,{v^{25}} +
   {\rm O}(v^{26})\,,
\label{HTsusc}
\end{eqnarray}
with $v=\tanh\beta$.
We notice that the generated series can be quite long 
(the typical order being $\sim 25$), and that they have been computed 
for many different physical
systems and lattice arrangements; an extensive list of such computations
can be found for example in \cite{PeVi2002}.

However, despite their length HT series usually show poor convergence, and 
sophisticated methods must be employed to analyze their content in terms of
RG quantities. A detailed account of many of the techniques which may be 
used (not only in relation to the case of HT series, but also for generic 
similar statistical-mechanical problems) can be found in \cite{Gu1989}.
The typical model problem, as derived from the predictions of the RG theory
or statistical mechanics and considered in this reference, is how to fit the
obtained HT series $S(z)=\sum_{n=0}^\infty S_n z^n$ to a function of the 
following form:
\begin{equation}
S(z)
\sim
A\left(1-\frac{z}{z_{\rm c}}\right)^{-\lambda}\left[
  1+
  A_1\left(1-\frac{z}{z_{\rm c}}\right)^{-\Delta_1}+
  A_2\left(1-\frac{z}{z_{\rm c}}\right)^{-\Delta_2}+
  \ldots
\right] \,,
\label{ModelFunction}
\end{equation}
for $z\rightarrow z_{\rm c}$ from below,
with $0<\Delta_1<\Delta_2<\ldots$. The constants $A$, $z_{\rm c}$ and 
$\lambda$ are called the critical amplitude, the critical point
and the critical exponent, respectively. The terms of the form
$\left(1-z/z_{\rm c}\right)^{-\Delta_i}$, which may be present
or not, and eventually be replaced by logarithmic singularities like
$\left|\ln\left(1-z/z_{\rm c}\right)\right|^{\delta_j}$,
are called corrections to scaling, and if one or more of the $\Delta_i$
(or the $\delta_j$) are non-integral we say we are in the presence of
confluent singularities. It is perhaps interesting to remark 
that for some special systems
even more complicated forms for the critical singularities are actually 
predicted by RG theory, and observed experimentally.

As shown in \cite{Gu1989}, the efficient extraction 
of the values for $A$, $z_{\rm c}$ and $\lambda$
(plus eventually the $A_i$ and the $\Delta_i$) 
from series like Eq.~(\ref{HTsusc}) is far from being straightforward.
The most frequently used approaches fall in two main broad classes:
\begin{enumerate}
\item[(i)] Methods consisting in fitting/extrapolating the sequence of the 
ratios of the terms of the series, that is the new series
$R_n\equiv{S_n}/{S_{n-1}}$. The main reason for employing the ratio series
lies in the fact that it has better convergence properties. In addition, under
reasonable assumptions about the functions $A$ and $A_i$ appearing in 
Eq.~(\ref{ModelFunction}), it is possible to rewrite the ratio terms as 
(see \cite{Gu1989})
\begin{eqnarray}
R_n\equiv\frac{S_n}{S_{n-1}}
&\sim&
\label{ExpConfl}
\frac{1}{z_{\rm critical}}\left[
  1+\frac{\lambda-1}{n}+\frac{c}{n^{1+\Delta}}
  +\frac{d}{n^{1+2\Delta}}+\frac{e}{n^{1+3\Delta}}+
\right.\\
\nonumber
&&+\left.
  {\rm O}(n^{-2})+ {\rm O}(n^{-2-\Delta})+
  {\rm O}(n^{-2-2\Delta})+{\rm O}(n^{-2-3\Delta})+\ldots
\ _{_{_{_{}}}}\right]\,,
\end{eqnarray}
which is a form particularly suitable for a fit versus $1/n$. (Actually, the 
last formula is valid in the case of one confluent singularity, but it may be 
generalized to more complicated instances. The easiness of the fit can be 
readily checked for the simpler case of no confluent singularity, which is 
recovered by setting $\Delta = 1$.)
\item[(ii)] 
Methods derived from Pad{\'e} approximants. Notably, a generalization
of Pad{\'e} approximants called differential approximants ---which 
has been proposed in \cite{GuJo1972} and subsequently developed 
in Refs.~\cite{HuBa1979,FiAY1979} and other
works--- has been derived just for the specific purpose of extracting the 
values of critical quantities, and stands nowadays as the technique of choice 
for the analysis of series like Eq.~(\ref{HTsusc}).
\end{enumerate}
 
Here we will content ourselves with a few
observations more about the results obtained by using the methods included
in class (i) above, to make a connection to Sec.~\ref{seqtr}.
Interestingly, many if not all of the ``classical'' acceleration 
schemes presented in Sec.~\ref{ThetaAlgorithm} (Richardson extrapolation, 
$\epsilon$-algorithm, $\theta$-algorithm, Levin transforms, and so on) have 
been applied in the literature to the study of ratio series of the type of
Eq.~(\ref{ExpConfl}). The careful and detailed analysis presented in 
{\cite{Gu1989} ---carried out comparing many different methods both for
some artificial test series and for some series derived from real problems 
in statistical mechanics---
shows how some of the examined acceleration schemes (notably the 
$\theta$-algorithm) 
behave more favourably than others, consistently with the observation
that for the case of ratio methods a logarithmic convergence should be
expected (see below); even more consistently, methods mainly suitable to 
accelerating linear convergence, like the $\epsilon$-algorithm, score poorly 
in such a situation.

One may now ask how it is possible to 
show that the convergence of the ratio sequence $R_n$ should be 
logarithmic. The answer
lies in the fact that the terms of a ratio series can be
seen as originating from a partial summation of a series $P_n$ defined as
\begin{equation}
P_n\equiv R_n-R_{n-1}=\frac{S_n}{S_{n-1}}-\frac{S_{n-1}}{S_{n-2}}=
\frac{S_n S_{n-2}-\left(S_{n-1}\right)^2}{S_{n-1}S_{n-2}}\,.
\end{equation}
Assuming the general form~(\ref{ExpConfl}) for the ratios, it can be checked that
\begin{equation}
P_n\sim\frac{1}{z_{\rm critical}}\left[
  -\frac{\lambda-1}{n\left(n-1\right)}
  -\frac{c\left(1+\Delta\right)}{n\left(n-1\right)^{1+\Delta}}
  +\ldots
\right]\sim\frac{\rm constant}{n^2}
\end{equation}
so that $P_n/P_{n-1}\rightarrow 1$, 
and we have logarithmic convergence indeed.

This example is instructive since it shows that there are real-life situations
where the validity of Eq.~(\ref{convergence_classification}) can be directly 
tested.

%
%
\subsubsection{Approach at Fixed Dimension $d=3$}
\label{fixeddim}

In QED, the fine-structure constant as the fundamental 
coupling parameters is small, and the computation of more
terms in the perturbative expansions is usually effective to get more precise
estimates of physical quantities even if the series we start with would not
in principle be convergent.
However, in the case of the determination of critical quantities by means of
many field-theoretical methods, the addition of terms of the expansion derived
from higher loops is simply not enough even when it is feasible: the coupling 
constant is so large in the relevant region where the phase transition occurs
that the additional terms do not by themselves provide any increase 
in the precision obtained for the results.

This consideration explains why in the latter case the loop calculations must
be complemented by a resummation prescription in order
to determine critical exponents.
In the case of the method at fixed dimension $d=3$, 
this observation is even more legitimate, since in such a framework 
the Borel summability of scalar $O(N)$ theories has been rigorously proven 
{\cite{EcMaSe1975}.
In this light, it is not any more surprising that
from an historical point of view, the computation of critical exponents by 
field-theoretical methods has represented a testbed for the application 
of many new resummation schemes; in particular, many of the devices implying 
Borel summation have been introduced in physics or invented just with the 
purpose of producing better estimates for critical exponents.

Typical results obtained for the $\beta$-function and the 
anomalous dimensions 
$\eta\!\left(g\right)$ and $\eta_2\!\left(g\right)$ 
in the $O(2)$-symmetric three-dimensional $(\phi^2)^2$-theory 
read~\cite{BaNiMe1978}
\begin{subequations}
\label{coeffs}
\begin{eqnarray}
\beta^{N=2}\!\left(u\right)&=&
-u+u^2 -0.4029629630\,u^3+0.314916942\,u^4 \nonumber\\
& & -0.31792848\,u^5+0.3911025\,u^6-0.552448\,u^7+
{\rm O}(u^8)\,,
\label{beta}\\[2ex]
\eta^{N=2}\!\left(u\right)&=&
0.0118518519\,u^2-0.0009873601\,u^3+0.0018368107\,u^4 \nonumber\\
& & -0.00058633\,u^5+0.0012514\,u^6+
{\rm O}(u^7)\,,
\label{eta}\\[2ex]
\eta_2^{N=2}\!\left(u\right)&=&
-\frac{2}{5}\,u+\frac{4}{50}\,u^2-0.0495134445\,u^3+
0.0407881056\,u^4\nonumber\\
& & -0.04376196\,u^5+0.0555573\,u^6+
{\rm O}(u^7)\,. \label{eta2}
\end{eqnarray}
\end{subequations}
Analogous series have been obtained for other values of $N$.
Following RG theory, a different universality class (that is, a different set
of physical systems) corresponds to each value of this parameter. For example,
the case $N=2$ is found to describe superfluid {\rm He}$^4$ at the 
$\lambda$-point and $XY$-ferromagnets, the case $N=1$ labels Ising-like 
ferromagnets and liquid-to-vapour transitions, the case $N=5$ has been proposed
as descriptive of some high-$T_c$ superconductors \cite{dG1972}, and the 
limit $N\rightarrow 0$ models polymers and self-avoiding walks \cite{Zh1997}.

These series describe the coefficients of the Callan-Symanzik equation
\begin{equation}
\left( m\frac{\partial}{\partial m}
+\beta\!\left(g\right) \, \frac{\partial}{\partial g}
-\left(\frac{a}{2}-b\right)\,\eta\!\left(g\right)-
b \,\eta_2\!\left(g\right)
\right)\Gamma^{\left(a,b\right)}=\Delta\Gamma^{\left(a,b\right)}\approx0
\label{CS}
\end{equation}
for the so-called three-dimensional $O\!\left(N\right)$-vector model,
which is defined by the bare Euclidean action (the index ${\rm B}$ labels 
the bare quantities)
\begin{equation}
\mathcal{S}_{\rm B}(\underline{\phi}_{\rm B})  \equiv \int {\rm d}^3x
\left(
\frac{1}{2}(\nabla \underline{\phi}_{\rm B})^2
+\frac{m^2_{\rm B}}{2} \underline{\phi}_{\rm B}^2
+\frac{g_{\rm B}}{4!}(\underline{\phi}_{\rm B}^2)^2
\right)\,.
\end{equation}
Here, $\underline{\phi}_{\rm B}$ is a vector of $N$ scalar fields, 
$\Gamma^{\left(a,b\right)}$ are the one-particle
irreducible Green functions (correlators) 
for the renormalized theory, with $a$ external vertices and $b$ insertions of 
$\underline{\phi}_{\rm B}^2$ operators. The
quantities $g$ and $u$ are related by a change of normalization.
We recall that the right-hand side of Eq.~(\ref{CS})
depends in general on the renormalization scheme adopted
[see again \cite{BaNiMe1978} for the precise definition of a scheme
leading to a right-hand side as in (\ref{CS})].

\begin{table}
\begin{center}
\begin{minipage}{14cm}
\begin{center}
\begin{tabular}{@{}c|ccccccccc|ccc}
\hline\hline
\vspace{-2mm}
&\multispan{9}{\hfill Incremental number\hfill }\vline&\multispan{3}{\hfill Total number\hfill}\\
\vspace{-0.5mm}
&\multispan{9}{\hspace*{\tabcolsep}\hrulefill\hspace*{\tabcolsep}}\vline
&\multispan{3}{\hspace*{\tabcolsep}\hrulefill\hspace*{\tabcolsep}}\\
Loop number& 0&1&2&3&4&5&6&7&8 & 6&7&8\\
\hline\hline
$\Gamma^{(2,0)^{^{^{}}}}_{_{_{}}}$& 0&0&1&2&6&19&75&317&1622 & 103&420&2042\\
\hline
$\Gamma^{(4,0)^{^{^{}}}}_{_{_{}}}$& 1&1&2&8&27&129&660&3986&26540 & 828&4814&31354\\
\hline\hline
\end{tabular}
\caption{\label{NumberOfDiagrams} Number of diagrams 
(one-particle irreducible and without tadpoles) contributing to
the two-point and four-point function up to $8$ loops in perturbative order.}
\end{center}
\end{minipage}
\end{center}
\end{table}

We would like to emphasize once again that the numerical coefficients in 
Eqs.~(\ref{beta})---(\ref{eta2}) are deceptively simple quantities. 
In fact, to obtain just each last term of these series,
on the order of $1000$ Feynman amplitudes had
to be integrated numerically, 
with a precision of $8$---$10$ digits each. 
Such amplitudes were originally known in terms of integral representations
in $18$ dimensions, and only a cleverly arranged set of computational
simplifications allowed for a reformulation~\cite{BaNiMe1978}
in terms
of equivalent amplitudes of lower dimensionality ($\leq 6$), thus opening
the way to a successful high-precision numerical evaluation.

The goal of calculating critical quantities in this computational scheme
is not straightforward, even if the coefficients in 
Eq.~(\ref{coeffs}) are assumed to be computable numerically.
Until the end of this subsection, we examine in detail all the necessary
steps to be performed.

(i) {\em The diagrammatic expansion.}
First of all, we have to generate all the diagrammatic expansion
for the Green functions $\gammat$, $\gammaf$ and $\gammato$, together with
that of $\dgammat$.
%
%
The knowledge of these correlators is indispensable to have 
enough information to be able to extract RG functions $\beta$, $\eta$ and
$\eta_2$ from Eq.~(\ref{CS}), respecting at the same time the renormalization
prescriptions set by the renormalization scheme employed.

A measure of the complexity of the generation stage can be found in 
Table \ref{NumberOfDiagrams}. It should be noticed that 
the number of diagrams for $\dgammat$ is excluded from the Table,
the corresponding raw number of diagrams being empirically intermediate between
that of $\gammat$ and that of $\gammaf$. On the other hand, 
since our renormalization scheme is such that the correlators must be evaluated
at zero external momenta, the diagrams of $\gammato$
can be considered a subset of those of $\gammaf$.

For each amplitude, the corresponding combinatorial/symmetry factor,
and the corresponding $O\!\left(N\right)$ polynomial must be computed too.

(ii) {\em Finding the best parameterizations for the needed amplitudes.}
At a given loop level $L$, the natural 
representation of each amplitude entails a $3L$-dimensional
momentum-space integral. Although other
representations are in principle possible, 
it was realized in~\cite{BaNiMe1978} that, starting from the standard
momentum representation of diagrams, one can systematically find and replace
in the amplitudes one-loop functions by analytic expressions
which are known in $d=3$ dimensions. 
This procedure lowers the number of needed final integrations.

As an example, we will show in some detail how the parameterization of the
diagram in Fig.~\ref{haus_diag}(a) 
can be carried out. 

\begin{figure}[tb!]
\begin{center}
\begin{minipage}{14cm}
\begin{center}
{\list{}{\itemindent 2\listparindent}\item[]
\begin{center}\begin{tabular}{@{}ccc}
\begin{minipage}[c]{0.192\textwidth} 
\vspace*{-5mm}\includegraphics[width=\textwidth]{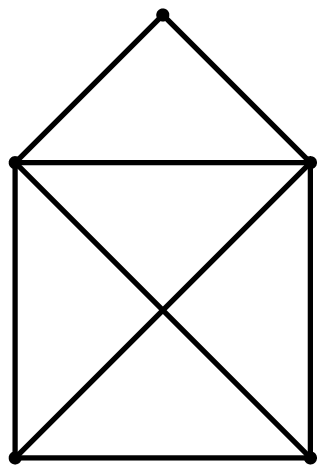}
\end{minipage}
&
\begin{minipage}[c]{0.228\textwidth} 
\includegraphics[width=\textwidth]{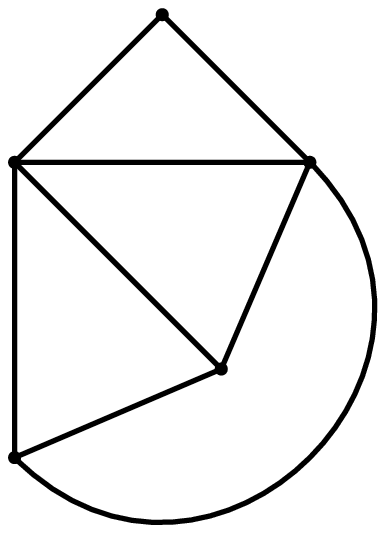}
\end{minipage}
&
\begin{minipage}[c]{0.228\textwidth} 
\includegraphics[width=\textwidth]{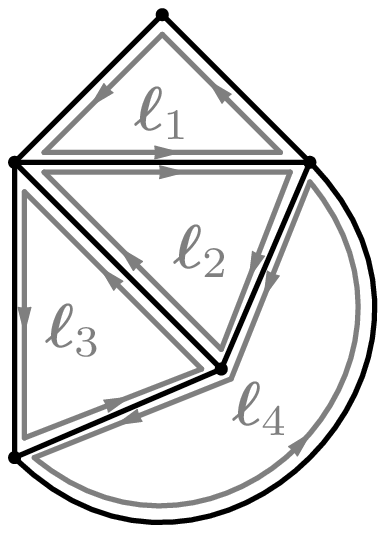}
\end{minipage}
\\
(a)
&
(b)
&
(c)
\\
\begin{minipage}[c]{0.24\textwidth} 
  \includegraphics[width=\textwidth]{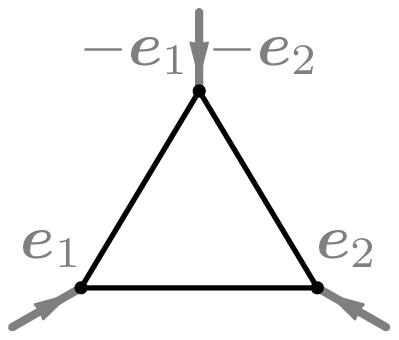}
\end{minipage}
&
\begin{minipage}[c]{0.24\textwidth} 
  \includegraphics[width=\textwidth]{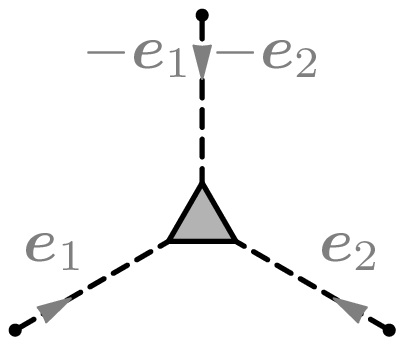}
\end{minipage}
&
\begin{minipage}[c]{0.192\textwidth} 
  \includegraphics[width=\textwidth]{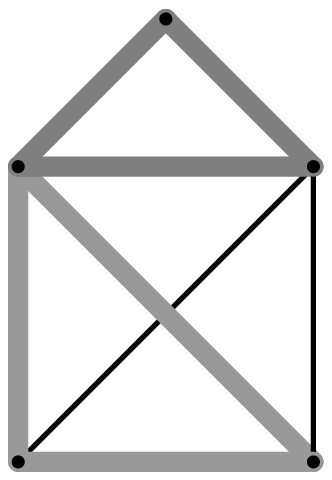}
\end{minipage}
\\
(d)
&
(e)
&
(f)
\\
\begin{minipage}[c]{0.192\textwidth} 
  \includegraphics[width=\textwidth]{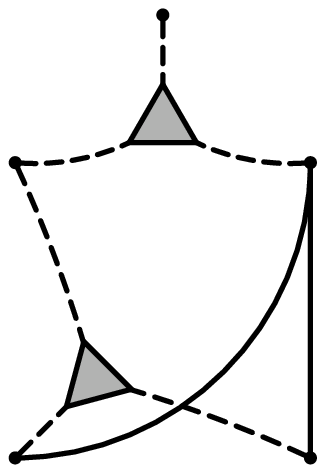}
\end{minipage}
&
\begin{minipage}[c]{0.21\textwidth} 
  \includegraphics[width=\textwidth]{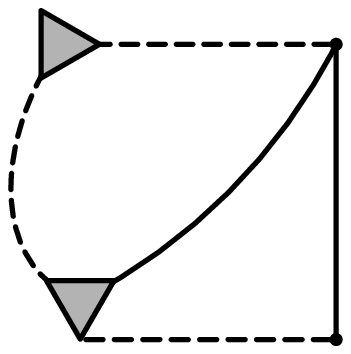}
\end{minipage}
&
\begin{minipage}[c]{0.21\textwidth} 
  \includegraphics[width=\textwidth]{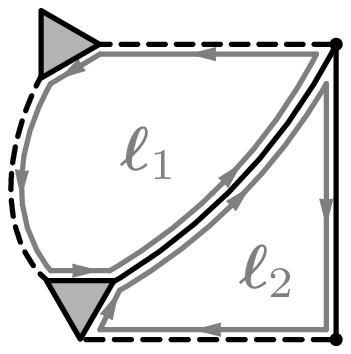}
\end{minipage}
\\
(g)
&
(h)
&
(i)
\end{tabular}\end{center}{\endlist}}
\caption{\label{haus_diag} Parameterization of a 
four-loop diagram in terms of ``triangles''
[diagrams (a), (b) and (c)], and
various steps in the parameterization
of the diagram in Fig.~\ref{haus_diag}(a) in terms of effective triangle
vertices [diagrams (d)--(i)]. 
Further explanations are given in the text.}
\end{center}
\end{minipage}
\end{center}
\end{figure}

The diagram in Fig.~\ref{haus_diag}(a)
has four loops. After redrawing it as in Fig.~\ref{haus_diag}(b) 
to highlight its planar structure, and assigning loop momenta as in
Fig.~\ref{haus_diag}(c), we can make use of standard Feynman rules 
for scalar theories in $3$ dimensions, arriving at the formula
\begin{eqnarray}
\nonumber\mathcal{A}\propto\int
{\rm d}^3\boldsymbol\ell_1 \,{\rm d}^3\boldsymbol\ell_2 \,
{\rm d}^3\boldsymbol\ell_3 \,{\rm d}^3\boldsymbol\ell_4 \,&&\hspace*{-7mm}
\;\mathrm{Pr}^2\!\left(\boldsymbol\ell_1\right)
\;\mathrm{Pr}\!\left(\boldsymbol\ell_3\right)
\;\mathrm{Pr}\!\left(\boldsymbol\ell_4\right)\\&&\hspace*{-7mm}
\times
\;\mathrm{Pr}\!\left(\boldsymbol\ell_1+\boldsymbol\ell_2\right)
\;\mathrm{Pr}\!\left(\boldsymbol\ell_2+\boldsymbol\ell_3\right)
\;\mathrm{Pr}\!\left(\boldsymbol\ell_2+\boldsymbol\ell_4\right)
\;\mathrm{Pr}\!\left(\boldsymbol\ell_3-\boldsymbol\ell_4\right)\,,
\label{Naive}
\end{eqnarray}
which expresses the amplitude at zero external momenta
(as specified by the renormalization scheme adopted
in \cite{Pa1980} and \cite{BaNiMe1978}) in terms 
of an integral representation of dimensionality $12$.
The propagator $\mathrm{Pr}\!\left(\boldsymbol\ell\right)$ here is 
the scalar Euclidean kernel
\begin{equation}
\mathrm{Pr}\!\left(\boldsymbol\ell\right)=\frac{1}{\ell^2+m^2}\,.
\end{equation}
The renormalized mass $m$ is set to one by the renormalization schemes
used in \cite{Pa1980} and \cite{BaNiMe1978}.
Needless to say, starting from Eq.~(\ref{Naive}),
it would be extremely difficult, if not impossible,
to give a precise numerical estimate for the amplitude corresponding
to the diagram in Fig.~\ref{haus_diag}(a).

Now, we can formulate a much more efficient strategy
to evaluate the diagram in Fig.~\ref{haus_diag}(a),
if we suppose for example to know,
with its complete dependence on external momenta 
${\boldsymbol e}_1$ and ${\boldsymbol e}_2$,
the analytic formula for the one-loop $3$-dimensional integral
corresponding to the ``triangle'' diagram of Fig.~\ref{haus_diag}(d), that is,
\begin{equation}
\mathrm{Tri}\!\left({\boldsymbol e}_1,{\boldsymbol e}_2\right) \equiv 
\int d^3\boldsymbol\ell \;
\;\mathrm{Pr}\!\left(\boldsymbol\ell\right)
\;\mathrm{Pr}\!\left(\boldsymbol\ell+{\boldsymbol e}_1\right)
\;\mathrm{Pr}\!\left(\boldsymbol\ell+{\boldsymbol e}_1+{\boldsymbol e}_2
\right)\,.\label{Triangle}
\end{equation}
In fact, in this case we can identify in the right-hand side 
of Eq.~(\ref{Naive}) 
suitable subintegrals being of the form of the right-hand side of 
Eq.~(\ref{Triangle}),
replacing them with the analytic formula for the function $\mathrm{Tri}$;
this operation can even be performed diagrammatically, by systematically
substituting all occurrences of Fig.~\ref{haus_diag}(d) 
with the ``effective vertex'' of Fig.~\ref{haus_diag}(e), 
where the dashed lines attached to triangle vertices 
are used to automatically implement momentum conservation.

This method leads, after the simple graphical manipulations 
of Figs.~\ref{haus_diag}(g), (h) and the choice of momenta of 
Fig.~\ref{haus_diag}(i), 
to the new integral representation
\begin{equation}
\mathcal{A}\propto\int {\rm d}^3\boldsymbol\ell_1 \;{\rm d}^3\boldsymbol\ell_2 \;
\;\mathrm{Tri}\!\left(\boldsymbol\ell_1,-\boldsymbol\ell_1\right)
\;\mathrm{Tri}\!\left(\boldsymbol\ell_1,\boldsymbol\ell_2\right)
\;\mathrm{Pr}\!\left(\boldsymbol\ell_1+\boldsymbol\ell_2\right)
\;\mathrm{Pr}\!\left(\boldsymbol\ell_2\right)\,.
\label{Clever}\end{equation}
After the explicit parameterization of momentum space 
${\rm d}^3\boldsymbol\ell_1\;{\rm d}^3\boldsymbol\ell_2$, 
this representation turns out to have a dimensionality of ``only'' $3$.
In Ref.~\cite{NiMeBa1977} a value of $0.1120344277(1)$ is quoted for this
diagram (we notice that a suitable normalization factor therein described 
must be inserted into Eq.~(\ref{Clever}) to obtain 
this numerical result.)

As the four-loop diagram in Fig.~\ref{haus_diag}(a) 
constitutes a relatively simple example, we would like to add that
much more complex effective vertices than the triangle alone 
must be used to make this technique effective for the large sets
of diagrams which are needed in the six- and seven-loop
orders. In addition, the way of
replacing effective vertices is not unique,
possibly leading to
many different integral representations, each one with a different
final dimensionality.
For example, as Fig.~\ref{haus_diag}(f)
already suggests, many other triangles than the two highlighted
ones could be identified at different positions in the diagram 
(however, it should also be noticed that in this example 
the maximal number of triangles
which can be replaced at the same time is always two,
no matter which choice of triangles is considered).
The identification of the most favorable 
parameterization defines a very complicated optimization
problem, which, if connected to the very large number of diagrams
to be evaluated, calls in turn for an automated way of finding out
the most favorable parameterizations. A strategy to solve this problem
in a more general and detailed way, even considering other possible analytic 
simplifications, is described in \cite{GuRi2006}.

It should be noticed that in this approach the renormalization procedure 
can be built into the step of amplitude parameterization.
The reason is that
the $O\left(N\right)$ vector model is superrenormalizable in $3$ dimensions
(see \cite{GuRi2006} for further details).
  
(iii) {\em Numerical evaluation of the needed amplitudes.}
Each integral representation obtained during the last stage must be
numerically evaluated at zero external momenta, following the 
renormalization-scheme prescriptions originally formulated 
in Ref.~\cite{Pa1980}. When combined with the symmetry factors and the 
$O\left(N\right)$ factors, as described above,
this operation leads to
the perturbative knowledge (up to the desired loop order) of RG functions
$\beta$, $\eta$ and $\eta_2$ appearing in Eq.~(\ref{CS}).

As previously emphasized,
the difficulty implied by this stage should not be underestimated.
For example, no published algorithm for multidimensional numerical integration
is able to guarantee that the needed number of significant digits
($7$--$8$) will be obtained after a reasonable number of function 
evaluations for all needed $9$-dimensional integral representations
at the level of $7$ perturbative loops.
However, an innovative algorithm described in \cite{GuRiPARIS}, 
which has been recently designed with this problem
in mind, should finally allow to obtain an asymptotic
behaviour of the integration error nice enough to reach 
the high precision required for this multi-dimensional computation.

(iv) {\em Resummation of the obtained series.}
In order to complete the analysis of the critical
exponents, we recall that the non-trivial value of the coupling 
$u$ such that the RG $\beta$-function in Eq.~(\ref{beta}) goes to zero
must be identified numerically
(this corresponds to the so-called Wilson-Fisher fixed point
of the $O\left(N\right)$ vector model).
One can then extract, for example, the value for 
the critical exponent $\alpha$
from the following formulas
\begin{subequations}
\begin{eqnarray}
\beta\!\left(u^{*}\right)&=&0\,,\\
\alpha&=&
2-\frac{d}{2-\eta\!\left(u^{*}\right)+\eta_2\!\left(u^{*}\right)}\,.
\end{eqnarray}
\end{subequations}
In this physical situation, the critical value of the coupling 
is non-perturbatively large (we have $g^*\approx 1.4$ here), and
therefore a suitable resummation of the series for the scaling functions 
is required to get meaningful results for the desired physical quantities.
In addition, and as mentioned before, the mathematical property of Borel 
summability has been rigorously established in \cite{EcMaSe1975} for the 
$\left(\phi^2\right)^2$ theory in the case $d=3$, so applying the procedure of
Borel resummation to this physical problem is definitely legitimate. 

From an historical perspective, it is noteworthy that the first precise 
numerical evaluation of critical quantities has been derived 
in~\cite{BaNiMe1978}
by means of a Borel--Pad{\'e} summation as described in 
Sec.~\ref{subsubsection_borel_pade}.
However, a spectacular theoretical breakthrough was achieved in the seminal 
paper \cite{LGZJ1977,LGZJ1980} 
when some additional information, 
derived from the large-order behavior analysis of perturbation theory
for scalar quantum field theories carried out in \cite{BrLGZJ1977}, 
has been conveyed to the Borel resummation of these series (a detailed review 
of the subject with additional results can also be found in 
\cite{BrPa1978}). Such analysis states that the coefficients of the 
perturbative series for any critical quantity $S(g)$ behave at large 
perturbative orders $n$ as
\begin{equation}
S_n = c (-a)^n n^b n! \left(1-O\left(\frac{1}{n}\right)\right),
\label{LargeOrderBehavior}
\end{equation}
with $a$, $b$ and $c$ being
computable numbers which depend on the critical function
to be evaluated. In particular, $a$ turns out to be independent of the series
under consideration and equal to
\[ a = 0.147774232 (1) \frac{9}{N + 8} \]
for the case $d = 3$.
It should be noticed how this discovery legitimates {\it a posteriori} the use
of the Borel--Pad{\'e} 
summation in~\cite{BaNiMe1978}, which had been proposed in
that context as a mere guess.

If we introduce a generalized Borel-Leroy transform $B_S^{(1, b')}(u)$ of $S(g)$, 
defined as
\begin{equation}
B_S^{(1, b')}(u)=\sum_{n=0}^\infty\frac{S_n}{\Gamma\left(n+b'+1\right)}u^n,
\label{Borel_b_transform}
\end{equation}
(which differs from the formulas presented in Sec.~\ref{borel_general_formulas}
only for the presence of an arbitrary parameter $b'$),
Eq.~(\ref{LargeOrderBehavior}) shows that the closest singularity of 
$B_S^{(1, b')}(u)$ (called instanton singularity) is located at the point
$-1/a$ (we refer the reader to \cite{ZJ1996} for a more detailed 
discussion about both this fact and the used terminology).

If the additional hypothesis is formulated that all the finite-distance
singularities of the Borel transform of the series come from 
subleading instantonic contributions (which are located further away on the 
negative axis of the complex plane), and the Borel transform is thus
supposed to be analytic in the whole complex plane except for the cut 
going from $-1/a$ to $-\infty$ (this statement 
is also known as the maximal analyticity hypothesis), then
a conformal mapping specially tailored to the analytic structure of the series
can be used, leading to a remarkable enhancement in the accuracy of the 
obtained resummed quantities. It should be noticed that, contrary to the 
property of Borel summability, the property of maximal analyticity has not been
rigorously proven for the case of $d=3$; however, arguments for its 
plausibility are given in \cite{ZJ1996}.

We then arrive again at a method based on the conformal mapping of the Borel 
plane, which we have already discussed in detail in 
Sec.~\ref{subsubsection_borel_pade-conformal}: by introducing the transformation
\begin{equation}
  z\left( u \right) = \frac{\sqrt{1 + au} - 1}{\sqrt{1 + au} + 1} \, ,
  \label{conformal_mapping_a}
\end{equation}
we map the Borel plane, cut at the instanton singularity $u = - 1 / a$, 
onto a circle in the $z$ plane in such a way to enforce maximal analyticity,
and thus to enhance the rate of convergence. With the help of 
Eq.~(\ref{conformal_mapping_a}), we can now rewrite the Borel transform 
(\ref{Borel_b_transform}) upon expansion of the $u$ 
in powers of $z$:
\begin{equation}
  B^{(1, b')}_S\left( u (z) \right) = \sum_{n = 0}^{\infty} \
  s^{[b']}_n \, z^n \, .
  \label{Borel_transform_in_z}
\end{equation}
where the coefficients $s^{[b']}_n$ can be uniquely determined as function
of $S_n$. Finally, the full physical solution $S(g)$ can be reconstructed 
by evaluating the Laplace--Borel integral:
\begin{eqnarray}
  S\!\left( g \right) &=& \frac{1}{g^{b' + 1}} \,
  \int\limits_{0}^{\infty} u^{b'} \exp(-u/g)\,
  B^{(1, b')}_S\left( u (z) \right) \, {\rm d} u \nonumber \\
  &=& \frac{1}{g^{b' + 1}} \,
  \sum_{n = 0}^{\infty} s^{(b')}_n \int_0^{\infty} u^{b'}
  \exp(-u/g) \left[ z(u) \right]^n {\rm d} u \, . 
  \label{Borel-Leroy-transform-exponents}
\end{eqnarray}
Interestingly, it is possible to connect the so far arbitrary parameter
$b'$ to $b$, and use it as a variational check to validate the method; in
addition, introducing more complicated prescriptions the knowledge of $c$ 
can be incorporated in the resummation algorithm as well (the reader
interested in this point may wish to consult \cite{LGZJ1980}). In this way,
one can properly take advantage of all the available information about the 
analytical properties of the critical quantities. Finally, additional 
refinements of the method described here are possible, like considering an 
additional homographic transformation to displace possible 
complex singularities in the $g$-plane \cite{GuZJ1998bx}.

However, we would like to 
mention that many other resummation methods have been proposed and 
implemented in this context. For example, an alternative approach
which does not make use neither of Borel summation nor of Pad{\'e} approximant
is that of order-dependent mapping given in~\cite{SeZJ1979} 
and~\cite{LGZJ1983}; to the best of our knowledge,
this method too has been invented for the particular case
of the resummation of critical exponents.

Anyway, let alone the problem of finding the correct value for the resummed 
series, we would like to emphasize how an even greater care should be spent 
in giving an accurate evaluation of the final
uncertainty of the resummation. 
In fact, the error in critical quantities basically comes from two sources, 
the summation error at fixed $g^{\ast}$, $\Delta S$, and the error induced 
by the error in $g^{\ast}$, $\Delta \tilde{g}^{\ast}$:
\begin{equation}
S = S^{\ast} \pm \Delta S \pm \left( \frac{d S}{d \tilde{g}}
   \right)_{\tilde{g}^{\ast}} \Delta \tilde{g}^{\ast}. 
\end{equation}
In this respect, the computation of some auxiliary quantities like the
shifted series
\begin{equation}
S_s \left( z \right) = \frac{1}{z^s} \left( S \left( z \right) - \sum_{k =
   0}^{s - 1} S_k z^k \right), s\in{1,2}, 
\end{equation}
or the series $1 / S$,
can give precious hints about the spread of the numerical results, thus
ultimately leading to a reasonable indication about the summation errors. 
A sophisticated analysis of the error bounds, as well as a careful review of 
the values for critical quantities obtained by Borel resummation from quantum
field theoretical techniques can be found in \cite{GuZJ1998bx}
(see also the earlier work~\cite{LGZJ1977}).

%
%
\subsubsection{Concluding Remarks on Renormalization Group--Related Quantities}
\label{rgremarks}

Since the determination of critical exponents is a particularly
prominent application of numerical convergence ``accelerators''
and summation techniques for divergent series, we would like to 
include here a few remarks regarding possible foreseeable progress in this
field. Indeed, there is some hope that the computations presented in the two 
last subsections can be partly extended in a not too distant future.

As discussed in \cite{NiRe1990}, in the case of the HT series the main
computational problem is apparently a graph-theoretical one. In particular, 
the algorithm of 
generation of the set of graph required for the LCE is very inefficient,
being plagued by the so-called problem of isomorphic copies:
every known algorithm for graph enumeration generates some subset of the 
needed graphs more than once, so that the effective number of graphs to be 
scanned during the generation stage can be orders of magnitude larger than 
the dimension of the desired final set. In addition, storage problems are quite
common during the computation of the series 
(see Refs.~\cite{CaEtAl2001,CaEtAl2002,CaEtAl2006}).
As a matter of fact, the reached level of $20$-$25$ terms of HT series seems
to be close to the saturation of presently available computational resources.

Similar considerations can be carried out in the case of the approach at fixed
dimension: even if promising steps have been done in~\cite{GuRi2006} towards 
the full extension to $7$ perturbative loops of the original $6$-loop 
computation of \cite{BaNiMe1978}, the method does not appear to be easily
extensible to a higher perturbative level. The most relevant problem with such
an approach lies in the fact that there is no a priori proof of its validity: 
no theorem exists stating that at a given loop order, for a given set of 
available effective vertices, some minimal reduction in the number of residual 
integrations will be obtained uniformly for all graphs needed for the 
evaluation of the desired physical quantities; the effectiveness of the method
can in fact be demonstrated only a posteriori, by an explicit inspection 
carried out diagram by diagram.
Indeed, such a proof has been recently given in~\cite{GuRi2006}
for the level of $7$ perturbative loops; however, it is unlikely that the 
method can be extended much further, since each loop more raises 
the upper bound for the dimensionality of the residual integration which must 
be carried out for each Feynman diagram, thus quickly making things 
computationally too expensive even if the considered set of effective vertices
is quite large.

Other computational schemes for RG quantities which are less suffering
from this explosive growth in computational complexity with the order of the 
approximation do exist --- notably Monte-Carlo methods. As mentioned before,
it is possible with these methods to reach a precision comparable to that
of the experimental results so far available; however, also the scaling of 
these algorithms is actually very bad, requiring as well a formidable
consumption of computational resources (for example, the sophisticated
theoretical estimates quoted in \cite{CaHaPeVi2006} required about
$20$ years of total CPU time).

In any case, while some progress can certainly be achieved with all 
three of the above mentioned methods,
it appears unlikely that any of the theoretical techniques known 
so far will be able to reach a precision comparable to that foreseen for the
next generation of experiments (which should be at least one order of
magnitude better than the accuracy reached so far).
Perhaps some theoretical breakthrough could change this picture.

%
%
\section{Conclusions}
\label{Conclu}

First, let us give a synopsis of the convergence 
acceleration methods discussed in Sec.~\ref{seqtr}
of this review. We can summarize a few important 
aspects as follows:
\begin{enumerate}
\item[(i)] As has been emphasized in~\cite{We1989} and here, 
our theoretical knowledge of the properties of many of the algorithms
is incomplete, although an important classification can be given 
in terms of alternating and nonalternating, as well as 
linearly and logarithmically convergent series.
We can say that a theory of their properties, if it is to be considered
as satisfactory from a practical point of view, should provide a 
theoretical estimate of the numerical stability properties
in addition to the convergence acceleration properties of the transforms.
\item[(ii)] None of the convergence accelerators discussed here
is linear in the sense that the $n$th-order transform of the sum of two 
input series is equal to the sum of the $n$th-order transforms
of the two input series. Linearity is thus not provided. 
However, a weaker requirement called translation under translation
can in general be imposed [see, e.g., Eq.~(3.1-4) of Ref.~\cite{We1989}].
It is generally acknowledged nowadays that the 
nonlinear transformations discussed in this review potentially
lead to much better numerical results that typical linear convergence
acceleration schemes.
\item[(iii)] Logarithmically convergent sequences are 
rather difficult to accelerate; in
particular, they constitute a set with the property of remanence
(see Ref.~\cite{BrRZ1991}), thus satisfying the hypothesis of a
``non-universality'' theorem~\cite{DeGB1982}. This implies
that no universal sequence transformation exists which is able to
accelerate all the sequences belonging to the set; in other words,
a given method will work only for some subset of 
the logarithmic sequences, and will not work for all the others.
However, a few potentially very powerful algorithms can be 
indicated (see, e.g., Sec.~\ref{OtherTraditional} and 
Table~\ref{master}).
\item[(iv)] It would be a hopeless task to attempt an exhaustive 
overview of all available convergence acceleration methods in this
review. As an example of an algorithm which 
is not discussed here, we only mention the Overholt process
and the ACCES algorithms 
which are described in detail in Ref.~\cite{BrRZ1991},
or the Euler transformation and the Euler--Maclaurin
formula which are discussed in many standard reference
books on mathematical methods for the physics community 
(e.g.,~Ref.\cite{AbSt1972,Ol1974}).
We have only attempted to provide a general overview of some
available methods, to introduce a rather universally 
applicable construction principle via the $E$-algorithm
(see Sec.~\ref{ThetaAlgorithm}), and to illustrate these 
general concepts by considering some examples. 
The interested reader will find further information in the 
more specialized literature indicated in Sec.~\ref{seqtr}.
\item[(v)] If one is simply interested in numerical results, 
there may be a strong motivation to skip the theoretical phase
of the algorithm-finding process sketched
in Sec.~\ref{seqtr} to a large extent, and to investigate the 
performance of convergence accelerators experimentally.
It is undeniable that this approach if far from 
satisfactory from a theoretical
point of view, but it can lead, nevertheless,
to numerically spectacular results.
In the case of no available analytic information, experiments are 
even unavoidable. In this context, one might note that
according to Refs.~\cite{SmFo1979,SmFo1982}, the 
nonlinear sequence transformation listed in Table~\ref{master},
notably the $d$ and $\delta$
transformations discussed in Sec.~\ref{NonlinearST}
below, are among the most
powerful and most versatile sequence transformations for 
alternating series that are currently known.
\item[(vi)] The notion of a certain ``complementarity'' appears to 
be rather universally applicable to the field of 
convergence acceleration algorithms, in various contexts.
E.g., the $\epsilon$ algorithm is known to be an
efficient accelerator for linearly convergent and even factorially
divergent input series, but fails for logarithmically
convergent input data. The $\rho$ algorithm is known to 
be an efficient accelerator for logarithmically convergent 
input series, but fails for linearly convergent and 
factorially divergent input data. Both the $\epsilon$ and $\rho$ algorithms
are numerically rather stable and rather universally applicable to 
series in their respective domains, but are in many cases outperformed 
by special sequence transformations which profit from 
explicit remainder estimates. On the other hand, the sequence transformation
with remainder fails completely if the assumptions about the 
structure of the remainder term are not met by a given input 
series. 
\item[(vii)] The notion of ``complementarity''
also affects the ``iteratability'' of the transformations:
Relatively simple sequence transformations like the $\Delta^2$ process,
which make only very crude assumptions about the structure of the 
remainder term, can typically be iterated, leading to improvements
in higher orders of the iteration process. In sharp contrast, 
the elimination of a complex remainder structure as implied by
a sequence transformation with a remainder estimate intuitively 
leads to the conjecture that a further iteration of the 
process cannot increase the apparent convergence any further. 
This is indeed confirmed in numerical studies performed 
by the authors, and illustrates the general complementarity aspect 
of the convergence accelerators once more: the power of a 
specialized algorithm has to be traded for a disadvantage in the
sense that further iterations typically cannot increase the 
numerical gain of a single iteration.
\end{enumerate}
The perhaps most important application of the convergence 
acceleration techniques is the calculation of the hydrogen Green 
function, described in Sec.~\ref{divcomp}.
Namely, while the 
so-called $Z\alpha$-expansion (``perturbative bound-state QED'') in general
converges better for low nuclear charges
$Z$ than for high $Z$, it is necessary, at the
current level of accuracy, to even include nonperturbative effects at low
$Z$. This surprising observation is especially important for the one-loop
self-energy correction, which is the dominant radiative correction in
hydrogenlike ions. To give an example, the difference between all known
terms of the $Z\alpha$-expansion up to the order $\alpha(Z\alpha)^6 m c^2$
(Ref.~\cite{Pa1993}) and the nonperturbative,
higher-order remainder for the 1S state~\cite{JeMoSo1999}
amounts to $28 \, {\rm kHz}$. This is about 700 times
larger than the current best measurement in atomic hydrogen, which has an
accuracy of about $40 \, {\rm Hz}$~\cite{NiEtAl2000,FiEtAl2004}.
Thus, while corrections of smaller absolute magnitude can still be described
perturbatively at low $Z$ (e.g., two-photon effects), the dominant
one-photon corrections have to be calculated nonperturbatively to match
current experimental precision, and from this point of view, atomic
hydrogen ($Z=1$) constitutes a ``high-Z system'' whose description would be
incomplete without a calculation of the nonperturbative effects.
An essential ingredient in that direction is given by the methods
discussed in Sec.~\ref{divcomp}.

Regarding divergent series and Borel summation treated in 
Sec.~\ref{Chap:DivBorel}, 
one important conclusion is that 
even the best summation method cannot reconstruct
a complete physical answer from perturbation theory alone,
if there are instanton-related contributions,
nonperturbatively suppressed, which must be added to the
perturbative expansion. This has been seen in one-dimensional
model problems, notably in the case of double-well like 
potentials treated in Sec.~\ref{DivMATdwf}.
In the calculation~\cite{SuLuZJJe2006}, an attempt
has been made to use resurgent expansions at
moderate and large couplings, for a few nontrivial
cases of one-dimensional potentials which,
nevertheless, admit a treatment via generalized 
quantization conditions. The conclusion has been that 
while the resurgent expansion can be useful to 
explore nonperturbative effects at small coupling, it cannot 
be used for accurate quantitative methods in the asymptotic
strong-coupling region. Roughly speaking, the triple resummation
of the resurgent series in $n$, $k$ and $l$ 
in Eq.~(\ref{resurgent_expansion_dw})
is too hard for any available summation algorithms to handle 
even if one enters directly into the quantization condition (\ref{qcond_dw})
and directly resums the $A_{\rm dw}$ and $B_{\rm dw}$ function given in 
Eqs.~(\ref{B_expansion_dw}) and~(\ref{A_expansion_dw}).
Nevertheless, the accurate understanding of nonperturbative
effects in quantum potentials with instanton contributions
via resurgent expansions at small and moderate coupling
could be seen as a significant consequence of the 
understanding gained by the path integral formalism.
Let us also mention that important connections of 
the quantum oscillators to field theoretical problems 
exist, as explained in detail in Ref.~\cite{ZJ1996}.

We should perhaps mention that our treatment of the even and odd
anharmonic oscillators in 
Secs.~\ref{even_oscillator_Borel-summation} and~\ref{DivMATodd} 
is entirely based on the concept of distributional
Borel summation and thus far from exhaustive.
One aspect left out in this review 
concerns renormalized strong coupling expansions~\cite{We1996b,We1996d}
and related methods
(see also~\cite{We1997,WeCiVi1993,We1992,We1990,WeCi1990,%
WeCiVi1991,We1996d,RoBhBh1998}) which make possible 
the determination of very accurate numerical energy eigenvalues
using only few perturbative coefficients, and which cover
both small and large couplings. Expansions of this kind,
however, fail for double-well like potentials and 
also cannot reproduce the width of resonances in the 
case of an odd perturbation.

To conclude this review, let us indicate some open problems and
directions for further research.
(i) There is a regrettable absence of a
general mathematical proof for the convergence
of some of the nonlinear sequence transformations described 
in Sec.~\ref{seqtr}. This might lead to a 
direction of research for mathematics.
Only for some special model problems could rigorous convergence
proofs be obtained (see \cite{Si1979,Si1980,Si1986} or Sections 13
and 14 of \cite{We1989}).
(ii) New sequence transformations with improved remainder 
estimates could be a significant direction of research for both physics and 
mathematics. In this context, one might mention
hyperasymptotics (see Ref.~\cite{Ol1974} and a couple
of preparatory calculations, e.g., Refs.~\cite{NeSc1975,Be1977}.
(iii) For nonalternating divergent series,
the next term omitted from the partial sum of the asymptotic series is 
typically not a good simple remainder estimate.
One may indicate the unsolved problem of finding a 
practically useful remainder estimates for nonalternating 
divergent series, in particular those which give rise to an imaginary part
in the sense of a distributional Borel summation.
(iv) Let us also indicate a possibly interesting 
further direction in the field-theoretical direction
which is based on the following observation:
In general, a perturbation potential or a field interaction 
determines the large-order behaviour of the generated perturbation series.
However, the interaction potential also determines the large-coupling 
expansion of the related quantities.
This means that there should be a connection between the 
large-coupling expansion
and the large-order behaviour of the perturbation series.
This notion and related ideas about connections of weak- and 
strong-coupling regimes have quite recently been explored in 
Refs.~\cite{Kl1998,Kl1998add,Kl1999,%
Su2001phi4,Su2001,Su2002,Su2005rev}, but further work
in this direction is definitely needed.
(v) Finally, there is a possible connection of the ``prediction''
or extrapolation of perturbative coefficients as described
here in Sec.~\ref{Accuracyrelations}, and Brodsky--Lepage--Mackenzie 
(BLM) scale setting~\cite{BrLeMa1983} which is based on the idea that,
roughly speaking,
seemingly wildly divergent perturbative expansions in field
theory can sometimes be formulated as not so wildly divergent 
series if the expansion variable (coupling constant) 
is taken at a momentum scale appropriate to the problem.
This approach is illustrated in Ref.~\cite{BrLeMa1983}
using the two-loop correction to the muon anomalous magnetic 
as an example, and a related approach has recently 
been generalized to higher orders in Ref.~\cite{KaSt1995}.
However, it seems that there is still room for improvement
in the renormalization-group-inspired approaches 
to the extrapolation of perturbative expansions
to higher-order loop corrections in field theory.

%
%
\subsection*{Acknowledgements}

First and foremost, the authors would like to 
acknowledge elucidating and insightful discussion with 
Professor Ernst Joachim Weniger. 
His seminal contributions to the field 
have been a significant motivation and 
inspiration to write the current review.
In addition, the authors acknowledge helpful discussions with 
Sergej V. Aksenov, Jens Becher, and Professor Sandro Graffi
as well as Professor Jean Zinn--Justin. 
P.R.~acknowledges the stimulating atmosphere at 
Gesellschaft f\"{u}r Schwerionenforschung (GSI).
U.D.J.~acknowledges support by the 
Deutsche Forschungsgemeinschaft (Heisenberg program). 
Finally, we would include an apology for the
necessarily incomplete list of references of this review;
and this goes out to all the authors whose work we failed to cite here.

\appendix

%
%
\section{Basic Terminology for the Spectra of Hamilton Operators}
\label{appbase}

%
%
\subsection{Resolvent, Spectrum, Green's Function}
\label{appbaseres}

In this subsection we shortly recall, for the convenience of the reader,
some of the basic mathematical definitions concerning  spectral theory of
Schr\"odinger operators, including resonances. A standard reference for
these topics is \cite{ReSi1978}.

\begin{definition}
Given a closed operator $H$ in a Hilbert space $\H$ with
domain $D(H)$, its resolvent set $\rho(H)$ is the set of all $\lambda\in\C$
such that the operator $H-\lambda I$ (where $I$ is the identity operator)  is
invertible and its range
$R(H-\lambda I)$  is the whole of $\H$.
\end{definition}

By the closed graph theorem (see e.g.~\cite{Ka1966}, Theorem
III.5.20) the inverse operator
$R(\lambda,H) =(H-\lambda I)^{-1}$, which by the above definition
exists and has domain $\H$, is bounded, or, equivalently, 
continuous, namely there is $C(\lambda)>0$ such that, for all $u\in\H$:
\begin{equation}
\|R(\lambda,H)u\|\leq C(\lambda)\|u\|\,.
\end{equation}
$\rho(H)$ is called resolvent set because the linear inhomogeneous
equation
$(H-\lambda I)u=f$ in the unknown $u\in\H$ can  always be solved: it has
indeed the unique solution 
$\displaystyle u=R(\lambda,H)f$ for any datum $f\in\H$ if $\lambda\in\rho(H)$. 

\begin{definition}
\label{spectrum}
The complement of $\rho(H)$ with respect to $\C$, i.e. the set of all
$\lambda\in\C$ such that either $R(\lambda,H)$ does not exist (in the
sense that the operator $H-\lambda I$ is not invertible), or, if it does, its
domain is not $\H$, is the spectrum of
$H$, denoted $\sigma(H)$. 
\end{definition}

The following examples can be given:
\begin{enumerate}
\item Let $H$ be a $n\times n$ matrix acting on $\H=\C^n$. Its spectrum
$\sigma(H)$ is clearly the set of the eigenvalues of $H$, because $H-\lambda I$ is
not invertible if and only if $Hu=\lambda u$ for some $u\neq 0$, i.e.
${\rm det}(H-\lambda I)=0$. On the other hand, if $H-\lambda I$ is invertible,
its range is always $\H$ because $\H$ has finite dimension.
\item Likewise, in general any eigenvalue $E$ of $H$ belongs to
$\sigma(H)$, because for any corresponding eigenvector $\psi$ one has
$(H-E I)\psi=0$.  Therefore the spectrum always contains the eigenvalues.
\item
If the Hilbert space is infinite dimensional, however, the
spectrum may have a totally different nature.
Let for instance  $H=X$ be the position
operator in quantum mechanics, namely the maximal multiplication operator
by $x$ in
$L^2(\R)$. Its action and domain are:
\begin{eqnarray}
(Xu)(x) =xu(x)\,,\qquad D(X) = \left\{u\in L^2(\R):
xu(x)\in L^2(\R) \right\} =
\nonumber\\
\left\{ u(x):\int_{\R}|u(x)|^2\,{\rm d}x<+\infty;
\int_{\R}|xu(x)|^2\,{\rm d}x<+\infty \right\} \,.
\end{eqnarray}
Then $\sigma(X)=\R$. We obviously have, indeed:
\begin{equation}
((R(\lambda,X) u)(x)=\frac{u(x)}{x-\lambda} \,.
\end{equation}
Now, for any given $\lambda \in \R$, pick $u(x)\in L^2(\R)$ continuous and 
nonzero in an open interval around $\lambda$. Then $\int_{\R}
\frac{|u(x)|^2}{|x-\lambda|^2}\,{\rm d}x$ does not exist and hence 
$R(X,\lambda)$ is not
defined on the whole of $\H$. 
\end{enumerate}
The most important example in the present context is represented by the
Schr\"odinger operator $H=T+V$  acting in $L^2(\R^n)$. Here $T$ is
the kinetic energy operator, and $V$ the (real-valued) potential.
In units with $\hbar=1$, these read
\begin{equation}
(Tu)(x)=- \Delta u(x)\,, \quad u\in D(T)\,, \qquad 
(Vu)(x)=V(x)u(x)\,,\quad  u\in D(V)\,.
\end{equation}
Here, $D(T)$ and $D(V)$ are the maximal domains of $T$ and $V$, namely
\begin{equation}
D(T) =\{u\in L^2(\R^n): \Delta u\in L^2(\R^n)\}, \qquad 
D(V) =\{u\in L^2(\R^n): V u\in L^2(\R^n)\} \,.
\end{equation}
We take as the domain $D(H)$ the maximal one,
\begin{equation}
D(H) =\{u\in L^2(\R^n): -\Delta u+Vu\in L^2(\R^n)\} \,.
\end{equation}
It is easy to see that $T$ and $V$ are self-adjoint operators,
namely $T=T^{\ast}$, $V=V^{\ast}$, where $A^\ast$ denotes the adjoint of the
operator $A$. Under suitable conditions on
$V$ (see Ref.~\cite{ReSi1978}, Chap.~10), also $H$
is self-adjoint (with the proof being often far from trivial). 
Concerning self-adjointness, we recall that:  
\begin{enumerate}
\item If $A$ is any self-adjoint operator in $\H$, then
$\sigma(A)\subset\R$.
\item The symmetry property $\langle Au | v \rangle=\langle u | A v\rangle$, $\forall\,(u,v)\in D(A)$ is not
sufficient to guarantee self-adjointness because it implies only
$A^{\ast}u=Au$ for
$u\in D(A)$ but it does not exclude $ D(A)\neq  D(A^{\ast})$. The
difference is quite substantial because while the spectrum of a
self-adjoint operator is real, the spectrum of a symmetric but not
self-adjoint operator is the whole of $\C$. 
\end{enumerate}
By its very nature, any Schr\"odinger operator is a differential operator 
acting on a Hilbert space of functions. This makes it possible to
characterize  its resolvent and spectrum in terms of the solutions
of the time-independent Schr\"odinger equation 
$(-\Delta + V ) \, \psi=E \, \psi$. 

\begin{definition}
\label{fund}
For any $\lambda\in\C$, let  $\G(x;\lambda)$ be a distributional solution of the
inhomogeneous equation
\begin{equation}
-\Delta \G(x;\lambda)+[V(x)-\lambda]\G(x;\lambda)=\delta(x), \quad
x\in\R^n \,,
\end{equation}
with $\delta(x)$ being the Dirac distribution. Under very general conditions on
$V$, it can be proven that $\G(x;\lambda)$ is unique, and it is called
fundamental solution, or Green function, of the Schr\"odinger operator
$H=T+V$.
\end{definition}

The Green function yields an explicit representation of the resolvent
$[H-\lambda I]^{-1}$. Consider indeed the integral operator in
$L^2(\R^n)$ of kernel $\G(x-y;\lambda)$, namely the operator acting
on functions $u\in L^2$ in the following way:
\begin{equation}
(G(\lambda) \, u)(x)=\int_{\R^n}\G(x-y;\lambda) \, u(y)\,{\rm d}y\,.
\end{equation}
We have, for any $u\in L^2$ (the interchange between differentiation and
integration can be justified):
\begin{eqnarray}
(H-\lambda \, I)(G(\lambda) \, u)(x) =
[-\Delta +V(x)-\lambda] \, 
\int_{\R^n} \G(x-y;\lambda) \, u(y)\,{\rm d}y =
\\
=\int_{\R^n}
[-\Delta +V(x)-\lambda] \, \G(x-y;\lambda)\, u(y)\,{\rm d}y =
\int_{\R^n}\delta (x-y)\, u(y)\,{\rm d}y = u(x) \,.
\end{eqnarray}
This shows that $G(\lambda)$ is the inverse of $[H-\lambda I]$, and
therefore coincides with the resolvent operator. Hence the above formula
yields  the explicit action of the resolvent $\Rl$:
\begin{equation}
\Rl \, u(x)=\int_{\R^n}\G(x-y;\lambda)\,u(y)\,{\rm d}y\,.
\end{equation}
As an example,
consider the free particle energy operator in one dimension,
\begin{equation}
H = T = -\frac{{\rm d}^2}{{\rm d}x^2}, \qquad 
D(T)=\{u: u \in L^2, u'\in L^2, u''\in L^2\}\,.
\end{equation}
The differential equation for the Green function is therefore
\begin{equation}
- \G''(x;\lambda) - \lambda \, \G (x;\lambda)=\delta(x)\,.
\end{equation}
To solve this equation, we switch to Fourier space.
Let $\hat{\G}(p;\lambda)$ be the
Fourier transform of $\G (x;\lambda)$:
\begin{equation}
\hat{\G}(p;\lambda) =
\frac{1}{\sqrt{2\pi}} \,
\int_\R \G (x;\lambda) \, {\rm e}^{-{\rm i} p x}\,{\rm d}x \,.
\end{equation}
Then (integrating twice by parts) we have
\begin{equation}
\frac{1}{\sqrt{2\pi}}\int_\R\G"(x;\lambda)\,
{\rm e}^{-{\rm i} p x}\,{\rm d}x=
p^2\hat{\G}(p;\lambda)\,.
\end{equation}
Since $\int_\R\delta(x){\rm e}^{-{\rm i} p x}\,{\rm d}x=1$,
the differential equation becomes
\begin{equation}
\label{Rp}
[p^2-\lambda]\hat{\G}(p;\lambda) = \frac{1}{\sqrt{2 \pi}}
\quad \Longrightarrow \quad
\hat{\G}(p;\lambda)=\frac{1}{\sqrt{2 \pi} (p^2-\lambda)}\,.
\end{equation}
If we now invert the Fourier transform (using for instance complex
integration) we readily obtain
\begin{equation}
\G(x;\lambda)={\rm e}^{\sqrt{-\lambda} \,|x|} \,.
\end{equation}
Hence, we have 
\begin{equation}
R(T,\lambda)u = [T-\lambda I]^{-1} \, u(x) =
\int_{\R} {\rm e}^{\sqrt{-\lambda} \, |x-y|} \, u(y)\,{\rm d}y\,.
\end{equation}
From this representation, we immediately see that $\sigma(T)=[0,+\infty)$
as it should be. If indeed $\lambda\in\C, \lambda \notin\R_+$, 
then we have for all $u\in L^2$, with
$\lambda=|\lambda| \, \exp({\rm i}\phi)$:
\begin{eqnarray}
\|R(T,\lambda)u\|^2 \leq
\int_\R\int_\R |{\rm e}^{2\sqrt{-\lambda} \, (x-y)}| \, 
|u(y)|^2\,{\rm d}y \, {\rm d}x \leq
\int_\R |u(y)|^2\, {\rm d}y \, \int_\R 
{\rm e}^{-2\sqrt{|\lambda|} \, \sin{(\phi/2)}) \, |x-y|}\,{\rm d}x
\nonumber\\
\leq \frac{1}{\sqrt{|\lambda|} \, \sin{(\phi/2)}} \, 
\int_\R |u(y)|^2\,{\rm d}y = C(\lambda) \, \|u\|^2 \,,
\qquad 
C(\lambda)=\frac{1}{\sqrt{|\lambda|}\sin{(\phi/2)}} \,,
\end{eqnarray}
whence the boundedness of $R(\lambda,T)$ 
because $0 < C(\lambda) < +\infty$ as long as
$0<\phi<2\pi$, {\rm i}.e. $\lambda\notin [0,+\infty)$.
In three dimensions with $x=(x_1,x_2,x_3)$, in exactly the same way
we obtain for the Green function of $T=-\Delta$:
\begin{equation}
\G(x;\lambda)=\frac{{\rm e}^{\sqrt{-\lambda} \, \|x\|}}{\|x\|},\quad
\|x\| =\sqrt{x_1^2+x_2^2+x_3^2}\,.
\end{equation}
As above, we can check that $\sigma(T)=[0,+\infty)$.

In the above examples, the spectrum is continuous. The question emerges
how to characterize the spectrum in  terms of  the
solutions of the Schr\"odinger equation.
We recall that by definition, $E$ is an eigenvalue of the Schr\"odinger
operator
$H$ (and hence a quantum bound state) if there is a non-zero function
$\psi(x,E)\in L^2(\R^n)$ 
(and hence normalizable: in other words, 
$\int_{\R^n}|\psi(x,E)|^2\,{\rm d}x=1$) such that
$H\psi =E\psi$. Then,
$\psi(x,E)$ is called an eigenfunction corresponding to $E$.
The number $m$ of linearly independent $L^2$ solutions of $H\psi=E\psi$
is the multiplicity of the eigenvalue $E$.  By the normalization
condition, all eigenfunctions  corresponding to any bound state  
vanish as $\|x\|\to +\infty$. Hence $E$ is an eigenvalue of the Schr\"odinger
operator if and only if the Schr\"odinger equation $H\psi=E\psi$
admits at least one normalizable solution.  It is well known that
eigenfunctions corresponding to different eigenvalues  of a
symmetric operator are orthogonal:
\begin{equation}
\langle \psi(x,E) | \psi(x,E')\rangle =
\int_{\R^n}\psi(x,E) \, \overline{\psi}(x,E')\,{\rm d}x=0, \quad E\neq E'\,.
\end{equation}
Linearly independent eigenfunctions corresponding to the same
eigenvalue can always be chosen to be orthogonal by the Gram-Schmidt
orthogonalization procedure when $\H$ is separable, i.e.~when $\H$
contains a countable basis. It follows that the eigenfunctions
corresponding to the bound states form an orthonormal system. If
the spectrum consists entirely of eigenvalues, this system is 
complete, namely every vector in $\H$ can be expanded in Fourier series
of the eigenvectors.

Under very general conditions on
$V$, it can actually be proven that any eigenfunction is
continuous and vanishes  at least exponentially (pointwise), 
i.e.~there is $K(E)>0$ such that 
\begin{equation}
|\psi(x,E)|\leq {\rm e}^{-K(E)\|x\|}, \quad \|x\|\to\infty\,.
\end{equation}
The points belonging to the continuous spectrum of $H$ can be
characterized in terms of the solutions of the stationary Schr\"odinger
equation  $H\psi=E\psi$ having a totally different behaviour at infinity:

{\it A point $\lambda\in\R$ belongs to the continuous spectrum of $H$ if
and only if there is a solution $f(x,\lambda)\notin L^2$ of the Schr\"odinger
equation $H \, f= \lambda \, f$ which fulfills the 
pointwise bound $|f(x,\lambda)|\leq
K(\lambda)$ for some $K(\lambda)>0$ and all $x\in\R^n$. }

Again, it is useful to consider an example. 
Consider once again the kinetic energy $T$ in one
dimension. Setting $\lambda=k^2, k\in\R$, the Schr\"odinger
equation  $T\psi=E\psi$ becomes 
\begin{equation}
-\frac{d^2}{{\rm d}x^2}\psi(x,k)=k^2 \psi(x,k)
\end{equation}
with the well known plane wave solutions
\begin{equation}
\psi(x,k)={\rm e}^{\pm {\rm i}kx}\,,
\end{equation}
which are not in $L^2$ because $|\psi(x,k)|=1$ if and only if $k\in\R$; 
however $\psi(x,k)$ is obviously bounded 
if and only if $k\in\R$, and hence
$\lambda \geq 0$. The functions $\psi(x,k)$ are often called continuum
eigenfunctions or sometimes generalized eigenfunctions.  Since
$\int_\R {\rm e}^{{\rm i}kx}\,{\rm d}x=2 \pi \delta(k)$, where $\delta(k)$ is the Dirac
distribution, they fulfill the continuum orthogonality relations
\begin{equation}
\int_\R\psi(x,k) \, \overline\psi (x,k')\,{\rm d}x =\delta(k-k')\,.
\end{equation}
Sometimes, this relation is written under the form $\langle
\psi(x,k) | \psi(x,k')\rangle = \delta(k-k')$ even if the scalar
product in the right-hand side does not make sense because $\psi(x,k)\notin L^2$.

Solutions of the Schr\"odinger equation which are 
both non-normalizable and unbounded have no relation with the spectrum.

We recall that an orthonormal system (ON) in a (separable) Hilbert
space $\H$ is a (finite or countable) set of vectors $\psi_n$,
with $n \in \mathbbm{N}_0$, 
such that $\langle\psi_n | \psi_m\rangle=\delta_{m,n}$.
Given any vector $u\in\H$, its Fourier coefficients with respect
to the given ON system are by definition the scalar products
\begin{equation}
a_n = \langle u | \psi_n\rangle \,.
\end{equation}
The ON system is complete when any vector in $\H$ can be expanded
in Fourier series, namely when
\begin{equation}
\label{F1}
u = \sum_{n=0}^{\infty}a_n\psi_n \,,
\end{equation}
where the convergence takes place in the norm of $\H$. 

If the spectrum of the Schr\"odinger operator $H$ is discrete,
namely it contains only isolated eigenvalues of finite multiplicity, then
the set of all its eigenfunctions can be selected in such a way that it
forms a complete ON system. The same property holds in general for any
self-adjoint operator with discrete spectrum. 

Let us consider again some example.
Let $H$ be the quantum harmonic oscillator, 
$H=-\frac{{\rm d}^2}{{\rm d}x^2}+x^2$ defined on its maximal domain. Then its
spectrum is discrete, because it can be proven that it contains only the
simple eigenvalues
$E_n= 2 n+1$ with $n \in \mathbbm{N}_0$. 
The corresponding eigenfunctions are the
Hermite functions 
$\psi_n(x)=\frac{1}{\sqrt{2^n \, n! \,\sqrt{\pi}}} \,
{\rm e}^{-x^2} \, H_n(x)$,
where $H_n(x)$ is the $n$th Hermite polynomial. The system $\psi_n$ is
orthonormal and complete:
\begin{equation}
\label{FH}
u(x)=\sum_{n=0}^{\infty} a_n \psi_n(x), \qquad 
a_n=\frac{1}{\sqrt{2^n \, n! \,\sqrt{\pi}}} \,
\int_\R u(x) \, {\rm e}^{-x^2}H_n(x)\,{\rm d}x \,.
\end{equation}
The convergence takes place in the $L^2$ norm (or quadratic mean), namely
the above equality sign means
\begin{equation}
\lim_{N\to\infty}
\int_{\R}|u(x)-\sum_{n=0}^{N} a_n \, \psi_n(x)|^2\,{\rm d}x= 0\,,
\end{equation}
and neither implies pointwise convergence nor
uniform convergence.

If the spectrum contains a continuous component, the ON set formed by the
eigenfunctions is not complete even when it is countable. For
example, the eigenfunctions of the Hydrogen atom are not complete
in $L^2(\R^3)$, because the spectrum of the corresponding Schr\"odinger
operator $ H=-\Delta-\frac{Z}{r}$ has a continuous component along
$[0,+\infty)$ in addition to the bound states $E_n=-\frac{Z^2}{n^2}$.

When the spectrum contains a continuous component the completeness
statement  requires the consideration of all eigenfunctions, namely
also the continuum eigenfunctions, which are the only ones when the
spectrum is purely continuous. 

Assume for the sake of simplicity that the continuous component stretches
over $[0,+\infty)$, so that setting $k=\sqrt{-E}$ we can label the
corresponding eigenfunctions by 
$\phi(x,k) = {\rm e}^{{\rm i}kx}/\sqrt{2\pi}$
for $k\in\R$. Then the 
continuum Fourier coefficient of $u(x)\in L^2$ is by definition the
integral transform
\begin{equation}
c(k)= \int_{\R}u(x) \, \overline{\phi(x,k)}\,{\rm d}x\,.
\end{equation}
If, as above, we denote by $\psi_n(x)$ the eigenfunctions corresponding to
the discrete eigenvalues $E_n$, the completeness theorem for the
eigenfunctions  is expressed by the formula
\begin{equation}
\label{compl}
u(x) = \sum_{n=0}^{\infty} a_n \, \psi_n(x) +
\int_{\R} c(k) \, \phi(x,k)\, {\rm d}k \,.
\end{equation}

Two examples illustrate the above statements.
(i) Consider once again the kinetic energy $T$ in one
dimension. There are no eigenvalues, and as we know the spectrum is
purely continuous along $[0,+\infty)$. 
Then, the completeness theorem provides nothing
else than the Fourier transform. We have indeed
\begin{equation}
c(k)= \frac{1}{\sqrt{2 \pi}}\,
\int_{\R} u(x) \, {\rm e}^{-{\rm i} k x}\,{\rm d}x=\hat{u}(k)\,,
\end{equation}
where $\hat{u}(k)$ is the Fourier transform of $u$. The above completeness
formula is here just the Fourier inversion formula
\begin{equation}
u(x) = \frac{1}{\sqrt{2 \pi}}\,
\int_{\R}c(k) \, \phi(x,k)\,{\rm d}k
= \frac{1}{\sqrt{2 \pi}}\,
\int_\R \hat{u}(k) \, {\rm e}^{{\rm i} k x}\,{\rm d}k \,,
\end{equation}
and the orthogonality relation 
$\int_\R\psi(x,k) \, \overline\psi (x,k')\,{\rm d}x =\delta(k-k')$ 
immediately yields the Fourier-Plancherel formula:
\begin{equation}
\int_\R |u(x)|^2\,{\rm d}x= \int_\R |\hat{u}(k)|^2\, {\rm d}k \,.
\end{equation}
(ii) Consider the radial equation for the Hydrogen atom, 
$H=-\Delta-\frac{Z}{r}$. Then Eq.~(\ref{compl}) holds for any function
$f(r)\in L^2(\R_+)$ with $\psi_n(x)$ the Laguerre functions of argument
$\frac{r}{n}$ and $\phi(r,k)$ a hypergeometric function as
defined, e.g.~in \cite{LaLi1958}. 

%
%
\subsection{Analytic Continuation of the Resolvent and Resonances}
\label{appbaseana}

Quantum mechanical  resonances are usually defined by complex,
second-sheet poles of the scattering amplitude. This means what follows:
one first proves that the scattering amplitude is an analytic function of
the energy in the whole complex plane cut along the positive real axis.
Then, if the potential is regular enough, one proves also that the
scattering amplitude admits a meromorphic continuation across the cut.
The complex poles of the continuation are the resonances. The real part of
the pole is the position of the resonance, and the imaginary part the
width. It can be shown, under fairly general conditions, that 
it suffices to prove these properties for the
resolvent of the   Schr\"odinger operator. Namely, one proves by the
Lippmann-Schwinger equation (see, e.g., Ref.~\cite{Si1973}) that
the existence of complex poles of the analytic continuation of the
resolvent (shortly, second sheet poles of the resolvent) implies that the
scattering amplitude admits the very same second sheet poles.  

Let us examine more closely the second sheet poles of the resolvent. 
First recall that, given any closed operator $H$, the resolvent $\Rl$ is
an analytic operator-valued function of $\lambda\in\rho(H)$ (shortly, the
resolvent is analytic for  $\lambda\in\rho(H)$). This means that for
any $\psi\in\H$ the scalar product
\begin{equation}
R_\psi(\lambda)=\langle\psi | \Rl \, \psi\rangle
\end{equation}
is an analytic function of $\lambda\in\rho(H)$. If $H$ is self-adjoint,
$R_\psi(\lambda)$ is in general analytic 
for $\lambda\notin\R$. If $\sigma(H)$ is
discrete, $R_\psi(\lambda)$ is meromorphic with simple poles along the real
axis at the eigenvalues of $H$. 
Let us illustrate these situations by
means of two examples from quantum mechanics.

(i) The position operator $X$. We have already seen that $R(\lambda,X)$ is
just the multiplication by $\frac{1}{x-\lambda}$. Hence:
\begin{equation}
R(\lambda,X)\psi(x)=\frac{\psi(x)}{x-\lambda}
\quad 
\Longrightarrow 
\quad 
\langle \psi |  R(\lambda,X) \, \psi\rangle=
\int_{\R}\frac{|\psi(x)|^2}{x-\lambda}\,{\rm d}x, 
\quad
\lambda\notin\R \,.
\end{equation}
The above formula for $R_\psi(\lambda)=\langle \psi | R(\lambda,X)\psi\rangle$ 
in general defines two distinct analytic
functions, one in the upper half-plane $\Im\lambda >0$ denoted by
$R^+_\psi(\lambda)$ and the other one, denoted $R^-_\psi(\lambda)$, in the lower half-plane
$\Im\lambda <0$. They are not the analytic continuation of each other,
unless $\psi$ vanishes in some open interval. Moreover, the possibility 
of continuing $R^-_\psi(\lambda)$ to the lower half-plane or $R^-_\psi(\lambda)$
to the upper half-plane depends on the regularity properties of $\psi$. 
Consider for instance the case $ \psi(x)= \exp(-x^2/2)$, and let us prove
that  $R^{+}_\psi(\lambda)$, a priori analytic for $\Im\lambda>0$, admits 
an analytic
continuation to the whole half-plane $\Im\lambda \leq 0$. In the same way, 
$R^{-}_\psi(\lambda)$, {\em a priori} 
analytic for $\Im\lambda<0$, admits an analytic
continuation to the whole half-plane $\Im\lambda \geq 0$. We have
\begin{equation}
R^{+}_\psi(\lambda)=\int_{\R}\frac{{\rm e}^{-x^2}}{x-\lambda}\,{\rm d}x, \quad 
\Im\lambda>0 \,.
\end{equation}
Since the integrand is analytic in $x$ for $\Im x\leq 0$ and
vanishes faster than exponentially when $|\Re x|\to+\infty$, we can
shift the integration down by $-a$, namely we can perform the change of
variable $x\to x-{\rm i}\,a$, $a>0$. Then
\begin{equation}
R^{+}_\psi(\lambda)=\int_{\R}
\frac{{\rm e}^{-(x-ia)^2}}{x-{\rm i} a-\lambda}\,{\rm d}x\,.
\end{equation}
The integrand in the right-hand side is analytic in $\lambda$ for $\Im\lambda>-a$,
and the integral converges absolutely and uniformly for
the same values of $\lambda$. Hence it represents the analytic continuation of 
$R^{+}_\psi(\lambda)$ to the half-plane 
$\Im\lambda >-a$. Since $a$ is arbitrary, 
$R^{+}_\psi(\lambda)$ thus admits a 
continuation to the whole of $\C$. A symmetric 
argument holds for $R^{-}_\psi(\lambda)$.
 
(ii) For the harmonic oscillator, let us verify directly that if $H$ is
the harmonic oscillator, the function $R_\psi(\lambda)$ is, for any 
$\psi\in L^2$, meromorphic with simple poles, which may occur only at the 
points at $\lambda_n=2n+1$ with $n \in \mathbbm{N}_0$.
Recall that if $\psi_n$ is a normalized eigenvector of $H$ corresponding
to the eigenvalue $\lambda_n=2n+1$, then
\begin{equation}
\Rl\psi_n=[H-\lambda I]^{-1}\psi_n=\frac{1}{\lambda_n-\lambda} \, \psi_n\,,
\end{equation} 
whence
\begin{equation}
R_{\psi_n}=\frac{1}{\lambda_n-\lambda}\langle\psi_n | \psi_n\rangle =
\frac{1}{2n+1-\lambda} \,.
\end{equation}
Let now $\psi\in L^2$ be arbitrary. By Eqs.~(\ref{F1}) and~(\ref{FH}),
we can write
\begin{equation}
\psi=\sum_{n=0}^{\infty}a_n \, \psi_n, \quad a_n=
\langle \psi | \psi_n\rangle\,,
\end{equation}
whence
\begin{subequations}
\begin{equation}
\Rl \, \psi=\Rl \, \sum_{n=0}^{\infty}a_n \, \psi_n =
\sum_{n=0}^{\infty} a_n \, \Rl \, \psi_n=
\sum_{n=0}^{\infty} \frac{a_n}{2n+1-\lambda} \, \psi_n \,,
\end{equation}
so that
\begin{eqnarray}
R_\psi(\lambda) &=& \langle \psi | \Rl \, \psi\rangle =
\sum_{m,n=0}^{\infty}
\langle a_m \, \psi_m | \frac{a_n}{2n+1-\lambda} \, \psi_n\rangle 
\nonumber \\
&=& \sum_{m,n=0}^{\infty}\frac{a_m\overline{a}_n}{2n+1-\lambda}
\langle \psi_m | \psi_n\rangle =
\sum_{n=0}^{\infty}\frac{|a_n|^2}{2n+1-\lambda}\,.
\end{eqnarray}
\end{subequations}
The last equality follows by the orthonormality relation
$\langle\psi_m | \psi_n\rangle=\delta_{m,n}$.  Hence $\R_\psi(\lambda)$ is a
meromorphic function with simple poles which may occur only at  the
points
$\lambda=2n+1$ with $n \in \mathbbm{N}_0$.  
Notice that the pole at $\lambda=2k+1$  is present
if and only if the  Fourier coefficient $a_k=\langle\psi | \psi_k\rangle$ 
is non-zero.

We can now give the definition of resonance associated with a given
Schr\"odinger operator
$H=T+V$ having continuous spectrum. For the sake of the simplicity we
will admit  that the continuous spectrum stretches either from zero to
$+\infty$ (which is always the case when $V\to 0$ as $\|x\|\to +\infty$)
or from $-\infty$ to $+\infty$ (as in the Stark effect). 

\begin{definition}
\label{resonance}
A complex number $E=E_1+{\rm i}\, E_2$, $E_2<0$, is a resonance of the
Schr\"odinger operator $H$ as above if there exists a dense set 
${\cal Q}$ in $\H$ such that
\begin{enumerate}
\item For any $\psi\in{\cal Q}$, the function $R_\psi(\lambda) =
\langle \psi | R(\lambda,H)\psi\rangle$, 
a priori holomorphic for $\Im\lambda >0$,  
admits a meromorphic continuation to $\Im\lambda \leq 0$ 
across the cut generated
by the continuous spectrum;
\item $E$ is a simple pole of any continuation $R_\psi(\lambda)$.
\end{enumerate}
\end{definition}

We remark that the assumption $E_2<0$ is standard and corresponds 
formally to a resonant state being a decaying state in the
future. Consider indeed the time-dependent Schr\"odinger equation 
$H\psi={\rm i}\partial_t\psi$. Assume that there exists $\phi(x)$ not in $L^2$ 
which however is a pointwise solution of the time-independent
Schr\"odinger equation
$H\phi=E\phi$ with $E$ as above. Then $\phi(x,t) 
=\phi(x) {\rm e}^{-{\rm i}Et}$
solves $H\psi={\rm i}\partial_t\psi$ and 
$ |\phi(x,t)|=|\phi(x)|{\rm e}^{E_2t} \to 0$
exponentially fast for any $x$ as
$t\to +\infty$ if $E_2<0$.  This is also the reason why the resonance
width is the inverse life-time of the resonant state.

In proving the existence of resonances for a given Schr\"odinger operator
the above definition requires the solution of two mathematical problems.
The first is represented by the actual construction of the meromorphic
continuation of the matrix elements $R_\psi(\lambda)$ across the cut for any
$\psi$ belonging to a suitable dense set in $\H$. The second is the proof
that the continuation actually has poles.
The most powerful method to construct the continuation is the 
dilation analyticity of Balslev and Combes~\cite{BaCo1971} 
(and its variants). It requires the analyticity of
the potential $V$, and works also for the Stark effect~\cite{GrGr1978}. 
Let us shortly review  how it works.

Given $\psi\in L^2(\R^N)$, and $\theta\in\R$ the corresponding 
dilated vector is
\begin{equation}
\psi_\t(x)={\rm e}^{N\theta/2}\psi({\rm e}^\t x)\,.
\end{equation}
Notice that $\psi_\t$ and $\psi$ have the same norm 
in $\R^N$ because 
\begin{equation}
\|\psi_\t(x)\|^2 =
{\rm e}^{N\t} \int_{\R^N}|\psi({\rm e}^\t x)|^2\,{\rm d}x =
\int_{\R^N}|\psi( x)|^2\,{\rm d}x=\|\psi\|^2 \,.
\end{equation}
The dilated position is clearly ${\rm e}^{\t} x$, and the dilated momentum 
${\rm e}^{-\t} \, \nabla_x$. Hence the dilated \Sc\ equation is
\begin{equation}
H(\t)=-{\rm e}^{-2\t}\Delta+V({\rm e}^\t x)\,.
\end{equation}
If we add a linear potential $F\,x_1$, with $F>0$, we get the \Sc\ 
equation of the Stark effect
\begin{equation}
H_F= -\Delta+V( x) + F \, x_1 \,,
\end{equation}
whose dilated form is
\begin{equation}
H_F(\t)=-{\rm e}^{-2\t}\Delta+V({\rm e}^\t x) + 
{\rm e}^\t \, F \, x_1 \,.
\end{equation}

Let now $\t$ be complex, $0<\Im\t<\pi/4$. The following
statements can be proven:
\begin{enumerate}
\item {\it The complex-dilated operator $H(\t)$ has the same real eigenvalues
as $H$. For $0\leq {\rm arg} \, \lambda < 2 \, {\rm arg} \, \t$,
the spectrum of $H(\t)$ consists at most of eigenvalues, 
which do not depend on $\t$.}
\item {\it The complex dilated operator $H_F(\t)$ has discrete spectrum, 
i.e.~it consists at most of isolated eigenvalues with finite multiplicity, with
negative imaginary part.}
\item {\it Let $\psi\in L^2$ be a dilation analytic vector. This means
that the function $\langle \psi_\t, \, \phi\rangle$ is an analytic
function of $\t$ for $|\Im\,\t|<\pi/4$ for any $\phi\in L^2$. The 
set of the dilation
analytic vectors is dense in $L^2$. Then one has:}
\begin{subequations}
\begin{eqnarray}
\label{Cont1}
& \left< \psi \left| \frac{1}{H-\lambda I} \right| \psi\right> = 
\left< \psi_{\overline \t} \left| \frac{1}{H(\t) - \lambda I} 
\right| \psi_\t \right>\,,
\qquad |\Im \, \t|<\frac{\pi}{4} \,,
\\[2ex]
\label{Cont2}
& \left< \psi \left| \frac{1}{H_F -\lambda I} \right| \psi\right> = 
\left< \psi_{\overline \t} \left| \frac{1}{H_F(\t) - \lambda I} \right|
\psi_\t\right> \,,
\qquad 0>\Im \, \t>-\frac{\pi}{4} \,.
\end{eqnarray}
\end{subequations}
\end{enumerate}

The above formulae yield the explicit analytic continuations of the
matrix elements 
$ R_\psi(H)=\langle \psi | \, [H-\lambda I]^{-1}\psi\rangle$ and 
$ R_\psi(H_F)=\langle \psi | \, [H_F-\lambda I]^{-1}\psi\rangle$, 
respectively. Moreover, we see that the poles of the
continuation must coincide with the complex eigenvalues of $H(\t)$ and
$H_F(\t)$, respectively, which are independent of $\t$. This last
characterization of the resonances, namely as eigenvalues of the 
non self-adjoint operator $H(\t)$ uniquely associated with the given
operator $H$ by the complex dilation is particularly useful. 

The above result settles the question of the definition of resonances and
of the construction of the analytic continuation of the resolvents (when
the potentials are analytic). Concerning the actual existence of
resonances, their characterization in terms of complex eigenvalues of the
complex dilated operators $H(\t)$ and $H_F(\t)$ has made 
it possible to obtain  a rigorous existence proof in the physically most relevant  
cases (atomic systems, Stark effect~\cite{Si1973,GrGr1978b}).

Finally, we remark that the resonance $E=E_1+ {\rm i} E_2$ is an 
eigenvalue of the complex dilated operator $H(\t)$. Therefore it 
corresponds to an eigenvector $\psi(x,E,\t)$, which depends on
$\theta$. If $\t\to 0$, $\psi(x,E,\t)$ becomes a solution of the
stationary Schr\"odinger equation, $H \, \psi(x,E)=E \, \psi(x,E)$, which is
neither $L^2$ nor oscillatory because $E$ is complex. In one dimension
typically it exhibits exponential increase in one direction and
exponential decrease in the opposite one. 

%
%
\section{Distributional Borel Summability}
\label{appdist}

%
%
\subsection{Mathematical Foundations of Distributional Borel Summability}
\label{appdistmath}

In Sec.~\ref{borel_general_formulas}, 
we have already seen that the Borel method cannot be directly
applied to the divergent expansion $\sum_{n=0}^{\infty} n! \, g^n$ because
its Borel transform has a pole on the real
positive axis and thus prevents the convergence of the Borel-Laplace
integral [see the discussion following Eq.~(\ref{defA})].  
The same consideration applies to any divergent
expansion $F(g) \sim \sum_{n=0}^{\infty} a_n \, g^n$ where the 
$a_n\sim \Gamma(k \,n+1)$ display a nonalternating sign pattern. 
Without loss of generality, we may assume $a_n > 0$ for all $n$. 
The Borel transform $B_F^{(k)}(g)$, defined according to 
Eq.~(\ref{BorelTransP_k}),
has a positive radius of convergence. However, since the coefficients of
its expansion are all positive, we can state that by the 
Vivanti-Pringsheim theorem~\cite{Pr1895,Pr1896,Vi1897}, the
function $B_F^{(k)}(g)$ has a singularity at the point where the convergence
circle meets the positive real axis. 

As we have seen in Sec.~\ref{borel_general_formulas}, the series 
$\sum_{n=0}^{\infty}n! \, g^n$ should be  associated to the
function ${\cal N}(g)= \int_0^{\infty} {\rm d}t\, 
{\rm e}^{-t}/(1-g \, t)$ for
$g>0$, where the integral can either be interpreted as a Cauchy principal
value integral or as an integral encircling the 
pole at $t = 1/g$ in one of the possible directions. 
Intuitively, one might conjecture that at least the result of the
principal-value integration, given this particular integration  
prescription, should be uniquely associated to the divergent input series,
whereas the sign of the imaginary part is ambiguous.
The main problem is to determine the uniqueness
conditions of the association of the divergent input series
to the value of the principal-value integral, which in turn 
may be referred to as a ``distributional'' Borel sum,
for reasons to be discussed below. The problem of 
clarifying the association is solved by the
method of distributional Borel summability~\cite{CaGrMa1986}.
Essentially, this concept extends the Nevanlinna criterion to boundary values
of analytic functions, which are in general distributions, whence
the name. The formal definition is as follows:

\begin{definition}
\label{DB}
We say that the formal power series $f(g) \sim \sum_{n=0}^{\infty}a_n \, g^n$ 
is Borel summable in the distributional sense to
$f(g)$ for $0 \leq g < R$, with $R>0$, if the following conditions are
satisfied:
\begin{itemize} 
\item[({\rm i})] In accordance with Eq.~(\ref{BorelTransP}), we set 
\begin{equation}
\label{BofT1}
B_f(t) = \sum_{n=0}^{\infty}\frac{a_n}{\Gamma(n+1)} \, t^n\,.
\end{equation}
Then $B(t)$, which is a priori  holomorphic in some circle $|t|< \Lambda$, 
admits a holomorphic continuation to the intersection 
of some neighbourhood of 
$\R_+ \equiv \{t\in{\R}\,:\,t >0\}$ with 
${\C}^+ \equiv \{t\in{\C}\,:\,\Im t >0\}$.
\item[(ii)]
The boundary value distributions 
$B(t\pm {\rm {\rm i}}\,0) \equiv \lim_{\epsilon \to 0}B(t\pm {\rm {\rm i}}\, \epsilon)$ 
exist $\forall t\in\R_+$ and the  following
representation holds, for $g$ belonging to the Nevanlinna disk
$C_R=\{g\,:\,\Re g^{-1}>R^{-1}\}$,  
\begin{equation}
\label{1.33}
f(g) = \frac{1}{g}\, 
\int_0^{\infty}\; {\rm PP}(B(t)) \; 
\exp\left( -\frac{t}{g} \right)\,{\rm d}t\,,
\end{equation} 
where ${\rm PP}(B(t))= \frac12\,(B(t+{\rm {\rm i}}\, 0)+
\overline{B(t+{\rm {\rm i}}\, 0)})$.
\end{itemize}
\end{definition}

As for the ordinary Borel sum, the representation (\ref{1.33}) is
unique among all real functions admitting the prescribed formal
power series expansion and fulfilling suitable analyticity
requirements and remainder estimates. In the ordinary Borel sum
case these are the Nevanlinna conditions of Theorem \ref{N}. 
Their generalization to the distributional case 
is as given in the following Theorem~\cite{CaGrMa1986}.

\begin{theorem}
\label{T.5.1}
Let $f(g)$ be bounded and analytic in the Nevanlinna disk 
$C_R = \{g : \Re{g^{-1}} > R^{-1}\}$, and let 
$f(g) = \frac12\,(\phi(g) + \overline{\phi(g)})$, 
with $\phi(g)$ analytic in $C_R$ so that
\begin{equation}
\label{5.1}
\left |\phi(g) - \sum_{n=0}^{N-1} a_n\,g^n\right | \leq 
A \, \sigma(\epsilon)^N \, N! \, |g|^N\,, \qquad \forall N=1,2,\dots
\end{equation}
uniformly in 
$C_{R,\epsilon} = \{g\in C_R : \arg{g}\geq -\pi/2 + \epsilon\}$, 
$\forall \epsilon>0$. 
Then the series $\sum_{n=0}^{\infty} (a_n/n!)u^n$
is convergent for small $|u|$,
and its sum admits an analytic continuation 
$B(u) = B_1(u) +  B_2(u)$. 
Here, $B_1(u)$ is analytic in 
$C_d^1 = \{ u: |\Im u| < d^{-1} \}$,
and $B_2(u)$ is analytic in 
$C_d^2 = \{u: (\Im u>0, \, \Re u > -d^{-1}) \;{\rm or}\; |u|<d^{-1}\}$ 
for some $d>0$. 
$B(u)$ satisfies
\begin{equation}
\label{5.2} 
|B(t+ {\rm {\rm i}} \, \eta_0)| \leq 
\frac{A'}{\eta_0} \, \exp\left( \frac{t}{R} \right)
\end{equation}
uniformly for $t>0$, for any $\eta_0$ such that $0<\eta_0<d^{-1}$. 
Moreover, setting
${\rm PP}(B(t)) = \frac12\, (B(t+{\rm {\rm i}}\,0) +
\overline{B(t+{\rm {\rm i}}\, 0)})$, 
the distributional Borel sum $f(g)$ admits the integral representation:
\begin{equation}
\label{5.3}
f(g)= 
\frac{1}{g} 
\int_0^{\infty} {\rm PP}(B(t)) \, \exp\left( -\frac{t}{g} \right)\,{\rm d}t\,,
\qquad 
g\in C_R \, , 
\end{equation}
{\rm i}.e. $f(g)$ is the distributional Borel sum of
$\sum_{n=0}^{\infty} a_n\,g^n$ for  $0\leq g <R$ in the sense of
Definition \ref{DB}.
Conversely, if $B(u) = \sum_{n=0}^{\infty} (a_n/n!)u^n$ is
convergent for $|u| < d^{-1}$
and admits the decomposition $B(u) = B_1(u) + B_2(u)$ with the above quoted
properties, then the function defined by (\ref{5.3}) 
is real-analytic in $C_R$ and
$\phi(g) = \int_0^{\infty} B(t+{\rm {\rm i}}\,0) \, \exp\left(-t/g\right) \, {\rm d}t$ 
is analytic and satisfies (\ref{5.1}) in $C_R$.  
\end{theorem}

{\it Proof.} See~\cite{CaGrMa1986}.

It is convenient to include the following
remark. The function 
$\phi(g) = g^{-1} \, \int_0^{\infty} 
B(t+{\rm {\rm i}}\,0) \, \exp(-t/g) \, {\rm d}t$
is called the "upper sum" and 
$\overline{\phi(\overline{g})} = g^{-1} \,
\int_0^{\infty} \overline{B(t+{\rm {\rm i}}\, 0)} \,
\exp(-t/g) \, {\rm d}t$ the
``lower sum'' of the series $\sum_{n=0}^\infty a_n \, g^n$. 
It follows that, for $g>0$, $f(g)
= \Re{\phi(g)}$. On the other hand, with this method, we can
single out a unique function with zero asymptotic power series
expansion, that is the ``discontinuity''
\begin{equation}
\label{defdisc}
d(g) = \frac{1}{g} \,
\int_0^{\infty}
\left[ B(t+{\rm {\rm i}}\,0) - \overline{B(t+{\rm {\rm i}}\,0)} \right] \,
\exp\left( -\frac{t}{g} \right)\,{\rm d}t = 
\phi(g) - \overline{\phi(\overline{g})}\,. 
\end{equation}
Thus, $d(g) = 2 {\rm {\rm i}}\,\Im{\phi(g)}$, for $g>0$.

Important example cases, where the concept of 
distributional Borel summability finds applications,
are being discussed in Secs.~\ref{DivMATodd}
(odd anharmonic oscillator) and below
in Sec.~\ref{appdiststar} (Stark effect).

Exactly as for the ordinary Borel summability
[we recall Eq.~(\ref{LaplaceIntP_k}) and Theorem~\ref{NL}], 
the distributional case can also be extended to perform the summation 
of power series diverging as fast as $n!^k$,
$k \in \mathbb{N}$. The corresponding generalization of 
Definition~\ref{DB} is:

\begin{definition} 
\label{defDNL}
Let $k$ be a rational number.
We say that the formal power series 
$F(g) \sim \sum_{n=0}^{\infty}a_n \, g^n$ is Borel-Leroy summable of order $k$ in
the distributional sense to $F(g)$ for $0\leq g<R$, with $R>0$, if the
following conditions are satisfied: 
\begin{itemize} 
\item[(1)] In accordance with Eq.~(\ref{BorelTransP_k}), we set
\begin{equation}
\label{BofT2}
B^{(k)}_F(t) =
\sum_{n=0}^{\infty} \frac{a_n}{\Gamma (k\,n+1)} \, t^n \, .
\end{equation}
Then $B(t)$, a priori holomorphic in some circle $|t|< \Lambda $, 
admits a holomorphic continuation to the intersection of 
some neighbourhood of
$\R_+ \equiv \{t\in{\R}\,:\,t >0\}$ with 
${\C}^+ \equiv \{t\in{\C}\,:\,\Im t >0\}$. 
\item[(ii)]
The boundary value distributions 
$B(t\pm {\rm {\rm i}}\,0) = \lim_{\epsilon \to 0}B(t\pm {\rm i}\epsilon) $ 
exist 
$\forall t\in\R_+$ and the  following
representation holds: 
\begin{equation}
\label{1.3}
F(g)=\frac1{kg} \, \int_0^{\infty} 
{\rm PP}\left(B^{(k)}_F(t)\right) \,
\exp\left[ - \left(\frac{t}{g}\right)^{1/k} \right] \,
\left(\frac{t}{g}\right)^{1/k-1}\,{\rm d}t
\end{equation} 
for $g$ belonging to the Nevanlinna disk of the $g^{1/k}$-plane,
which is $C_R \equiv \{g\in\C\,:\,{\rm Re}\,g^{-1/k}>R^{-1}\}$, and
where ${\rm PP}(B_j(t))=\frac12 \, (B^{(k)}_F(t+ {\rm {\rm i}}\,0)+
\overline{B^{(k)}_F(t+{\rm {\rm i}}\,0)})$.
\end{itemize}
\end{definition}

In this case, a criterion analogous to Theorem \ref{T.5.1} holds (see 
\cite{CaGrMa1986} for details). 
The generalization implied by Definition~\ref{defDNL}
is necessary, for instance, to prove the
distributional Borel summability of the odd anharmonic oscillators
$x^{2m+1}$ with $m>1$.

%
%
\subsection{Distributional Borel Summability of the Stark Effect}
\label{appdiststar}

The Hydrogen Stark effect is observed when a single-electron atom is placed in
an external constant electric field of strength $F$. If $Z$ is
the atomic number and $\mu$ the electron mass,
then  the Schr\"odinger equation becomes (with $\hbar = e = 1$)
\begin{equation}
\label{S1}
\left( -\frac{1}{2\mu}\Delta\psi-\frac{Z}{r}\psi +F x_3
\right) \psi=E \, \psi, \quad
\psi=\psi(x_1,x_2,x_3),\quad r=\sqrt{x_1^2+x_2^2+x_3^2}\,,
\end{equation}
where without loss of generality the electric field is directed along the 
$x_3$ axis. Observe that in contrast to the 
discussion in Sec.~\ref{appbaseana},
we are manifestly working in three dimensions here.
The perturbation theory yields a subtle problem whose solution
has direct connections to the odd anharmonic oscillators 
treated in Sec.~\ref{DivMATodd}.
In the presence of an  electric field, the ${\rm SO}(4)$ symmetry of
the hydrogen atom is broken;  if however the field is constant the 
${\rm SO}(2)$ symmetry around the direction of the field is conserved and the
Schr\"odinger equation is separable in parabolic coordinates. Therefore the
parabolic quantum numbers of the Hydrogen atom,
$n_1$, $n_2$ and $m$, are used for the classification of the atomic
states~\cite{LaLi1958}. Explicitly, let us introduce the 
``squared'' parabolic coordinates $(u,v,\phi)$, with
$0\leq u,v<+\infty$ and $0\leq \phi<2\pi$:
\begin{subequations}
\label{sqpar}
\begin{eqnarray}
u=r+x_3, \quad v=r-x_3, \quad \phi={\rm \arctan}\,{\frac{x_2}{x_1}}  \,,
\end{eqnarray}
inverted as
\begin{eqnarray}
x_1=uv\cos{\phi}, \quad x_2=uv\sin{\phi}, \quad x_3=\frac12(u^2-v^2) \,.
\end{eqnarray}
\end{subequations}
Now, express the Laplace operator in the squared parabolic coordinate (see
e.g.~\cite{LaLi1958}); then exploit the 
cylindrical symmetry around the $x_3$ axis to
separate out the $\phi$ dependence writing
$f(u,v,\phi)$ under the form 
$f(u,v,\phi)=f(u,v){\rm e}^{im\phi}$, $m=0,\pm 1,
\pm 2,\ldots$  and finally write $f(u,v)$ under the form
$f(u,v)=\xi(u)\eta(v)$. Setting from now on $\mu=1$,
we see that the Schr\"odinger equation (\ref{S1}) then reduces to
a coupled system of ordinary differential equations:
\begin{eqnarray}
\label{S2a}
-\frac{1}{2}\frac{{\rm d}^2\xi(u)}{{\rm d}u^2}+\frac{m^2-1/4}{u^2} \, \xi(u)
- 2Eu^2 \, \xi(u)+Fu^4 \, \xi(u) = Z_1\xi(u)
\\
\label{S2b}
-\frac{1}{2}\frac{{\rm d}^2\eta(v)}{{\rm d}v^2}+\frac{m^2-1/4}{v^2} \, \eta(v)
- 2Eu^2 \, \eta(v)-F \, v^4\eta(v)=Z_2\eta(v) \,,
\end{eqnarray}
where $Z_1+Z_2=Z$ and the boundary conditions $\xi(0)=\eta(0)=0$ hold for
$m=0$. The spectrum of (\ref{S1}) is recovered in the following way: let
$Z_1^{n_1;m}(E,F)$ with $n,m \in \mathbbm{N}_0$ be the eigenvalues of
Eq.~(\ref{S2a}), and
$Z_2^{n_2;m}(E,F)$  with $n,m \in \mathbbm{N}_0$ those of 
Eq.~(\ref{S2b}).  Then
the equation
\begin{equation}
\label{Sev}
Z_1^{n_1;m}(E,F)+Z_2^{n_2;m}(E,F)=Z
\end{equation}
implicitly defines, for any triple $(n_1,n_2,m)$, a family of functions
$E(n_1,n_2,m,F)$ to be identified with the resonances of the Stark effect,
which has no bound states as already recalled. In this context 
it should be remarked that the minus sign appearing
in front of the quartic term in equation (\ref{S2b}) requires some care in
the definition of the corresponding operator and thus of its eigenvalues. Note
moreover  that for $F=0$ and $E<0$, the hydrogenic eigenvalues are recovered 
because in this case 
$Z_1^{n_1;m}(E,0)=\sqrt{-2E}\,(n_1+|m|+1/2)$ and
$Z_2^{n_2;m}(E,0)=\sqrt{-2E}\,(n_2+|m|+1/2)$. Hence the condition $Z_1+Z_2=Z$
yields
\begin{equation}
E(n_1,n_2,m)=-\frac{Z^2}{2(n_1+n_2+|m|+1)^2}.
\end{equation}
Thus, $m$ is the standard magnetic quantum number; $n_1$ and $n_2$ are called
parabolic quantum numbers, and the principal quantum number $n$
has the expression $n=n_1+n_2+|m|+1$ in terms of the parabolic quantum numbers.

As already mentioned, no hydrogen
eigenvalue persists no matter how small  the intensity $F$ of the applied
electric field is. We have indeed:
 
\begin{theorem}
\label{SpettroS}
For $F\neq 0$ the Schr\"odinger operator $H(F)$ whose domain and action are
defined as follows
\begin{eqnarray} 
\label{SA1}
D(H(F)) = C_0^{\infty}(\R^3); \qquad
H(F) \, u = \left( -\frac{1}{2\mu}\Delta-\frac{Z}{r} +
F \, x_3 \right) \,
u, \qquad u\in D(H(F))\,,
\end{eqnarray}
is essentially self-adjoint, i.e.~it has one and only one self-adjoint
extension. Its spectrum (identified with the spectrum of the self-adjoint
extension) extends from $-\infty$ to $+\infty$ and is absolutely continuous.
\end{theorem}
 
{\it Proof.} The first proof goes back to Titchmarsh, and relies on the
separability in squared parabolic coordinates. 
A more recent and direct proof can be found in~\cite{GrGr1978b}.

The absolute continuity of the spectrum in particular excludes the occurrence
of eigenvalues embedded in the continuum.  All eigenvalues of the hydrogen
atom are thus expected to turn into resonances, and this physical intuition can
actually be proven.

\begin{theorem}
Let $ E(n_1,n_2,m) = E(n_1,n_2,m;0) = -Z^2/(n_1+n_2+m+1)^2$ be any
eigenvalue of the hydrogen atom, classified by the parabolic quantum numbers
$n_1\geq 0, n_2\geq 0$ and by the magnetic quantum number 
$m \in \mathbbm{N}_0$. Then for $|F|\neq 0$ it 
turns into a resonance $E(n_1,n_2,m;F)$, namely:
\begin{enumerate}
\item Let $0<{\rm arg} \, F<\pi$. Then the operator family $H(F)$ defined by
(\ref{SA1}) has discrete spectrum. 
\item There is $B(n_1,n_2,m)>0$ such that the eigenvalue 
$E(n_1,n_2,m)$ of the hydrogen atom
is stable as an eigenvalue $E(n_1,n_2,m;F)$ of $H(F)$ for 
$|F|<B(n_1,n_2,m)$, $\forall\,0<{\rm arg} \, F<\pi$. 
\item The eigenvalues $E(n_1,n_2,m;F)$ are holomorphic functions of $F$ in the
complex sector $\Omega(n_1,n_2,m) = \{F\in\C\;:\,|F| < B(n_1,n_2,m)\,;\,
0<{\rm arg}\,F<\pi\}$.  They admit an analytic continuation up to any sector
$\{F\in\C\; : \,|F| < B_\epsilon(n_1,n_2,m)\, ; \,
-\pi/2+\epsilon < {\rm arg}\, F < 3\pi/2-\epsilon\}$, 
$\forall\,\epsilon >0$, through the real axis. 
\item For $F\in\R$, $F\neq 0$, any function $E(n_1,n_2,m;F)$ is complex valued,
with the properties 
\begin{subequations}
\begin{eqnarray}
{\rm Re} \, E(n_1,n_2,m;F+{\rm i}\,0) &=& 
{\rm Re} \, E(n_1,n_2,m;F-{\rm i}\,0),\\ 
{\rm Im} \, E(n_1,n_2,m;F+{\rm i}\,0) &=& 
-{\rm Im} \, E(n_1,n_2,m;F-{\rm i}\,0) \,. 
\end{eqnarray}
\end{subequations}
Any complex-valued function $E(n_1,n_2,m;F)$, $F\in\R$  is a resonance of the
Stark effect. More specifically,  ${\rm Re} E(n_1,n_2,m;F)$ is the location of
the resonance, and ${\rm Im} E(n_1,n_2,m;F)$ its width, i.e. its inverse
life-time.
\end{enumerate}
\end{theorem}
{\it Proof.} See~\cite{GrGr1978b}.

We remark that the functions $E(n_1,n_2,m;F)$ are resonances of the Stark effect
in the strongest possible sense; namely they verify the following properties
\begin{enumerate}
\item They are second-sheet poles of the resolvent. This means 
the following: Consider the resolvent operator $ R(F,E)=[H(F)-E]^{-1}$.  Since the
spectrum of $H(F)$ is the whole of $\R$, the function 
$R_{\psi}(F,E)=\langle \psi | R(F,E)\psi\rangle$ is a holomorphic function
of $E$ in the upper half-plane ${\rm Im} \, E >0$ for any $\psi\in L^2$. Then  
$R_{\psi}(F,E)$ has a meromorphic continuation to ${\rm Im} \, E \leq 0$. The
poles of the continuation coincide with the functions $E(n_1,n_2,m;F)$.
\item They are the eigenvalues of the complex dilated operator uniquely associated
with $H(F)$ by the Balslev-Combes analytic dilation procedure (see
e.g.~\cite{ReSi1978}, XIII.10). More precisely: for $\theta\in\R$ let 
\begin{equation}
(U(\theta)\psi)(x) = {\rm e}^{3\theta/4}\psi({\rm e}^{\theta/2}x); \quad
(U(\theta)^{-1}\psi)(x) = {\rm e}^{-3\theta/4}\psi({\rm e}^{-\theta/2}x)
= (U(-\theta)\psi)(x)
\end{equation}
be the unitary dilation operator in $L^2(\R^3)$. Let
$H(F,\theta)=U(\theta) \, H(F) \, U(\theta)^{-1}$ be the unitary image
of $H(F)$, expressed by
\begin{equation}
H(F,\theta)=-\frac{{\rm e}^{-2\theta}}{2 \mu}\Delta
-\frac{Z{\rm e}^{-\theta}}{r}+F{\rm e}^{\theta} x_3\,.
\end{equation}
Then the operator family $H(F,\theta)$ can be analyzed also for $\theta$
complex, $0 < {\rm Im} \, \theta < \pi$. It can be proven that for such values of
$\theta$, the operator $H(F,\theta)$ has discrete spectrum. Its eigenvalues do not depend on
$\theta$ and coincide with the functions $E(n_1,n_2,m;F)$.
\item They are second-sheet poles of the scattering matrix  $S(E)$. This 
means that the scattering matrix for the Stark effect  $S(E,F)$ exists, is
unitary and has  the same analyticity properties of the resolvent. Namely,
it can be analytically continued from the upper half-plane ${\rm Im} \, E >0$ to a
meromorphic function to the lower half-plane ${\rm Im} \, E \leq 0$. The
poles of the continuation coincide with the functions $E(n_1,n_2,m;F)$.
\end{enumerate}

{\it Proof.}  Assertions 1, 2, 3 and 4 of the theorem have been proven in
\cite{GrGr1978b} through the separability of the problem in squared parabolic
coordinates recalled above,  and in
\cite{He1979} by a direct analysis of the operator $H(F,\theta)$. The complete
equivalence between the two methods is proven in
\cite{GrGrSi1970}. Assertion 5 is proven in \cite{BeGrHaSi1979}. 
An important issue
about the equivalence between the separated 
Eqs.~(\ref{S2a}) and (\ref{S2b})
and the operator $H(F)$ for complex values of $F$ is the interpretation of
(\ref{S2b}) as the analytic continuation of  (\ref{S2a}) from $F>0$ to
$-F={\rm e}^{{\rm i}\pi}F$.
Remark 3 is proven in~\cite{GrGr1978b}. 

Another point completely clarified in this example is the meaning
of divergent perturbation theory. 
For the Stark effect, the perturbative
expansion of the energy eigenvalue $E(n_1,n_2,m,F)$ reads (see
Eq.~(59) of~\cite{Si1978}),
\begin{equation}
\label{PertSer}
E(n_1,n_2,m,F) \sim \sum_{N=0}^{\infty}
E^{(N)}_{n_1 n_2 m} \, F^N\,,
\end{equation}
where $F$ is the electric field strength. 

\begin{theorem} Consider the perturbation series (\ref{PertSer}) for the
resonance of quantum numbers $(n_1,n_2;m)$:
\begin{equation}
E(n_1,n_2,m,F) \sim \sum_{N=0}^{\infty} E^{(N)}_{n_1 n_2 m} \, F^N\,.
\end{equation}
Then:
\begin{enumerate}
\item If $0<{\arg F}<\pi$, $0\leq |F|\leq B(n_1,n_2,m)$, the series is Borel
summable  to the eigenvalue $E(n_1,n_2,m,F)$ of $H(F)$ which exists for such
complex values of the field;
\item If $F\in\R$, $0\leq |F|\leq B((n_1,n_2,m)$ the series is Borel summable 
in the distributional sense  to the resonance $E(n_1,n_2,m,F)$. 
\end{enumerate}
\end{theorem}

{\it Proof.} Assertion 1 is proven in \cite{GrGr1978b},
assertion 2 in~\cite{CaGrMa1993}.
 
Finally, we remark that the eigenvalues $E(n_1,n_2,m,F)$, $0<{\arg F}<\pi$
can be continued across the real axis, and that their continuation to $F$ real
from above yields the resonances.  Since both the Borel sum and the analytic
continuation are unique, we can say that the Borel summability along the 
complex directions of the field $F$  is already  sufficient to determine the
resonances existing for $F$ real. However, the direct summation of the original,
real series  to the resonances requires the distributional Borel procedure.

The proof of the existence of resonances and of the (ordinary) Borel
summability of their divergent perturbation expansion has been extended to the
case of the Stark effect of an arbitrary atomic system, namely a system of
$N$ quantum particles interacting through two-body Coulomb potentials and under
the action of a constant electric field (see Refs.~\cite{GrGr1981,HeSi1981}). 

%
%
\section{Perturbation Series in Quantum Chromo{\rm d}ynamics}
\label{OtherFields}

We would like to summarize the results of the
investigations in Refs.~\cite{StMHMaSc1996,MHEtAl1996,MHSc1999,KnMHSo2002}
in the light of the concepts described in the current review.
Typically, in quantum chromodynamics (QCD), the problem is
to estimate the nonperturbative contribution to 
physical 
quantities ${\cal P}$ which, on the perturbative level, can be expressed
as
\begin{equation}
{\cal P} \sim \sum_{n=0}^{\infty} A_n \, a_s^n(Q)\,,
\end{equation}
where the strong coupling $a_s^n(Q)$ is a 
running coupling constant that depends the momentum 
scale $Q$. Infrared (IR) renormalons lead to 
singularities along the integration path of the 
Laplace--Borel integral which have been amply 
discussed in Sec.~\ref{Chap:DivBorel}. In the light
of a resurgent expansion like Eq.~(\ref{resurgent_expansion_dw}),
which has been conjectured for QCD perturbation series
(see Ref.~\cite{St2002,St2004}), it is natural to associate
the smallest term of a perturbation series with a
fundamental uncertainty that has to be ascribed to the perturbative
expansion due to nonperturbative effects.
This prescription is also consistent with the 
numerical data in Table~\ref{table_splitting_dw_2},
where the one-instanton series for the splitting of 
energy levels of the double well finds a natural 
barrier at the level of two-instanton shifts.

The leading nonalternating 
divergence of perturbative expansions in QCD 
is commonly referred to as the leading IR-renormalon 
pole. Suppose that we can reformulate, using the 
$\beta$ function of QCD,
\begin{equation} \label{ren2.8}
\exp\left(-\frac{t}{a_s(Q)} \right)
\sim
\exp\left(-t b_0\, \ln\left(\frac{Q^2}{\Lambda^2 {\rm e}^{-C}}\right)\right)
\sim
\left(\frac{Q^2}{\Lambda^2 {\rm e}^{-C}}\right)^{-t b_0}
\sim
\left(\frac{\Lambda^2 {\rm e}^{-C}}{Q^2}\right)^{t b_0}
\quad,
\end{equation}
where $\Lambda$ denotes the Landau-pole and $C$ corrects for the
renormalization scheme. 
Suppose, furthermore, that $n_0$ is the smallest term 
of the perturbation series, and that the renormalon
poles are located at positive integer $k$ as a function
of the argument of the Borel transform, defined according 
to Eq.~(\ref{BorelTransP}).
We then have as an estimate for the fundamental uncertainty
of the perturbative expansion,
\begin{eqnarray} \label{ren2.11}
A_{n_0} \, a_s^{n_0}(Q)
&\approx&
{\cal P} - \sum_{n=0}^{n_0-1} B^{(n)} a_s^n(Q)
\nn
&\approx&
\sum_{k=1}^\infty
\int_0^\infty {\rm d}t \, B_{\cal P}(t) \, {\rm e}^{-t/a_s(Q)}\,
\delta\left(t-\frac{k}{b_0}\right)
\nn
&\sim&
\sum_{k=1}^\infty
\int_0^\infty {\rm d}t \, B_{\cal P}(t) \, 
\left(\frac{\Lambda^2 {\rm e}^{-C}}{Q^2}\right)^{tb_0} \,
\delta\left(t-\frac{k}{b_0}\right)
\nn
&\sim&
\sum_{k=1}^\infty C^{(k)} \,
\left(\frac{\Lambda^2 {\rm e}^{-C}}{Q^2}\right)^k\,.
\end{eqnarray}
The nonperturbatively corrected perturbation series
thus has the structure,
\begin{equation} \label{ren2.12}
{\cal P} \;=\; 
\sum_{n=0}^{n_0-1} A_n \, s^n
+ \sum_{k=1}^\infty C^{(k)}
\left(\frac{\Lambda^2 {\rm e}^{-C}}{Q^2}\right)^k \,.
\end{equation}
In some cases the $k=1$ IR-renormalon is absent and the
leading IR-renormalon appears at $k=2$ only,
as for example in the case of the photon 
two-point-function~\cite{Za1992}. From the vast number of 
related field-theoretical investigations, which profit from variants
of the Borel method, we restrict ourselves to indicating the
articles and books corresponding to
Refs.~\cite{Da1982,LGZJ1990,Ra1991prd,Sh1994,ZJ1996,Be1999,Pi1999,St2002prep}.


\begin{thebibliography}{100}

\bibitem{LiLuSh1995}
C.~B. Liem, T. L\"{u}, and T.~M. Shih, {\em The Splitting Extrapolation Method}
  (World Scientific, Singapore, 1995).

\bibitem{Br1991a}
C. Brezinski, {\em History of Continued Fractions and {P}ad\'{e} Approximants}
  (Springer, Berlin, 1991).

\bibitem{Ri1927}
L.~F. Richardson, Phil. Trans. Roy. Soc. London A {\bf 226},  229  (1927).

\bibitem{Ai1926}
A.~C. Aitken, Proc. Roy. Soc. Edinburgh {\bf 46},  289  (1926).

\bibitem{Kn1964}
K. Knopp, {\em Theorie und {A}nwendung der unendlichen {R}eihen} (Springer,
  Berlin, 1964).

\bibitem{Ha1949}
G.~H. Hardy, {\em Divergent Series} (Clarendon Press, Oxford, UK, 1949).

\bibitem{Pe1966}
G.~M. Petersen, {\em Regular Matrix Transformations} (McGraw-Hill, London,
  1966).

\bibitem{Pe1969}
A. Peyerimhoff, {\em Lectures on Summability} (Springer, Berlin, 1969).

\bibitem{ZeBe1970}
K. Zeller and W. Beekmann, {\em Theorie der {L}imitierungsverfahren} (Springer,
  Berlin, 1970).

\bibitem{Wi1981}
J. Wimp, {\em Sequence Transformations and Their Applications} (Academic Press,
  New York, NY, 1981).

\bibitem{Wi1984}
A. Wilansky, {\em Summability through Functional Analysis} (North-Holland,
  Amsterdam, 1984).

\bibitem{PoSh1988}
R.~E. Powell and S.~M. Shah, {\em Summability Theory and Its Applications}
  (Prentice-Hall of India, New Delhi, 1988).

\bibitem{Bo2000}
J. Boos, {\em Classical and Modern Methods of Summability} (Oxford University
  Press, Oxford, UK, 2000).

\bibitem{Wy1956eps}
P. Wynn, Math. Tables Aids Comput. {\bf 10},  91  (1956).

\bibitem{Wy1956b}
P. Wynn, Proc. Camb. Phil. Soc. {\bf 52},  663  (1956).

\bibitem{Br1971}
C. Brezinski, C. R. Acad. Sc. Paris {\bf 273},  727  (1971).

\bibitem{Ba1975}
G.~A. Baker, {\em Essentials of Pad\'{e} approximants} (Academic Press, New
  York, 1975).

\bibitem{Br1977}
C. Brezinski, {\em Acc\'el\'eration de la Convergence en Analyse Num\'erique}
  (Springer, Berlin, 1977).

\bibitem{Br1978}
C. Brezinski, {\em Algorithmes d'Acc\'el\'eration de la Convergence -- \'Etude
  Num\'erique} (Editions Technip, Paris, 1978).

\bibitem{BeOr1978}
C.~M. Bender and S.~A. Orszag, {\em Advanced Mathematical Methods for
  Scientists and Engineers} (McGraw-Hill, New York, NY, 1978).

\bibitem{Br1980pade}
C. Brezinski, {\em Pad\'{e}-type Approximation and General Orthogonal
  Polynomials} (Birkh\protect{\"{a}}user, Basel, 1980).

\bibitem{MaSh1983}
G.~I. Marchuk and V.~V. Shaidurov, {\em Difference Methods and Their
  Extrapolations} (Springer, New York, 1983).

\bibitem{CuWu1987}
A. Cuyt and L. Wuytack, {\em Nonlinear Methods in Numerical Analysis}
  (North-Holland, Amsterdam, 1987).

\bibitem{De1988}
J.-P. Delahaye, {\em Sequence Transformations} (Springer, Berlin, 1988).

\bibitem{BrRZ1991}
C. Brezinski and M. Redivo-Zaglia, {\em Extrapolation Methods} (North-Holland,
  Amsterdam, 1991).

\bibitem{Wa1996}
G. Walz, {\em Asymptotics and Extrapolation} (Akademie-Verlag, Berlin, 1996).

\bibitem{BaGr1996}
G.~A. Baker and P. Graves-Morris, {\em Pad\'{e} Approximants}, 2nd ed.
  (Cambridge University Press, Cambridge, UK, 1996).

\bibitem{Si2003}
A. Sidi, {\em Practical Extrapolation Methods} (Cambridge University Press,
  Cambridge, 2003).

\bibitem{BoLaWaWa2004}
F. Bornemann, D. Laurie, S. Wagon, and J. Waldvogel, {\em The SIAM 100-Digit
  Challenge: A Study in High-Accuracy Numerical Computing} (Society of
  Industrial Applied Mathematics, Philadelphia, 2004).

\bibitem{Gu1989}
A.~J. Guttmann,  in {\em Phase Transitions and Critical Phenomena 13}, edited
  by C. Domb and J.~L. Lebowitz (Academic Press, London, 1989), pp.\ 3--234.

\bibitem{We1989}
E.~J. Weniger, Comput. Phys. Rep. {\bf 10},  189  (1989).

\bibitem{Ho2000}
H.~H.~H. Homeier, J. Comput. Appl. Math. {\bf 122},  81   (2000).

\bibitem{Br2000}
C. Brezinski, J. Comput. Appl. Math. {\bf 122},  1   (2000).

\bibitem{JeMoSoWe1999}
U.~D. Jentschura, P.~J. Mohr, G. Soff, and E.~J. Weniger, Comput. Phys. Commun.
  {\bf 116},  28  (1999).

\bibitem{JeMoSo1999}
U.~D. Jentschura, P.~J. Mohr, and G. Soff, Phys. Rev. Lett. {\bf 82},  53
  (1999).

\bibitem{Br2004}
C. Brezinski, Numer. Algor. {\bf 36},  309  (2004).

\bibitem{Br1997}
C. Brezinski, {\em Projection Methods for Systems of Equations} (Elsevier,
  Amsterdam, 1997).

\bibitem{SiFoSm1986}
A. Sidi, W.~F. Ford, and D. Smith, SIAM J. Numer. Anal. {\bf 23},  178  (1986).

\bibitem{HoRaKr1995}
H.~H.~H. Homeier, S. Rast, and H. Krienke, Comput. Phys. Commun. {\bf 92},  188
   (1995).

\bibitem{Ev1993}
G. Evans, {\em Practical Numerical Integration} (WILEY, Chichester, 1993).

\bibitem{SaHo2003}
H. Safouhi and P.~E. Hoggan, Mol. Phys. {\bf 101},  19  (2003).

\bibitem{El1995}
E. Elizalde, {\em Ten Physical Applications of Spectral Zeta Functions}
  (Springer, Berlin, 1995).

\bibitem{PaKa2001}
R.~B. Paris and D. Kaminsky, {\em Asymptotics and Mellin-Barnes Integrals}
  (Cambridge University Press, Cambridge, 2001).

\bibitem{ReSi1978}
M. Reed and B. Simon, {\em Methods of Modern Mathematical Physics IV: Analysis
  of Operators} (Academic Press, New York, 1978).

\bibitem{Bo1899}
E. Borel, Ann. sci. Ec. norm. sup. Paris {\bf 16},  9  (1899).

\bibitem{Bo1928}
E. Borel, {\em Le\c{c}on sur les S\'{e}ries Divergentes}, 2nd ed.
  (Gauthiers-Villars, Paris, 1928).

\bibitem{Bo1975}
E. Borel, {\em Lectures on Divergent Series}, {\em Translation LA-6140-TR} (Los
  Alamos Scientific Laboratory, Los Alamos, 1975).

\bibitem{Dy1952}
F.~J. Dyson, Phys. Rev. {\bf 85},  631  (1952).

\bibitem{Li1976Lett}
L.~N. Lipatov, Pis'ma Zh. \'{E}ksp. Teor. Fiz. {\bf 24},  179  (1976), [JETP
  Lett. {\bf 24}, 157 (1976)].

\bibitem{Li1976}
L.~N. Lipatov, Zh. \'{E}ksp. Teor. Fiz. {\bf 71},  2010  (1976), [JETP {\bf
  44}, No.~6 (1977)].

\bibitem{Li1977Lett}
L.~N. Lipatov, Pis'ma Zh. \'{E}ksp. Teor. Fiz. {\bf 25},  116  (1977).

\bibitem{Li1977}
L.~N. Lipatov, Zh. \'{E}ksp. Teor. Fiz. {\bf 72},  411  (1977), [JETP {\bf 45},
  216 (1977)].

\bibitem{BrKr2000}
D. Broadhurst and D. Kreimer, Phys. Lett. B {\bf 475},  63  (2000).

\bibitem{BrKr2001}
D. Broadhurst and D. Kreimer, Nucl. Phys. B {\bf 600},  403  (2001).

\bibitem{BeWu1969}
C.~M. Bender and T.~T. Wu, Phys. Rev. {\bf 184},  1231  (1969).

\bibitem{BeWu1971}
C.~M. Bender and T.~T. Wu, Phys. Rev. Lett. {\bf 27},  461  (1971).

\bibitem{BeWu1973}
C.~M. Bender and T.~T. Wu, Phys. Rev. D {\bf 7},  1620  (1973).

\bibitem{ArFeCa1990}
G.~A. Arteca, F.~M. Fern\'{a}ndez, and E.~A. Castro, {\em Large Order
  Perturbation Theory and Summation Methods in Quantum Mechanics} (Springer,
  Berlin, 1990).

\bibitem{LGZJ1990}
J.~C. LeGuillou and J. Zinn-Justin, {\em Large-Order Behaviour of Perturbation
  Theory} (North-Holland, Amsterdam, 1990).

\bibitem{Ad1994}
B.~G. Adams, {\em Algebraic Approach to Simple Quantum Systems} (Springer,
  Berlin, 1994).

\bibitem{Fi1994}
J. Fischer, Fortschr. Physik {\bf 42},  665  (1994).

\bibitem{DeDi1991}
E. Delabaere and H. Dillinger, Ph.D. thesis, University of Nice, Nice, 1991
  (unpublished).

\bibitem{Ph1989}
F. Pham, C. R. Acad. Sc. Paris {\bf 309},  999  (1989).

\bibitem{CaNoPh1993}
B. Candelpergher, J.~C. Nosmas, and F. Pham, {\em Approche de la
  R\'{e}surgence} (Hermann, Paris, 1993).

\bibitem{Bo1994}
L. $\mathrm{Boutet~de~Monvel}$ (Ed.), {\em M\'{e}thodes R\'{e}surgentes}
  (Hermann, Paris, 1994).

\bibitem{Su2001phi4}
I.~M. Suslov, Zh. \'{E}ksp. Teor. Fiz. {\bf 120},  5  (2001), [JETP {\bf 93}, 1
  (2001)].

\bibitem{Su2001}
I.~M. Suslov, Pis'ma v. Zh. \'{E}ksp. Teor. Fiz. {\bf 74},  211  (2001), [JETP
  {\bf 74}, 191 (2001)].

\bibitem{Su2002}
I.~M. Suslov, Pis'ma v. Zh. \'{E}ksp. Teor. Fiz. {\bf 76},  387  (2002), [JETP
  {\bf 76}, 327 (2002)].

\bibitem{Su2005rev}
I.~M. Suslov, Zh. \'{E}ksp. Teor. Fiz. {\bf 127},  1350  (2005), [JETP {\bf
  100}, 1188 (2005)].

\bibitem{Jo1965}
C. Jordan, {\em Calculus of Finite Differences} (Chelsea, New York, 1965).

\bibitem{MT1981}
L.~M. Milne-Thomson, {\em The Calculus of Finite Differences} (Chelsea, New
  York, 1981).

\bibitem{Po1886}
H. Poincar\'{e}, Acta Math. {\bf 8},  295   (1886).

\bibitem{Sh1955}
D. Shanks, J. Math. and Phys. (Cambridge, Mass.) {\bf 34},  1  (1955).

\bibitem{ClGrAd1969}
W.~D. Clark, H.~L. Gray, and J.~E. Adams, J. Res. Natl. Bur. Stand. B {\bf 73},
   25  (1969).

\bibitem{Ed1974}
H.~M. Edwards, {\em Riemann's zeta function} (Academic Press, New York and
  London, 1974).

\bibitem{Ol1974}
F.~W.~J. Olver, {\em Asymptotics and Special Functions} (Academic Press, New
  York, NY, 1974).

\bibitem{Ti1986}
E.~C. Titchmarsh, {\em The Theory of the Riemann Zeta-Function} (Clarendon
  Press, Oxford, UK, 1986).

\bibitem{BeWe2001}
C.~M. Bender and E.~J. Weniger, J. Math. Phys. {\bf 42},  2167   (2001).

\bibitem{DeGB1982}
J.~P. Delahaye and B. Germain-Bonne, SIAM J. Numer. Anal. {\bf 19},  840
  (1982).

\bibitem{Br1980E}
C. Brezinski, Numer. Math. {\bf 35},  175  (1980).

\bibitem{Ha1979E}
T. Havie, B.I.T. {\bf 19},  204  (1979).

\bibitem{Wo1988}
S. Wolfram, {\em Mathematica-A System for Doing Mathematics by Computer}
  (Addison-Wesley, Reading, MA, 1988).

\bibitem{PrFlTeVe1993}
W.~H. Press, B.~P. Flannery, S.~A. Teukolsky, and W.~T. Vetterling, {\em
  Numerical Recipes in {C}: The Art of Scientific Computing}, 2 ed. (Cambridge
  University Press, Cambridge, UK, 1993).

\bibitem{FeFoSm1983}
T. Fessler, W.~F. Ford, and D.~A. Smith, ACM Trans. Math. Soft. {\bf 9},  346
  (1983).

\bibitem{JeHomeHD}
See the URL http://www.mpi-hd.mpg.de/\~{}ulj.

\bibitem{Gi1973}
J. Gilewicz,  in {\em Pad\'{e} Approximants and Their Applications}, edited by
  P.~R. Graves-Morris (Academic Press, London, 1973), p.\ 99.

\bibitem{Gi1978}
J. Gilewicz, {\em Approximants de Pad\'e -- Lecture Notes in Mathematica 667}
  (Springer, Berlin, 1978).

\bibitem{Ba1965}
G.~A. Baker, Adv. Theor. Phys. {\bf 1},  1  (1965).

\bibitem{BaGa1970}
{\em The Pad\'e Approximant in Theoretical Physics}, edited by G.~A. Baker and
  J.~L. Gammel (Academic Press, New York, 1970).

\bibitem{Ba1972}
J.~L. Basdevant, Fortschr. Physik {\bf 20},  283  (1972).

\bibitem{Br1976}
C. Brezinski,  in {\em Pad\'e Approximation Method and Its Application to
  Mechanics}, edited by H. Cabannes (Springer, Berlin, 1976).

\bibitem{Ba1990}
G.~A. Baker, {\em Quantitative theory of critical phenomena} (Academic Press,
  San Diego, 1990).

\bibitem{Ma1997}
D.~A. Macdonald, Math. Comput. {\bf 66},  1619  (1997).

\bibitem{Wy1966}
P. Wynn, SIAM J. Numer. Anal. {\bf 3},  91  (1966).

\bibitem{Br1975}
C. Brezinski, C. R. Acad. Sc. Paris {\bf 280A},  729  (1975).

\bibitem{Br1973}
C. Brezinski, C. R. Acad. Sc. Paris {\bf 276A},  305  (1973).

\bibitem{Ku1837}
E.~E. Kummer, J. Reine Angew. Math. {\bf 16},  206  (1837).

\bibitem{Dr1976}
J.~E. Drummond, J. Austral. Math. Soc. {\bf 19},  416  (1976).

\bibitem{JaOB1978}
M.~J. Jamieson and T.~H. O'Beirne, J. Phys. B {\bf 11},  L31  (1978).

\bibitem{Sa1987}
P. Sablonniere, J. Comput. Appl. Math. {\bf 19},  55  (1987).

\bibitem{Ca1977}
B.~C. Carlson, {\em Special Functions of Applied Mathematics} (Academic Press,
  New York, 1977).

\bibitem{WeCiVi1993}
E.~J. Weniger, J. {\v C}{\'\i}{\v z}ek, and F. Vinette, J. Math. Phys. {\bf
  34},  571  (1993).

\bibitem{Br1985anm}
C. Brezinski, Appl. Numer. Math. {\bf 1},  457  (1985).

\bibitem{GaGu1974}
D.~S. Gaunt and A.~J. Guttmann,  in {\em Phase Transitions and Critical
  Phenomena 3}, edited by C. Domb and M.~S. Green (Academic Press, London,
  1974), p.\ 181.

\bibitem{Le1973}
D. Levin, Int. J. Comput. Math. B {\bf 3},  371  (1973).

\bibitem{SmFo1979}
D.~A. Smith and W.~F. Ford, SIAM J. Numer. Anal. {\bf 16},  223  (1979).

\bibitem{Je2003habil}
U.~D. Jentschura, Quantum Electrodynamic Bound--State Calculations and
  Large--Order Perturbation Theory, e-print hep-ph/0306153.

\bibitem{We1992}
E.~J. Weniger, Numer. Algor. {\bf 3},  477  (1992).

\bibitem{We2004}
E.~J. Weniger, J. Math. Phys. {\bf 45},  1209  (2004).

\bibitem{CiZaSk2003}
J. $\check{\rm C}{\rm i}\check{\rm z}$ek, J. Zamastil, and L. Sk\'{a}la, J.
  Math. Phys. {\bf 44},  962  (2003).

\bibitem{RoBhBh1998}
D. Roy, R. Bhattacharya, and S. Bhowmick, Comput. Phys. Commun. {\bf 113},  131
   (1998).

\bibitem{WeCiVi1991}
E.~J. Weniger, J. {\v C}{\'\i}{\v z}ek, and F. Vinette, Phys. Lett. A {\bf
  156},  169  (1991).

\bibitem{ElStChMiSp1998}
V. Elias, T.~G. Steele, F. Chishtie, R. Migneron, and K. Sprague, Phys. Rev. D
  {\bf 58},  116007  (1998).

\bibitem{StEl1998}
T.~G. Steele and V. Elias, Mod. Phys. Lett. A {\bf 13},  3151  (1998).

\bibitem{ChElMiSt1999}
F.~A. Chishtie, V. Elias, V.~A. Miransky, and T.~G. Steele, Pad\'{e}-Summation
  Approach to QCD Beta-Function Infrared Properties, e-print hep-ph/9905291.

\bibitem{ChElSt1999a}
F. Chishstie, V. Elias, and T.~G. Steele, Phys. Lett. B {\bf 446},  267
  (1999).

\bibitem{ChElSt1999b}
F. Chishstie, V. Elias, and T.~G. Steele, Phys. Rev. D {\bf 59},  105013
  (1999).

\bibitem{ChElSt2000}
F. Chishstie, V. Elias, and T.~G. Steele, J. Phys. G {\bf 26},  93  (2000).

\bibitem{ElChSt2000}
V. Elias, F.~A. Chishtie, and T.~G. Steele, e-print hep-ph/0004140.

\bibitem{SaLi1991}
M.~A. Samuel and G. Li, Phys. Rev. D {\bf 44},  3935  (1991).

\bibitem{SaLiSt1993}
M.~A. Samuel, G. Li, and E. Steinfelds, Phys. Rev. D {\bf 48},  869  (1993).

\bibitem{SaLi1994c}
M.~A. Samuel and G. Li, Int. J. Theor. Phys. {\bf 33},  1461  (1994).

\bibitem{SaLiSt1994}
M.~A. Samuel, G. Li, and E. Steinfelds, Phys. Lett. B {\bf 323},  188  (1994).

\bibitem{SaLi1994a}
M.~A. Samuel and G. Li, Phys. Lett. B {\bf 331},  114  (1994).

\bibitem{SaLi1994b}
M.~A. Samuel, G. Li, and E. Steinfelds, Phys. Lett. B {\bf 323},  188  (1994).

\bibitem{SaElKa1995}
M.~A. Samuel, J. Ellis, and M. Karliner, Phys. Rev. Lett. {\bf 74},  4380
  (1995).

\bibitem{Sa1995a}
M.~A. Samuel, Int. J. Theor. Phys. {\bf 34},  903  (1995).

\bibitem{Sa1995b}
M.~A. Samuel, Int. J. Theor. Phys. {\bf 34},  1113  (1995).

\bibitem{SaLiSt1995}
M.~A. Samuel, G. Li, and E. Steinfelds, Phys. Rev. E {\bf 51},  3911  (1995).

\bibitem{ElGaKaSa1996}
J. Ellis, E. Gardi, M. Karliner, and M.~A. Samuel, Phys. Lett. B {\bf 366},
  268  (1996).

\bibitem{ElGaKaSa1996a}
J. Ellis, E. Gardi, M. Karliner, and M.~A. Samuel, Phys. Lett. B {\bf 366},
  268  (1996).

\bibitem{ElGaKaSa1996b}
J. Ellis, E. Gardi, M. Karliner, and M.~A. Samuel, Phys. Rev. D {\bf 54},  6986
   (1996).

\bibitem{Ga1997}
E. Gardi, Phys. Rev. D {\bf 56},  68  (1997).

\bibitem{ElKaSa1997}
J. Ellis, M. Karliner, and M.~A. Samuel, Phys. Lett. B {\bf 400},  176  (1997).

\bibitem{SaAbYu1997}
M.~A. Samuel, T. Abraha, and J. Yu, Phys. Lett. B {\bf 394},  165  (1997).

\bibitem{JaJoSa1997}
I. Jack, D.~R.~T. Jones, and M.~A. Samuel, Phys. Lett. B {\bf 407},  143
  (1997).

\bibitem{ElJaJoKaSa1998}
J. Ellis, I. Jack, D.~R.~T. Jones, M. Karliner, and M.~A. Samuel, Phys. Rev. D
  {\bf 57},  2665  (1998).

\bibitem{BuAbSaStMa1996}
P. Burrows, T. Abraha, M. Samuel, E. Steinfelds, and H. Masuda, Phys. Lett. B
  {\bf 272},  223  (1996).

\bibitem{CvKo1998}
G. Cveti\v{c} and R. K\"{o}gerler, Nucl. Phys. B {\bf 522},  396  (1998).

\bibitem{Cv1998a}
G. Cveti\v{c}, Phys. Rev. D {\bf 57},  R3209  (1998).

\bibitem{CvYu2000}
G. Cveti\v{c} and J.~Y. Yu, e-print hep-ph/0003241.

\bibitem{JeWeSo2000}
U.~D. Jentschura, E.~J. Weniger, and G. Soff, J. Phys. G {\bf 26},  1545
  (2000).

\bibitem{JeBeWeSo2000}
U.~D. Jentschura, J. Becher, E.~J. Weniger, and G. Soff, Phys. Rev. Lett. {\bf
  85},  2446  (2000).

\bibitem{vW1965}
A. van Wijngaarden,  in {\em Cursus: Wetenschappelijk Rekenen B, Process
  Analyse} (Stichting Mathematisch Centrum, Amsterdam, 1965), pp.\ 51--60.

\bibitem{NP1961}
C.~W. Clenshaw, E.~T. Goodwin, D.~W. Martin, G.~F. Miller, F.~W.~J. Olver, and
  J.~H. Wilkinson, {\em (National Physical Laboratory), Modern Computing
  Methods, Notes on Applied Science}, 2 ed. (H. M. Stationary Office, London,
  1961), Vol.~16.

\bibitem{Da1969}
J.~W. Daniel, Math. Comput. {\bf 23},  91  (1969).

\bibitem{PeKi1998}
P.~J. Pelzl and F.~E. King, Phys. Rev. E {\bf 57},  7268  (1998).

\bibitem{JeAkMoSaSo2003}
U.~D. Jentschura, S.~V. Aksenov, P.~J. Mohr, M.~A. Savageau, and G. Soff,  in
  {\em Proceedings of the Nano\-tech 2003 Conference (Vol. 2)}, edited by M.
  Laudon and B. Romanowicz (Computational Publications, Boston, 2003), pp.\
  535--537.

\bibitem{BrDeGB1983}
C. Brezinski, J.~P. Delahaye, and B. Germain-Bonne, SIAM J. Numer. Anal. {\bf
  20},  1099  (1983).

\bibitem{GrAt1968}
H. Gray and T. Atchison, Math. Comput. {\bf 22},  595  (1968).

\bibitem{GB1978}
B. Germain-Bonne, Ph.D. thesis, University of Lille, Lille, 1978 (unpublished).

\bibitem{PeKi2002}
P.~J. Pelzl, G.~J. Smethells, and F.~E. King, Phys. Rev. E {\bf 65},  036707
  (2002).

\bibitem{DeKePaGl1984}
J.~P. Dempsey, L.~M. Keer, N.~B. Patel, and M.~L. Glasser, ASME J. Appl. Mech.
  {\bf 51},  324  (1984).

\bibitem{BaGa1995}
P. Baratella and R. Gabutti, J. Comput. Appl. Math. {\bf 62},  181  (1995).

\bibitem{BoDe1992}
J. Boersma and J.~P. Dempsey, Math. Comput. {\bf 59},  157  (1992).

\bibitem{Ga1991}
W. Gautschi, Math. Comput. {\bf 57},  325  (1991).

\bibitem{DeLiDe1990}
K.~M. Dempsey, D. Liu, and J.~P. Dempsey, Math. Comput. {\bf 55},  693  (1990).

\bibitem{AkSaJeBeSoMo2003}
S.~V. Aksenov, M.~A. Savageau, U.~D. Jentschura, J. Becher, G. Soff, and P.~J.
  Mohr, Comput. Phys. Commun. {\bf 150},  1  (2003).

\bibitem{Be1999master}
J. Becher, Numerical Convergence Acceleration Techniques and Their Application
  in Theoretical Physics (in German), master thesis, Technische Universit\"{a}t
  Dresden, 1999 (unpublished).

\bibitem{Ba1953vol1}
H. Bateman, {\em Higher Transcendental Functions} (McGraw-Hill, New York, NY,
  1953), Vol.~1.

\bibitem{AbSt1972}
M. Abramowitz and I.~A. Stegun, {\em Handbook of Mathematical Functions}, 10
  ed. (National Bureau of Standards, Washington, D. C., 1972).

\bibitem{JeMoSo2001pra}
U.~D. Jentschura, P.~J. Mohr, and G. Soff, Phys. Rev. A {\bf 63},  042512
  (2001).

\bibitem{JeMo2004pra}
U.~D. Jentschura and P.~J. Mohr, Phys. Rev. A {\bf 69},  064103  (2004).

\bibitem{JeMo2005p}
U.~D. Jentschura and P.~J. Mohr, Phys. Rev. A {\bf 72},  014103  (2005).

\bibitem{SuKoFr2005}
A. Surzhykov, P. Koval, and S. Fritzsche, Phys. Rev. A {\bf 71},  022509
  (2005).

\bibitem{Ro1961}
M.~E. Rose, {\em Relativistic Electron Theory} (J. Wiley \& Sons, New York, NY,
  1961).

\bibitem{SwDr1991b}
R.~A. Swainson and G.~W.~F. Drake, J. Phys. A {\bf 24},  95  (1991).

\bibitem{EiMe1995}
J. Eichler and W. Meyerhof, {\em Relativistic Atomic Collisions} (Academic
  Press, San Diego, 1995).

\bibitem{VaMoKh1988}
D.~A. Varshalovich, A.~N. Moskalev, and V.~K. Khersonskii, {\em Quantum Theory
  of Angular Momentum} (World Scientific, Singapore, 1988).

\bibitem{Mo1974a}
P.~J. Mohr, Ann. Phys. (N.Y.) {\bf 88},  26  (1974).

\bibitem{MoPlSo1998}
P.~J. Mohr, G. Plunien, and G. Soff, Phys. Rep. {\bf 293},  227  (1998).

\bibitem{Mo1974b}
P.~J. Mohr, Ann. Phys. (N.Y.) {\bf 88},  52  (1974).

\bibitem{WhWa1944}
E.~T. Whittaker and G.~N. Watson, {\em A Course of Modern Analysis} (Cambridge
  University Press, Cambridge, UK, 1944).

\bibitem{MaObSo1966}
W. Magnus, F. Oberhettinger, and R.~P. Soni, {\em Formulas and Theorems for the
  Special Functions of Mathematical Physics} (Springer, New York, 1966).

\bibitem{Sl1960}
L.~J. Slater, {\em The Confluent Hypergeometric Function} (Cambridge University
  Press, Cambridge, UK, 1960).

\bibitem{LGZJ1980}
J.~C. LeGuillou and J. Zinn-Justin, Phys. Rev. B {\bf 21},  3976  (1980).

\bibitem{ZJ1996}
J. Zinn-Justin, {\em Quantum Field Theory and Critical Phenomena}, 3rd ed.
  (Clarendon Press, Oxford, 1996).

\bibitem{ZJ1981npb}
J. Zinn-Justin, Nucl. Phys. B {\bf 192},  125  (1981).

\bibitem{ZJJe2004i}
J. Zinn-Justin and U.~D. Jentschura, Ann. Phys. (N.Y.) {\bf 313},  197  (2004).

\bibitem{ZJJe2004ii}
J. Zinn-Justin and U.~D. Jentschura, Ann. Phys. (N.Y.) {\bf 313},  269  (2004).

\bibitem{JeZJ2004plb}
U.~D. Jentschura and J. Zinn-Justin, Phys. Lett. B {\bf 596},  138  (2004).

\bibitem{Sh1994}
M. Shifman, {\em Instantons in Gauge Theories} (World Scientific, Singapore,
  1994).

\bibitem{Fo2000primer}
H. Forkel, A Primer on Instantons, e-print hep-ph/0009136.

\bibitem{Fo2001}
H. Forkel, Instantons and Glueballs, e-print hep-ph/0103204.

\bibitem{Fo2002}
H. Forkel, Direct instantons, topological charge screening and QCD glueball sum
  rules, e-print hep-ph/0312049.

\bibitem{St2002}
M. Stingl, Field--Theory Amplitudes as Resurgent Functions, e-print
  hep-ph/0207049.

\bibitem{St2004}
M. Stingl, Ann. Fac. Sci. Toulouse {\bf 13},  659  (2004).

\bibitem{Be1999}
M. Beneke, Phys. Rep. {\bf 317},  1  (1999).

\bibitem{CaGrMa1986}
E. Caliceti, V. Grecchi, and M. Maioli, Commun. Math. Phys. {\bf 104},  163
  (1986).

\bibitem{CaGrMa1993}
E. Caliceti, V. Grecchi, and M. Maioli, Commun. Math. Phys. {\bf 157},  347
  (1993).

\bibitem{BaCo1971}
E. Balslev and J.~C. Combes, Commun. Math. Phys. {\bf 22},  280  (1971).

\bibitem{ShWa1994}
B. Shawyer and B. Watson, {\em Borel's Method of Summability} (Oxford
  University Press, Oxford, 1994).

\bibitem{Ka1993a}
M. Kaku, {\em Quantum Field Theory} (Oxford University Press, New York, 1993).

\bibitem{ItZu1980}
C. Itzykson and J.~B. Zuber, {\em Quantum Field Theory} (McGraw-Hill, New York,
  1980).

\bibitem{NeOr1988}
J.~W. Negele and H. Orland, {\em Quantum Many-Particle Systems}
  (Addison-Wesley, Reading, MA, 1988).

\bibitem{Pa1988a}
G. Parisi, {\em Statistical Field Theory} (Addison-Wesley, Reading, MA, 1988).

\bibitem{Ri1991}
V. Rivasseau, {\em From Perturbative to Constructive Renormalization}
  (Princeton University Press, Princeton, 1991).

\bibitem{Kl1995path}
H. Kleinert, {\em Path Integrals in Quantum Mechanics, Statistics, and Polymer
  Physics}, 2nd ed. (World Scientific, Singapore, 1995).

\bibitem{DuHa1999}
G.~V. Dunne and T.~M. Hall, Phys. Rev. D {\bf 60},  065002  (1999).

\bibitem{GrGrTu1971}
S. Graffi, V. Grecchi, and G. Turchetti, Nuovo Cim. B {\bf 4},  313  (1971).

\bibitem{GrMa1984a}
V. Grecchi and M. Maioli, Ann. Inst. Henri Poincar\'e A {\bf 41},  37  (1984).

\bibitem{GrMa1984b}
V. Grecchi and M. Maioli, J. Math. Phys. {\bf 25},  3439  (1984).

\bibitem{Ca1926}
T. Carleman, {\em Les Fonctions Quasi-Analytiques} (Gauthiers-Villars, Paris,
  1926).

\bibitem{Ne1919}
F. Nevanlinna, Ann. Acad. Sci. Fenn. {\bf 12}, No.~3 (1918-1919).

\bibitem{So1980}
A.~D. Sokal, J. Math. Phys. {\bf 21},  261  (1980).

\bibitem{GrGrSi1970}
S. Graffi, V. Grecchi, and B. Simon, Phys. Lett. B {\bf 32},  631  (1970).

\bibitem{FrGrSi1985}
V. Franceschini, V. Grecchi, and H.~J. Silverstone, Phys. Rev. A {\bf 32},
  1338  (1985).

\bibitem{Je2000prd}
U.~D. Jentschura, Phys. Rev. D {\bf 62},  076001  (2000).

\bibitem{Je2001pra}
U.~D. Jentschura, Phys. Rev. A {\bf 64},  013403  (2001).

\bibitem{SeZJ1979}
R. Seznec and J. Zinn-Justin, J. Math. Phys. {\bf 20},  1398  (1979).

\bibitem{LGZJ1983}
J.~C. LeGuillou and J. Zinn-Justin, Ann. Phys. (N.Y.) {\bf 147},  57  (1983).

\bibitem{Gr1979}
R.~E. Grundy, J. Comp. Phys. {\bf 33},  290  (1979).

\bibitem{Ma1987a}
M.~F. Marziani, Nuovo Cim. B {\bf 99},  145  (1987).

\bibitem{Ma1984}
M.~F. Marziani, J. Phys. A {\bf 17},  547  (1984).

\bibitem{JeSo2001}
U.~D. Jentschura and G. Soff, J. Phys. A {\bf 34},  1451  (2001).

\bibitem{HuPi1983}
W. Hunziker and C.~A. Pillet, Commun. Math. Phys. {\bf 90},  219  (1983).

\bibitem{Si1982}
B. Simon, Int. J. Quantum Chem. {\bf 21},  3  (1982).

\bibitem{Si1970}
B. Simon, Ann. Phys. (N.Y.) {\bf 58},  76  (1970).

\bibitem{GrGr1978}
S. Graffi and V. Grecchi, J. Math. Phys. {\bf 19},  1002  (1978).

\bibitem{LoMa1970}
J. Loeffel and A. Martin, CERN preprint TH-1031 (1970).

\bibitem{LoMaSiWi1969}
J. Loeffel, A. Martin, B. Simon, and A.~S. Wightman, Phys. Lett. B {\bf 30},
  656  (1969).

\bibitem{GrGr1975}
S. Graffi and V. Grecchi, Lett. Nuovo Cimento {\bf 12},  425  (1975).

\bibitem{vW1968}
B.~L. van~der Waerden, {\em Sources of Quantum Mechanics} (Dover, New York,
  1968).

\bibitem{CaGrMa1980}
E. Caliceti, V. Graffi, and M. Maioli, Commun. Math. Phys. {\bf 75},  51
  (1980).

\bibitem{Ca2000}
E. Caliceti, J. Phys. A {\bf 33},  3753  (2000).

\bibitem{Ca2006}
E. Caliceti, Ann. Henri Poincar\'{e} {\bf 7},  563  (2006).

\bibitem{Be2004}
C.~M. Bender, Czech. J. Phys. {\bf 184},  13  (2004).

\bibitem{CaGrSj2005}
E. Caliceti, S. Graffi, and J. Sj\"{o}strand, J. Phys. A {\bf 18},  185
  (2005).

\bibitem{Sh2002cmp}
K.~C. Shin, Commun. Math. Phys. {\bf 229},  543  (2002).

\bibitem{DoDuTa2001i}
P. Dorey, C. Dunning, and R. Tateo, J. Phys. A {\bf 34},  L391  (2002).

\bibitem{DoDuTa2001ii}
P. Dorey, C. Dunning, and R. Tateo, J. Phys. A {\bf 34},  5679  (2002).

\bibitem{DoDuTa2005}
P. Dorey, C. Dunning, and R. Tateo, J. Phys. A {\bf 38},  1305  (2005).

\bibitem{Ti1957}
E.~C. Titchmarsh, {\em Eigenfunction Expansions, Vol. II} (Cambridge University
  Press, Cambridge, UK, 1957).

\bibitem{Op1928}
J.~R. Oppenheimer, Phys. Rev. {\bf 31},  66  (1928).

\bibitem{YaEtAl1978}
R. Yaris, J. Bendler, R.~A. Lovett, C.~M. Bender, and P.~A. Fedders, Phys. Rev.
  A {\bf 18},  1816  (1978).

\bibitem{AlCa2000}
G. Alvarez and C. Casares, J. Phys. A {\bf 33},  5171  (2000).

\bibitem{LaLi1958}
L.~D. Landau and E.~M. Lifshitz, {\em Quantum Mechanics (Volume 3 of the Course
  of Theoretical Physics)} (Pergamon Press, London, 1958).

\bibitem{Ha1978}
E. Harrell, Commun. Math. Phys. {\bf 60},  73  (1978).

\bibitem{CaGrMa1988}
E. Caliceti, V. Grecchi, and M. Maioli, Commun. Math. Phys. {\bf 113},  625
  (1988).

\bibitem{CaGrMa1996}
E. Caliceti, V. Grecchi, and M. Maioli, Commun. Math. Phys. {\bf 176},  1
  (1996).

\bibitem{GrGr1984}
S. Graffi and V. Grecchi, Commun. Math. Phys. {\bf 92},  397  (1984).

\bibitem{JLMaSc1981}
G. Jona-Lasinio, F. Martinelli, and I. Scoppola, Commun. Math. Phys. {\bf 80},
  223  (1981).

\bibitem{Si1985}
B. Simon, J. Funct. Anal. {\bf 63},  123  (1985).

\bibitem{HeSi1978plb}
I.~W. Herbst and B. Simon, Phys. Lett. B {\bf 78},  304  (1978).

\bibitem{ZJ1981jmp}
J. Zinn-Justin, J. Math. Phys. {\bf 22},  511  (1981).

\bibitem{ZJ1983npb}
J. Zinn-Justin, Nucl. Phys. B {\bf 218},  333  (1983).

\bibitem{ZJ1984jmp}
J. Zinn-Justin, J. Math. Phys. {\bf 25},  549  (1984).

\bibitem{SuLuZJJe2006}
A. Surzhykov, M. Lubasch, J. Zinn-Justin, and U.~D. Jentschura, Phys. Rev. B
  {\bf 74},  205317  (2006).

\bibitem{Wi1975}
K.~G. Wilson, Rev. Mod. Phys. {\bf 47},  773  (1975).

\bibitem{Wi1983}
K.~G. Wilson, Rev. Mod. Phys. {\bf 55},  583  (1983).

\bibitem{CaHaPeVi2006}
M. Campostrini, M. Hasenbusch, A. Pelissetto, and E. Vicari, e-print
  cond-mat/0605083.

\bibitem{LiEtAl1996}
J.~A. Lipa, D.~R. Swanson, J. Nissen, T.~C.~P. Chui, and U.~E. Israelson, Phys.
  Rev. Lett. {\bf 76},  944  (1996).

\bibitem{MISTE}
MISTE-Experiment, home page http://miste.jpl.nasa.gov.

\bibitem{GuZJ1998bx}
R. Guida and J. Zinn-Justin, J. Phys. A {\bf 31},  8103  (1998).

\bibitem{Kl1999}
H. Kleinert, Phys. Rev. D {\bf 60},  085001  (1999).

\bibitem{LaEtAl2004}
C. L\"{a}mmerzahl, G. Ahlers, N. Ashby, M. Barmatz, P.~L. Biermann, H. Dittus,
  V. Dohm, R. Duncan, K. Gibble, J. Lipa, N. Lockerbie, N. Mulders, and C.
  Salomon, Gen. Relativ. Gravit. {\bf 36},  615  (2004).

\bibitem{PaSwWaWi1984}
G.~S. Pawley, R.~H. Swendsen, D.~J. Wallace, and K.~G. Wilson, Phys. Rev. B
  {\bf 29},  4030  (1984).

\bibitem{BaFeMMMu1996}
H.~G. Ballesteros, L.~A. Fernandez, V. Martin-Mayor, and A.~M. Sudupe, Phys.
  Lett. B {\bf 387},  125  (1996).

\bibitem{CaEtAl2001}
M. Campostrini, M. Hasenbusch, A. Pelissetto, P. Rossi, and E. Vicari, Phys.
  Rev. B {\bf 63},  214503  (2001).

\bibitem{CaEtAl2002}
M. Campostrini, M. Hasenbusch, A. Pelissetto, P. Rossi, and E. Vicari, Phys.
  Rev. B {\bf 65},  144520  (2002).

\bibitem{CaEtAl2006}
M. Campostrini, M. Hasenbusch, A. Pelissetto, and E. Vicari, Phys. Rev. B {\bf
  74},  144506  (2006).

\bibitem{FeMoWo1971}
M. Ferer, M.~A. Moore, and M. Wortis, Phys. Rev. B {\bf 4},  3954  (1971).

\bibitem{BuCo1997}
P. Butera and M. Comi, Phys. Rev. B {\bf 56},  8212  (1997).

\bibitem{WiFi1972}
K.~G. Wilson and M.~E. Fisher, Phys. Rev. Lett. {\bf 28},  240  (1972).

\bibitem{Wi1972vs}
K.~G. Wilson, Phys. Rev. Lett. {\bf 28},  548  (1972).

\bibitem{WiKo1974}
K.~G. Wilson and J.~B. Kogut, Phys. Rep. {\bf 12},  75  (1974).

\bibitem{GoLaTkCh1983}
S.~G. Gorishnii, S.~A. Larin, F.~V. Tkachov, and K.~G. Chetyrkin, Phys. Lett. B
  {\bf 132},  351  (1983).

\bibitem{KlSF2001jpa}
H. Kleinert and V. Schulte-Frohlinde, J. Phys. A {\bf 34},  1037  (2001).

\bibitem{KlNeSFChLa1991}
H. Kleinert, J. Neu, V. Schulte-Frohlinde, K.~G. Chetyrkin, and S.~A. Larin,
  Phys. Lett. B {\bf 272},  39  (1991), [Erratum Phys. Lett. B {\bf 319}, 545
  (1993)].

\bibitem{Fi1998}
M.~E. Fisher, Rev. Mod. Phys. {\bf 70},  653  (1998).

\bibitem{ZJ2001}
J. Zinn-Justin, Phys. Rep. {\bf 344},  159  (2001).

\bibitem{BaNiGrMe1976}
G.~A. Baker, B.~G. Nickel, M.~S. Green, and D.~I. Meiron, Phys. Rev. Lett. {\bf
  36},  1351  (1976).

\bibitem{NiMeBa1977}
B.~G. Nickel, D.~I. Meiron, and G.~A. Baker, Compilation of {$2$}-point and
  {$4$}-point graphs for continuous spin models, University of Guelph Report
  1977 (unpublished).

\bibitem{BaNiMe1978}
G.~A. Baker, B.~G. Nickel, and D.~I. Meiron, Phys. Rev. B {\bf 17},  1365
  (1978).

\bibitem{ChDo2004}
X.~S. Chen and V. Dohm, Phys. Rev. E {\bf 70},  056136  (2004).

\bibitem{BaBe2001}
C. Bagnuls and C. Bervillier, Phys. Rep. {\bf 348},  91  (2001).

\bibitem{CaDeMoVi2003}
L. Canet, B. Delamotte, D. Mouhanna, and J. Vidal, Phys. Rev. B {\bf 68},
  064421  (2003).

\bibitem{WeHo1973}
F.~J. Wegner and A. Houghton, Phys. Rev. A {\bf 8},  401  (1973).

\bibitem{vGWe2001}
G. v.~Gersdorff and C. Wetterich, Phys. Rev. B {\bf 64},  054513  (2001).

\bibitem{ParisiTwoLoops}
G. Parisi, Carg\`{e}se Lecture Notes, Columbia University preprint 1973
  (unpublished).

\bibitem{Pa1980}
G. Parisi, J. Stat. Phys. {\bf 23},  49  (1980).

\bibitem{BrLGZJ1973}
E. Br{\'e}zin, J.~L. Guillou, and J. Zinn-Justin, Phys. Rev. D {\bf 8},  434
  (1973).

\bibitem{KlSF2001}
H. Kleinert and V. Schulte-Frohlinde, {\em Critical Properties of
  {$\phi^4$}--Theories} (World Scientific, Singapore, 2001).

\bibitem{DombAndGreen}
{\em Phase Transitions and Critical Phenomena}, edited by C. Domb and M.~S.
  Green (Academic Press, London, 1972--2000), Vol.~1--20.

\bibitem{SyEsHeHi1966}
M.~F. Sykes, J.~W. Essam, B.~R. Heap, and B.~J. Hiley, J. Math. Phys. {\bf 7},
  1557  (1966).

\bibitem{ItDr1989}
C. Itzykson and J.~M. Drouffe, {\em Statistical Field Theory} (Cambridge
  University Press, Cambridge, UK, 1989).

\bibitem{Wo1974}
M. Wortis,  in {\em Phase Transitions and Critical Phenomena 3}, edited by C.
  Domb and M.~S. Green (Academic Press, London, 1974), p.\ 113.

\bibitem{Ca2001}
M. Campostrini, J. Stat. Phys. {\bf 103},  369  (2001).

\bibitem{PeVi2002}
A. Pelissetto and E. Vicari, Phys. Rep. {\bf 368},  549  (2002).

\bibitem{GuJo1972}
A.~J. Guttmann and G.~S. Joyce, J. Phys. A {\bf 5},  L81  (1972).

\bibitem{HuBa1979}
D.~L. Hunter and G.~A. Baker, Phys. Rev. B {\bf 19},  3808  (1979).

\bibitem{FiAY1979}
M.~E. Fisher and H. Au-Yang, J. Phys. A {\bf 12},  1677  (1979).

\bibitem{EcMaSe1975}
J.~P. Eckmann, J. Magnen, and R. S{\'e}n{\'e}or, Commun. Math. Phys. {\bf 39},
  251  (1975).

\bibitem{dG1972}
P.~G. de~Gennes, Phys. Lett. A {\bf 38},  339  (1972).

\bibitem{Zh1997}
S.~C. Zhang, Science {\bf 275},  1089  (1997).

\bibitem{GuRi2006}
R. Guida and P. Ribeca, J. Stat. Mech. (2006) P02007, e-print cond-mat/0512222.

\bibitem{GuRiPARIS}
R. Guida and P. Ribeca, PARIS: improving the DCUHRE integrator in an
  intermediate number of dimensions, in preparation.

\bibitem{LGZJ1977}
J.~C. LeGuillou and J. Zinn-Justin, Phys. Rev. Lett. {\bf 39},  95  (1977).

\bibitem{BrLGZJ1977}
E. Br\'{e}zin, G. Parisi, and J. Zinn-Justin, Phys. Rev. D {\bf 15},  1544
  (1977).

\bibitem{BrPa1978}
E. Br\'{e}zin and G. Parisi, J. Stat. Phys. {\bf 19},  269  (1978).

\bibitem{NiRe1990}
B.~G. Nickel and J.~J. Rehr, J. Stat. Phys. {\bf 61},  1  (1990).

\bibitem{SmFo1982}
D.~A. Smith and W.~F. Ford, Math. Comput. {\bf 38},  481  (1982).

\bibitem{Pa1993}
K. Pachucki, Ann. Phys. (N.Y.) {\bf 226},  1  (1993).

\bibitem{NiEtAl2000}
M. Niering, R. Holzwarth, J. Reichert, P. Pokasov, {Th.~Udem}, M. Weitz, T.~W.
  H\"{a}nsch, P. Lemonde, G. Santarelli, M. Abgrall, P. Laurent, C. Salomon,
  and A. Clairon, Phys. Rev. Lett. {\bf 84},  5496  (2000).

\bibitem{FiEtAl2004}
M. Fischer, N. Kolachevsky, M. Zimmermann, R. Holzwarth, {Th.~Udem}, T.~W.
  H\"{a}nsch, M. Abgrall, J. Gr\"unert, I. Maksimovic, S. Bize, H. Marion, F.
  Pereira Dos~Santos, P. Lemonde, G. Santarelli, P. Laurent, A. Clairon, C.
  Salomon, M. Haas, U.~D. Jentschura, and C.~H. Keitel, Phys. Rev. Lett. {\bf
  92},  230802  (2004).

\bibitem{We1996b}
E.~J. Weniger, Ann. Phys. (N.Y.) {\bf 246},  133  (1996).

\bibitem{We1996d}
E.~J. Weniger, Phys. Rev. Lett. {\bf 77},  2859  (1996).

\bibitem{We1997}
E.~J. Weniger, Phys. Rev. A {\bf 56},  5165  (1997).

\bibitem{We1990}
E.~J. Weniger, J. Comput. Appl. Math. {\bf 32},  291  (1990).

\bibitem{WeCi1990}
E.~J. Weniger and J. {\v C}{\' \i}{\v z}ek, Comput. Phys. Commun. {\bf 59},
  471  (1990).

\bibitem{Si1979}
A. Sidi, Math. Comput. {\bf 33},  315  (1979).

\bibitem{Si1980}
A. Sidi, Math. Comput. {\bf 35},  833  (1980).

\bibitem{Si1986}
A. Sidi, SIAM J. Math. Anal. {\bf 17},  1222  (1986).

\bibitem{NeSc1975}
W. Neuhaus and W. Schottl\"{a}nder, Computation {\bf 15},  41  (1975).

\bibitem{Be1977}
L. Berg, Computation {\bf 18},  361  (1977).

\bibitem{Kl1998}
H. Kleinert, Phys. Rev. D {\bf 57},  2264  (1998).

\bibitem{Kl1998add}
H. Kleinert, Phys. Rev. D {\bf 58},  107702  (1998).

\bibitem{BrLeMa1983}
S.~J. Brodsky, G.~P. Lepage, and P.~B. Mackenzie, Phys. Rev. D {\bf 28},  228
  (1983).

\bibitem{KaSt1995}
A.~L. Kataev and V.~S. Starshenko, Phys. Rev. D {\bf 52},  402  (1995).

\bibitem{Ka1966}
T. Kato, {\em Perturbation Theory for Linear Operators} (Springer, Berlin,
  1966).

\bibitem{Si1973}
B. Simon, Ann. Math. {\bf 97},  247  (1973).

\bibitem{GrGr1978b}
S. Graffi and V. Grecchi, Commun. Math. Phys. {\bf 62},  83  (1978).

\bibitem{Pr1895}
A. Pringsheim, Sitzungsberichte d. K. bay. Akad. d. Wiss. {\bf 25},  75
  (1895).

\bibitem{Pr1896}
A. Pringsheim, Math. Ann. {\bf 17},  121  (1896).

\bibitem{Vi1897}
G. Vivanti, Math. Ann. {\bf 18},  457  (1897).

\bibitem{He1979}
I.~W. Herbst, Commun. Math. Phys. {\bf 64},  279  (1979).

\bibitem{BeGrHaSi1979}
L. Benassi, V. Grecchi, E. Harrell, and B. Simon, Phys. Rev. Lett. {\bf 42},
  704  (1979).

\bibitem{Si1978}
H. Silverstone, Phys. Rev. A {\bf 18},  1853  (1978).

\bibitem{GrGr1981}
S. Graffi and V. Grecchi, Commun. Math. Phys. {\bf 79},  91  (1981).

\bibitem{HeSi1981}
I.~W. Herbst and B. Simon, Commun. Math. Phys. {\bf 80},  181  (1981).

\bibitem{StMHMaSc1996}
E. Stein, M. Meyer-Hermann, L. Mankiewicz, and A. Sch\"afer, Phys. Lett. B {\bf
  376},  177  (1996).

\bibitem{MHEtAl1996}
M. Meyer-Hermann, M. Maul, L. Mankiewicz, E. Stein, and A. Sch\"afer, Phys.
  Lett. B {\bf 383},  463  (1996), [Erratum Phys. Lett. B {\bf 383}, 487
  (1996)].

\bibitem{MHSc1999}
M. Meyer-Hermann and A. Sch\"{a}fer, Eur. Phys. J. A {\bf 5},  91  (1999).

\bibitem{KnMHSo2002}
A. Knauf, M. Meyer-Hermann, and G. Soff, Phys. Lett. B {\bf 549},  109  (2002).

\bibitem{Za1992}
V. Zakharov, Nucl. Phys. B {\bf 385},  452  (1992).

\bibitem{Da1982}
F. David, Nucl. Phys. B {\bf 209},  433  (1982).

\bibitem{Ra1991prd}
P.~A. R\c{a}czka, Phys. Rev. D {\bf 43},  R9  (1991).

\bibitem{Pi1999}
M. Pindor, e-print hep-th/9903151.

\bibitem{St2002prep}
M. Steinhauser, Phys. Rep. {\bf 364},  247  (2002).

\end{thebibliography}
\end{document}